\documentclass[11pt]{article}

\usepackage{amssymb,amsfonts,amsthm}
\usepackage{graphicx}
\usepackage{subfigure}
\usepackage{amssymb}
\usepackage{amsthm}
\usepackage{amsmath}
\usepackage{bm}             %% boldmath-command: \bm{}
\usepackage[mathscr]{eucal}

\usepackage{cancel}
\usepackage{outbraces}

\usepackage{psfrag}

\usepackage{wasysym}

\usepackage{fancyheadings}

\renewcommand{\sectionmark}[1]%
        {\markboth%
                {}%
                {{\rm\bfseries\thesection}\quad{\sc #1}}}
\newcommand{\sE}[2]{e_{#1}{}^{#2}}
\newcommand{\sEE}[2]{e^{#1}_{\:\: #2}}
\newcommand{\E}[2]{E_{#1}{}^{#2}}
\newcommand{\EE}[2]{E^{#1}_{\:\: #2}}
\newcommand{\N}[2]{\mathcal{N}^{#1}{}_{#2}}
\newcommand{\bN}[2]{\bar{\mathcal{N}}^{#1}{}_{#2}}

\newcommand{\dN}[2]{\dot{\mathcal{N}}^{#1}_{\:\:#2}}

\renewcommand{\P}[2]{\mathcal{P}^{#1}_{\:\:#2}}

\newcommand{\weg}{\:\,}
\newcommand{\wweg}{\:\;}

\newenvironment{details}%
{\begin{minipage}[t]{12.5cm}\begingroup \mdseries \itshape}%
{\endgroup\end{minipage}}

\newenvironment{subsections}%
{\begin{minipage}[t]{11.5cm}\begingroup \mdseries
\begin{itemize}\setlength{\itemsep}{-1ex}}%
{\end{itemize}\endgroup\end{minipage}}

\newcommand{\onpage}[1]{\hfill\pageref{#1}}
\newcommand{\textfrac}[2]{{\textstyle \frac{#1}{#2}}}

\newcommand{\barn}{\bar{n}}

\newcommand{\cT}{\mathcal{T}}
\newcommand{\cS}{\mathcal{S}}
\newcommand{\cA}{\mathcal{A}}
\newcommand{\cB}{\mathcal{B}}
\newcommand{\cO}{\mathcal{O}}

\newcommand{\li}{l_{\mathrm{i}}}
\newcommand{\lf}{l_{\mathrm{f}}}
\newcommand{\lin}{l_{\mathrm{in}}}
\newcommand{\lout}{l_{\mathrm{out}}}
\newcommand{\uin}{u^{\mathrm{in}}}
\newcommand{\uout}{u^\mathrm{out}}
\newcommand{\uinn}[1]{\uin_{\,#1}}
\newcommand{\uoutt}[1]{\uout_{\,#1}}

\theoremstyle{plain}

\newtheorem*{proposition}{Proposition}
\newtheorem*{conjecture}{Conjecture}

\theoremstyle{remark}

\newtheorem*{remark}{Remark}
\newcommand{\parb}{\pmb{\partial}}

\newcommand{\delzero}{\pmb{\partial}_{\boldsymbol{0}}}

\newcommand{\udot}{\dot{u}}
\newcommand{\Udot}{\dot{U}}

\newcommand{\cn}{{\cal N}}
\newcommand{\cg}{{\cal G}}
\newcommand{\nt}{\tilde{N}}

\newcommand{\bptl}{\bar{\partial}}

\newcommand{\la}{\langle}
\newcommand{\ra}{\rangle}

\newcommand{\sfrac}[2]{{\textstyle \frac{#1}{#2}}}

\begin{document}

\thispagestyle{empty}

%%%%%%%%%%%%%%%%%%%%%%%%%%%%%%%%%%%%%%%%%%%%%%%%%%%%%%%%%%%%%%%%%%%%
\title{\bfseries The cosmological billiard attractor}
%%%%%%%%%%%%%%%%%%%%%%%%%%%%%%%%%%%%%%%%%%%%%%%%%%%%%%%%%%%%%%%%%%%%

\author{\sc
J.\ Mark Heinzle$^{1}$\thanks{Electronic address: {\tt
Mark.Heinzle@univie.ac.at}}\ ,\  Claes Uggla$^{2}$\thanks{Electronic
address: {\tt Claes.Uggla@kau.se}}\ ,\ and Niklas
R\"ohr$^{2}$\thanks{Electronic address: {\tt
Niklas.Rohr@kau.se}}\\
$^{1}${\small\em Gravitational Physics, Faculty of Physics, University of Vienna,}\\
{\small\em %W\"ahringerstra\ss e 17,
A-1090 Vienna, Austria}\\
$^{2}${\small\em Department of Physics, University of Karlstad,}\\
{\small\em S-651 88 Karlstad, Sweden}}

%%%%%%%%%%%%%%%%%%%%%%%%%%%%%%%%%%%%%%%%%%%%%%%%%%%%%%%%%%%%%%%%%%%
\date{}%\normalsize{February 20, 2007}}
%%%%%%%%%%%%%%%%%%%%%%%%%%%%%%%%%%%%%%%%%%%%%%%%%%%%%%%%%%%%%%%%%%%
\maketitle
%%%%%%%%%%%%%%%%%%%%%%%%%%%%%%%%%%%%%%%%%%%%%%%%%%%%%%%%%%%%%%%%%%%

\thispagestyle{empty}

%%%%%%%%%%%%%%%%%%%%%%%%%%%%%%%%%%%%%%%%%%%%%%%%%%%%%%%%%%%%%%%%%%%
\begin{abstract}
%%%%%%%%%%%%%%%%%%%%%%%%%%%%%%%%%%%%%%%%%%%%%%%%%%%%%%%%%%%%%%%%%%%

\begin{center}
\begin{minipage}{11cm}
This article is devoted to a study of the asymptotic dynamics of
generic solutions of the Einstein vacuum equations toward a generic
spacelike singularity. Starting from fundamental assumptions
%that formulate the basic properties of
about the nature of generic spacelike singularities we derive in a
step-by-step manner the cosmological billiard conjecture: we show
that the generic asymptotic dynamics of solutions is represented by
(randomized) sequences of heteroclinic orbits on the `billiard
attractor'. Our analysis rests on two pillars: (i) a dynamical
systems formulation based on the conformal Hubble-normalized
orthonormal frame approach expressed in an Iwasawa frame; (ii)
stochastic methods and the interplay between genericity and
stochasticity. Our work generalizes and improves the level of rigor
of previous work by Belinskii, Khalatnikov, and Lifshitz;
furthermore, we establish that our approach and the Hamiltonian
approach to `cosmological billiards', as elaborated by Damour,
Hennaux, and Nicolai, can be viewed as yielding `dual'
representations of the asymptotic dynamics.

\begin{center}
\end{center}

\end{minipage}
\end{center}

%%%%%%%%%%%%%%%%%%%%%%%%%%%%%%%%%%%%%%%%%%%%%%%%%%%%%%%%%%%%%%%%%%
\end{abstract}
%%%%%%%%%%%%%%%%%%%%%%%%%%%%%%%%%%%%%%%%%%%%%%%%%%%%%%%%%%%%%%%%%%%

\newpage

\thispagestyle{empty}

\begin{center}
\begin{Large}
\textbf{List of Contents}
\end{Large}
\end{center}

\vspace{4ex}

\newcounter{listenzaehler}

\begingroup\bfseries

\begin{list}{\arabic{listenzaehler}}{\labelsep1em\leftmargin3em}
\usecounter{listenzaehler}
\item Introduction \onpage{introduction}
\item The Hamiltonian billiard approach \onpage{Hamilton} \\
\begin{details}
A brief review of the Hamiltonian approach to generic asymptotic
dynamics. The asymptotic dynamical evolution is described as motion
in `cosmological billiards'.
\end{details}
\item The dynamical systems approach \onpage{confsec} \\
\begin{details}
An introduction to conformal Hubble-normalized variables and the associated
dynamical systems approach to
the Einstein field equations. We specialize to an Iwasawa decomposition
of conformal orthonormal frames and give
the definition of the state space associated with the Einstein
equations in the chosen gauge.
\end{details}

\begin{subsections}
\item Conformal Hubble-normalization \onpage{confHubb}
\item Iwasawa frame variables \onpage{Iwasframe}
\end{subsections}
\item Asymptotic silence, locality, and the silent boundary \onpage{asympsil} \\
\begin{details}
Discussion of the basic concepts and assumptions about
generic spacelike singularities in the dynamical systems approach.
Due to the spatial decoupling of the equations
we can introduce a finite dimensional state space that is
expected to capture the essential asymptotic dynamics---the
so-called SH silent boundary.
\end{details}

\begin{subsections}
\item Asymptotic silence and asymptotic locality \onpage{asysilasyloc}
\item The silent boundary \onpage{silbound}
\end{subsections}
\item Kasner circle stability and the oscillatory subset \onpage{prebilliard} \\
\begin{details}
Discussion of important structures associated with the state space.
We investigate the Kasner circle and the associated generalized Kasner
solutions, and we reduce the SH silent boundary system
to a system on an invariant subset: the oscillatory dynamical system.
\end{details}

\begin{subsections}
\item {\small The Kasner circle, stable variables, and the oscillatory subset} \onpage{Kascirc}
\item The oscillatory dynamical system \onpage{theosci}
\end{subsections}

\thispagestyle{empty}

\item Dynamics on the components of the oscillatory subset \onpage{dynosc} \\
\begin{details}
Investigation of the state space and the dynamics of the oscillatory
dynamical system. We analyze step-by-step the invariant subsets of
the oscillatory system, which are associated with Bianchi types I,
II, $VI_0$, and $VII_0$. The orbits that turn out to be of
prime importance for our purposes are 'transitions',
i.e., heteroclinic orbits
that connect fixed points on the Kasner circle.
\end{details}

\begin{subsections}
\item The silent Kasner subset (Bianchi type I subset) \onpage{silentBI}
\item The silent Bianchi type II subset \onpage{silentBII}
\item The silent Bianchi type VI$_0$ and VII$_0$ subsets \onpage{silentBVIVII}
\end{subsections}

\newpage

\item Sequences, eras, and phases \onpage{transition} \\
\begin{details}
Definition and analysis of the fundamental objects for generic asymptotic
dynamics of solutions: sequences of transitions.
A sequence of transitions is defined to be an infinite concatenation
of heteroclinic orbits on the oscillatory subset.
Each (generic) sequence is naturally associated with a sequence of Kasner states
that is determined by a recursive map, the so-called Kasner map.
We investigate the basic properties of sequences of transitions
and Kasner sequences; in particular we define the concepts
of large and small curvature phases.
\end{details}

\begin{subsections}
\item Sequences of transitions and Kasner sequences \onpage{seqtransKas}
\item Eras, large curvature phases, and small curvature phases \onpage{eralcp}
\end{subsections}
\item Stochastic analysis of Kasner sequences \onpage{stat} \\
\begin{details}
Introduction to the stochastic aspects of our analysis.
Kasner sequences possess a probabilistic interpretation, since they
are associated with a probability density on a compact interval.
We investigate Kasner sequences from this stochastic point of view.
\end{details}

\thispagestyle{empty}

\item Growth \onpage{growth} \\
\begin{details}
Investigation of growth rates.
Along sequences of transitions several quantities associated with
the state space variables exhibit well-defined growth rates.
We give a detailed discussion of growth in terms of the growth function.
\end{details}

\begin{subsections}
\item Auxiliary differential equations and the growth function \onpage{auxand}
\item Properties of the growth function \onpage{propgrow}
\end{subsections}
\item Asymptotic shadowing \onpage{asysha} \\
\begin{details}
Discussion of a central issue: sequences of transitions act as the
asymptotic limits of generic solutions of the Einstein equations with
a spacelike singularity.
However, not all sequences are admissible;
by establishing the asymptotic suppression of certain variables, we restrict
the set of possible `attractors' to billiard sequences.
In this context stochastic methods are essential.
\end{details}

\begin{subsections}
\item Shadowing---asymptotic sequences of orbits \onpage{sha---}
\item $R_2\rightarrow 0$ and $N_2\rightarrow 0$ as $\tau\rightarrow \infty$ \onpage{R2N2go}
\item Exclusion of $R_2$ and $N_2$ \onpage{excluR2}
\item Randomized Kasner sequences \onpage{randomiK}
\end{subsections}
\item The billiard attractor \onpage{billiardattractor}\\
\begin{details}
Definition and discussion of the billiard attractor.
We reduce billiard sequences to attractor sequences, which
are infinite concatenations of
transitions of merely three types.
We show that attractor sequences describe the asymptotic
dynamics of generic solutions with a spacelike singularity.
This section contains a brief summary of the main steps
of the analysis.
\end{details}

\begin{subsections}
\item The billiard subset \onpage{billsubse}
\item The asymptotic suppression of multiple transitions \onpage{asysuppre}
\item The billiard attractor \onpage{thebillatt}
\end{subsections}

\newpage

\thispagestyle{empty}

\item `Duality' of Hamiltonian and dynamical systems billiards \onpage{dual} \\
\begin{details}
Discussion of the connection between the Hamiltonian approach and
the dynamical systems approach. We show that the Hamiltonian
cosmological billiard and the billiard attractor can be viewed as
dual representations of generic asymptotic dynamics.
\end{details}
\item Models with symmetries in an Iwasawa frame \onpage{sym}\\
\begin{details}
Specialization of our treatment to models with symmetries.
\end{details}

\begin{subsections}
\item $G_1$ models \onpage{G1mo}
\item $G_2$ models \onpage{G2mo}
\item Generic Bianchi type VI$_{-1/9}$ models \onpage{genexpB}
\end{subsections}

\thispagestyle{empty}

\item Gauge considerations \onpage{gauge} \\
\begin{details}
Discussion of issues connected with gauge and analysis
of gauge-invariant quantities.
\end{details}
\item Concluding remarks \onpage{concl} \\
\begin{details}
A brief recapitulation and discussion of some basic results.
We also give an outlook and make contact with related fields of research.
\end{details}
\quad

\vspace{2ex}

\item[] Appendices \onpage{bappe} \\
\begin{details}
The paper concludes with a number of appendices that provide
additional information on special issues. Some appendices contain
rather technical computations that are omitted in the main text,
while others are devoted to independent material: in particular, in
Appendix~\ref{Kasner} we derive the BKL generalized Kasner metric
within the dynamical systems formulation, and in
Appendix~\ref{asympconst} we establish asymptotic freezing, i.e.,
that certain variables converge to temporal constants in the
asymptotic limit.
\end{details}

\begin{subsections}
\item[A] Iwasawa variables and useful equations \onpage{relations}
\item[B] Deriving the generalized Kasner line element of BKL \onpage{Kasner}
\item[C] Kasner solutions in a rotating frame \onpage{KasnernonFermi}
\item[D] Multiple transitions \onpage{multiple}
\item[E] Behavior of an auxiliary quantity \onpage{Bbehave}
\item[F] Convergence of sums \onpage{convergenceofsums}
\item[G] Hitting intervals stochastically \onpage{hitting}
\item[H] Asymptotic constants of the motion \onpage{asympconst}
\item[I] %Relationship between work on
AVTD singularities and the dynamical systems approach \onpage{rel}
\end{subsections}

\vspace{3ex}

\item[] References \onpage{bibbegin}
\end{list}

\endgroup

\thispagestyle{empty}

\newpage

%%%%%%%%%%%%%%%%%%%%%%%%%%%%%%%%%%%%%%%%%%%%%%%%%%%%%%%%%%%%%%%%%%%
\section{Introduction}
\label{introduction}
%%%%%%%%%%%%%%%%%%%%%%%%%%%%%%%%%%%%%%%%%%%%%%%%%%%%%%%%%%%%%%%%%%%

Remarkable developments took place in the sixties, seventies, and
early eighties, that molded our understanding of singularities in
General Relativity (GR). On the one hand, the singularity theorems
of Hawking and Penrose~\cite{hawkingpenrose} proved the
inevitability of spacetime singularities under rather general
conditions. On the other hand, in a series of papers, Lifshitz,
Khalatnikov and Belinskii, \cite{lk63,bkl70,bkl82} and references
therein, set out to give a description of the actual nature of
generic singularities---henceforth we will refer to these authors
and their work as BKL. These authors performed a heuristic analysis
that eventually resulted in the claim that a generic spacelike
singularity for the Einstein field equations with a perfect fluid
with a radiation equation of state as the matter source is
\textit{spacelike}, \textit{local}, \textit{vacuum dominated}, and
\textit{oscillatory}. BKL obtained this picture by (i) using certain
spatially homogeneous (SH) metrics, (ii) replacing constants with
spatial functions, (iii) inserting the resulting expressions into
Einstein's field equations and making a perturbative expansion, (iv)
checking if this ad hoc procedure yielded a consistent result and
thereby completing the basic viability test. Furthermore, BKL
employed synchronous coordinates, i.e., Gaussian normal coordinates,
such that the singularity occurred simultaneously. The BKL approach
led to the conjecture that the time evolution of a generic solution
in the vicinity of a generic singularity is schematically described
by a sequence of generalized Kasner solutions (i.e., vacuum Bianchi
type I solutions where constants are replaced with spatially
dependent functions), where the transitions are `mediated' by
generalized vacuum Bianchi type II solutions through the so-called
Kasner map. In subsequent work, this map was shown to be associated
with chaotic behavior~\cite{khaetal85,bar82,chebar83}.

The BKL picture obtained further heuristic support from the
Hamiltonian approach developed by Misner and
Chitr\'e~\cite{mis69,grav73,chi72}, originally for asymptotic
Bianchi type IX dynamics. In one variety of this approach the
dynamics was described in terms of a free motion in an abstract flat
Lorentzian \mbox{(minisuper-)} space surrounded by potential walls,
or alternatively by means of a spatial projection leading to a free
motion inside a potential well described by moving walls; for
further developments of this picture, see~\cite{jan01}
and~\cite[Chapter 10]{waiell97}. In another variety of the
Hamiltonian approach an intrinsic time variable was introduced,
which led to a description of the asymptotic dynamics in terms of a
projected `billiard' motion in a region of hyperbolic space bounded
by infinitely high straight stationary walls, see
e.g.~\cite{grav73}. This work on `cosmological billiards' was later
generalized in order to deal with general inhomogeneous cases, which
culminated in the recent work by Damour, Hennaux, and Nicolai,
see~\cite{dametal03,damnic05} and references therein. Apart from
these studies, special inhomogeneous spacetimes with non-oscillatory
singularities have also been investigated by means of Hamiltonian
methods---notably by Moncrief, Isenberg, Berger, and collaborators,
who also obtained numerical support for the general basic BKL
picture, see~\cite{beretal98}, and~\cite{ber02} for a review and
additional references.

Despite the ingenuity of the BKL and the Hamiltonian methods, there
has been, unfortunately, little progress as regards a desired
sharpening of the heuristic arguments in order to turn conjectures
into rigorous
mathematical statements.%
\footnote{Note, however, the successful rigorous treatment
  of special non-oscillatory cases in~\cite{ber02,andren01},
  which is partly based on Fuchsian methods.}
The requirement of mathematical rigor is met in a different approach
to cosmological singularities: the dynamical systems approach.
During the last decade, based on dynamical systems formulations,
there has been considerable progress in the SH case as regards
theorems about dynamical behavior, largely in connection with the
book ``Dynamical Systems in Cosmology''~\cite{waiell97}; notably,
Ringstr\"om obtained the first mathematical theorems in the context
of oscillatory behavior for Bianchi type VIII and, more
substantially, type IX models~\cite{rin00,rin01}. In an attempt to
extend the dynamical systems approach to the inhomogeneous context,
Uggla et al.~\cite{uggetal03} introduced a dynamical systems
formulation for the Einstein field equations without any
symmetries---in the following, we will refer to this work as UEWE.
The results were: a detailed description of the generic attractor;
concisely formulated conjectures about the asymptotic dynamic
behavior toward a generic spacelike singularity; a basis for a
numerical investigation of generic singularities---following UEWE
numerical results yielded additional support for the expected
generic picture as well as the discovery of new phenomena and
subsequent refinements~\cite{gar04,andetal05,limetal06}.

The purpose of the present paper is two-fold: first, we establish a
link between the Hamiltonian picture as described by Damour et
al.~\cite{dametal03,damnic05} and the dynamical systems approach to
inhomogeneous cosmologies as initiated in UEWE. In particular, we
demonstrate that the `Hamiltonian billiards' and the corresponding
dynamical systems description can be viewed as yielding dual
representations of the generic asymptotic dynamics (`configuration
space description' vs.\ `momentum space description'). To avoid
excessive clutter that would obscure the main ideas we simplify our
presentation by confining ourselves to the four-dimensional vacuum
case. Inclusion of matter sources such as perfect fluids, see
e.g.~\cite{uggetal03,limetal04}, and to more general cases like
those considered by Damour et al.~\cite{dametal03,damnic05}
should---in principle---be straightforward.

The second and main purpose of this paper is to \textit{derive} and
give rigor to some of the key conjectures formulated in the BKL and
Hamiltonian approach, starting from `first principles' connected
with the full state space picture of our dynamical systems approach.
This derivation does \textit{not} constitute a mathematical proof of
the conjectures; however, in many respects we go beyond what has
been accomplished previously: we identify the `billiard attractor'
as the attractor for the asymptotic dynamics toward a generic
spacelike singularity; we give decay rates that describe the
approach to the attractor; we show how asymptotic constants of the
motion, which were obtained from the Hamiltonian billiard approach
initially, arise. An important ingredient in our treatment is the
use of stochastic methods: we emphasize the connection between
genericity and stochasticity. The detailed statements and
conjectures we obtain are accessible to numerical experiments and
can be compared with results for special models.

The present paper is to a large extent self-contained. First, we
give a condensed review of the Hamiltonian approach of Damour et
al.~\cite{dametal03,damnic05} in Section~\ref{Hamilton}. Second, we
present the conformal Hubble-normalized orthonormal frame approach
leading to the dynamical systems formulation in
Sections~\ref{confsec} and~\ref{asympsil}. Endowed with the basic
techniques and the interrelation between the two approaches, we then
derive the billiard attractor, which then serves as the starting
point for further discussions, developments, and concluding remarks.

%%%%%%%%%%%%%%%%%%%%%%%%%%%%%%%%%%%%%%%%%%%%%%%%%%%%%%%%%%%%%%%%%%%%%%%%%%%%%%%%%%%%%
\section{The Hamiltonian billiard approach}
\label{Hamilton}
%%%%%%%%%%%%%%%%%%%%%%%%%%%%%%%%%%%%%%%%%%%%%%%%%%%%%%%%%%%%%%%%%%%%%%%%%%%%%%%%%%%%%

In~\cite{dametal03}, Damour, Henneaux, and Nicolai used a
Hamiltonian approach to study the dynamics of the
Einstein-dilaton-$p$-form system in the neighborhood of a generic
spacelike singularity. The authors described the asymptotic behavior
of the fields by a `billiard' motion in a region of hyperbolic space
bounded by straight `walls'. The techniques used in~\cite{dametal03}
represent generalizations of methods that are due to Chitr\'e and
Misner~\cite{mis69,grav73,chi72}, which have been applied and
extended by many authors, see the references in~\cite{dametal03}. In
the following we give a brief review of the Hamiltonian `billiard
approach' restricted to the four-dimensional vacuum case in GR; the
presentation is based on~\cite{dametal03}, but the notation is
tailored to our later purposes. In particular, for frame indices we
use $\alpha,\beta,\ldots$, as in~\cite{waiell97,uggetal03}, and for
spatial coordinate indices we use $i,j,\ldots$, as
in~\cite{dametal03} and~\cite{waiell97,uggetal03}.

In~\cite{dametal03}, the metric is written in 3+1 form
with vanishing shift vector,
\begin{equation}\label{ds2}
ds^2 = -N^2\,(dx^0)^2 + g_{i j}\,dx^i\,dx^j\:.
\end{equation}
Let $g^{ij}$ denote the inverse of the spatial three-metric $g_{ij}$
and $g = \det g_{ij}$. We introduce
\begin{equation}
\label{NDeW}
\cg^{ijkl} :=
g^{i(k}\,g^{l)j} - g^{ij}\,g^{kl}\, \quad\text{and}\quad
\cg_{ijkl} :=
g_{i(k}\,g_{l)j} - \textfrac{1}{2}g_{ij}\,g_{kl}\, \:;
\end{equation}
$\cg^{ijkl}$ is the DeWitt metric~\cite{dew67} multiplied with
$g^{-1/2}$ so that it becomes a tensor instead of a tensor density;
$\cg_{ijmn}\,\cg^{mnkl} = \delta_{ij}{}^{kl} = \delta_{(i}^{(k} \delta_{j)}^{l)}$.

The Lagrangian (density) associated with the reduced
Einstein-Hilbert action reads
\begin{equation}\label{Lagrange}
\mathcal{L} = \textfrac{1}{4} \sqrt{g} \, N^{-1} \, \cg^{ijkl} \,\dot{g}_{i j}
\dot{g}_{k l} + \sqrt{g}\, N \;{}^{3}\!R \:,
\end{equation}
where a dot refers to the partial derivative w.r.t.\ $x^0$ and
${}^{3}\!R$ denotes the spatial three-curvature associated with $g_{i j}$.
To obtain the Hamiltonian $\mathcal{H}$ we introduce
the conjugate momenta
\begin{equation*}
\pi^{i j} = \partial \mathcal{L}/\partial {\dot g}_{i j} =
\textfrac{1}{2} \nt^{-1}\, \cg^{ijkl} \dot{g}_{kl}\:,
\end{equation*}
where we define
\begin{equation}\label{Ntildedef}
\nt =  g^{-1/2}\,N \:.
\end{equation}
This leads to
\begin{equation*}
\mathcal{L} =
%\pi^{i j} \dot{g}_{i j} - N g^{-1/2} \mathcal{H} =
\pi^{i j} \dot{g}_{i j} - {\cal H} = \pi^{i j} \dot{g}_{i j} - \nt
\left[ \cg_{ijkl}\,\pi^{i j}\pi^{kl} - g\:{}^{3}\!R\right]\:.
\end{equation*}
Variation of ${\cal H}$ w.r.t.\ ${\tilde N}$ yields the Hamiltonian
constraint ${\cal H}=0$, while the momentum constraints are
introduced separately, see~\cite{dametal03}.

There exists a unique oriented orthonormal spatial coframe
$\{\omega^{\alpha}\,|\,{\alpha}=1\ldots 3\}$,
$\omega^{\alpha} =
e^{\alpha}_{\:\,i}\, dx^i$, such that $e^{\alpha}_{\:\,i} =
\sum_\beta\mathcal{D}^\alpha_{\:\,\beta}\, \N{\beta}{i}$, where
$\mathcal{D}$ is a diagonal matrix,  which
we choose to express as $\mathcal{D} =
\mathrm{diag}\big[\exp(-b^1),\exp(-b^2),\exp(-b^3)\big]$, and
where $\N{{\alpha}}{i}$ a unit upper triangular matrix,
\begin{equation}\label{IwasawaNdef}
\Big(\,\N{{\alpha}}{i}\,\Big) =
\begin{pmatrix}
1 & \N{1}{2} & \N{1}{3} \\
0 & 1 & \N{2}{3} \\
0 & 0 & 1
\end{pmatrix} =
\begin{pmatrix}
1 & n_1 & n_2 \\
0 & 1 & n_3 \\
0 & 0 & 1
\end{pmatrix} \:.
\end{equation}
This choice of frame
leads to the so-called \textit{Iwasawa decomposition} of the
spatial metric $g_{i j}$:
\begin{equation*}
g_{i j} = \sum_{\alpha} \exp(-2 b^{\alpha}) \N{{\alpha}}{i}
\N{{\alpha}}{j}\:.
\end{equation*}

Existence and uniqueness of the frame $\{\omega^{\alpha}\}$ is
associated with the theorem on the uniqueness of the QR
decomposition in linear algebra; $\N{\alpha}{i}$ can also be viewed
as representing the Gram-Schmidt orthogonalization of the spatial
coordinate coframe $\{d x^i\}$. The Iwasawa decomposition
corresponds to a Cholesky decomposition of a symmetric matrix $A$
into a product $R^T R$, where the diagonal elements of the
triangular matrix $R$ are factored out and parametrized as
$\exp(-b^\alpha)$.

To avoid confusion with the Greek frame indices $\alpha,\beta,\ldots$,
we prefer to use $b$ as the kernel letter for the
diagonal degrees of freedom instead of $\beta$, which was used by
Damour et al.~\cite{dametal03}. Note that the negative sign in the
exponentials is in agreement with the conventions of~\cite{dametal03}, but
contrary to the conventions of e.g.~\cite{jan01,waiell97}. For
representations that are adapted to metric anisotropies, see
Misner~\cite{mis69} and e.g.~\cite{grav73,waiell97}.
Following the summation convention of~\cite{dametal03}, summation of
pairs of coordinate indices $i, j,...$ is understood, whereas sums
over the frame indices $\alpha, \beta,...$ are written out
explicitly in this section.

The frame $\{e_{\alpha}\}$ that is
dual to $\{\omega^{\alpha}\}$ is given by
$e_{\alpha} = \sE{\alpha}{i} \partial_{x^i} =
\exp(b^{\alpha})\,\bN{i}{{\alpha}}\partial_{x^i}$, where the matrix
$(\bN{i}{\alpha})$ is the inverse of $(\N{\alpha}{i})$,
\begin{equation}\label{inverseoffdiag}
%\Big(\,\mathcal{N}^{-1}\,\Big)_{i,{\alpha}} =
\Big(\,\bN{i}{\alpha}\,\Big) =
\begin{pmatrix}
1 & \bN{1}{2} & \bN{1}{3} \\
0 & 1 & \bN{2}{3} \\
0 & 0 & 1
\end{pmatrix}
=
\begin{pmatrix}
1 & \bar{n}_1 & \bar{n}_2 \\
0 & 1 & \bar{n}_3 \\
0 & 0 & 1
\end{pmatrix}
=
\begin{pmatrix}
1 & -n_1 & n_1 n_3 - n_2 \\
0 & 1 & -n_3 \\
0 & 0 & 1
\end{pmatrix} \:.
\end{equation}

Expressing the Lagrangian~\eqref{Lagrange} in the Iwasawa frame variables
$b^{\alpha}$ and $\N{{\alpha}}{i}$ yields
\begin{equation*}
%\sum_{\alpha} (\dot{b}^{{\alpha}})^2 - \Big(\sum_{\alpha} \dot{b}^{\alpha}\Big)^2 +
\mathcal{L} = \sqrt{g}\,N^{-1} \left[ \sum_{{\alpha}, \beta} \cg_{{\alpha}
\beta} \dot{b}^{\alpha} \dot{b}^{\beta} + \textfrac{1}{2}
\sum\limits_{{\alpha} < \beta} \exp(2(b^{\beta}-b^{\alpha}))
\left(\dN{{\alpha}}{i}\, \bN{i}{\beta}\right)^2 \right] +  \sqrt{g}\,N\;{}^{3}\!R\:,
\end{equation*}
where we have introduced the reduced metric $\cg_{\alpha\beta}$, %from $\cg^{ijkl}$,
which is associated with the diagonal degrees of freedom, i.e.,
`$b^{\alpha}$-space':
\begin{equation*}
\sum_{{\alpha},\beta} \cg_{{\alpha} \beta} v^{\alpha} w^{\beta}  :=
-\sum_{\gamma\neq \delta} v^{\gamma} w^{\delta} = \sum_{\alpha}
v^{\alpha} w^{\alpha} - \Big( \sum_{\alpha} v^{\alpha} \Big) \Big(
\sum_{\beta} w^{\beta} \Big)\:;
\end{equation*}
when `$b^{\alpha}$-space' is endowed with the metric $\cg_{\alpha\beta}$,
it is isometric to a $(2+1)$-dimensional Minkowski space, since the signature of
$\cg_{\alpha\beta}$ is $(- + +)$.
The inverse metric $\cg^{{\alpha} \beta}$ of $\cg_{{\alpha} \beta}$
is given by
\begin{equation*}
\sum_{\alpha,\beta} \cg^{\alpha \beta} v_\alpha w_\beta =
\sum_{\alpha,\beta} \delta^{\alpha \beta} v_\alpha w_\beta -
\textfrac{1}{2} \sum_{\alpha,\beta} v_\alpha w_\beta =
\sum_\gamma v_\gamma w_\gamma -\textfrac{1}{2} \Big(\sum_\alpha v_\alpha\Big) \Big(\sum_\beta w_\beta\Big)\:.
\end{equation*}

The momenta $\pi_{\alpha}$ that are conjugate to $b^{\alpha}$ and
the momenta $\P{i}{{\alpha}}$ ($i > \alpha$) that are conjugate to $\N{{\alpha}}{i}$ read
%\begin{subequations}\label{dotmom}
%\begin{align}
%\pi_{\alpha} & = 2\frac{\sqrt{g}}{N}\sum_{\beta} G_{{\alpha} \beta}
%\dot{b}^{\beta}\,, &
%\dot{b}^{\alpha} & = \frac{N}{2\sqrt{g}} \sum_{\beta} G^{{\alpha} \beta}\pi_{\beta} \:,\\
%\P{i}{{\alpha}} & = \frac{\sqrt{g}}{N} \sum_{\beta}
%\exp(2(b^{\beta}-b^{\alpha}))\dN{{\alpha}}{j}\N{j}{\beta}
%\N{i}{\beta}\,, & \N{{\alpha}}{i} &= \frac{N}{\sqrt{g}}\sum_{\beta}
%\exp(-2(b^{\beta}-b^{\alpha}))\P{j}{{\alpha}}\N{\beta}{i}\N{\beta}{j}
%\:,
%\end{align}
%\end{subequations}
%
\begin{equation}\label{momentarel}
\pi_{\alpha} = 2\nt^{-1}\sum_{\beta} \cg_{{\alpha} \beta}
\dot{b}^{\beta}\,; \qquad  \P{i}{{\alpha}} = \nt^{-1} \sum_{\beta}
e^{2(b^{\beta}-b^{\alpha})}\dN{{\alpha}}{j}\bN{j}{\beta}
\bN{i}{\beta} \quad (i > \alpha)\,,
\end{equation}
while $\P{i}{{\alpha}}=0$ when $i \leq \alpha$; accordingly,
$\P{i}{{\alpha}}$ can be viewed as a lower triangular matrix, whose
components we for convenience call ${\cal P}_i$:
\begin{equation}\label{momentamatrix}
\Big(\,\P{i}{{\alpha}}\,\Big) =
\begin{pmatrix}
0 & 0 & 0 \\
\P{2}{1} & 0 & 0 \\
\P{3}{1} & \P{3}{2} & 0
\end{pmatrix} =
\begin{pmatrix}
0 & 0 & 0 \\
{\cal P}_1 & 0 & 0 \\
{\cal P}_2 & {\cal P}_3 & 0
\end{pmatrix} \:.
\end{equation}
Inverting the relationships~\eqref{momentarel} yields
\begin{subequations}\label{momentarelinv}
\begin{align} \label{def}
\dot{b}^{\alpha} & = \textfrac{1}{2}\nt \sum_{\beta} \cg^{{\alpha}
\beta}\pi_{\beta}\:  ,&
\dot{n}_2 & = \nt\, e^{2(b^1-b^2)}\,\left(n_3\,{\cal
P}_1 + (e^{2(b^2-b^3)} + n_3^2)\,{\cal P}_2\right)\:, \\
\label{def2}
\dot{n}_1 & = \nt
\,e^{2(b^1-b^2)}\,\left({\cal P}_1 + n_3\,{\cal P}_2\right)\:, &
\dot{n}_3 &
= \nt \, e^{2(b^2-b^3)} \, {\cal P}_3\: ,
\end{align}
\end{subequations}
which leads to
\begin{equation}\label{Hamilto}
\mathcal{L} = \sum_{\alpha}\left( \pi_{\alpha} \dot{b}^{\alpha} +
 \mathcal{P}^i_{\weg {\alpha}} \dN{{\alpha}}{i} \right) - \nt
\left[ \textfrac{1}{4} \sum_{{\alpha},\beta}\cg^{{\alpha} \beta}
\pi_{\alpha} \pi_{\beta} + \textfrac{1}{2} \sum_{{\alpha}<\beta}
e^{-2(b^{\beta}-b^{\alpha})} (\mathcal{P}^i_{\weg \alpha}
\N{{\beta}}{i})^2 - g R \right]\:,
\end{equation}
where the Hamiltonian $\mathcal{H}$ can be read off easily.

We assume that the spacetime with metric~\eqref{ds2} possesses a
spacelike singularity in the past; then, as argued
in~\cite{dametal03}, $b^{\alpha}$ is expected to be timelike in the
vicinity of this singularity, i.e., $\sum_{\alpha,\beta}
\cg_{{\alpha} \beta} b^{\alpha} b^{\beta} <0$. Based on these
considerations, it is possible to replace the metric variables
$b^{\alpha}$ by new variables: we introduce $\rho^2 =
-\sum_{\alpha,\beta} \cg_{{\alpha} \beta} b^{\alpha} b^{\beta}$ and
`orthogonal' angular variables, collectively denoted by $\gamma$,
i.e., $\partial_\gamma$ is orthogonal to $\partial_\rho$\,. In these
variables the line element associated with $\cg_{\alpha\beta}$ can
be written in the form
\begin{equation*}
d \sigma^2 = \sum_{{\alpha},\beta} \cg_{{\alpha} \beta} d b^{\alpha}
d b^{\beta} = -d \rho^2 + \rho^2 d\Omega_h^2\:,
\end{equation*}
where $d\Omega_h^2$ is the standard metric on hyperbolic space.
Making a further variable change according to
\begin{equation}\label{lambdarho}
\lambda = \log \rho = \textfrac{1}{2} \log
\Big(-\sum_{{\alpha},\beta} \cg_{{\alpha} \beta} b^{\alpha}
b^{\beta} \Big)
\end{equation}
yields $d\sigma^2 = \rho^2 (-d\lambda^2 +  d\Omega_h^2 )$ and thus in particular
\begin{equation*}
\sum_{{\alpha},\beta}\cg^{{\alpha} \beta} \pi_{\alpha} \pi_{\beta} =
-\pi_\rho^2 + \rho^{-2} \pi_\gamma^2 = \rho^{-2} \left[
-\pi_\lambda^2 + \pi_\gamma^2 \right]\:,
\end{equation*}
cf.~\eqref{Hamilto}.

Consequently, choosing the rescaled lapse according to ${\tilde
N}=\rho^2$ leads to the Hamiltonian
\begin{equation}\label{Hami}
\mathcal{H} = \textfrac{1}{4} \left[ -\pi_\lambda^2 + \pi_\gamma^2
\right] + \rho^2 \sum\nolimits_A c_A\,e^{-2 \rho\, w_A(\gamma)}\:;
\end{equation}
here, the sum is over a number of terms of the same type: $c_A$ are
functions of spatial derivatives of the metric, off-diagonal metric
variables, and momenta; $w_A(\gamma)$ denote linear forms of the
variables $\gamma^\alpha$, i.e., $w_A(\gamma) = \sum_{\alpha}\,
(w_{A})_\alpha \:\gamma^{\beta}$.

The Hamiltonian approach relies on the expectation
that $\rho\rightarrow +\infty$ in the approach to the singularity.
As a consequence of this assumption, each term $\rho^2 \exp[-2\rho
w_A(\gamma)]$ becomes a sharp wall, i.e., an infinitely high
potential which is described by an infinite step function
$\Theta_\infty(x)$ that vanishes for $x<0$ and is infinite when
$x\geq 0$. In the GR vacuum case there exist three `dominant' terms
in the sum, which is the minimum number of terms required to define
the `billiard table'. The coefficients $c_A$ of the dominant terms
are non-negative functions of the variables, but `generically' they
are positive, see~\cite{dametal03}. The three dominant terms $\rho^2
c_A \exp[-2\rho w_A(\gamma)]$ thus generate `dominant' walls
$\Theta_\infty(-2 w_A(\gamma))$. Only the dominant walls are assumed
to be of importance for the generic asymptotic dynamics; the other,
`subdominant', terms of~\eqref{Hami}, whose exponential
$b^\alpha$-dependence can be obtained by multiplying dominant terms,
are dropped accordingly. We thus obtain an asymptotic Hamiltonian of
the form
\begin{equation}\label{limiHami}
{\cal H}_{\infty} =  \textfrac{1}{4} \left[ -\pi_\lambda^2 +
\pi_\gamma^2 \right] + \sum_{A=1}^3 \Theta_\infty(-2 w_A(\gamma))\:,
\end{equation}
where the sum is over the three dominant terms.

This limiting Hamiltonian is believed to describe generic asymptotic
dynamics. It is independent of $\N{{\alpha}}{i}$, $\P{i}{{\alpha}}$,
and $\lambda$, which suggests that $\P{i}{{\alpha}}$,
$\N{{\alpha}}{i}$ and $\pi_\lambda$ are asymptotic constants of the
motion. The remaining non-trivial dynamics resides in hyperbolic
space, i.e., in the variables $\gamma$. The asymptotic dynamics can
be described via~\eqref{limiHami} as a geodesic motion in hyperbolic
space, which is constrained by the existence of three sharp
reflective walls, i.e., by a `billiard motion'. The asymptotic
dynamics is thus given by a `cosmological billiard,' see
Figure~\ref{cosmobilliard}.

%%%%%%%%%%%%%%%%%%%%%%%%%%%%%%%%%%%%%%%%%%%%%%%%%%%%%%%%%%%%%%%%%%%%%%%%%%%%%%%%%%%%%
\section{The dynamical systems approach}
\label{confsec}
%%%%%%%%%%%%%%%%%%%%%%%%%%%%%%%%%%%%%%%%%%%%%%%%%%%%%%%%%%%%%%%%%%%%%%%%%%%%%%%%%%%%%

%--------------------------------------------------------------------------------------
\subsection*{Conformal Hubble-normalization}
\label{confHubb}
%--------------------------------------------------------------------------------------

Physics is a science of scales: in many problems there exists a
variable scale of particular importance, which can be factored out
of the equations in order to obtain a simpler mathematical
description. In GR there exists a single dimensional unit which is
chosen to be `length' (or, equivalently, `time'). The geometrical
way of factoring out a scale in GR is by means of a conformal
transformation: choose a conformal factor that depends on the
variable scale so that the conformal factor carries the dimension
and the conformal metric becomes dimensionless. Then, for
dimensional reasons, the Einstein field equations split into
decoupled equations for the conformal factor and a coupled system of
dimensionless equations for quantities
associated with the dimensionless conformal metric.%
\footnote{Logarithmic derivatives of
  the conformal factor occur in the equations; however, one of the
  evolution equations (which is the Raychaudhuri equation in the present
  Hubble-normalized case) can be used to algebraically solve for the
  logarithmic time derivative,
  and the constraints can be used to
  generically solve for the logarithmic spatial derivatives.}
This reduced dimensionless system carries the essential information
about the problem, since one can solve for the decoupled equations
for the conformal factor once it has been understood, and thus
recover the physical metric.

The problem of generic spacelike singularities and the dynamics of
solutions in a neighborhood thereof is one example where one has a
variable scale: it is provided by the affine parameter of
inextendible causal geodesics, or the expansion along such
geodesics. Here we consider non-rotating timelike congruences that
may or may not be geodesic; the latter is of no particular
relevance, since the generic behavior toward a spacelike singularity
is similar, as will be discussed later. One equation plays an
essential role in the singularity theorems---the Raychaudhuri
equation for the expansion. This makes a suitable function of the
expansion an excellent candidate for a conformal factor. By
factoring out the expansion, which blows up toward the singularity,
we will obtain a system of \textit{regular}
dimensionless equations.%
\footnote{For historical reasons we will
choose the Hubble variable instead of the expansion, but since they
are proportional to each other this makes no essential difference.}
Let us begin by introducing the basic set-up and nomenclature:

Consider a four-dimensional Lorentzian manifold (the `physical
spacetime') with a metric $\mathbf{g}$ of signature $(-+++)$ that
satisfies the Einstein vacuum equations. By convention we set $c =
1$ and $8 \pi G =1$, which leaves `length' (or, equivalently,
`time') as the single remaining unit. Let us consider a timelike
non-rotating congruence in the physical spacetime whose tangential
vector field we denote by $\mathbf{u}$. We choose an orthonormal
frame $\mathbf{e}_a$ ($a = 0\ldots 3$) such that $\mathbf{e}_0 =
\mathbf{u}$, and introduce a orthonormal coframe dual to
$\mathbf{e}_a$ which we denote by $\boldsymbol{\omega}^a$, i.e.,
$\boldsymbol{\omega}^a(\mathbf{e}_b) = \delta^a_{\weg b}$. Our
choice of coordinates is that of a coordinate $3+1$ decomposition
associated with $\mathbf{u}$, where we set the shift vector to zero
and let $\partial_{x^0} = N \mathbf{u}$, so that the flow lines of
$\mathbf{u}$ are the timelines. Accordingly,
\begin{equation*}
\mathbf{e}_0 = N^{-1} \partial_{x^0} \:,
\quad \mathbf{e}_\alpha = \sE{\alpha}{i} \partial_{x^i}
\qquad\text{and}\qquad
\boldsymbol{\omega}^0 = N dx^0 \:,
\quad\boldsymbol{\omega}^\alpha = \sEE{\alpha}{i} dx^i
\end{equation*}
with $\alpha =1,2,3$ and $i =1,2,3$.
%The components $\sEE{\alpha}{i}$ are dual to $\sE{\alpha}{i}$,
%i.e., $\sEE{\alpha}{i} \sE{\beta}{i} = \delta^\alpha_{\weg\beta}$.
The metric thus reads
\begin{equation*}
\mathbf{g} = \eta_{a b}\, \boldsymbol{\omega}^a \boldsymbol{\omega}^b =
-N^2 (d x^0)^2 + \delta_{\alpha \beta} \sEE{\alpha}{i} \sEE{\beta}{j} dx^i dx^j \:,
\end{equation*}
where $\eta_{a b} = \mathrm{diag}(-1,1,1,1)$.
The Hubble scalar of the vector field $\mathbf{u} = \mathbf{e}_0$
is given by $H = \frac{1}{3} \boldsymbol{\triangledown}_a \mathbf{u}^a$
%$H = \frac{1}{3} \boldsymbol{\nabla}_a\,\mathbf{u}^a$,
where $\boldsymbol{\triangledown}$
%$\boldsymbol{\nabla}$
denotes the covariant derivative associated with $\mathbf{g}$, and
hence $H$ is related to the expansion $\theta$ of $\mathbf{u}$ by
$H=\frac{1}{3}\theta$. Note that $H^{-1}$ carries dimension `length'
(see~\cite{rohugg05,ear74} and compare with discussions about FRW
models where this is a frequently used fact).

The natural way of keeping track of dimensions in a parametrized
theory like GR is by means of using rigid measuring rods that carry
the dimension, which corresponds to assigning a dimension (`length')
to the orthonormal one-forms. To obtain Hubble-normalized variables
and associated Hubble-normalized equations, we factor out the Hubble
scale out of the frame vectors and dual one-forms, i.e., we go over
to a conformal Hubble-normalized orthonormal frame $\parb_a$ and its
dual coframe $\boldsymbol{\Omega}^a$:
\begin{equation*}
\parb_a = \frac{\mathbf{e}_a}{H}\:,\qquad
\boldsymbol{\Omega}^a = H \boldsymbol{\omega}^a\:.
\end{equation*}
The dimensionless `unphysical metric' $\mathbf{G}$ that is
associated with the conformal frame is defined as $\mathbf{G} =
\eta_{a b}\, \boldsymbol{\Omega}^a \boldsymbol{\Omega}^b$; it is
related to $\mathbf{g}$ through a conformal rescaling:
\begin{equation} \label{defconfon}
\mathbf{G} =  \eta_{a b}\, \boldsymbol{\Omega}^a \boldsymbol{\Omega}^b =
H^2 \eta_{a b} \boldsymbol{\omega}^a \boldsymbol{\omega}^b = H^2 \mathbf{g}\:.
\end{equation}
The conformal Hubble-normalized orthonormal frame satisfies
\begin{equation} \label{13cONfr}
\parb_0 =  \cn^{-1} \partial_{x^0} \:,
\quad \parb_\alpha = \E{\alpha}{i} \partial_{x^i}
\qquad\text{and}\qquad
\boldsymbol{\Omega}^0 = \cn dx^0 \:,
\quad\boldsymbol{\Omega}^\alpha = \EE{\alpha}{i} dx^i\:,
\end{equation}
where $\cn$ is the conformal lapse, $\E{\alpha}{i}$ are the
conformal spatial frame vector components, and $\EE{\alpha}{i}$ the
associated dual vector components, i.e.,
\begin{equation}\label{framedef}
\cn=H\,N\:,\qquad \E{\alpha}{i} = \frac{\sE{\alpha}{i}}{H} \:,\qquad
\EE{\alpha}{i} = H \sEE{\alpha}{i} \:.
\end{equation}
Since $\boldsymbol{\Omega}^\alpha$ is dual to $\parb_\alpha$, i.e.,
$\boldsymbol{\Omega}^\alpha(\parb_\beta) = \delta^\alpha_{\weg
\beta}$, the same is true for the components $\E{\alpha}{i}$ and
$\EE{\alpha}{i}$, i.e., $\EE{\alpha}{i} \E{\beta}{i} =
\delta^\alpha_{\weg\beta}$. The unphysical metric $\mathbf{G}$ can
be written
\begin{equation*}
\mathbf{G} = \eta_{a b}\, \boldsymbol{\Omega}^a \boldsymbol{\Omega}^b =
-\cn^2 (d x^0)^2 + \delta_{\alpha \beta} \EE{\alpha}{i} \EE{\beta}{j} dx^i dx^j \:.
\end{equation*}
For further details and discussions see~\cite{rohugg05}.
%where Hubble-normalized lapse $\cn$ is not to be confused with the
%Iwasawa matrix $\cn$.

The starting point for the conformal Hubble-normalized orthonormal
frame approach to the Einstein equations
is the decomposition of the commutators of $\parb_a$,
\begin{subequations}\label{commuta}
\begin{align}
\label{ccomts0}
[\,\parb_{0}, \parb_{\alpha}\,] & =
\dot{U}_{\alpha}\,\parb_{0} +
F_\alpha{}^\beta\,\parb_{\beta}\,,
&& F_\alpha{}^\beta = q\,\delta_{\alpha}{}^{\beta} -
\Sigma_{\alpha}{}^{\beta} -
\epsilon_{\alpha}{}^{\beta}{}_{\gamma}\,R^{\gamma}\, ,\\
\label{ccomt2} [\,\parb_{\alpha}, \parb_{\beta}\,] & =
(2A_{[\alpha}\,\delta_{\beta]}{}^{\gamma} +
\epsilon_{\alpha\beta\delta}\,N^{\delta\gamma})\,\parb_{\gamma}\, \:,
\end{align}
\end{subequations}
which can also be written in the alternative form
\begin{align}
\label{dcomts}
0 & = (\parb_{\alpha}+ \Udot_{\alpha})\parb_{0} -
(\delta_{\alpha}{}^{\beta}\,\parb_{0} -
F_{\alpha}{}^{\beta})\,\parb_{\beta}\, ,
&& F_\alpha{}^\beta = q\,\delta_{\alpha}{}^{\beta} -
\Sigma_{\alpha}{}^{\beta} -
\epsilon_{\alpha}{}^{\beta}{}_{\gamma}\,R^{\gamma}\, ,
\tag{\ref{ccomts0}${}^\prime$}
\\
0 & = \bm{C}_{\alpha}{}^{\beta}\,\parb_{\beta}\,,
&& \bm{C}_{\alpha}{}^{\beta}= \epsilon_{\alpha}{}^{\gamma\beta}\,
(\parb_{\gamma}-A_{\gamma}) - N_{\alpha}{}^{\beta}\:,
\tag{\ref{ccomt2}${}^\prime$}
\end{align}
see~\cite{rohugg05}. In~\eqref{ccomts0}, $\Udot^\alpha$ are the
frame components of the acceleration of $\parb_0$ w.r.t.\ the
conformal metric $\mathbf{G}$, i.e., $\Udot =
\boldsymbol{\nabla}_{\!\!\boldsymbol{\partial}_0} \parb_0$, where
$\boldsymbol{\nabla}$ is the covariant derivative associated with
$\mathbf{G}$. The variable $q$ is the negative Hubble scalar of the
vector field $\parb_0$ in the unphysical spacetime, i.e., $q =
-\frac{1}{3} \varTheta$, where $\varTheta= \boldsymbol{\nabla}_{\!a}
\parb_0^{\;a}$ is the expansion of $\parb_0$ w.r.t.\ the metric
$\mathbf{G}$.
%In terms of the physical spacetime, $q$ is the deceleration parameter
%associated with $\mathbf{e}_0 = \mathbf{u}$ and $\mathbf{g}$.
The object $\Sigma_{\alpha\beta}$ is the (trace-less) shear,
$R_\alpha$ is the Fermi rotation, which describes how the frame
rotates w.r.t.\ a Fermi propagated frame, associated with the vector
$\parb_0$ and the conformal metric $\mathbf{G}$. In~\eqref{ccomt2},
$N^{\alpha\beta}$ and $A_\alpha$ are spatial commutator functions
that describe the three-curvature of $\mathbf{G}$, see below; we
also refer to~\cite{waiell97,rohugg05} where the analogous
non-normalized objects are described.

Using the analog of~\eqref{ccomts0} in the physical spacetime,
i.e., $[\mathbf{e}_0, \mathbf{e}_\alpha] =
\dot{u}_\alpha \mathbf{e}_0 +f_{\alpha}^{\weg \beta} \mathbf{e}_\beta$,
where the trace of $f_{\alpha}^{\weg \beta}$ equals $-3 H$,
cf.~\cite{rohugg05}, it is straightforward to derive the equation
\begin{subequations}\label{Hubblenorm}
\begin{equation}
\parb_0 H = -(1+q) H\:.
\end{equation}
We thus see that, in terms of the physical spacetime, $q$ is
interpreted as the deceleration parameter associated with
$\mathbf{e}_0 = \mathbf{u}$ and $\mathbf{g}$. As we will describe
below, $q$ can be algebraically determined by the Raychaudhuri
equation. Analogously, we define the spatial Hubble gradient
$r_\alpha$ by
\begin{equation}\label{ralph}
\parb_\alpha H = -r_\alpha\,H\:,
\end{equation}
\end{subequations}
where $r_\alpha$ is generically determined by the Codazzi
constraint, see below. For dimensional reasons, the
equations~\eqref{Hubblenorm} for $H$ decouple from the remaining
dimensionless equations~\eqref{udotN}--\eqref{allconeq}, which
are given next.

We are now in a position to describe the dynamics of Einstein's
vacuum field equations in terms of the conformal Hubble-normalized
variables introduced above. From the Hubble-normalized commutator
equations, the Einstein field equations, and the Jacobi identities,
see~\cite{rohugg05}, we obtain a dimensionless system of coupled
equations; we split the system into gauge equations, evolution
equations, and constraint equations:

\textit{Gauge equation}:
\begin{equation} \label{udotN}
\hspace{-3em} \Udot_{\alpha} = \parb_{\alpha}\ln \cn\:.
\end{equation}

\textit{Evolution equations}:
\begin{subequations}\label{alleveq}
\begin{align}
\label{dl13comts}
\parb_{0}E_{\alpha}{}^{i} & =
F_{\alpha}{}^{\beta}\,E_{\beta}{}^{i}\,, \\
\label{dlsigdot}
\parb_{0}\Sigma_{\alpha\beta} & =
-(2-q)\Sigma_{\alpha\beta} -
2\epsilon^{\gamma\delta}{}_{\langle\alpha}\,\Sigma_{\beta\rangle\gamma}\,R_{\delta}
- {}^{3}{S}_{\alpha\beta} + (I_\Sigma)_{\alpha\beta}
\,, \\
\label{dlndot}
\parb_{0}N^{\alpha\beta} & =
(3q\,\delta_{\gamma}{}^{(\alpha}-2F_{\gamma}{}^{(\alpha}{})\,N^{{\beta}){\gamma}}
+ (I_N)^{\alpha\beta}\, ,\\
\label{dladot} \parb_{0}A_{\alpha} & =
F_{\alpha}{}^{\beta}\,A_{\beta} + (I_A)_\alpha\,.
\end{align}
\end{subequations}

\textit{Constraint equations}:
\begin{subequations}\label{allconeq}
\begin{align}
\label{dl13com}
0 & = \bm{C}_{\alpha}{}^{\beta}\,E_{\beta}{}^{i}\, ,\\
\label{dlgauss} 0 & = 1 - \Sigma^2 - \Omega_k - (I_G)\, ,\\
\label{dlcodazzi} 0 & =  -3A_{\beta}\Sigma_{\alpha}{}^{\beta}+
\epsilon_{\alpha}{}^{\beta\gamma}\Sigma_{\beta}{}^{\delta}N_{\delta\gamma} +
(I_C)_\alpha\, ,\\
\label{dljacobi1} 0 & =  A_{\beta}\,N_{\alpha}{}^{\beta} +
(I_J)_\alpha\, ,\\
\label{dljacobi2} 0 & = (I_{J\omega})_\alpha\, .
\end{align}
\end{subequations}
Here,  $(\ldots)$ and $\la \ldots\ra$ denote symmetrization and
trace free symmetrization, respectively; let us briefly comment on
this system: the equations~(\ref{udotN}), (\ref{dl13comts}),
and~(\ref{dl13com}) are obtained from the commutator equations;
(\ref{dlsigdot}) is the trace free Einstein evolution equations,
while the remaining Einstein evolution equation, which can be
written as the Raychaudhuri equation, yields the expression for $q$
that is given below in~\eqref{defofmuchsig2q}; (\ref{dlgauss}) is
the Gauss constraint; (\ref{dlcodazzi}) are the Codazzi constraint
equations; (\ref{dlndot}), (\ref{dladot}), (\ref{dljacobi1}),
(\ref{dljacobi2}) are obtained from the Jacobi identities.

In these equations we have employed the following definitions:
\begin{subequations}\label{defofmuch}
\begin{align}
\label{defofmuchsig2q}
\Sigma^2 & := \textfrac{1}{6}\Sigma_{\alpha\beta}\Sigma^{\alpha\beta}\, , &
q & = 2\Sigma^2 + (I_q)\, ,\\
\Omega_k & := -\textfrac{1}{6}\,{}^{3}{\cal R}\, ,& {}^{3}{\cal R} & =
- 6A^2 - \textfrac{1}{2}B_\alpha{}^\alpha + (I_R) \, ,\\
{}^{3}{\cal S}_{\alpha\beta} & =
2A_\gamma\,N_{\delta\langle\alpha}\epsilon_{\beta\rangle}{}^{\gamma\delta} +
B_{\langle\alpha\beta\rangle} + (I_S)_{\alpha\beta}\, , & B_{\alpha\beta} &
=  2N_{\alpha}{}^\gamma\,N_{\gamma\beta} -
N_\gamma{}^\gamma\,N_{\alpha\beta}\,;
\end{align}
\end{subequations}
here and in the following the norm of a spatial vector $V^\alpha$ is
written as $V_\alpha\,V^\alpha=V^2$; ${}^{3}{\cal R}, {}^{3}{\cal
S}_{\alpha\beta}$ are the conformal scalar curvature and trace free
Ricci curvature, respectively. Finally, the expressions for
$(I_{\ast})_{\cdot\cdot}$, which are zero in the symmetry adapted
spatially homogeneous (SH)
case, are given by
\begin{subequations}\label{theIs}
\begin{eqnarray}
\label{Isig}(I_\Sigma)_{\alpha\beta} & = &
-\epsilon^{\gamma\delta}{}_{\la\alpha}\,N_{\beta\ra\delta}\,(\Udot_{\gamma}-2r_{\gamma})
+ (\parb_{\la\alpha} + A_{\la\alpha})(\Udot_{\beta\ra} +
2r_{\beta\rangle}) + \Udot_{\la\alpha}\Udot_{\beta\ra} -
2r_{\langle\alpha}r_{\beta\rangle}\, ,\qquad\\
%
%\label{Isig}(I_\Sigma)_{\alpha\beta} & = &
%-\epsilon^{\gamma\delta}{}_{\la\alpha}\,N_{\beta\ra\delta}\,(\Udot_{\gamma}-2r_{\gamma})
%+ (\parb_{\la\alpha} + A_{\la\alpha} +
%\Udot_{\la\alpha})\Udot_{\beta\ra} + 2(\parb_{\la\alpha} +
%A_{\la\alpha}-r_{\langle\alpha})r_{\beta\rangle}\, ,\\
%
(I_q)  & = &  - \textfrac{1}{3}(\parb_{\alpha} + \Udot_{\alpha}
-2A_{\alpha}+2r_{\alpha})(\Udot^{\alpha} + r^{\alpha})\, ,\\
\label{IN}(I_N)^{\alpha\beta} & = &(\parb_{\gamma}+
\Udot_{\gamma})\epsilon^{\gamma\delta
(\alpha}F_{\delta}{}^{\beta)}\,, \mspace{90mu} (I_A)_{\alpha}  =
\textfrac{1}{2}\,(\parb_{\beta}+
\Udot_{\beta})(\,3q\,\delta_{\alpha}{}^{\beta}-F_{\alpha}{}^{\beta}\,)\, ,\\
(I_G)& = &
\textfrac{1}{3}\,(2\parb_{\alpha}-4A_{\alpha}+r_{\alpha})r^{\alpha}\,,
\mspace{63mu}  (I_C)_\alpha  =
\parb_{\beta}\Sigma_{\alpha}{}^{\beta} +
(2\delta_{\alpha}{}^{\beta} + \Sigma_{\alpha}{}^{\beta})r_{\beta}\, ,\\
(I_J)_\alpha & = &
-\textfrac{1}{2}(\parb_{\beta}\,N_{\alpha}{}^{\beta} +
\epsilon_\alpha{}^{\beta\gamma}\parb_{\beta}A_{\gamma})\,
,\mspace{30mu}
(I_{J\omega})_\alpha = \bm{C}_\alpha{}^\beta\,\Udot_\beta\, ,\\
(I_S)_{\alpha\beta} & = & \parb_{\la\alpha}\, A_{\beta\ra} -
\parb_\gamma \,N_{\delta\la\alpha}\epsilon_{\beta\ra}{}^{\gamma\delta}\, ,
\mspace{65mu}
(I_R) = 4\parb_{\alpha}A^{\alpha}\, .
\end{eqnarray}
\end{subequations}

The above dimensionless coupled system of partial differential
equations is associated with a \textit{state space} described by the
state vector
\begin{subequations}\label{Xstatesp}
\begin{align}
\bm{X} & = \left(E_{\alpha}{}^{i}, \Sigma_{\alpha\beta},A_\alpha,N_{\alpha\beta}\right)\:;
\intertext{we make the split}
\label{Xstatespsplit}
\bm{X} &= (E_{\alpha}{}^{i})\oplus\bm{S}\:, \quad\text{where}\qquad
\bm{S}=\left(\Sigma_{\alpha\beta},A_\alpha,N_{\alpha\beta}\right)\:.
\end{align}
\end{subequations}
The Fermi rotation variables $R_\alpha$, which describe how the
chosen frame rotates w.r.t.\ a Fermi propagated frame, are gauge
variables: thus $R_\alpha=0$ corresponds to choosing a Fermi frame
while $R_\alpha \neq 0$ yields a rotating frame. The variable
$\Udot_\alpha$ is regarded as a gauge variable determined by
$\mathcal{N}$ (or, equivalently, by $N$); the spatial Hubble
gradient $r_\alpha$ is generically determined by the Codazzi
constraint~\eqref{dlcodazzi}, however, for certain choices of
$\mathcal{N}$ one can derive an evolution equation for $r_\alpha$
and include $r_\alpha$ into the state vector $\bm{S}$, see UEWE;
sometimes it is also useful to elevate $q$ to a dependent variable,
see~\cite{gar04}.

%--------------------------------------------------------------------------------------
\subsection*{Iwasawa frame variables}
\label{Iwasframe}
%--------------------------------------------------------------------------------------

In this paper we choose the conformal orthonormal frame
$\parb_\alpha = \E{\alpha}{i}\partial_{x^i}$ to be a conformally
rescaled Iwasawa frame, i.e., $\E{\alpha}{i} = H^{-1}
\sE{\alpha}{i}$, where $\sE{\alpha}{i}$ is a lower triangular
matrix%
\footnote{Since $\sE{\alpha}{i} = \exp(b^\alpha) \bN{i}{\alpha}$, the
  components $\E{\alpha}{i}$
  form a lower triangular matrix when
  $\bN{i}{\alpha}$ is an upper triangular matrix, which is in accordance with
  the definitions of Section~\ref{Hamilton}.
  For the case of the coframe, both $\EE{\alpha}{i}$ and $\N{\alpha}{i}$
  are upper triangular matrices.}
with the diagonal entries written in an exponential
form, see Section~\ref{Hamilton}.

In Appendix~\ref{relations} we show that choosing a conformal
Iwasawa frame in the conformal Hubble-normalized dynamical systems
approach corresponds to choosing the gauge
\begin{equation*}
\Sigma_{23}= - R_1\:,\qquad \Sigma_{31} = R_2\:, \qquad
\Sigma_{12} = -R_3\:,
\end{equation*}
and setting
\begin{equation*}
N_3 = N_{33} = 0\:.
\end{equation*}
This makes it natural to introduce the notation
$\Sigma_{\alpha}=\Sigma_{\alpha\alpha}$, and to replace the
off-diagonal components of $\Sigma_{\alpha\beta}$ by $R_\alpha$
according to the above equation, i.e.,
\begin{equation}\label{sigmamatrix}
\begin{pmatrix}
\Sigma_{11} & \Sigma_{12} & \Sigma_{13} \\
\Sigma_{21} & \Sigma_{22} & \Sigma_{23} \\
\Sigma_{31} & \Sigma_{32} & \Sigma_{33}
\end{pmatrix} =
\begin{pmatrix}
\Sigma_{1} & -R_3 & R_2 \\
-R_3 & \Sigma_{2} & -R_1 \\
R_2 & -R_1 & \Sigma_{3}
\end{pmatrix} \:.
\end{equation}
By this choice of gauge we write the state vector $\bm{S}$ as
\begin{equation}\label{Sstatesp}
\bm{S} = \left(\Sigma_{\alpha}, R_{\alpha}, A_\alpha,N_{\alpha\beta}\right)\:.
\end{equation}
For further details we refer to Appendix~\ref{relations}, where we
give the relationships between the Hamiltonian Iwasawa variables and
the Hubble-normalized variables of the conformal Hubble-normalized
dynamical systems approach.

As we will see in this paper, the Iwasawa gauge has non-trivial
consequences for the description of the generic asymptotic dynamics
of solutions in a neighborhood of a generic spacelike singularity.
However, next we describe some central concepts that are independent
of the chosen frame.

%%%%%%%%%%%%%%%%%%%%%%%%%%%%%%%%%%%%%%%%%%%%%%%%%%%%%%%%%%%%%%%%%%%%%%%%%%%%%%%%%%%%%%%%%%
\section{Asymptotic silence, locality, and the silent boundary}
\label{asympsil}
%%%%%%%%%%%%%%%%%%%%%%%%%%%%%%%%%%%%%%%%%%%%%%%%%%%%%%%%%%%%%%%%%%%%%%%%%%%%%%%%%%%%%%%%%%

%--------------------------------------------------------------------------------------
\subsection*{Asymptotic silence and asymptotic locality}
\label{asysilasyloc}
%--------------------------------------------------------------------------------------

A generic \textit{spacelike} singularity is expected to be a scalar
curvature singularity%
\footnote{In contrast, a non-scalar curvature
  singularity is expected to require fine-tuning of initial data.}
associated with ultra strong gravity which increasingly focuses
light in all directions as the singularity is approached.
The resulting asymptotic collapse of the light cones%
\footnote{We may hence
  refer to this as an `anti-Newtonian' limit.}
causes the particle horizons to shrink to
zero size toward the singularity along any timeline. As a consequence,
communication
between different timelines is prohibited in the asymptotic limit,
and we therefore refer to this causal feature of shrinking particle
horizons as \textit{asymptotic silence}; the associated singularity
is said to be \textit{asymptotically silent}.

Evidence for the conjecture that generic singularities are
asymptotically silent is discussed in UEWE and~\cite{limetal04}; we
also refer to corroborative results from the context of spatially
homogeneous cosmologies, see~\cite{waiell97,rin00,rin01,grietal72}
and references therein. In Section 4 in UEWE it is shown that,
generically, asymptotic silence is connected with the property that
the Hubble-normalized spatial frame variables vanish asymptotically,
i.e.,
\begin{equation*}
\E{\alpha}{i} \rightarrow 0
\end{equation*}
toward the singularity; however, this is not sufficient for
asymptotic silence, as is illustrated by special examples
in~\cite[Section 4]{limetal06}. We also refer to~\cite{limetal06}
for examples of various non-generic asymptotically silent
singularities and for examples of singularities for which asymptotic
silence does not hold---a phenomenon referred to as
\textit{asymptotic silence-breaking}.

We define the dynamics along a timeline toward a singularity to be
\textit{asymptotically local} if
\begin{equation}
\label{asylocdef} E_{\alpha}{}^{i}\rightarrow 0 \:, \quad
\parb_{\alpha} (\bm{S}, r_\beta, \Udot_\beta) \rightarrow 0\:, \quad
(r_\alpha, \Udot_\alpha) \rightarrow 0 \:.
\end{equation}
toward the singularity. (Note that this definition slightly differs
from the one used in~\cite{andetal05}). Asymptotic silence and
asymptotic local dynamics are closely related concepts. The
shrinking of particle horizons and the associated loss of
communication between timelines suggests that inhomogeneities are
irrelevant for the asymptotic dynamics, since they are shifted
outside the shrinking horizons faster than they grow, i.e.,
$E_{\alpha}{}^{i}$ goes to zero at a rate that is faster than the
possible growth rate of $\partial_i \bm{S}$, $\partial_i \log(H)$,
$\partial_i \log({\cal N})$. This is supported for generic timelines
toward generic asymptotically silent spacelike singularities by
numerical experiments~\cite{gar04,andetal05}, but it may not be the
case for all timelines. This is an issue that will be discussed in
the concluding remarks in connection with `recurring spike
formation'~\cite{andetal05}; see also the discussion
in~\cite{limetal06}. However, apart from in the concluding remarks,
in the remainder of this paper we will be concerned with generic
timelines approaching generic asymptotically silent singularities
for which the dynamics is assumed to be asymptotically local.

%--------------------------------------------------------------------------------------
\subsection*{The silent boundary}
\label{silbound}
%--------------------------------------------------------------------------------------

Consider the Einstein field equations in the conformal Hubble-normalized
approach, i.e., the system of equations~\eqref{udotN}-\eqref{theIs} introduced
in Section~\ref{confsec}.
It follows from~\eqref{dl13comts}, in connection with~\eqref{dl13com},
that
\begin{equation}\label{silbou}
\E{\alpha}{i} = 0
\end{equation}
defines an invariant subset of the state space; we refer to this
invariant subset as the \textit{silent boundary}, which is
characterized by the state vector $\bm{S}$. Since $\parb_\alpha =
\E{\alpha}{i} \partial_i$, Equation~\eqref{silbou} implies
$\parb_\alpha \bm{S} = 0$ and $\parb_\alpha \Udot_{\beta} = 0$,
$\parb_\alpha r_\beta = 0$ in~\eqref{udotN}-\eqref{theIs}, so that
the equations \textit{on} the silent boundary reduce to a system of
ordinary differential equations. Note, however, that we obtain a
system of ODEs \textit{for each spatial point} $x^i$; hence the
silent boundary can be visualized as an infinite set of copies, one
for each spatial point, of a finite dimensional state space.
The equations on the silent boundary are identical to the equations
for the state vector $\bm{S}$ of spatially self-similar and
SH models. This is a direct consequence from the symmetry
assumptions, since spatial self-similarity implies that
$\parb_\alpha \bm{S} =0$ and $\parb_\alpha(\Udot_\alpha,r_\alpha)=0$
in a symmetry adapted
frame, which in turn entails that the equations for $E_\alpha{}^i$
decouple, so that one obtains a reduced coupled system of ODEs for
the variables $\bm{S}$.

The \textit{SH subset of the silent
boundary} (in brief: SH silent boundary) is defined as the invariant
subset of the silent boundary given by
\begin{equation}\label{SHsilentdef}
\E{\alpha}{i} =0 \quad\text{ and } \quad \Udot_\alpha = 0\,,\:
r_\alpha = 0\:, \tag{\ref{silbou}${}^\prime$}\:.
\end{equation}
Since this yields $(I_{\ast})_{\cdot\cdot}= 0$,
the equations on the SH silent boundary reduce
to a system of ordinary differential equations, which is obtained
by setting the quantities $(I_{\ast})_{\cdot\cdot}$
to zero in~\eqref{udotN}-\eqref{theIs}.
This system
is identical to the (reduced) system of equations for
SH models, since spatial homogeneity implies
$\parb_\alpha \bm{S} = 0$ and $\Udot_\alpha = r_\alpha =0$
in a symmetry adapted frame, which subsequently
leads to a decoupling of the quantities $E_\alpha{}^i$;
%and leaves a reduced system for $\bm{S}$;
%which is the coupled system of
%ordinary differential equations~\eqref{udotN}-\eqref{theIs}
%where $(I_{\ast})_{\cdot\cdot}= 0$;
see also~\cite{andetal04}. Note again that the SH silent boundary is
an infinite set of copies, parametrized by the spatial coordinates,
of a finite dimensional state space with $\bm{S}$ as state vector.

The particular importance of the SH silent boundary in the study of
generic spacelike singularities stems from its connection with
asymptotically local dynamics; compare~\eqref{asylocdef}
and~\eqref{SHsilentdef}. The reasoning is as follows: The
Hubble-normalized conformal orthonormal frame approach produces
field equations that are regular in the asymptotic limit toward a
generic spacelike singularity. This allows us to \textit{extend} the
state space to also include the silent boundary in our analysis of
the dynamical system. This is highly advantageous; in particular,
based on our expectation that generic spacelike singularities are
asymptotically local, see~\eqref{asylocdef}, we are now able to
conjecture that the asymptotic behavior of asymptotically local
solutions is reflected in the dynamics on the SH silent boundary;
see the discussion in Section~\ref{asysha}. Therefore, the SH silent
boundary and its neighborhood in the full state space will be the
main object of investigation in this paper.

In contrast to the spatially self-similar and SH models, the
variables $\bm{S}$ on the SH silent boundary depend on the spatial
coordinates, and the constants of integration are therefore
spatially dependent functions. The dynamical systems formalism thus
explains the first step in the heuristic ad hoc procedure by BKL,
which consists in replacing the constants in certain SH models with
spatially dependent functions. Moreover, the ODE structure on the
silent boundary in turn induces ODE structures associated with
perturbations thereof---in a series expansion spatial partial
derivatives only act on spatial functions associated with previous
lower order terms, thus only yielding temporal constants for the
perturbative equations; this feature naturally captures the BKL
statement that spatial derivatives only enters `passively' into the
equations (BKL~\cite[pp.656--657]{bkl82}). Note, however, that the
dynamical systems approach allows one to derive the results of BKL
and thus put them in the rigorous context of the regularized
dimensionless state space picture of Einstein's field equations, a
picture which also simplifies comparisons with analytical results
for special cases as well as with numerical investigations.

The asymptotic ODE structure induced by asymptotic local dynamics
can be exploited in several ways; in particular, it allows us to
reparametrize the {\em individual\/} timelines, for which the
spatial coordinates $x^i$ are fixed, so that $\parb_{0}f =
-df/d\tau$ along a given timeline, where $f(x^0,x^i)$ is any
variable occurring in an ODE, and $\tau(x^0,x^{i}):=
-\log(\ell/\hat{\ell})$ is a `local time function' directed toward
the singularity. In this context, $\ell = g^{1/6}$ and $g$ is the
determinant of the physical spatial metric. In our convention,
hatted objects refer to objects that are functions of the spatial
coordinates alone; note also that we choose a time direction toward
the singularity, which is in contrast to UEWE. To obtain the
solution in the chosen time coordinate $x^0$, one subsequently
integrates the relation $dx^0 = -\cn^{-1}d\tau$ so that
\begin{equation} \label{ttaurel} x^0 = {\hat x}^0(x^{i}) -
\int_{\tau_0}^{\tau}\cn^{-1}({\tau}',x^{i})d{\tau}'\, . \end{equation}

The reparametrization freedom allows us to introduce a dynamical
systems treatment of the equations on the silent boundary and its
neighborhood, in complete analogy with the treatment of the
so-called silent cosmological models from a dynamical systems
perspective, see~\cite{matetal94,bruetal95} and~\cite[Chapter
13]{waiell97}.\footnote{In the present context it is irrelevant that
the only nontrivial spatially inhomogeneous non-rotating exact
silent solutions without a cosmological constant turned out to be
the Szekeres dust models; see~\cite{hveetal97} and, for similar
topics and additional references,~\cite{wylber06}.} From now on we
will use the local reparametrization freedom induced by
asymptotically local dynamics and write
\begin{equation*}
\parb_{0} = -\partial_\tau\:,
\end{equation*}
to study the asymptotic dynamics along timelines by means of finite
dimensional dynamical systems techniques.\footnote{As a simple
illustration, we derive the generalized Kasner line element of BKL
in Appendix~\ref{Kasner}.}

%%%%%%%%%%%%%%%%%%%%%%%%%%%%%%%%%%%%%%%%%%%%%%%%%%%%%%%%%%%%%%%%%%%%%%%%%%%%%%%%%%%%%%%%%%
\section{Kasner circle stability and the oscillatory subset}
\label{prebilliard}
%%%%%%%%%%%%%%%%%%%%%%%%%%%%%%%%%%%%%%%%%%%%%%%%%%%%%%%%%%%%%%%%%%%%%%%%%%%%%%%%%%%%%%%%%%

It is suggestive that there exists an open set of vacuum models that
possess asymptotically silent singularities and obey asymptotically
local dynamics for a generic set of timelines. This follows from the
analysis of UEWE~\cite{uggetal03} in combination with the results
of~\cite{limetal04}, where a dynamical systems formulation was used
that added the null geodesic equations to the Einstein field
equations; further support comes from the numerical analysis
of~\cite{gar04} and from~\cite{andetal05,limetal06}. We therefore
conclude that at each spatial point the SH silent boundary---or
rather a subset of the SH silent boundary---constitutes a local
attractor for solutions in the full inhomogeneous state space
associated with the Einstein equations; note, however, that it can
only be a local attractor in the full state space since there
exists, e.g., an open set of solutions without any singularity at
all~\cite{chrkla93}; moreover, there may exist an open set of
solutions with weak null singularities, see~\cite{dafrod05} and
references therein, or an open set of solutions with a singularity
that consists of parts that are asymptotically silent and spacelike
and parts that are weak null singular; for special examples of
singularities of this latter type, see~\cite{limetal06}.

The main aim of the present paper is to identify the subset of the
SH silent boundary that acts as the attractor for generic asymptotic
dynamics. As suggested by the analysis of SH models, the subsets
that represent Bianchi type I and II models will be of crucial
importance. The common denominator of these models is the most
fundamental structure of the SH silent boundary state space: the
Kasner circle.

%The most general SH vacuum models are the Bianchi type VIII and IX
%models and the general type VI$_{-1/9}$ models.
%There exists heuristic and
%numerical evidence for the case of Bianchi type VI$_{-1/9}$ and VIII,
%see~\cite{hewetal03,ber02}, that the attractor consists of a union
%of the Bianchi type I (Kasner) and type II subsets; for Bianchi
%type IX models it has recently been rigorously proved
%that the attractor resides on this union of subsets~\cite{rin01}.
%The aim of the present paper
%to identify the subset of the SH silent boundary
%that acts as the attractor for generic asymptotic dynamics.
%As suggested by the analysis of SH models,
%Bianchi type I and II models will be of fundamental importance.
%However, there are numerous representations of
%Bianchi type I and II models and associated subsets.
%Based to the results in UEWE, we expect the
%attractor to only belong to a subset of their union, where this subset
%depends on the choice of frame, i.e., on our choice of
%Iwasawa gauge.
%We begin by investigating the most fundamental structure
%of the SH silent boundary state space: the Kasner circle.

%--------------------------------------------------------------------------------------
\subsection*{The Kasner circle, stable variables, and the oscillatory subset}
\label{Kascirc}
%--------------------------------------------------------------------------------------

The SH part of the silent boundary $\E{\alpha}{i} =0$ is given by
$\Udot_\alpha = r_\alpha=0$\,; it consists of an infinite number of
identical copies of the finite dimensional state space spanned by
the state vector $\bm{S} =
(\Sigma_{\alpha\beta},A_\alpha,N_{\alpha\beta})$; hereby, the
spatial variables can be regarded as acting as an index set. The
associated system of evolution equations and constraints is obtained
when the quantities $(I_{\ast})_{\cdot\cdot}$ are set to zero
in~\eqref{dlsigdot}-\eqref{dladot}
and~\eqref{dlgauss}-\eqref{dljacobi1}, respectively. This system of
equations possesses a one-parameter set of equilibrium points which
is fundamental for our analysis: the \textit{Kasner circle}
$\mathrm{K}^{\ocircle}$. It is determined by
\begin{equation}\label{Kasnercircledef1}
1-\Sigma^2 = A_{\alpha} = N_{\alpha\beta}= R_\alpha = 0 \,.
\end{equation}

Since $R_\alpha = 0$ corresponds to $\Sigma_{\alpha\beta} = 0$
($\alpha \neq \beta$), $\Sigma_{\alpha\beta}$ is diagonal for every
point on $\mathrm{K}^{\ocircle}$, i.e., $\Sigma_{\alpha\beta}=
\mathrm{diag}[\hat{\Sigma}_{1},\hat{\Sigma}_{2},\hat{\Sigma}_{3}]$,
where the numbers $\hat{\Sigma}_{\alpha}$ ($\alpha =1 \ldots 3$) are
constants. (All considerations apply to each of the (infinitely
many) copies of the finite dimensional SH state space. Since the
numbers $\hat{\Sigma}_{\alpha}$ ($\alpha =1 \ldots 3$) may differ
between the individual copies of the SH state space,
$\hat{\Sigma}_{\alpha}$ are in fact spatially dependent functions.
In this context recall that we employ hatted objects to denote
temporally constant spatially dependent functions.) It is standard
to represent the Kasner circle in terms of the (generalized, when
viewed as spatially dependent) Kasner exponents $p_\alpha$,
\begin{equation}\label{Kasnercircledef2}
\Sigma_{\alpha\beta}=\text{diag}[{\hat \Sigma}_{1},{\hat
\Sigma}_{2},{\hat \Sigma}_{3}]= \text{diag}[3p_1-1,3p_2-1,3p_3-1]\:,
\end{equation}
where we have omitted the hats on top of $p_\alpha$ in order to
agree with standard notation. Since $\mathrm{tr}\:
\Sigma_{\alpha\beta} = 0$ and $\Sigma^2 =1$, the Kasner exponents
satisfy the Kasner relations
\begin{equation*}
p_1+p_2+p_3=1\:, \qquad p_1^2 + p_2^2 + p_3^2 = 1\:.
\end{equation*}
In Appendix~\ref{Kasner} we derive the generalized Kasner line
element, see~\cite{bkl82},
\begin{equation*}
ds^2 = -dt^2 + \left( t^{2 p_1} l_i l_j +
t^{2 p_2} m_i m_j + t^{2 p_3} n_i n_j \right) dx^i dx^j
\end{equation*}
from the fixed point solution given
by~\eqref{Kasnercircledef1} and~\eqref{Kasnercircledef2}.

\begin{figure}[ht]
\centering
{\psfrag{a}[cc][cc]{$\Sigma_1$} \psfrag{b}[cc][cc]{$\Sigma_3$}
\psfrag{c}[cc][cc]{$\Sigma_2$} \psfrag{d}[cc][cc]{$T_1$}
\psfrag{e}[cc][cc]{$Q_2$} \psfrag{f}[cc][cc]{$T_3$}
\psfrag{g}[cc][cc]{$Q_1$} \psfrag{h}[cc][cc]{$T_2$}
\psfrag{i}[cc][cc]{$Q_3$} \psfrag{j}[cc][cc]{0}
\psfrag{k}[cc][cc]{$\textfrac{\pi}{3}$}\psfrag{l}[cc][cc]{$\textfrac{2\pi}{3}$}
\psfrag{m}[cc][cc]{$\pi$} \psfrag{n}[cc][cc]{$\textfrac{4\pi}{3}$}
\psfrag{o}[cc][cc]{$\textfrac{5\pi}{3}$}\psfrag{p}[cc][cc]{$2\pi$}
\psfrag{q}[cc][cc]{$u$} \psfrag{r}[cc][cc]{$\alpha$}
\psfrag{A}[cc][cc]{$N_2,R_3$} \psfrag{B}[cc][cc]{$N_1$}
\psfrag{C}[cc][cc]{$N_1,R_1$} \psfrag{D}[cc][cc]{$R_1,R_2$}
\psfrag{E}[cc][cc]{$R_1,R_2,R_3$}\psfrag{F}[cc][cc]{$N_2,R_2,R_3$}
\subfigure[Kasner sectors and unstable variables.]{
        \label{Ktrigg}
        \includegraphics[width=0.50\textwidth]{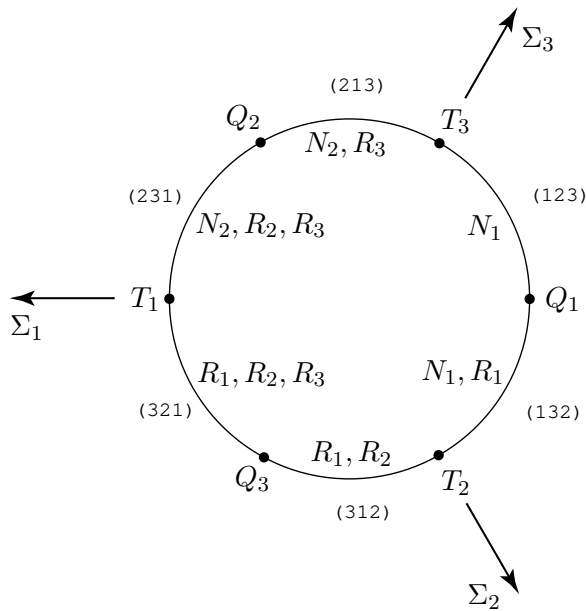}}} \\[4ex]
{\psfrag{a}[cc][cc]{$\Sigma_1$} \psfrag{b}[cc][cc]{$\Sigma_3$}
\psfrag{c}[cc][cc]{$\Sigma_2$} \psfrag{d}[cc][cc]{$T_1$}
\psfrag{e}[cc][cc]{$Q_2$} \psfrag{f}[cc][cc]{$T_3$}
\psfrag{g}[cc][cc]{$Q_1$} \psfrag{h}[cc][cc]{$T_2$}
\psfrag{i}[cc][cc]{$Q_3$} \psfrag{j}[cc][cc]{0}
\psfrag{k}[cc][cc]{$\textfrac{\pi}{3}$}\psfrag{l}[cc][cc]{$\textfrac{2\pi}{3}$}
\psfrag{m}[cc][cc]{$\pi$} \psfrag{n}[cc][cc]{$\textfrac{4\pi}{3}$}
\psfrag{o}[cc][cc]{$\textfrac{5\pi}{3}$}\psfrag{p}[cc][cc]{$2\pi$}
\psfrag{q}[cc][cc]{$u$} \psfrag{r}[cc][cc]{$\alpha$}
\psfrag{A}[cc][cc]{$N_2,R_3$} \psfrag{B}[cc][cc]{$N_1$}
\psfrag{C}[cc][cc]{$N_1,R_1$} \psfrag{D}[cc][cc]{$R_1,R_2$}
\psfrag{E}[cc][cc]{$R_1,R_2,R_3$}\psfrag{F}[cc][cc]{$N_2,R_2,R_3$}
\subfigure[Variation of the Kasner parameter $u$ along the Kasner circle.]{
        \label{uplot}
        \includegraphics[width=0.55\textwidth]{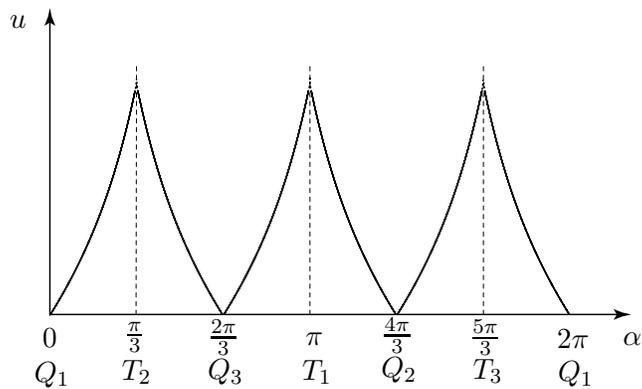}}}
        \caption{The Kasner circle of fixed points; sectors and unstable variables
        and relation to the Kasner parameter $u$. Note that $u=1$ at the $Q_\alpha$
        points and that $u=\infty$ at the Taub points $T_\alpha$.}
    \label{triggu}
\end{figure}

The Kasner circle $\mathrm{K}^{\ocircle}$ can be divided into six
sectors, where each sector is characterized by a typical ordered
sequence of the Kasner exponents $(p_\alpha)$; hence, sector $(123)$
is defined as the part of $\mathrm{K}^{\ocircle}$ where $p_1 < p_2 <
p_3$, whereas sector $(312)$ is characterized by the order
$p_3<p_1<p_2$, etc.; see Figure~\ref{Ktrigg}. Evidently, the six
sectors can be identified with each other through permutations of
the spatial axes. Of particular interest are the six special points
on the Kasner circle that are associated with solutions possessing
an additional symmetry, the so-called locally rotationally symmetric
(LRS) solutions. The points $Q_\alpha$ correspond to the three
equivalent LRS solutions whose intrinsic geometry is non-flat; for
$Q_1$ we have $(\hat{\Sigma}_1,\hat{\Sigma}_2,\hat{\Sigma}_3) =
(-2,1,1)$ and $(p_1,p_2,p_3) =
(-\frac{1}{3},\frac{2}{3},\frac{2}{3})$, cyclic permutations yield
the parameters for $Q_2$ and $Q_3$. Each point $T_\alpha$
corresponds to a flat LRS solution: the Taub representation of
Minkowski spacetime; the Taub points are given by
$(\hat{\Sigma}_1,\hat{\Sigma}_2,\hat{\Sigma}_3) = (2,-1,-1)$ or
$(p_1,p_2,p_3) = (1,0,0)$, and cyclic permutations; see
Figure~\ref{Ktrigg}.

It is convenient to parametrize the Kasner exponents in terms of the
\textit{Kasner parameter} $u$, which can be defined frame
invariantly through
\begin{equation}\label{detSigmau}
\det(\Sigma_{\alpha\beta})= 2 - \frac{27u^2(1+u)^2}{(1+u+u^2)^3},\qquad u\in [1,\infty]\:,
\end{equation}
see Appendix~\ref{KasnernonFermi} for details. Due to frame
invariance, the Kasner parameter naturally captures the equivalence
of the six sectors of $\mathrm{K}^{\ocircle}$, see
Figure~\ref{uplot}. On sector $(\alpha,\beta,\gamma)$, where
$p_\alpha < p_\beta < p_\gamma$, we set
\begin{equation}\label{ueq}
p_\alpha=\frac{-u}{1+u+u^2}\, ,\qquad p_\beta=\frac{1+u}{1+u+u^2}\,
,\qquad p_\gamma = \frac{u(1+u)}{1+u+u^2} \qquad u\in (1,\infty)\:.
\end{equation}
It is easy to check that $p_\alpha + p_\beta +
p_\gamma = 1$ and $p_\alpha^2 + p_\beta^2 + p_\gamma^2 = 1$. The
parameter $u=1$ describes the points $\mathrm{Q}_1, \mathrm{Q}_2,
\mathrm{Q}_3$, i.e., the three equivalent representations of the
non-flat LRS Kasner solution, while $u=\infty$ defines the Taub
points $\mathrm{T}_1, \mathrm{T}_2, \mathrm{T}_3$ and thus the Taub solution.%
\footnote{For the BKL
  definition of $u$, the order of $p_\alpha$ is fixed according
  to $p_1\leq p_2 \leq p_3$; we have found it convenient to
  use the above frame independent definition instead, and to permute the ordering
  of $p_\alpha$ according to the sector one considers when dealing
  with frame dependent matters.}

We proceed by performing a local dynamical systems analysis in a
neighborhood of the Kasner circle $\mathrm{K}^{\ocircle}$.
Linearization of the dynamical system at an arbitrary point
$(p_1,p_2,p_3)$ of $\mathrm{K}^{\ocircle}$ yields
\begin{subequations}
\begin{align}
\label{Edecay}
\partial_\tau E_\alpha{}^i & = -3 (1 - p_\alpha) E_\alpha{}^i & & ({\rm no\,\,sum\,\, over\,\,}\alpha) \\
\intertext{for the conformal frame variables, and}
\label{Adecay}
\partial_\tau A_\alpha & = -3 (1 - p_\alpha) A_\alpha & & ({\rm no\,\,sum\,\, over\,\,}\alpha) \\
\label{Nalphabetadecay}
\partial_\tau N_{\alpha \beta} & =
-6 ( 1 -p_\gamma ) N_{\alpha \beta} & & (\alpha \neq \beta \neq \gamma \neq \alpha)\,,  \\
\intertext{for the variables $A_\alpha$, $N_{\alpha\beta}$
$(\alpha\neq \beta)$ on the SH silent boundary; furthermore,}
\partial_\tau N_1 & = - 6 p_1 N_1 \,, &
& \partial_\tau N_{2} = -6 p_{2}\,N_{2} \\
\partial_\tau R_1 & = 3 (p_2 - p_3) R_1\,, &
& \partial_\tau R_2 =  -3(p_3 - p_1)R_2\,, &
& \partial_\tau R_3 = 3(p_1 - p_2)R_3\:,
\end{align}
\end{subequations}
where we set $N_1 = N_{11}$ and $N_2 = N_{22}$ (recall that $N_3 =
N_{33}=0$). We see that the variables $E_\alpha{}^i$,
$N_{\alpha\beta}$ ($\alpha \neq \beta$) and $A_\alpha$ belong to the
stable subspace of each fixed point of $\mathrm{K}^{\ocircle}$
(except for the Taub points). In contrast, the variables
$(R_1,R_2,R_3)$ and $(N_1, N_2)$ are stable or unstable depending on
the sector of $\mathrm{K}^{\ocircle}$ the point $(p_1,p_2,p_3)$ lies
in. Finally, the variables $\Sigma_\alpha = \Sigma_{\alpha\alpha}$
belong to the center subspace, i.e., they are constant to first
order. The analysis of the stability of the Kasner circle
$\mathrm{K}^{\ocircle}$ is summarized in Figure~\ref{Ktrigg}, where
the unstable variables are given for each sector of
$\mathrm{K}^{\ocircle}$.

We now decompose the state vector $\bm{S} = (\Sigma_{\alpha\beta},
A_{\alpha}, N_{\alpha\beta})$ of the SH silent boundary into a
`stable' and an `oscillatory' part: $\bm{S} =
\bm{S}_{\mathrm{stable}} \oplus \bm{S}_{\mathrm{osc}}$, where
$\bm{S}_{\mathrm{stable}} = (N_{\alpha\beta}, A_{\alpha})$ ($\alpha
\neq \beta$) and $\bm{S}_{\mathrm{osc}} = (\Sigma_\alpha,
R_{\alpha}, N_1, N_2)$. The subset determined by
$\bm{S}_{\mathrm{stable}} = 0$ on the SH silent boundary is an
invariant subset, which we call the \textit{oscillatory subset}
$\mathcal{O}$.

We conjecture that there exists an open set of solutions whose
behavior is governed by the Kasner states as $\tau\rightarrow
\infty$, i.e., we expect solutions to spend an increasing
amount of time close to
$\mathrm{K}^{\ocircle}$ as $\tau\rightarrow\infty$.
Equations~\eqref{Adecay} and~\eqref{Nalphabetadecay} suggest that
the variables $E_\alpha{}^i$ and the stable variables
$\bm{S}_{\mathrm{stable}} = (N_{\alpha\beta}, A_{\alpha})$ ($\alpha
\neq \beta$) decay rapidly, and that
$(E_\alpha{}^i,\bm{S}_{\mathrm{stable}})\rightarrow 0$ in the limit
$\tau\rightarrow \infty$; this motivates the notation `stable'. In
anticipation of results to come we note that the behavior of the
variables $\bm{S}_{\mathrm{osc}} = (\Sigma_\alpha, R_{\alpha}, N_1,
N_2)$ on the oscillatory subset, by contrast, can be best described
as `oscillatory', whence the chosen nomenclature. Since we conjecture
$(E_\alpha{}^i,\bm{S}_{\mathrm{stable}})\rightarrow 0$, it is the
complicated dynamics of $\bm{S}_{\mathrm{osc}}$ on the oscillatory
subset $\mathcal{O}$ that will play the decisive role in our
description of the generic asymptotic dynamics---we thus begin by
closely investigating the dynamical system on $\mathcal{O}$.

%--------------------------------------------------------------------------------------
\subsection*{The oscillatory dynamical system}
\label{theosci}
%--------------------------------------------------------------------------------------

Recall from Section~\ref{asympsil} that the silent boundary contains
an infinite number of identical copies of the SH state space and
thus of the oscillatory state space $\mathcal{O}$; hereby, the
spatial variables act as the index set. Although it is essential for
our ultimate aims to consider this infinite collection of spaces
$\mathcal{O}$, this is irrelevant for our proximate purposes: in the
present context we may regard the oscillatory subset $\mathcal{O}$
as one finite dimensional state space.

The evolution equations on the oscillatory subset $\mathcal{O}$ are
given by setting the stable variables to zero in
Eqs.~\eqref{dlsigdot}-\eqref{dlndot}, i.e.,
$\bm{S}_{\mathrm{stable}} = (N_{\alpha\beta}, A_{\alpha}) = 0$
($\alpha \neq \beta$). (Recall that the quantities
$(I_{\ast})_{\cdot\cdot}$ are zero because $\mathcal{O}$ is a subset
of the SH silent boundary determined by $\E{\alpha}{i} =0$,
$\Udot_\alpha = r_\alpha=0$.) Accordingly, the dynamical system on
$\mathcal{O}$ takes the form
\begin{subequations}\label{unstableeqs}
\begin{eqnarray}
\label{unstableeq3}
\partial_{\tau} \Sigma_1 & = &
2(1-\Sigma^2)\Sigma_1 - 2(R^2_2 + R^2_3) + {}^3\mathcal{S}_{11} \\
\label{unstableeq4}
\partial_{\tau} \Sigma_2 & = &
2(1-\Sigma^2)\Sigma_2 - 2(R^2_1 - R^2_3) + {}^3\mathcal{S}_{22} \\
\label{unstableeq5}
\partial_{\tau}\Sigma_3 & = &
2(1-\Sigma^2)\Sigma_3 + 2(R^2_1 + R^2_2) + {}^3\mathcal{S}_{33} \\
\label{unstableeq6}
\partial_{\tau} R_1 & = & 2(1-\Sigma^2)R_1 +
(\Sigma_2 - \Sigma_3)R_1 +2R_2R_3 \\
\label{unstableeq7}
\partial_{\tau} R_2 & = &
2(1-\Sigma^2)R_2 - (\Sigma_3 - \Sigma_1)R_2 \\
\label{unstableeq8}
\partial_{\tau} R_3 & = & 2(1-\Sigma^2)R_3 +
(\Sigma_1 - \Sigma_2)R_3 - 2R_1R_2 \\
\label{unstableeq9}
\partial_{\tau} N_{1} & = & -2(\Sigma^2 + \Sigma_{1})N_{1}\\
\label{unstableeq10}
\partial_{\tau} N_{2} & = & -2(\Sigma^2 + \Sigma_{2})N_{2} \: ,
\end{eqnarray}
where
\begin{align}
\nonumber
\Sigma^2 & = \textfrac{1}{6}\left(\Sigma_1^2 + \Sigma_2^2 + \Sigma_3^2
+
2 R_1^2 + 2 R_2^2 + 2 R_3^2\right) \\[0.5ex]
\label{someabbrevs}
{}^3\mathcal{S}_{11} & =  \textfrac{1}{3} (N_1 - N_2) (2 N_1
+ N_2) \:, \\
\nonumber
{}^3\mathcal{S}_{22} & =  -\textfrac{1}{3} (N_1 - N_2)
(N_1+ 2 N_2) \:,\qquad {}^3\mathcal{S}_{33} =  -\textfrac{1}{3} (N_1 -
N_2)^2\:.
\end{align}
The Gauss constraint~\eqref{dlgauss} becomes
\begin{equation}\label{gausscon}
1-\Sigma^2 - \textfrac{1}{12}(N_1-N_2)^2 = 0\, ,
\end{equation}
while the Codazzi constraints~\eqref{dlcodazzi} result in
\begin{equation}\label{codazzi}
N_1 R_2 = 0 \:, \qquad
N_1 R_3 = 0 \:, \qquad
N_2 R_1 = 0 \:, \qquad
N_2 R_3 = 0\:.
\end{equation}
\end{subequations}
Note that---a priori---the Codazzi constraints read
\begin{equation}\label{codazziap}
\tag{\ref{codazzi}$^\prime$}
N_1 R_2 = 0 \:, \qquad
N_2 R_1 = 0 \:, \qquad
(N_1 - N_2) R_3 = 0\:;
\end{equation}
however, unless $N_1 \equiv N_2 \neq 0$ on a finite $\tau$-interval,
the constraints~\eqref{codazzi} ensue. Now assume that $N_1 \equiv
N_2 \neq 0$, so that $1 -\Sigma^2 =0$; then $R_1 = 0$ and $R_2 =0$,
and the equations~\eqref{unstableeq3}--\eqref{unstableeq10} imply
$\Sigma_1 \equiv \Sigma_2$ (and $\partial_\tau \Sigma_1 =
\partial_\tau \Sigma_2$), whereby $R_3 \equiv 0$. Therefore, $N_1
R_3 = 0$ and $N_2 R_3 = 0$, and~\eqref{codazzi} holds in the case
$N_1 \equiv N_2 \neq 0$ as well.

The constraints determine the structure of the oscillatory state
space $\mathcal{O}$: it consists of a number of invariant subsets
(`components') that are connected with each other by parts of their
boundaries only. This partitioning of $\mathcal{O}$ is to be
understood in terms of the Bianchi classification: the invariant
subset given by $(N_1 =0) \wedge (N_2=0)$ is the Bianchi type I
state space, the components $(N_1 = 0) \wedge (N_2 \neq 0)$ and
$(N_1 \neq 0) \wedge (N_2 = 0)$ describe Bianchi type II states, and
the components $N_1 N_2 < 0$ and $N_1 N_2 > 0$ are the Bianchi type
$\mathrm{VI}_0$ and $\mathrm{VII}_0$ subsets, respectively. In other
words, the equation system~\eqref{unstableeqs}, when restricted to
the respective component, is identical to the equations for the
Bianchi type I, II, $\mathrm{VI}_0$, $\mathrm{VII}_0$ vacuum models.
In the following we more closely examine these `Bianchi components'
of $\mathcal{O}$.

The Bianchi type I component is defined by $(N_1 =0)$ and $(N_2=0)$,
and hence $(R_1, R_2, R_3)$ are arbitrary, but subject to the Gauss
constraint. Every orbit on the Bianchi type I component describes a
Kasner solution, since
\begin{equation}\label{bIk}
\Udot^2=r^2 = 0 \quad \text{and}\quad 1-\Sigma^2 =
A^2=N_{\alpha\beta}\,N^{\alpha\beta}=0
\end{equation}
on this set; we therefore denote the Bianchi type I component
alternatively as the \textit{silent Kasner subset} $\mathcal{K}$.
However, since $(R_1,R_2,R_3) \neq (0,0,0)$ (except for the fixed
points on $\mathrm{K}^{\ocircle}$), every orbit on $\mathcal{K}$ is
a representation of a Kasner solution in a frame that rotates
w.r.t.\ a Fermi propagated frame; this is a consequence of the
Iwasawa frame choice; see Appendix~\ref{KasnernonFermi} for details.
We distinguish several invariant Bianchi type I subsets which are
defined by the vanishing of different combinations of the three
variables $(R_1, R_2, R_3)$. We will use $\mathcal{B}$ as a kernel
letter for an invariant subspace and a subscript that denotes the
\textit{non-zero} variables associated with the different subsets.
Accordingly, $\mathcal{B}_{R_\alpha}$ is an invariant subset, where
$R_\alpha \neq 0$, $R_\beta = 0$, $R_\gamma = 0$
($\alpha\neq\beta\neq\gamma\neq\alpha$); for $\mathcal{B}_{R_1 R_3}$
we have $R_1 \neq 0$, $R_2 =0$, $R_3\neq 0$; the set
$\mathcal{B}_{R_1 R_2 R_3}$ is characterized by $R_1 R_2 \neq 0$ or
$R_2 R_3 \neq 0$ and thus comprises the case of all three variables
being non-zero. (It is easy to see that the
equations~\eqref{unstableeqs} prevent the set $(R_1 =0)$ and the set
$(R_3 = 0)$ from being invariant subsets.) The Kasner circle
$\mathrm{K}^{\ocircle}$ is a special subset of the Bianchi type I
component; here, $(R_1,R_2,R_3) = (0,0,0)$. In the standard notation
$\overline{\mathcal{B}}_{\ast}$ for the closure of a set, we find
that, e.g., $\overline{\mathcal{B}}_{R_1} = \mathcal{B}_{R_1} \cup
\mathrm{K}^{\ocircle}$, or, $\mathcal{K} =
\overline{\mathcal{B}}_{R_1 R_2 R_3} = \mathcal{B}_{R_1 R_2 R_3}
\cup \mathcal{B}_{R_1 R_3} \cup \mathcal{B}_{R_1} \cup
\mathcal{B}_{R_2} \cup \mathcal{B}_{R_3} \cup
\mathrm{K}^{\ocircle}$.

The Bianchi type II component is defined by either $N_1$ or $N_2$
being zero. The Codazzi constraints~\eqref{codazzi} enforce
$N_\alpha R_\beta = 0$ $(\alpha \neq \beta)$, i.e., the Bianchi type
II component consists of four invariant subsets: ${\cal B}_{N_1}$,
${\cal B}_{N_2}$, ${\cal B}_{N_1R_1}$, and ${\cal B}_{N_2R_2}$.
Solutions on the former subsets represent Bianchi type II models in
a Fermi propagated frame, since $R_\alpha =0$ for all $\alpha$;
solutions on ${\cal B}_{N_1R_1}$, ${\cal B}_{N_2R_2}$ are
representations in a frame that rotates w.r.t.\ a Fermi frame.

Finally, we denote the Bianchi type $\mathrm{VI}_0$ and $\mathrm{VII}_0$
subsets by ${\cal B}_{N_1N_2-}$ and ${\cal B}_{N_1N_2+}$,
respectively, where the subscript denotes the sign of $N_1 N_2$. The
Codazzi constraints~\eqref{codazzi} enforce $R_\alpha = 0$ for all
$\alpha$; hence all solutions on these components are represented in
a Fermi propagated frame.

\begin{figure}[Ht]
\psfrag{a}[cc][cc]{${\cal B}_{R_1R_2R_3}$} \psfrag{b}[cc][cc]{${\cal
B}_{R_3R_1}$} \psfrag{c}[cc][cc]{${\cal B}_{N_1R_1}$}
\psfrag{d}[cc][cc]{${\cal B}_{N_2R_2}$} \psfrag{e}[cc][cc]{${\cal
B}_{N_1N_2+}$} \psfrag{f}[cc][cc]{${\cal B}_{N_1N_2-}$}
\psfrag{g}[cc][cc]{${\cal B}_{R_1}$} \psfrag{h}[cc][cc]{${\cal
B}_{R_2}$} \psfrag{i}[cc][cc]{${\cal B}_{R_3}$}
\psfrag{j}[cc][cc]{${\cal B}_{N_1}$} \psfrag{k}[cc][cc]{${\cal
B}_{N_2}$} \psfrag{l}[cc][cc]{$\mathrm{K}^{\ocircle}$} \centering{
\includegraphics[height=0.45\textwidth]{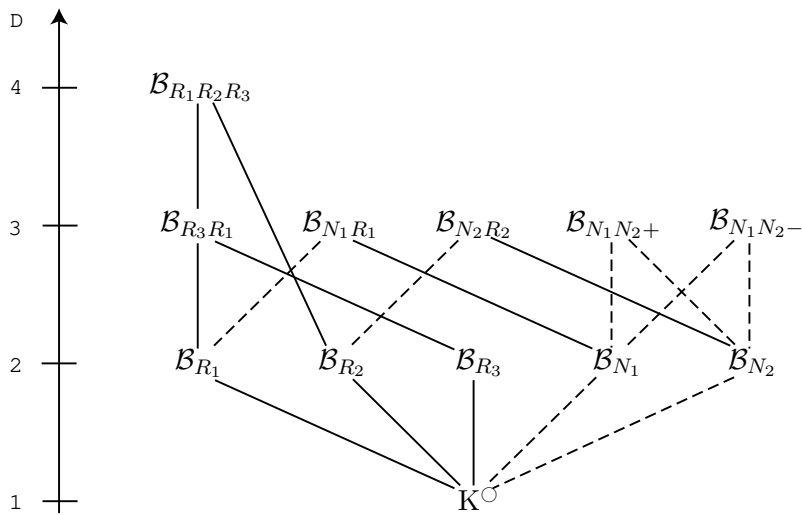}}
\caption{Subset contraction diagram of the oscillatory subset
$\mathcal{O}$. In our notation, the subscripts denote the non-zero
variables, and $\mathsf{D}$ hence describes the dimension of the
subsets $\mathcal{B}_{\ast}$. Solid lines correspond to setting
$R_\alpha$-variables to zero (frame rotation contractions); dashed
lines are associated with setting $N_1$ or $N_2$ to zero (Bianchi
type contractions). Note that, e.g., ${\cal B}_{N_1}$ includes the
two possible representations (one for each possible sign of $N_1$)
of the Fermi propagated subset; similar statements hold for the
other subsets that involve $N_1$ or $N_2$. Analogously, e.g., ${\cal
B}_{R_\alpha}$ is the union of two subsets that are characterized by
the sign of $R_\alpha$.}\label{contractionf}
\end{figure}

In Figure~\ref{contractionf} we present a diagram containing the
subsets $\mathcal{B}_{\ast}$ introduced above. We give the dimension
of the different subsets and show how they are related to each other
by setting variables to zero; the figure thus represents a
contraction diagram.

%As we will see in the next section, the Codazzi constraints~(\ref{codazzi}) prevent
%asymptotic oscillatory dynamics in the limit $\lim \tau \rightarrow +\infty$ from taking
%place, but we are not interested in the oscillatory system per se, but in its role for
%asymptotic dynamics in the full physical state space. As stated previously, we expect that
%$\bm{S}_{\mathrm{stable}}\rightarrow 0$ when $\tau\rightarrow 0$.
%However, at a given $\tau$ the stable variables are never exactly zero for
%a physical solution, no matter how far we are into the asymptotic regime,
%and hence the left hand sides of the Codazzi constraints~(\ref{codazzi})
%are not exactly zero either. This makes it possible for an orbit to oscillate between the
%different components, and this is what motives the name --- oscillatory subset.
%The building blocks for the oscillations are the orbits on the different components of
%$\mathcal{O}$, and hence an understanding of the dynamics of the components is essential.

We conclude this section by noting that models of Bianchi type VIII
and IX are not described by the equations on the SH silent boundary
we consider. This is due to the inherent incompatibility of the SH
frames of the Bianchi type VIII and IX models with the Iwasawa
frames we use throughout this paper; w.r.t.\ an Iwasawa frame
Bianchi type VIII and IX solutions appear as inhomogeneous solutions
and are thus associated with general solutions in the full interior
state space $\bm{X}$. In contrast, the SH frames of Bianchi type
VI$_{-1/9}$ models are Iwasawa compatible as we shall see in
Section~\ref{sym}. Since these models are expected to possess the
same attractor as generic timelines of generic asymptotically silent
models when expressed in an Iwasawa frame, it is rather these SH
models that are of interest in the present context.

%%%%%%%%%%%%%%%%%%%%%%%%%%%%%%%%%%%%%%%%%%%%%%%%%%%%%%%%%%%%%%%%%%%%%%%%%%%%%%%%%%%%%%%%%%
\section{Dynamics on the components of the oscillatory subset}
\label{dynosc}
%%%%%%%%%%%%%%%%%%%%%%%%%%%%%%%%%%%%%%%%%%%%%%%%%%%%%%%%%%%%%%%%%%%%%%%%%%%%%%%%%%%%%%%%%%

In the following we will describe the dynamics of the
system~\eqref{unstableeqs} on the oscillatory subset $\mathcal{O}$.
Since $\mathcal{O}$ decomposes into independent components, we will
analyze each component of $\mathcal{O}$ separately. We will see that
the solutions of~\eqref{unstableeqs} are heteroclinic orbits, which,
except for in the Bianchi type $\mathrm{VII}_0$ case, connect
different fixed points on the Kasner circle and thus provide
transitions between different Kasner states; accordingly we will
henceforth refer to such heteroclinic orbits as
\textit{transitions}.

%--------------------------------------------------------------------------------------
\subsection*{The silent Kasner subset (Bianchi type I subset)}
\label{silentBI}
%--------------------------------------------------------------------------------------

The silent Kasner subset, $\mathcal{K}$, is defined as the invariant
component of $\mathcal{O}$ given by $(N_1 = 0)$ and $(N_2 = 0)$ and
thus by the conditions~\eqref{bIk}. Consequently, solutions on
$\mathcal{K}$ represent Kasner states, though in general not in a
Fermi frame, since $(R_1, R_2, R_3) \neq (0,0,0)$. In this way, a
given Kasner state has several different representations on
$\mathcal{K}$.

We call the solutions of~\eqref{unstableeqs} on $\mathcal{K}$
\textit{frame transitions}. Depending on how many of the variables
$(R_1,R_2,R_3)$ are non-zero, we distinguish between single frame
transitions and multiple frame transitions.

\begin{figure}[Ht]
\centering
        \subfigure[$\mathcal{T}_{N_1}$]{
        \label{Kcurvtrans1}
        \includegraphics[height=0.28\textwidth]{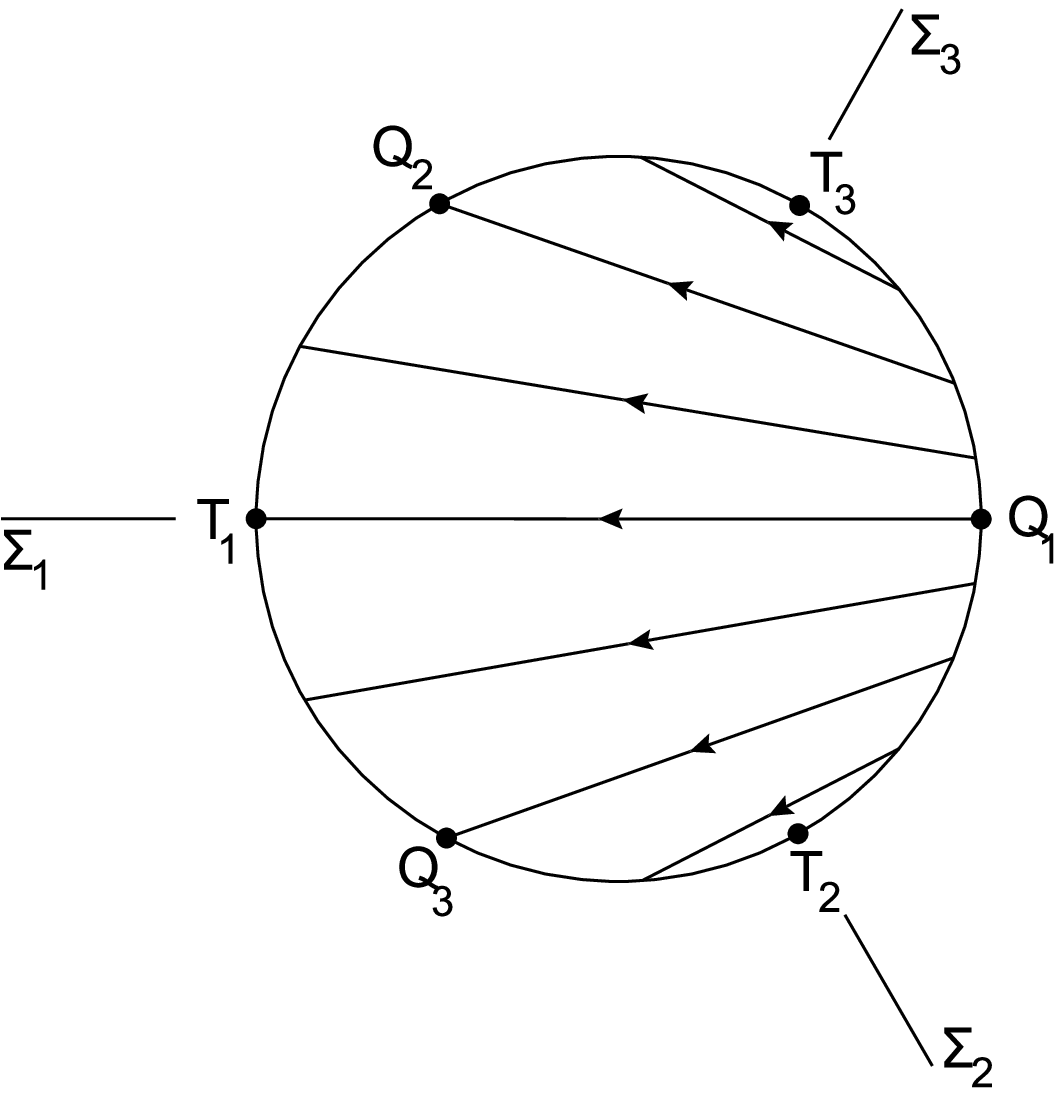}}\qquad
        \subfigure[$\mathcal{T}_{R_1}$]{
        \label{Kframetrans1}
        \includegraphics[height=0.28\textwidth]{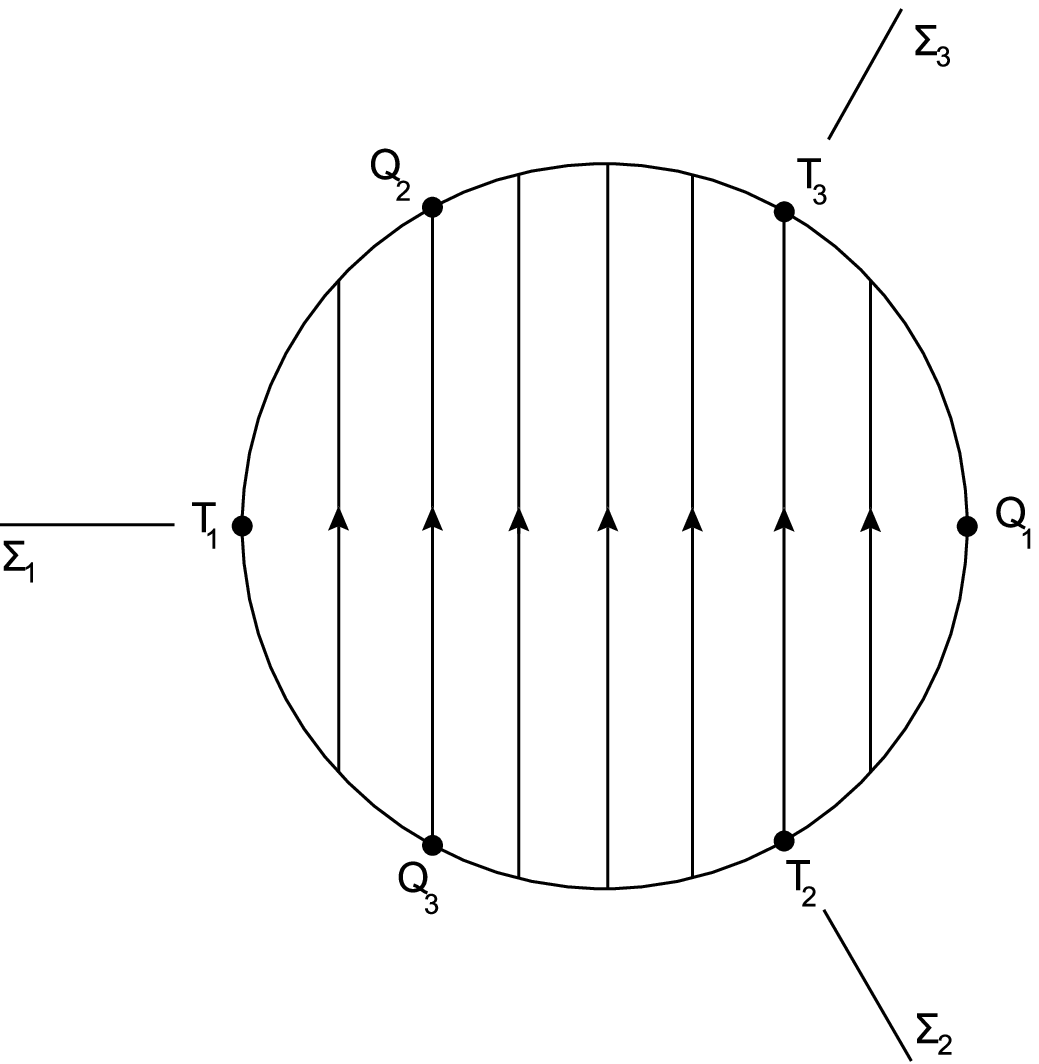}}\qquad
         \subfigure[$\mathcal{T}_{R_3}$]{
        \label{Kframetrans3}
        \includegraphics[height=0.28\textwidth]{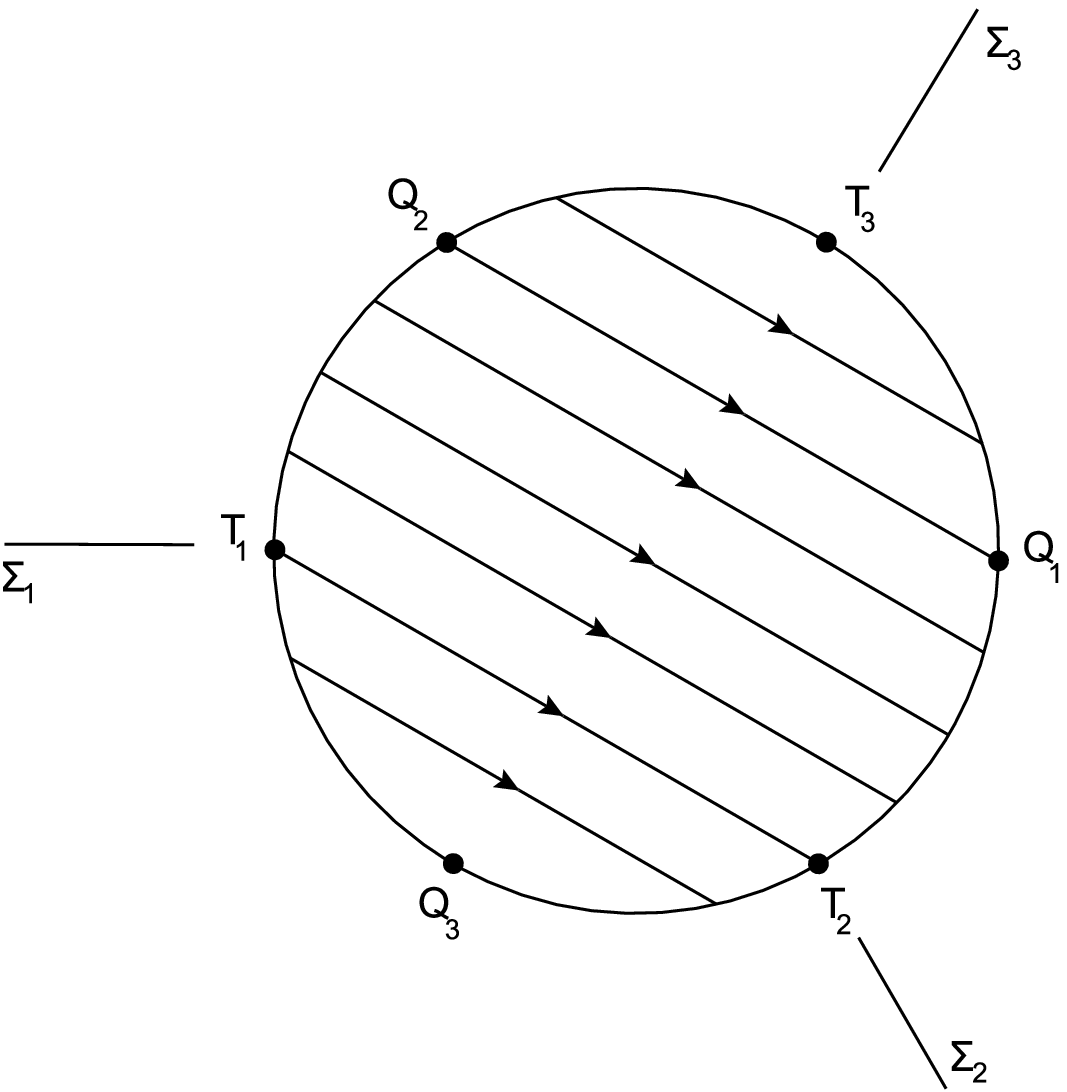}}\\
        \subfigure[$\mathcal{T}_{N_2}$]{
        \label{Kcurvtrans2}
        \includegraphics[height=0.28\textwidth]{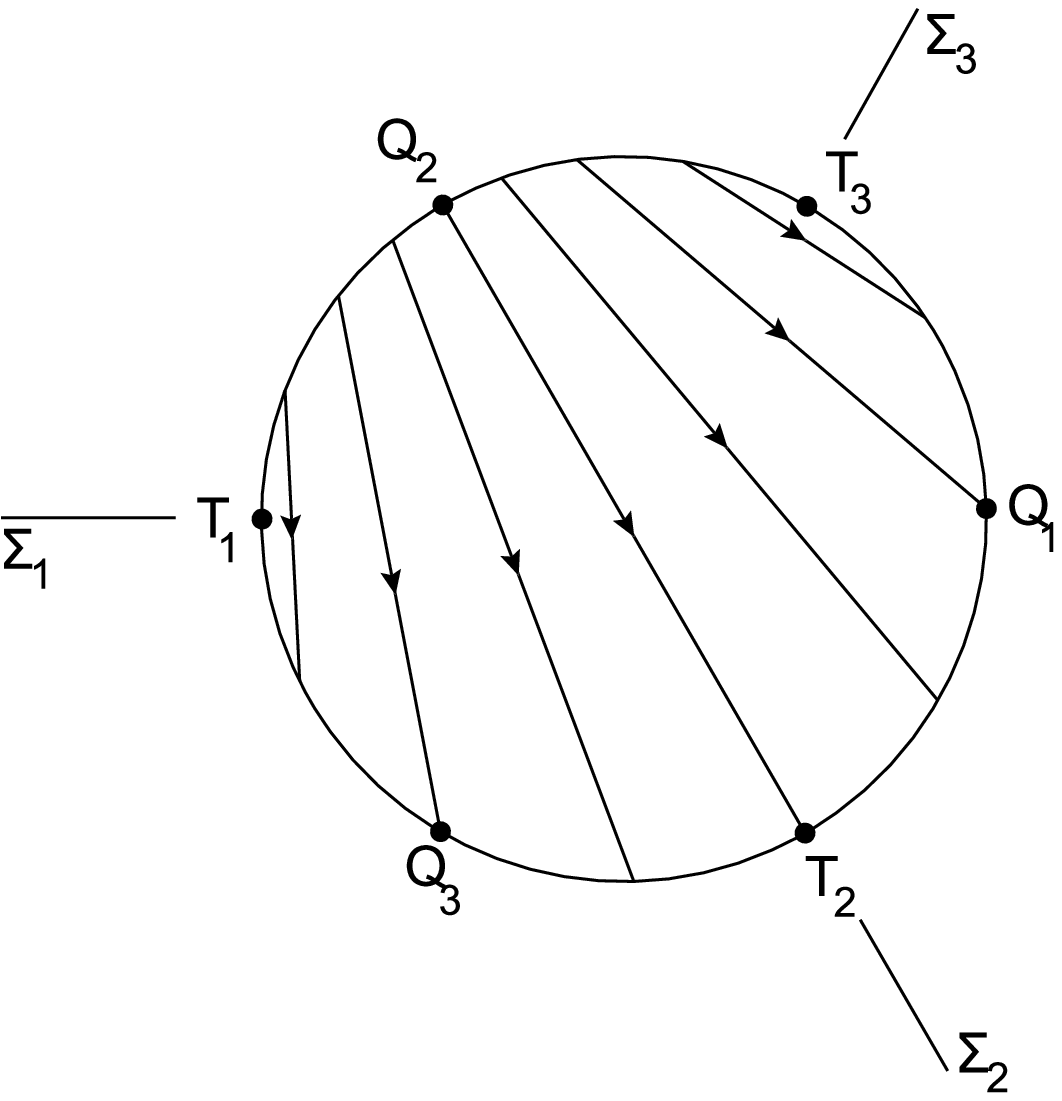}}\qquad
        \subfigure[$\mathcal{T}_{R_2}$]{
        \label{Kframetrans2}
        \includegraphics[height=0.28\textwidth]{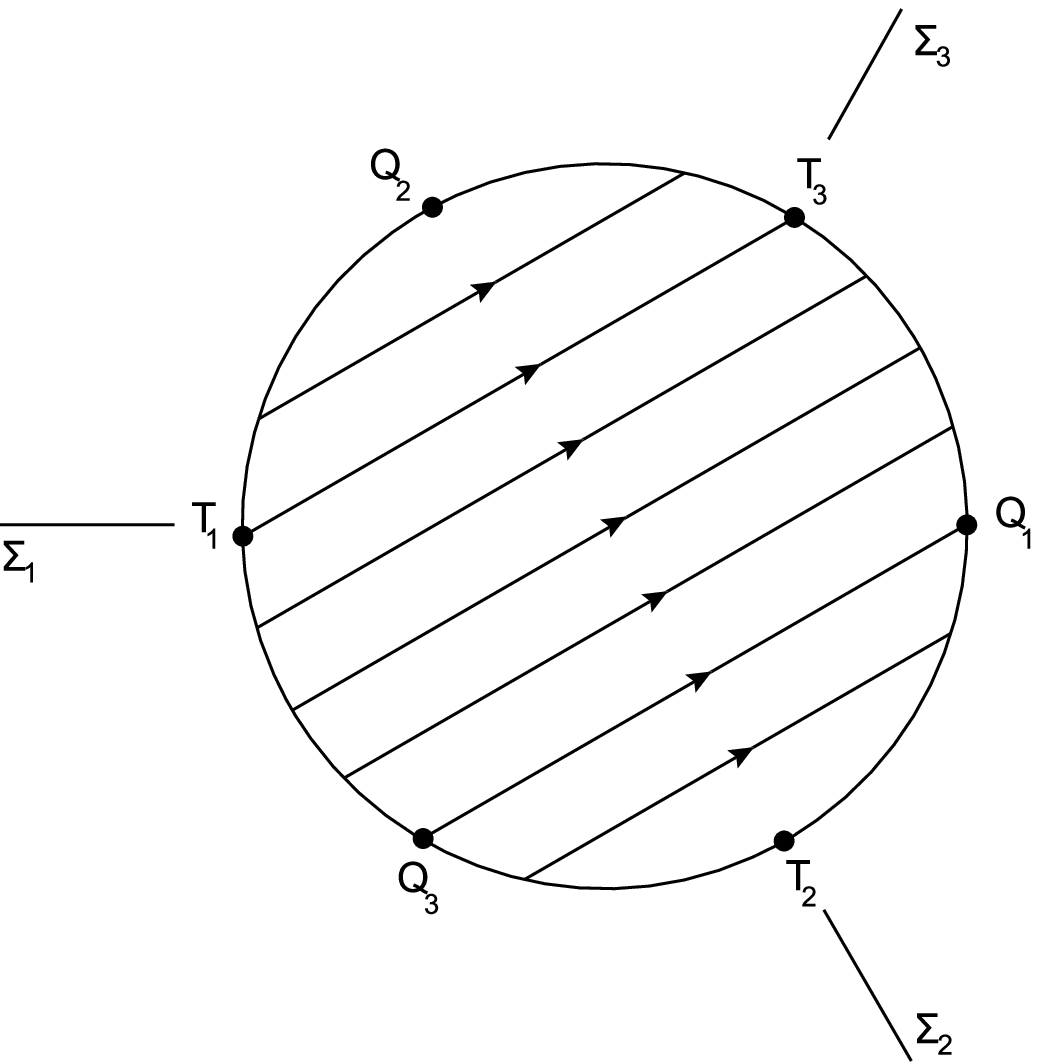}}
        \caption{Projections of single transitions for the Iwasawa frame onto diagonal $\Sigma_\alpha$-space.
        The first three transitions, $\mathcal{T}_{N_1}$, $\mathcal{T}_{R_1}$, $\mathcal{T}_{R_3}$,
        are the transitions relevant for the billiard attractor in Section~\ref{billiardattractor};
        the two other ones, $\mathcal{T}_{N_2}$, $\mathcal{T}_{R_2}$, are of relevance for the
        approach to the attractor, but asymptotically they are suppressed as shown in
        Section~\ref{asysha}.}
    \label{singletrans}
\end{figure}

There exist three families of \textit{single frame transitions},
denoted by $\cT_{R_1}$, $\cT_{R_2}$, $\cT_{R_3}$, which are
associated with the subsets ${\cal B}_{R_1}$, ${\cal B}_{R_2}$,
${\cal B}_{R_3}$. In the case of a $\cT_{R_\alpha}$ transition the
equations~\eqref{unstableeq3}--\eqref{unstableeq5} yield that
$\Sigma_\alpha=\mathrm{const}$ and that $\Sigma_\beta$,
$\Sigma_\gamma$ are monotonically increasing or decreasing,
see~Figure~\ref{singletrans}.
%\footnote{Both in UEWE and
%  here we use a frame choice such that $\Sigma_{23}=\epsilon_1\,R_1,
%  \Sigma_{31}=\epsilon_2\,R_2, \Sigma_{12}=\epsilon_3\,R_3$, with
%  $(\epsilon_1,\epsilon_2,\epsilon_3)= (1,1,1)$ in UEWE and
%  $(\epsilon_1,\epsilon_2,\epsilon_3)= (-1,1,-1)$ here; the only
%  effect of the difference in sign in the present context is to
%  reverse the direction of the flow of the orbits, but this eventually
%  affects the description of the dynamics of a generic timeline
%  considerably.}
The initial and final states of the transitions are equilibrium
points on the Kasner circle $\mathrm{K}^{\ocircle}$ that are related
by $(\Sigma_\alpha)_+ = (\Sigma_\alpha)_-$, $(\Sigma_\beta)_+ =
(\Sigma_\gamma)_-$, and $(\Sigma_\gamma)_+ = (\Sigma_\beta)_-$,
where the subscripts ${}_+$ and ${}_-$ denote the final and initial
states of the transition, i.e., $(\Sigma_\delta)_\pm =
\lim_{\tau\rightarrow \pm\infty} \Sigma_\delta(\tau)$. Consequently,
a $\cT_{R_\alpha}$ orbit corresponds to a Kasner solution viewed in
a frame that rotates around the $\alpha$-axes; in their final state,
the $\beta$- and $\gamma$-axes are rotated by $\pi/2$ w.r.t.\ their
initial position, which corresponds to an interchange of the
$\beta$-/$\gamma$-directions. Since the Kasner parameter $u$ is a
frame invariant, it is necessarily invariant during a frame
transition; in particular,
\begin{equation}\label{uconstinBI}
u = u_+ = u_- = \mathrm{const}\:;
\end{equation}
see Appendix~\ref{KasnernonFermi} for details. Expressed in $u$ we
obtain for $\cT_{R_\alpha}$,
\begin{equation}\label{DeltaSigma2s}
|\Delta \Sigma_\beta| = |(\Sigma_\beta)_+ - (\Sigma_\beta)_-| \, =
\, \left\{\begin{array}{l}
3 (1+ 2 u)(1 + u + u^2)^{-1} \\[0.5ex]
3 (2 u+u^2)(1 + u + u^2)^{-1} \\[0.5ex]
3 (-1+u^2)(1 + u + u^2)^{-1}
\end{array}\right. \, ,
\end{equation}
depending on whether the transition starts from sector
$(\beta,\gamma, \alpha)$, $(\beta,\alpha,\gamma)$, or $(\alpha, \beta, \gamma)$.
A simple consequence is $|\Delta \Sigma_\beta| \leq 2\sqrt{3}$.

For \textit{multiple frame transitions} more than one of the
variables $(R_1,R_2,R_3)$ is non-zero. There exists one class of
double frame transitions, $\cT_{R_3R_1}$, which is associated with
${\cal B}_{R_3R_1}$, and one class of triple frame transitions,
$\cT_{R_1R_2R_3}$, connected with ${\cal B}_{R_1 R_2 R_3}$; for the
latter, $R_\alpha \not\equiv 0$ $\forall \alpha$. The initial state
of a multiple frame transition is an equilibrium point on sector
$(321)$ of $\mathrm{K}^{\ocircle}$, while the final state lies on
$(123)$; for details see Appendix~\ref{multiple}. The Kasner
parameter $u$ is invariant under multiple frame transitions, since
it is a frame invariant; furthermore, we have $|\Delta\Sigma_2| = 0$
and $(\Sigma_1)_\pm = (\Sigma_3)_\mp$ (as for $\cT_{R_2}$
transitions). For examples of multiple frame transitions, see
Figure~\ref{TR1R3f}.

%--------------------------------------------------------------------------------------
\subsection*{The silent Bianchi type II subset}
\label{silentBII}
%--------------------------------------------------------------------------------------

Solutions of the oscillatory system~\eqref{unstableeqs} that satisfy
$N_\alpha \not \equiv 0$ for some $\alpha$ are located on the
Bianchi type II, VI$_0$, or VII$_0$ component. These solutions are
characterized by non-vanishing three-curvature, i.e.,
${}^3\mathcal{R} = - N_{\alpha\beta} N^{\alpha \beta} + \frac{1}{2}
(N_\alpha^{\weg\alpha})^2$, see~\eqref{defofmuch}. We denote
solutions of Bianchi type II and VI$_0$ as \textit{curvature
transitions}; for solutions of Bianchi type VII$_0$ we do not use
this terminology for reasons that will become clear below.
%since they do not represent transitions
%between points on $\mathrm{K}^{\ocircle}$; see below.
Curvature transitions (and Bianchi $\mathrm{VII}_0$ solutions)
differ from frame transitions in one important respect:
the Kasner parameter $u$ changes under the transition.
We begin by considering those curvature transitions that
are associated with the Bianchi type II subset.

The silent Bianchi type II subset is the subset of $\mathcal{O}$
determined by $(N_1 \neq 0) \wedge (N_2 =0)$ and $(N_1 = 0) \wedge
(N_2 \neq 0)$; it is the union of the invariant subsets ${\cal
B}_{N_1}$, ${\cal B}_{N_1R_1}$, and ${\cal B}_{N_2}$, ${\cal
B}_{N_2R_2}$.

The simplest solutions on the silent Bianchi type II subset are the
orbits on ${\cal B}_{N_1}$ or ${\cal B}_{N_2}$, denoted by
$\cT_{N_1}$ and $\cT_{N_2}$, which we call \textit{single curvature
transitions}; the absence of a third family, $\cT_{N_3}$ is due to
the peculiarities of the Iwasawa frame. In the following we consider
$\cT_{N_1}$ transitions; the treatment of $\cT_{N_2}$ transitions is
analogous. For $\cT_{N_1}$, the Gauss constraint~\eqref{gausscon}
yields $N_1^2/12 = 1 -\Sigma^2$. The resulting system for
$(\Sigma_1,\Sigma_2,\Sigma_3)$,
Eqs.~\eqref{unstableeq3}--\eqref{unstableeq5}, possesses solutions
whose initial states are equilibrium points on sectors $(123)$ or
$(132)$ of $\mathrm{K}^{\ocircle}$, see Figure~\ref{singletrans}.
Without loss of generality we consider $\cT_{N_1}$ transitions from
sector $(123)$ (as $\cT_{N_1}$ orbits that originate from sector
$(132)$ are easily obtained by a permutation). The orbits can be
parametrized by $u = u_-$, the Kasner parameter characterizing the
initial state; in terms of an auxiliary function $\zeta(\tau)$ we
obtain
\begin{subequations}\label{Zeqs}
\begin{equation}\label{Zeq}
\Sigma_{1} = -\,4 + (1 + u^2)\, \zeta \: , \qquad \Sigma_{2} = 2 -
u^2 \,\zeta \: , \qquad \Sigma_{3} = 2 - \zeta \:,
\end{equation}
where the evolution equation for $\zeta$ is given by
\begin{equation}\label{Zeq2}
\partial_{\tau} \zeta = 2(1-\Sigma^2)\,\zeta \qquad\text{with}\qquad
(1-\Sigma^{2}) = \frac{3}{\zeta_+ \zeta_-}\left(\zeta_+ -
\zeta\right)\left(\zeta-\zeta_-\right)\:,
\end{equation}
\end{subequations}
where $\zeta_\pm = 3/(1 \mp u + u^2)$; by definition, $ 0 < \zeta_-
< 1$, and $0 < \zeta_+ < 3$. The function $\zeta(\tau)$ interpolates
monotonically between $\zeta_-$ (as $\tau\rightarrow -\infty$) and
$\zeta_+$ (as $\tau\rightarrow \infty$). Letting $u=1$
in~\eqref{Zeqs} yields the orbit that connects $\mathrm{Q}_1$ with
$\mathrm{T}_1$, see Figure~\ref{Kcurvtrans1}; $u=\infty$ corresponds
to the point $\mathrm{T}_3$.

Evaluation of~\eqref{Zeq} at $\tau = \mp \infty$ (corresponding to
$\zeta_\mp$) yields the initial/final states of the $\cT_{N_1}$
transition, and thus a map that connects the initial and final
Kasner states, which are described by $u=u_-$ and $u_+$,
respectively:
\begin{equation}\label{BKLMap}
u_+  \:= \:
\left\{\begin{array}{ll}
u_- - 1 & \qquad \text{if}\quad u_- \geq 2 \\[1ex]
(u_- - 1)^{-1} &\qquad  \text{if} \quad 1 \leq u_- < 2
\end{array}\right.\:.
\end{equation}
This formula was first obtained by BKL via different methods; we
will refer to~\eqref{BKLMap} as the \textit{Kasner map}. Note that
this result holds for both $\cT_{N_1}$ and $\cT_{N_2}$ curvature
transitions.
%
%\begin{equation}
%g = \left\{\begin{array}{lll}
%    \frac{1+u_-+u_-^2}{1 + u_+ + u_+^2} &\quad
% {\rm if} &\quad
%u_- \geq 2
%    \\
%   \frac{1+u_- + u_-^2}{(u_- - 1)^2\,(1 + u_+ + u_+^2)}
%&\quad  {\rm if} &\quad 1 < u_- < 2
%\end{array}\right. \, .
%\end{equation}
%%

For later purposes we define the growth factor $g$ according to%
\footnote{The quantity $g$ should not be confused with the
determinant of the spatial metric---it should be clear from the
context which one is meant.}
\begin{equation}\label{geq}
g:=\frac{\zeta_+}{\zeta_-} = \frac{1+u+u^2}{1-u+u^2} = \frac{1 + u_-+u_-^2}{1-u_-+u_-^2} \:.
\end{equation}
Since $u = u_-\in [1,\infty)$
for $\cT_{N_1}$ orbits, we have $1<g<3$.

There exist two classes of \textit{mixed frame/curvature
transitions}: orbits on ${\cal B}_{N_1R_1}$ and orbits on ${\cal
B}_{N_2R_2}$, which we denote by $\cT_{N_1 R_1}$ and $\cT_{N_2
R_2}$, respectively. The initial state of a $\cT_{N_1 R_1}$ orbit is
an equilibrium point on sector $(132)$, and the final state lies on
$(213)$ or $(231)$; for $\cT_{N_2 R_2}$ transitions, the numbers $1$
and $2$ are interchanged; see Appendix~\ref{multiple}. In analogy to
single curvature transitions, the initial and the final Kasner state
are related by the Kasner map~\eqref{BKLMap}. This fact relies on
the frame invariance of the parameter $u$: $\cT_{N_1}$ and $\cT_{N_1
R_1}$ transitions are representations of the same Bianchi type II
solutions---they are merely expressed in different frames (Fermi
frame vs.\ rotating frame); the transformation law~\eqref{BKLMap} of
the frame invariant $u$ is thus unaffected. For more details on
mixed frame/curvature transitions, see Appendix~\ref{multiple}; for
examples, see Figure~\ref{TN1R1f}. Note that henceforth, for
brevity, we will refer to mixed frame/curvature transitions as mixed
curvature transitions.

\begin{figure}[Ht]
\psfrag{a}[cc][cc]{$\Sigma_1$} \psfrag{b}[cc][cc]{$\Sigma_3$}
\psfrag{c}[cc][cc]{$\Sigma_2$} \psfrag{d}[cc][cc]{$T_1$}
\psfrag{e}[cc][cc]{$Q_2$} \psfrag{f}[cc][cc]{$T_3$}
\psfrag{g}[cc][cc]{$Q_1$} \psfrag{h}[cc][cc]{$T_2$}
\psfrag{i}[cc][cc]{$Q_3$} \psfrag{m}[cc][cc]{$A_{+}$}
\psfrag{l}[lc][cc]{$A_{-}$} \psfrag{j}[cc][cc]{$B_{-}$}
\psfrag{k}[cc][cc]{$B_{+}$} \centering
     \subfigure[Double frame transitions ${\cal T}_{R_3R_1}$.]{
     \label{TR1R3f}
     \includegraphics[width=0.42\textwidth]{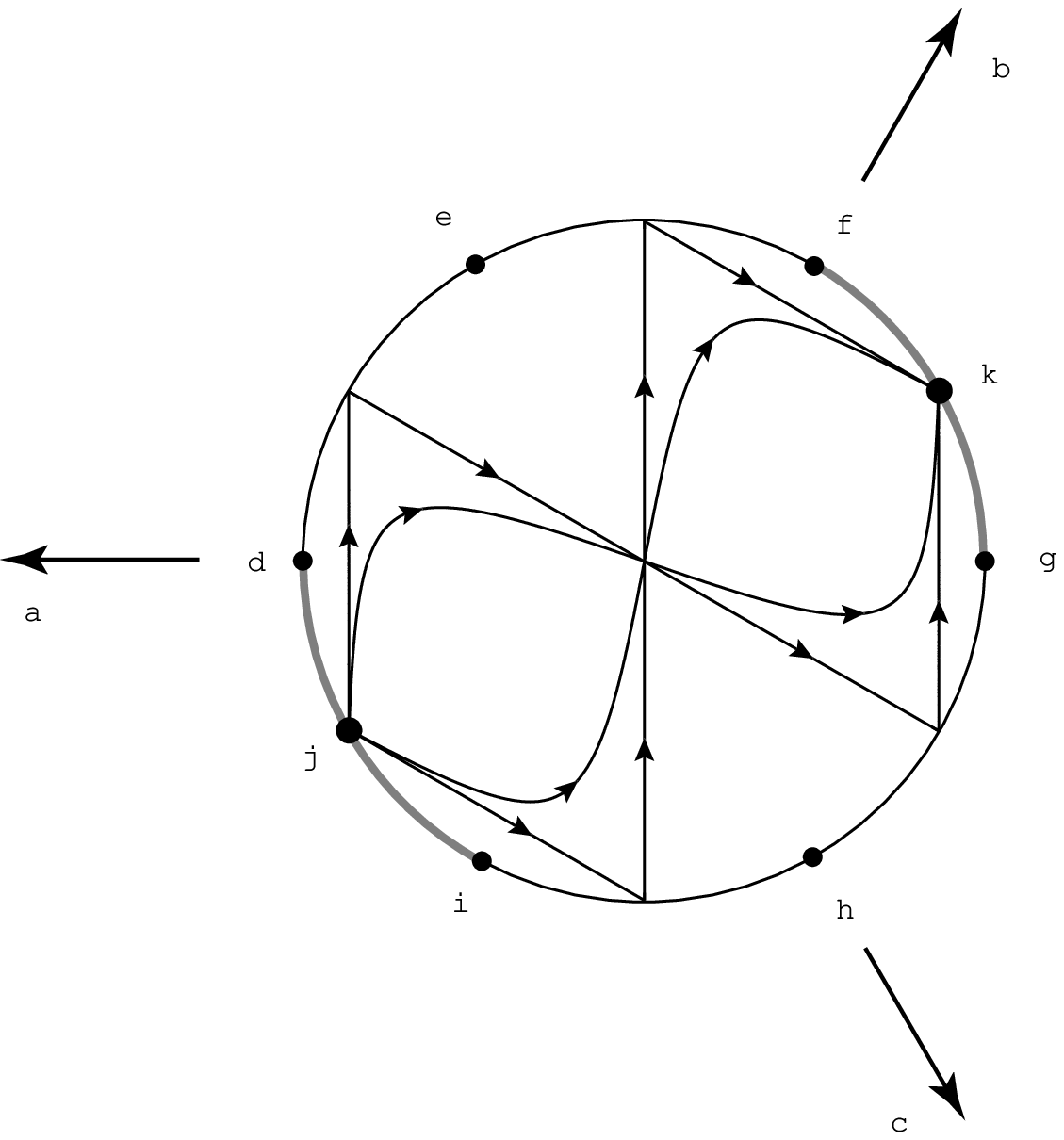}}
     \qquad
     \subfigure[Mixed frame/curvature transitions ${\cal T}_{N_1R_1}$.]{
     \label{TN1R1f}
     \includegraphics[width=0.42\textwidth]{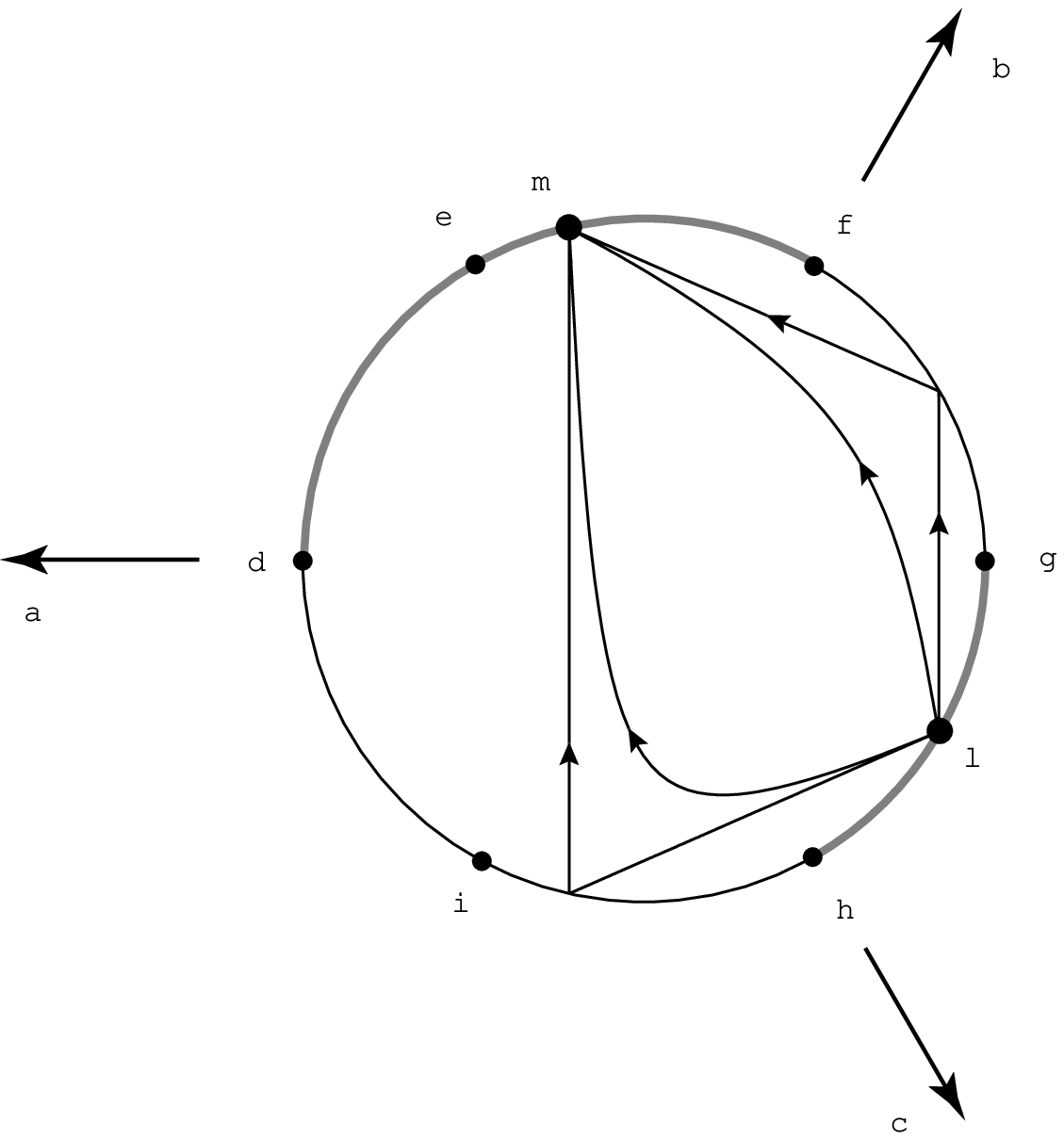}}
\caption{Figure (a) shows a projection of double frame transitions
onto $\Sigma_\alpha$-space; note that single transitions form the
boundary of this subset and that the final state of a double frame
transition can be obtained in terms of a series of single
transitions that are associated with the same initial Kasner state. Figure (b) shows
mixed frame/curvature transitions for which analogous remarks hold.}
     \label{doubletransition}
\end{figure}

%--------------------------------------------------------------------------------------
\subsection*{The silent Bianchi type VI$_0$ and VII$_0$ subsets}
\label{silentBVIVII}
%--------------------------------------------------------------------------------------

The silent Bianchi type $\mathrm{VI}_0$ subset is given by $N_1 N_2
<0$ and $R_{\alpha} =0$ for all $\alpha$; it is denoted by ${\cal
B}_{N_1N_2-}$. We call the orbits on ${\cal B}_{N_1N_2-}$
\textit{double curvature transitions}: $\cT_{N_1 N_2-}$, or simply
$\cT_{N_1 N_2}$. In Appendix~\ref{multiple} we show that the
$\alpha$-limit of each $\cT_{N_1 N_2}$ transition is the Taub point
$\mathrm{T}_3$; the $\omega$-limit is a fixed point on sector
$(312)$ or $(321)$ of $\mathrm{K}^\ocircle$. Accordingly, in terms
of the Kasner parameter $u$, we find
\begin{equation}\label{doublecurvu}
u_- = \infty \:\, \xrightarrow{\cT_{N_1N_2}\,(\text{Bianchi } \mathrm{VI}_0)} \:\,
u_+ \in [1,\infty) \:;
\end{equation}
note that for each $u\in [1,\infty)$ there exist double curvature
transitions that map the Taub state $u_- =\infty$ to exactly that
final Kasner state, i.e.,  $u = u_+$. It is evident that the
behavior of double curvature transitions is fundamentally different
from the behavior of Bianchi type II transitions (single and mixed
curvature transitions); compare~\eqref{doublecurvu} with the Kasner
map~\eqref{BKLMap}. We observe that there do not exist double
curvature transitions with arbitrarily small amplitudes: there
exists $\epsilon > 0$ such that $\max_\tau |N_1(\tau)| \geq
\epsilon$ (and $\max_\tau |N_2(\tau)| \geq \epsilon$) uniformly for
all double curvature transitions; for a proof see
Appendix~\ref{multiple}.

Finally, let us now consider the silent Bianchi type
$\mathrm{VII}_0$ subset ${\cal B}_{N_1N_2+}$, which is defined by
$N_1 N_2 >0$ and $R_{\alpha} =0$ for all $\alpha$. The state space
contains a line of fixed points, $\mathcal{L}_3^\pm$, given by the
conditions $(\Sigma_1,\Sigma_2,\Sigma_3) = (1,1,-2)$ and $N_1 =
N_2$; see Appendix~\ref{multiple} for details. The $\alpha$-limit of
each Bianchi $\mathrm{VII}_0$ orbit orbit is a point on
$\mathcal{L}_3^\pm$, the $\omega$-limit is a fixed point on sector
$(312)$ or $(321)$ of $\mathrm{K}^\ocircle$. Hence, since Bianchi
type $\mathrm{VII}_0$ orbits do not connect two fixed points on
$\mathrm{K}^\ocircle$, and we therefore refrain from calling them
transitions. For this reason we will not introduce the nomenclature
$\cT_{N_1 N_2+}$.

%%%%%%%%%%%%%%%%%%%%%%%%%%%%%%%%%%%%%%%%%%%%%%%%%%%%%%%%%%%%%%%
\section{Sequences, eras, and phases}
\label{transition}
%%%%%%%%%%%%%%%%%%%%%%%%%%%%%%%%%%%%%%%%%%%%%%%%%%%%%%%%%%%%%%%

In the previous section we have described the dynamics on the
oscillatory subset by analyzing the solutions (transitions) on the
different components of $\mathcal{O}$. These transitions form pieces
of a `heteroclinic orbit puzzle' that governs asymptotically local
dynamics; in the following we describe and characterize its key
features.

%--------------------------------------------------------------------------------------
\subsection*{Sequences of transitions and Kasner sequences}
\label{seqtransKas}
%--------------------------------------------------------------------------------------

We define a \textit{sequence of transitions} as a heteroclinic
sequence on the oscillatory subset $\mathcal{O}$, i.e., as an
infinite concatenation of transition orbits (heteroclinic orbits) on
$\mathcal{O}$. (Note that, by definition, Bianchi type VII$_0$
orbits are excluded from these sequences; since the initial states
of type VII$_0$ orbits do not lie on $\mathrm{K}^{\ocircle}$, but on
$\mathcal{L}_3^\pm$, they cannot be concatenated with transition
orbits.) Accordingly, a sequence of transitions, which we denote by
$\cS_\cT$, is described by
$\cS_\cT = (\cT_i)_{i\in\mathbb{N}} = (\cT_0, \cT_1,\cT_2, \ldots)$,%
\footnote{In order
  to agree with the conventions of~\cite{bkl70} we define
  the first transition of $\cS_\cT$ to carry the index number $0$.}
where each transition $\cT_i$ is either a frame transition
$\cT_{R_1}$, $\cT_{R_2}$, $\cT_{R_3}$, $\cT_{R_3 R_1}$,
$\cT_{R_1 R_2 R_3}$, or a single or mixed curvature transition
$\cT_{N_1}$, $\cT_{N_2}$, $\cT_{N_1 R_1}$, $\cT_{N_2 R_2}$, or a
double curvature transition $\cT_{N_1 N_2}$.
It is understood that the $\omega$-limit of each transition $\cT_i$
(which is a fixed point on $\mathrm{K}^\ocircle$) coincides
with the $\alpha$-limit of the transition $\cT_{i+1}$.

The initial state of each individual transition $\cT_i$ is a fixed
point on $\mathrm{K}^\ocircle$ (which coincides with the final state
of $\cT_{i-1}$), and is thus associated with a particular value of
the Kasner parameter $u$. Therefore, every sequence of transitions
$\cS_\cT$ generates a series of Kasner parameters in a natural way.
Since frame transitions do not induce any change in $u$,
see~\eqref{uconstinBI}, the series of Kasner states is not
associated with the sequence $\cS_\cT$ directly, but rather with the
sequence of curvature transitions of $\cS_\cT$, which may be
regarded as its supporting `skeleton'. This makes it convenient to
introduce a second running index that consecutively numbers
curvature transitions: $l=0,1,2,\ldots\,$. The index $l$ can be
viewed as a function $l:\mathbb{N}\rightarrow \mathbb{N}$ such that
$l(i) = l_i$ denotes the number of curvature transitions among the
transitions $\{\cT_0,\cT_1,\ldots,\cT_{i-1}\}$.

Let $u_l$ denote the initial Kasner state of the
$l$\raisebox{0.7ex}{\small th} curvature transition; then the
transition maps $u_l$ to $u_{l+1}$,
\begin{equation*}
u_l \:\,\xrightarrow{\;\text{$l$\raisebox{0.5ex}{th} curvature
transition}\;}\:\, u_{l+1}\:.
\end{equation*}
For single and mixed curvature transitions the Kasner
map~\eqref{BKLMap} applies:
\begin{equation}
\label{useq} u_{l+1}  \:= \:
\begin{cases}
u_l - 1 & \quad\text{if}\quad u_l \geq 2 \\
(u_l - 1)^{-1} &  \quad\text{if} \quad 1 \leq u_l < 2\:.
\end{cases}
\end{equation}
A double curvature transition, on the other hand,
maps a Kasner state characterized by $u=\infty$ to
a Kasner state with $u\in (1,\infty)$.

Two cases occur: (i) The initial state $u_0$ is an irrational
number, which is the generic case since $\mathbb{Q}$ is a set of
measure zero in $\mathbb{R}$. Then the sequence
$(u_l)_{l\in\mathbb{N}}$ is given through the
recursion~\eqref{useq}; in particular, for all $l$, $u_l \in
(1,\infty)$. A sequence $(u_l)_{l\in\mathbb{N}}$ of this type we
denote as a \textit{Kasner sequence}. Note that a Kasner sequence is
associated with a sequence of transitions $\cS_\cT$ that does not
contain any double curvature transitions $\cT_{N_1 N_2}$. (ii) The
initial state $u_0$ is a rational number. Then there exists $k <
\infty$ such that $u_{k-1} = 1$, whereby $u_{k} = \infty$. At this
point the recursion defined by the Kasner map~\eqref{useq} is
interrupted. A double curvature transition $\cT_{N_1 N_2}$ must
follow, which yields $u_{k+1} \in [1,\infty)$. Generically the
Kasner parameter $u_{k+1}$ is an irrational number, hence the series
$(u_l)_{l > k}$ is a Kasner sequence given by the
recursion~\eqref{useq}; by a shift of origin (i.e., by redefining
$u_0 := u_{k+1}$) we obtain again a standard Kasner sequence. If the
Kasner parameter $u_{k+1}$ is a rational number, which is the
non-generic case, we are back to square one, at the beginning of the
loop.

%These arguments entail that that sequences that contain double
%curvature transitions emerge from repeated fine-tuning.
From this discussion we are led to classify possible sequences of
transitions. Sequences with a finite number of double curvature
transitions are innocuous: they are standard Kasner sequences by a
simple shift of origin and thus do not require special treatment.
Sequences that contain infinitely many double curvature transitions
are qualitatively different, but they are highly non-generic since
they emerge from repeated fine-tuning: for such sequences the final
state of each individual $\cT_{N_1 N_2}$ transition must be
represented by a rational value of the Kasner parameter. This
indicates that these sequences are not to be treated on an equal
footing with generic sequences, i.e., sequences that are free from
double curvature transitions. Henceforth, unless otherwise stated, a
sequence of transitions $\cS_\cT$ always denotes a sequence where
double curvature transitions are excluded, i.e., it is an infinite
concatenation of frame transitions and single and mixed curvature
transitions. It is sequences of this type that are intrinsically
tied to Kasner sequences $(u_l)_{l\in\mathbb{N}}$ which are defined
through the recursion~\eqref{useq}.

%--------------------------------------------------------------------------------------
\subsection*{Eras, large curvature phases, and small curvature phases}
\label{eralcp}
%--------------------------------------------------------------------------------------

Here we investigate qualitative properties of Kasner sequences
$(u_l)_{l\in\mathbb{N}}$; in particular, we introduce a partition of
the sequence into phases of small curvature and phases of large
curvature; however, we begin by defining the concepts of an epoch
and an era.

Consider a Kasner sequence $(u_l)_{l\in\mathbb{N}}$ as given by the
recursion~\eqref{useq}. A Kasner \textit{epoch} is simply defined as
an individual Kasner state $u_l$ of the sequence. The sequence
$(u_l)_{l\in\mathbb{N}}$ possesses a natural partition into eras: an
\textit{era} is a set $[\lin,\lout]\ni l$ such that $u_l$ is
monotonically decreasing from a maximal value $\uin =
u_{\lin}=(u_{\lin-1}-1)^{-1}$ to a minimal value $1 < \uout =
u_{\lout} < 2$. The length of an era is given by the number of
Kasner epochs it contains: $L = \lout - \lin+1$. Since $u_{l+1} =
u_l -1$ for all $l$ of an era $[\lin,\lout]$, we have $L
=\mathrm{int}(\uin)$, where $\mathrm{int}(x)$ is a function that
gives the integer part of $x$, see~\cite{khaetal85}. Note that an
era can be, and many will be, of length one.

We now introduce a somewhat more flexible concept that comprises the
concept of an era as a special case. Let $\mathbb{R} \ni \eta_u \geq
1$; we define a phase of small curvature---a \textit{small curvature
phase}---as a set $[\lin,\lout] \ni l$ such that $u_l$ monotonically
decreases from a maximal value $\uin = u_{\lin}$ to a minimal value
$\uout = u_{\lout}$ according to
\begin{equation}\label{scpdef}
\uin = u_{\lin}= (u_{\lin-1}-1)^{-1}> \eta_u \geq 1
\quad\xrightarrow{l \in [\lin,\lout]}\quad
\eta_u+1 > \uout = u_{\lout} > \eta_u\:.
\end{equation}

When $\eta_u$ is chosen to be equal to $1$ the definition of a small
curvature phase reduces to the definition of an era; however, the
usefulness of the concept of a small curvature phase stems from the
possibility of choosing $\eta_u \gg 1$, and this is the choice we
will typically make. In this case a small curvature phase contains
only large values of $u$. Accordingly, the Kretschmann scalar
$\mathcal{W}_1$ associated with the metric $\mathbf{G}$ is small,
since $\mathcal{W}_1 = 27 u^2(1+u)^2/(1 + u +u^2)^3 = 27 u^{-1} ( 1
+ u^{-1} + O(u^{-2}))$, cf.~Appendix~\ref{KasnernonFermi}. Since
this quantity determines the magnitude of the curvature during
a Bianchi type II transition, see Section~\ref{gauge} and
Figure~\ref{W1W2weyl}, the terminology `phase of small curvature'
suggests itself. Finally, note that the length of a small curvature
phase is defined in analogy to the length of an era as the number of
Kasner epochs (i.e., Kasner parameters) it comprises:
\begin{equation}\label{capphaselength}
L = \lout-\lin+1=\mathrm{int}(\uin - \eta_u + 1)\approx \uin -\eta_u  \approx \uin\:,
\end{equation}
where the last approximation holds only in the special case $\uin \gg \eta_u$.

The complement of a phase of small curvature is a phase of large curvature.
A set $[\li,\lf] \ni l$ is called a \textit{large curvature phase}
if
\begin{equation*}
u_{\li -1}> \eta_u \:, \qquad
u_l \leq \eta_u \quad {\rm for\; all}\quad l\in[\li,\lf]\:, \qquad
u_{\lf+1} > \eta_u\:.
\end{equation*}
Accordingly, the length $L$ of a large curvature phase is at least
$\mathrm{int}(u_{\li}) \geq \mathrm{int}(\eta_u-1)$, since $u_{\li}
\in (\eta_u-1,\eta_u)$. In a large curvature phase, the values of
the Kasner parameter are comparatively small, i.e., small in
comparison with the $u$-values of a small curvature
phase~\eqref{scpdef}; it follows that curvature is comparatively large
in the case of a large curvature phase, see Section~\ref{gauge},
which motivates the nomenclature `large curvature phase'. Note that
the concept of a large curvature phase does not exist when $\eta_u$
is chosen to be equal to $1$; recall that in this case the concepts
era/small curvature phase are identical.

Small curvature phases and large curvature phases occur
alternately. A large curvature phase continues and ends the era
that began simultaneously with the small curvature phase preceding
the large curvature phase. When $1< u_l <2$, the Kasner parameter
number $(l+1)$ begins a new era: if $u_{l+1} \leq \eta_u$, then that
entire era belongs to the continuing large curvature phase; if, and
only if, $u_{l+1} >\eta_u$, then a large curvature phase has ended
and a new small curvature phase begins. Consequently, a large curvature
phase can contain an arbitrary number of eras; In the following
example, where the choice $\eta_u = 3.5$ has been made, the large
curvature phase contains two and a half eras.
%\begin{equation}
%\overunderbraces{& \br{1}{\text{\scriptsize captured}} &
%\br{3}{\text{\scriptsize free}}
%&\br{1}{\text{\scriptsize captured}}}%
%{ & \quad 6.29 \quad 5.29 \quad 4.29 \;\;\,& \;\;\, 3.29 \quad 2.29
%\quad 1.29 \;\;\,& \;\;\, 3.45 \quad 2.45 \quad 1.45 \;\;\,& \;\;\,
%2.23 \quad 1.23 \;\;\,& \;\;\, 4.33 \;\;\, & \ldots}%
%{ &\br{2}{\text{\scriptsize era}}  & \br{1}{\text{\scriptsize era}}
%& \br{1}{\text{\scriptsize era}} & \br{2}{\text{\scriptsize era}}}
%\end{equation}
%
\begin{small}
\begin{equation}\label{phases}
\overunderbraces{& \br{1}{\text{\scriptsize small curvature phase}}
& \br{3}{\text{\scriptsize large curvature phase}}
&\br{1}{\text{\scriptsize s.c.p.}}}%
{& 6.29 \rightarrow 5.29 \rightarrow 4.29 \rightarrow & 3.29
\rightarrow 2.29 \rightarrow 1.29 \rightarrow & 3.45 \rightarrow
2.45 \rightarrow 1.45 \rightarrow & 2.23
\rightarrow 1.23 \rightarrow & 4.33 \rightarrow & \ldots}%
{ &\br{2}{\text{\scriptsize era}}  & \br{1}{\text{\scriptsize era}}
& \br{1}{\text{\scriptsize era}} & \br{2}{\text{\scriptsize era}}}
\end{equation}
\end{small}

\begin{figure}[Ht]
\psfrag{a}[cc][cc]{$\Sigma_1$} \psfrag{b}[cc][cc]{$\Sigma_3$}
\psfrag{c}[cc][cc]{$\Sigma_2$} \psfrag{d}[cc][cc]{$T_1$}
\psfrag{e}[cc][cc]{$Q_2$} \psfrag{f}[cc][cc]{$T_3$}
\psfrag{g}[cc][cc]{$Q_1$} \psfrag{h}[cc][cc]{$T_2$}
\psfrag{i}[cc][cc]{$Q_3$} \centering{
  \includegraphics[height=0.40\textwidth]{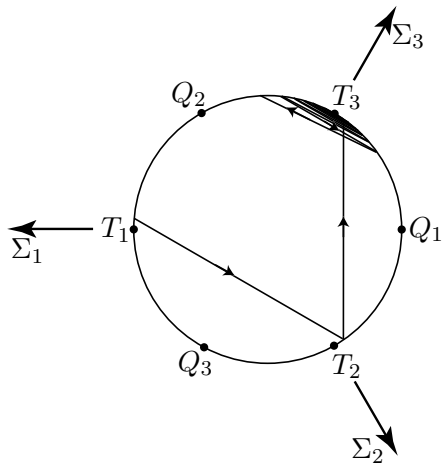}}
\caption{A small curvature phase. Note that this particular small
curvature phase starts with a $\cT_{R_3}$ transition followed by a
$\cT_{R_1}$ transition and a series of alternating $\cT_{N_1}$ and
$\cT_{R_3}$ transitions; the latter oscillatory behavior is of
particular relevance for the billiard attractor discussed in
Section~\ref{billiardattractor}.}
     \label{smallcurvphase}
\end{figure}

Let us now briefly discuss how curvature phases are represented in
the state space $\mathcal{O}$. Consider a sequence of transitions
$\cS_\cT = (\cT_i)_{i\in\mathbb{N}}$ on $\mathcal{O}$ and the
associated Kasner sequence $(u_l)_{l\in\mathbb{N}}$. Suppose that
$l\in[\lin,\lout]$ is a small curvature phase (associated with a
value $\eta_u \gg 1$) of the Kasner sequence
$(u_l)_{l\in\mathbb{N}}$. Since the inverse of the Kasner parameter,
i.e., $u^{-1}$, is a measure for the (angular) distance (on
$\mathrm{K}^\ocircle$) from $\mathrm{T}_3$ (and equivalently from
$\mathrm{T}_1$, $\mathrm{T}_2$), the condition $\uin  > \eta_u$
characterizing the small curvature phase is equivalent to
$(\uin)^{-1} < \eta_u^{-1}$, i.e., to the statement that the
(angular) distance of the transitions from $\mathrm{T}_3$
($\mathrm{T}_1$, $\mathrm{T}_2$) be less than a given distance
$\eta_u^{-1}$. Suppose that the curvature transition number $\lin$
(which is the first curvature transition of the phase) is a
$\cT_{N_1}$ transition that takes place in the
$\eta_u^{-1}$-neighborhood of the Taub point $\mathrm{T}_3$. In this
neighborhood of $\mathrm{T}_3$ only three types of transitions are
possible: $\cT_{N_1}$, $\cT_{N_2}$, and $\cT_{R_3}$,
see~Figure~\ref{singletrans}; it is thus immediate from the figures
that the initial $\cT_{N_1}$ transition generates a sequence where
$\cT_{N_1}$ transitions and $\cT_{N_2}$ or $\cT_{R_3}$ transitions
occur alternately while the distance $u^{-1}$ from $\mathrm{T}_3$
slowly increases (as $u$ decreases by $u\mapsto u-1$). For some
time, therefore, the sequence is `captured' in the
$\eta_u^{-1}$-neighborhood of the point $\mathrm{T}_3$; since the
curvature is small in this neighborhood, the phase is a `phase of
small curvature'. From~\eqref{capphaselength} we have that the
length of a captured phase is given by $L \approx \uin$ when $\uin
\gg \eta_u$; in other words, the number of curvature transitions
needed to leave the given $\eta_u^{-1}$-neighborhood of
$\mathrm{T}_3$ is inversely proportional to the initial distance
$(\uin)^{-1}$ from $\mathrm{T}_3$. For an example of a small
curvature phase in the state space picture, see
Figure~\ref{smallcurvphase}.

%We conclude this section by briefly
%%
%\marginalnote{jmh: should we have this remark at all? or
%move it to growth section?}
%%
%considering the non-generic case of
%sequences of transitions that contain infinitely
%many double curvature transitions $\cT_{N_1 N_2}$.
%Following the remarks of the previous subsection
%we choose to view these transitions as repeated interruptions
%of a regular Kasner sequence. Thereby the concepts
%introduced above (i.e., eras, small/large curvature phases)
%are still relevant; however, two successive eras
%are no longer ``causally connected'': the initial Kasner parameter
%$\uin$ of an era corresponds to the final state of
%a $\cT_{N_1 N_2}$ transition and is thus not related
%to the Kasner parameters of the preceding era.
%
%This suggests that we treat sequences of transitions that contain
%infinitely many double curvature transition
%as a concatenation of pieces of standard Kasner sequences
%glued together by $\cT_{N_1 N_2}$ transitions.
%In accordance with this viewpoint we henceforth
%adopt the convention that the indices $i$ (transition counting)
%and $l$ (epoch counting) are stalled while the
%$\cT_{N_1 N_2}$ transitions take places.

In our description of generic asymptotic dynamics
it will turn out to be important
to characterize Kasner sequences from a stochastic point of view.
In order to do so, in the subsequent section, we will
consider the space of all possible Kasner
sequences.

%%%%%%%%%%%%%%%%%%%%%%%%%%%%%%%%%%%%%%%%%%%%%%%%%%%%%%%%%%%%%%%%%%
\section{Stochastic analysis of Kasner sequences}
\label{stat}
%%%%%%%%%%%%%%%%%%%%%%%%%%%%%%%%%%%%%%%%%%%%%%%%%%%%%%%%%%%%%%%%%%

The recursion formula~\eqref{useq} that defines Kasner sequences
typically generates sequences of pairwise different values $u_l$.
However, there also exist Kasner sequences $(u_l)_{l\in\mathbb{N}}$
that are periodic; for instance, the sequence
\begin{equation}\label{periodicu}
u_0 = \frac{5 + \sqrt{13}}{\sqrt{13}-1} \,\longrightarrow\,
u_1 = \frac{6}{\sqrt{13}-1} \,\longrightarrow\,
u_2 = \frac{7-\sqrt{13}}{\sqrt{13}-1} \,\longrightarrow\,
u_3 = u_0 = \frac{5 + \sqrt{13}}{\sqrt{13}-1} \,\longrightarrow \ldots
\end{equation}
is a Kasner sequence with period $3$, see~\cite{ma88} or~\cite[p.\
236]{waiell97}. Note that this Kasner sequence is associated with a
heteroclinic cycle (or rather, a set of entangled heteroclinic
cycles) in the state space description of sequences.

Consider an arbitrary Kasner sequence $(u_l)_{l\in\mathbb{N}}$. As
discussed in Section~\ref{transition}, this sequence possesses a
natural partition into eras $j=1,2,\ldots$, where each era $j$ is
associated with an initial value $u = \uin>1$ of the Kasner
parameter, which we denote by $\uinn{j}$. The Kasner
map~\eqref{useq} thus generates a sequence
$(\uinn{j})_{j\in\mathbb{N}}$ obeying the recursion formula
\begin{equation}\label{Tt}
\uinn{j+1} = \frac{1}{\mathrm{frac}(\uinn{j})}\:,
\end{equation}
where $\mathrm{frac}(\cdot)$ is a function that gives the fractional
part of its argument. It is known that the transformation~\eqref{Tt}
is associated with exponential instability: the distance between two
points (whose initial distance is small) grows exponentially with
the number of iterations $j$, see~\cite{khaetal85}.

For the following we regard the sequence
$(\uinn{j})_{j\in\mathbb{N}}$ as a distribution of points (given by
the values of $\uinn{j}$) in the interval $(1,\infty)\subseteq
\mathbb{R}$. By adopting this viewpoint it becomes natural to ask
whether this distribution possesses a probabilistic description,
i.e., whether it can be modeled by a continuous probability
function. In general, this will not be the case, since there exist
Kasner sequences that generate finite distributions
$\{\upsilon_1,\ldots,\upsilon_n\}$, i.e., $\uinn{j} \in
\{\upsilon_1,\ldots,\upsilon_n \}$ for all $j \in \mathbb{N}$. In
the example~\eqref{periodicu} we even obtain $\uinn{j} = \upsilon =
(5 + \sqrt{13})/(\sqrt{13}-1)$ for all $j \in \mathbb{N}$.

However, the picture changes when we consider not the distribution
of one single sequence $(\uinn{j})_{j\in\mathbb{N}}$, but the
entirety of all distributions generated by the collection of all
possible Kasner sequences. This collective distribution is indeed
modeled by a continuous probability function, which arises as the
stationary limit from an arbitrary probability density via the map
$\uin \mapsto [\mathrm{frac}(\uin)]^{-1}$,
cf.~\eqref{Tt}; %which is induced on $\uin$ by the Kasner map,
see~\cite{khaetal85}. In other words, $\uin$ can be regarded as a
random variable with a specific probability density given by
\begin{equation}\label{probdensuin}
w(\uin) = \frac{1}{\log 2} \:\frac{1}{\uin(1+\uin)}\:.
\end{equation}

According to the probabilistic description with probability
density~\eqref{probdensuin}, when we choose an arbitrary $\uin$ from
an arbitrary era of an arbitrary Kasner sequence, the probability of
$\uin$ taking a value in $(\upsilon_1,\upsilon_2)$ is given by
\begin{equation*}
\mathrm{P}(\upsilon_1 < \uin < \upsilon_2) =  \frac{1}{\log 2}
\int_{\upsilon_1}^{\upsilon_2} \frac{1}{u(1+u)}\: d u\:.
\end{equation*}
In order to obtain a probability distribution on a bounded interval
the inverse of the Kasner parameter $\uin$ is used: we denote the
inverse of $\uin$ by $\varkappa$, i.e., $\varkappa = (\uin)^{-1}$;
by definition, $\varkappa \in (0,1)$. The probability density of the
random variable $\varkappa$ reads
\begin{equation}\label{probdens}
w(\varkappa) = \frac{1}{\log 2} \:\frac{1}{1 + \varkappa}\:,
\end{equation}
see~\cite{khaetal85}.

It is straightforward to adapt the probabilistic description to the
concept of small/large curvature phases. The Kasner sequence
$(u_l)_{l\in\mathbb{N}}$ consists of infinitely many small curvature
phases $j=1,2,\ldots$, where each small curvature phase is associated with
a value $\uinn{j} > \eta_u$ of the initial Kasner parameter (which
naturally coincides with the initial parameter of the era that begins
simultaneously with the small curvature phase $j$,
see~\eqref{phases}). Let $\varkappa_j$ be the inverse of $\uinn{j}$,
i.e., $\varkappa_j = (\uinn{j})^{-1} < \eta_u^{-1}$; the
distribution of the sequence $(\varkappa_{j})_{j\in\mathbb{N}}$ in
the interval $(0,\eta_u^{-1})$ is described by a random variable
$\varkappa$ with probability density
\begin{equation}\label{hatw}
\hat{w}(\varkappa) = \frac{1}{\log (1+\eta_u^{-1})} \:\frac{1}{1 +
\varkappa}\:;\qquad \int_0^{\eta_u^{-1}} \hat{w}(\varkappa)
d\varkappa = 1\:;
\end{equation}
evidently, for $\eta_u =1$ the distribution $\hat{w}(\varkappa)$ reduces
to~\eqref{probdens}. Note in particular that $\hat{w}(0) \neq 0$.

Consider a Kasner sequence $(u_l)_{l\in\mathbb{N}}$ whose
distribution of the parameters $(\uinn{j})_{j\in\mathbb{N}}$ has a
probabilistic description (in terms of the probability
density~\eqref{hatw}). For such a Kasner sequence small curvature
phases are expected to dominate over large curvature phases in the
following sense: the probability that an epoch of the sequence
chosen randomly among $l$ epochs belongs to a small curvature phase
tends to one as $l\rightarrow \infty$. The proof is based on
stochastic arguments:

We first compute the expectation value $\langle L \rangle$ of the
length of a small curvature phase. Since $L \approx \uin - \eta_u =
\varkappa^{-1} -\eta_u$, see~\eqref{capphaselength}, we obtain
\begin{equation*}
\langle L \rangle = \int_0^{\eta_u^{-1}} \hat{w}(\varkappa) L(\varkappa) d\varkappa
\approx \hat{w}(0) \int_0^{\eta_u^{-1}} (\varkappa^{-1} -\eta_u) d\varkappa = \infty \:,
\end{equation*}
i.e., $\langle L \rangle$ is infinite. Note that
this feature is independent of the probability density $\hat{w}(\varkappa)$
as long as $\hat{w}(0) > 0$.

Second, we compute the expectation value for the length of a large
curvature phase. A large curvature phase continues and ends the era
that started simultaneously with the preceding small curvature
phase, see~\eqref{phases}.
The probability for the large curvature phase
to contain another (complete) era is given by
\begin{equation*}
p= 1- \int_0^{\eta_u^{-1}} w(\varkappa) d\varkappa =
\int_{\eta_u^{-1}}^1 w(\varkappa) d\varkappa = 1 -
\frac{\log(1+\eta_u^{-1})}{\log 2} \approx 1 - \frac{\eta_u^{-1}}{\log 2} < 1\:;
\end{equation*}
the approximation assumes $\eta_u \gg 1$. Accordingly, the
probability that a large curvature phase contains at least another
$n$ (complete) eras is $p^n$;
the probability that a large curvature phase contains exactly
$n$ (complete) eras is $(1-p) p^n$. (Note here that the assumption $\eta_u
\gg 1$ ensures that there will be sufficiently many eras in one
large curvature phase to permit a probabilistic description.) The
expectation value for the length of an era that is contained in a
large curvature phase is
\begin{equation*}
\left(\int_{\eta_u^{-1}}^1 w(\varkappa) d\varkappa\right)^{-1}
\int_{\eta_u^{-1}}^1 \mathrm{int}(\varkappa^{-1}) w(\varkappa)
d\varkappa \,>\, \int_{\eta_u^{-1}}^1 \left(\varkappa^{-1}-1 \right) w(\varkappa)
d\varkappa
% = (\log 2)^{-1} \log (\eta_u +1) -1
\approx (\log 2)^{-1} \log \eta_u \:,
\end{equation*}
where again $\eta_u \gg 1$ is assumed.
Accordingly, a sequence of $n$ eras within a large curvature phase
is expected to be of length $n (\log 2)^{-1} \log\eta_u$.
Combining the results we find
that the expectation value for the length of a large curvature phase
is given by
\begin{equation*}
\eta_u + (\log 2)^{-1} (\log \eta_u) (1-p) \sum_{n=0}^\infty n p^{n}
\approx \eta_u + \eta_u \log \eta_u  \approx \eta_u \log\eta_u  < \infty \:.
\end{equation*}
Since the mean length of a large curvature phase is expected to be
finite, while the expectation value $\langle L \rangle$ of the
length of a small curvature phase is infinite, we conclude that,
asymptotically as $l\rightarrow \infty$, small curvature phases
dominate over large curvature phases.

Recall that, in the state space description, small curvature phases
are phases where the sequence of transitions $\cS_\cT$, which
generates the Kasner series $(u_l)_{l\in\mathbb{N}}$ we consider, is
captured in a small neighborhood of the Taub points, see
Figure~\ref{smallcurvphase}. Accordingly, dominance of small
curvature phases over large curvature phases means that the sequence
of transitions $\cS_\cT$ spends more and more `time' (as measured by
the number of epochs) in a neighborhood of the Taub points as
compared to the time spent close to ordinary Kasner states; above
all, the ratio diverges in the limit. Considerations of this type
quite naturally bring the Taub points into focus when we aim at
investigating generic asymptotic dynamics of solutions.

%%%%%%%%%%%%%%%%%%%%%%%%%%%%%%%%%%%%%%%%%%%%%%%%%%%%%%%%%%%%%%%%%%
\section{Growth}
\label{growth}
%%%%%%%%%%%%%%%%%%%%%%%%%%%%%%%%%%%%%%%%%%%%%%%%%%%%%%%%%%%%%%%%%%

In this section we analyze the analytic and stochastic properties
of a function that is naturally associated with a sequence of
transitions: the growth function.

%--------------------------------------------------------------------------------------
\subsection*{Auxiliary differential equations and the growth function}
\label{auxand}
%--------------------------------------------------------------------------------------

In our analysis we will encounter certain types of differential
equations; the following auxiliary equation represents the common
denominator:
\begin{equation} \label{A1eq}
\partial_\tau\, A =  2 (1-\Sigma^2) A\:.
\end{equation}
When a solution of~\eqref{unstableeqs} on the oscillatory subset
$\mathcal{O}$ is given, then the equation can be regarded as a
linear differential equation with a prescribed time-dependent
coefficient $2(1-\Sigma^2)$. Let us now integrate~\eqref{A1eq} along
the different Bianchi type I and II transitions.

Along frame transitions we obtain $A = \mathrm{const}$, since
\mbox{$(1-\Sigma^2)=0$}. Single and mixed curvature transitions
yield $A(\tau_0) \mapsto A(\tau) = A(\tau_0)
[\zeta(\tau)/\zeta(\tau_0)]$, which follows from~\eqref{Zeq2}; note
that $\zeta(\tau)/\zeta(\tau_0)>1$ for all $\tau > \tau_0$. Letting
$\tau_0 \rightarrow -\infty$ and $\tau\rightarrow \infty$ shows that
the initial value of $A$ (i.e., $\lim_{\tau\rightarrow -\infty}
A(\tau)$) and the final value of $A$ (i.e., $\lim_{\tau\rightarrow
\infty} A(\tau)$) are related by the \textit{growth factor} $g$;
recall that $g$ is given by
\begin{equation*}%\label{geqagain}
g = \frac{1+u+u^2}{1-u+u^2}\:,
\end{equation*}
where $u$ is the (initial) Kasner parameter of the curvature
transition; see~\eqref{geq}. In condensed form, we write
\begin{equation}\label{Agro}
A \mapsto g A
\end{equation}
for the effect of a curvature transition on the quantity $A$. Recall
that, since $u\in (1,\infty)$ for single and mixed curvature
transitions, we have $1<g<3$.

\begin{figure}[ht]
\psfrag{a}[cc][cc]{$u_{l-2}$} \psfrag{b}[cc][cc]{$G_{l-2}$}
\psfrag{c}[cc][cc]{$l-2$} \psfrag{d}[cc][cc]{$u_{l-1}$}
\psfrag{e}[cc][cc]{$G_{l-1}$} \psfrag{f}[cc][cc]{$l-1$}
\psfrag{g}[cc][cc]{$u_{l}$} \psfrag{h}[cc][cc]{$G_{l}$}
\psfrag{i}[cc][cc]{$l$} \psfrag{j}[cc][cc]{$u_{l+1}$}
\psfrag{k}[cc][cc]{$G_{l+1}$} \psfrag{l}[cc][cc]{$l+1$}
\psfrag{m}[cc][cc]{$u_{l+2}$} \psfrag{n}[cc][cc]{$G_{l+2}$}
\psfrag{o}[cc][cc]{$l+2$} \psfrag{p}[cc][cc]{$g_{l-2}$}
\psfrag{q}[cc][cc]{$g_{l-1}$} \psfrag{r}[cc][cc]{$g_{l}$}
\psfrag{s}[cc][cc]{$g_{l+1}$} \psfrag{t}[cc][cc]{$g_{l+2}$}
\centering{
  \includegraphics[width=0.98\textwidth]{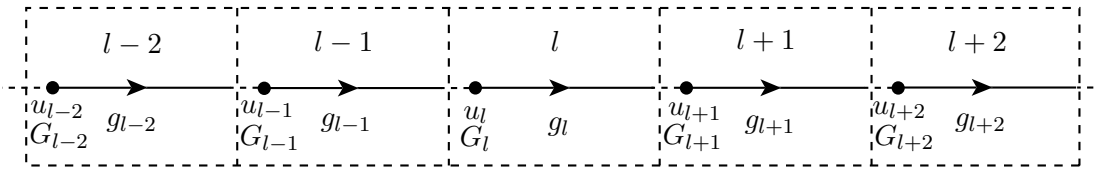}}
\caption{Transition counting conventions.}\label{utrans}
\end{figure}

Solving Eq.~\eqref{A1eq} along a sequence of transitions $\cS_\cT$
on $\mathcal{O}$ amounts to iterating~\eqref{Agro}, which we will
elaborate on in the following. (Recall that, in our convention,
$\cS_\cT$ contains only Bianchi type I and II transitions, unless
otherwise stated.) For the $l$\raisebox{0.7ex}{\small th} curvature
transition of a sequence $\cS_\cT$ the growth factor is given by
\begin{equation}\label{grofacl}
g_{l} := \frac{1 + u_{l} + u_{l}^2}{1 - u_{l} + u_{l}^2} = \frac{1 +
u_{l} + u_{l}^2}{1 + u_{l+1} + u_{l+1}^2} \:\,r_l^2 \:,
\qquad\text{where }\,
r_{l} := \big(\min\{ u_{l}-1 ,1\}\big)^{-1} \:;
\end{equation}
the quantity $r_l$ satisfies $r_l \geq 1$, where $r_l = 1$ whenever
$u_l > 2$ ---this is the case when $u_{l+1} = u_l -1$,
see~\eqref{BKLMap}. We define the \textit{growth function} as the
product of the growth factors: we let $G:\mathbb{N} \rightarrow
\mathbb{R}$ be defined by $G(0)=G_0 =1$ and
\begin{equation*}%align}\label{Gf}
G(l)=G_l = \prod\limits_{j=0}^{l-1} g_{j} =
\prod\limits_{j=0}^{l-1} \left( \frac{1 + u_{j} + u_{j}^2}{1 +
u_{j+1} + u_{j+1}^2} \right) r_{j}^2 = \left( \frac{1 + u_0 +
u_0^2}{1 + u_l + u_l^2} \right) \prod\limits_{j=0}^{l-1} r_{j}^2\:.
\end{equation*}
Since $g_j>1, r_j \geq 1$ for all $j>1$ (and $g_j \not\rightarrow 1$
as $j\rightarrow \infty$), $G_l$ increases rapidly with $l$ and
$G_l\rightarrow \infty$ with $l\rightarrow \infty$. In our
convention, $G_l$ denotes the value of the growth function at the
beginning of the $l$\raisebox{0.7ex}{\small th} curvature
transition; see Figure~\ref{utrans}.

Iterating~\eqref{Agro} along $l$ curvature transitions of the sequence $\cS_\cT$
thus yields
\begin{equation}\label{A0mapstoGlA0}
A_0 \mapsto G(l)\, A_0 = G_l\, A_0 = \Big( \prod\limits_{j=0}^{l-1} g_{j}\Big) A_0\:;
\end{equation}
hence $A$ is a function whose rate of growth is described by the
growth function $G$. Finally, in order to write out
Eq.~\eqref{A0mapstoGlA0} in proper notation, let $A_i$ denote the
value of $A$ at the beginning of the $i$\raisebox{0.7ex}{\small th}
transition $\cT_i$ of the sequence $\cS_\cT =
(\cT_0,\cT_1,\cT_2,\ldots)$; then we can write
\begin{equation}\label{Ainc}
A_i = G_{l(i)}\, A_0 \:;
\end{equation}
recall that $l(i) = l_i$ is the curvature transition index
associated with $\cS_\cT$, i.e., the number of curvature transitions
among $\{\cT_0,\cT_1,\ldots,\cT_i\}$, see Section~\ref{transition}.
The use of this notation appears superfluous in the present context,
since the function $A$ does not change under frame transitions and
it would thus suffice to focus on the curvature transitions
$l=0,1,2,\ldots$ of $\cS_\cT$; however, next we will discuss a more
general differential equation whose solutions are affected by frame
transitions.

Consider the following auxiliary differential equation, which is
frame dependent:
\begin{equation}\label{A2eq}
\partial_\tau  B  =  2 (1- \Sigma^2) B + 2 R_3^2 -2 R_1^2\:;
\end{equation}
as with~\eqref{A1eq}, our aim is to integrate this equation
along a sequence of transitions $\cS_\cT$.
%in complete analogy with~\eqref{A1eq},
%when a solution of~\eqref{unstableeqs} is given,
%we are able to integrate~\eqref{A2eq}
%by viewing $2(1-\Sigma^2)$ and $2 R_3^2 -2 R_1^2$ as given
%time-dependent coefficients.

For simplicity, let $\cS_\cT = (\cT_0, \cT_1, \cT_2, \ldots)$ be a
sequence of transitions that does not contain $\cT_{N_2}$ and
$\cT_{N_2 R_2}$ transitions; hence $N_2 \equiv 0$ along $\cS_\cT$,
so that
\begin{equation*}
\partial_\tau (\Sigma_2 - 2) =
2 (1-\Sigma^2) (\Sigma_2 - 2) + 2 R_3^2 -2 R_1^2
\end{equation*}
holds for each transition $\cT_i$, which follows
from~\eqref{unstableeq4} and~\eqref{gausscon}. Consequently,
$(\Sigma_2 -2)$ is a particular solution of~\eqref{A2eq}.

The general solution for $B$ along $\cS_\cT$ can thus be constructed
explicitly as the superposition of the particular solution
$(\Sigma_2 - 2)$ and the general solution of the homogeneous
equation. Since the latter is given by~\eqref{Ainc} we obtain
\begin{equation}\label{Binc}
B_i \:=\: \big(\Sigma_2(i)-2\big) + G_{l(i)}\, \Big[ B_0 -
(\Sigma_2(0) -2)\Big]\:,
\end{equation}
where $\Sigma_2(i)$ denotes the value of $\Sigma_2$ after $i$
transitions (which, by the convention illustrated in
Figure~\ref{utrans}, is the value at the beginning of the transition
with index $i$). In the generic case, i.e., $B_0 \neq
(\Sigma_2(0)-2)$, we therefore find
\begin{equation}\label{Bincgen}
%B_i \:\sim\: G_{l(i)}\, [B_0 - \mathrm{const}] \qquad
%( i\rightarrow \infty)\:;
B_i \:\sim\: \mathrm{const}\:G_{l(i)} \qquad
( i\rightarrow \infty)
\end{equation}
with $\mathrm{const} \neq 0$;
in particular, $|B_i| \rightarrow \infty$ as $i\rightarrow \infty$.
The result~\eqref{Binc} is slightly modified when $\cT_{N_2}$ and
$\cT_{N_2 R_2}$ transitions are taken into account, see
Appendix~\ref{Bbehave}; however, the generic behavior of $B$ as
described by~\eqref{Bincgen} remains unchanged.

In Section~\ref{transition} we have argued that generically
sequences of transitions do not contain double curvature
transitions; in accordance with these considerations we have hence
focused our analysis on sequences $\cS_\cT$ of this type. In the
remainder of this subsection, however, we discuss non-generic
sequences, in particular in view of the growth associated with
double curvature transitions.

From Section~\ref{dynosc} and Appendix~\ref{multiple} we recall that
the equation $\partial_\tau (2-\Sigma_3) = 2 (1-\Sigma^2)
(2-\Sigma_3)$ holds along double curvature transitions (i.e., Bianchi
type $\mathrm{VI}_0$ orbits) and analogously along Bianchi type
$\mathrm{VII}_0$ orbits. Consequently, we find that $A(\tau) =
A(\tau_0) (2 -\Sigma_3)(\tau)/(2-\Sigma_3)(\tau_0)$; the behavior of
$B$ is identical, since setting $(R_1, R_2, R_3) = 0$
reduces~\eqref{A2eq} to~\eqref{A1eq}. Since the $\alpha$-limit of
all orbits is a fixed point with $\Sigma_3 = 2$, the expression $(2
-\Sigma_3)(\tau)/(2-\Sigma_3)(\tau_0)$ diverges as
$\tau_0\rightarrow -\infty$. The growth factor $g$ associated with a
Bianchi type $\mathrm{VI}_0$/$\mathrm{VII}_0$ orbit can thus be
considered as being infinite. It is immediate from these
considerations that the growth function is ill-defined along
sequences of transitions that contain double curvature transitions;
it is natural, however, to consider a `piecewise' growth function:
following the remarks of Section~\ref{transition} we choose to view
double curvature transitions as repeated interruptions of a regular
Kasner sequence. Thereby the concepts introduced in
Section~\ref{transition} (i.e., eras, small/large curvature phases)
are still relevant; the interruptions, however, correspond to two
successive eras not being `causally connected'---the initial Kasner
parameter $\uin$ of an era that arises from the final state of a
$\cT_{N_1 N_2}$ transition is not related to the Kasner parameter of
the preceding era. This suggests that we treat sequences of
transitions that contain double curvature transition as a
concatenation of `pieces' of standard Kasner sequences glued
together by $\cT_{N_1 N_2}$ transitions.
%
%``In accordance with this viewpoint we henceforth
%adopt the convention that the indices $i$ (transition counting)
%and $l$ (epoch counting) are stalled while the
%$\cT_{N_1 N_2}$ transitions take places.''
%
Along these fragments the growth function is defined in the standard
way; however, joining the pieces is not possible, since this is
connected with a divergent growth along the joining $\cT_{N_1 N_2}$
transition.

In anticipation of our considerations in Section~\ref{asysha} we
conclude these remarks by considering orbits that are not exactly of
Bianchi type $\mathrm{VI}_0$/$\mathrm{VII}_0$, but arbitrarily close
(in a higher dimensional state space where the
$\alpha$-/$\omega$-limits of the $\mathrm{VI}_0$/$\mathrm{VII}_0$
orbits are saddles). It is immediate that the growth of $A$ and $B$
along such orbits can be arbitrarily large depending on the degree
of the type $\mathrm{VI}_0$/$\mathrm{VII}_0$ approximation.

%--------------------------------------------------------------------------------------
\subsection*{Properties of the growth function}
\label{propgrow}
%--------------------------------------------------------------------------------------

Since the growth function is the key element in the description of
the behavior of $A$ and $B$, it is appropriate to study this
function in more detail. Let us begin by investigating the
properties of the growth function $G(l)$ along a sequence of
transitions $\cS_\cT$ with regard to the concepts of small and large
curvature phases, and eras.

Consider first a small curvature phase $[\lin,\lout]$, where we
assume $\eta_u > 2$. For all $l$ the growth factor $g_l$ is given
by
\begin{equation}\label{grofacscp}
g_{l} = \frac{1 + u_{l} + u_{l}^2}{1 - u_{l} + u_{l}^2} = \frac{1 +
u_{l} + u_{l}^2}{1 + u_{l+1} + u_{l+1}^2} \:,
\end{equation}
see~\eqref{grofacl}. The growth function satisfies the recursive
relation $G_{l+1} = G_l g_l$, so that we obtain
\begin{equation}\label{gfcap}
\frac{G(\lout+1)}{G(\lin)} = \prod_{l=\lin}^{\lout} g_l =
%\frac{1+\uin + (\uin)^2}{1-u_{\lout} + (u_{\lout})^2}
\frac{1+\uin + (\uin)^2}{1+u_{\lout+1} + (u_{\lout+1})^2} \approx
\frac{1+\uin + (\uin)^{2}}{1+\eta_u+\eta_u^2} \approx
\frac{(\uin)^{2}}{\eta_u^2}\:,
\end{equation}
where the approximation requires $\eta_u \gg 1$.
Eq.~\eqref{gfcap} describes the growth of $G(l)$ during
one small curvature phase.

In a large curvature phase $[\li,\lf]$, since $u_l \leq \eta_u$ for
all $l\in[\li,\lf]$, we have
\begin{equation}\label{glfree}
g_l = \frac{1 + u_{l} + u_{l}^2}{1 - u_{l} + u_{l}^2} \geq \frac{1 +
\eta_u + \eta_u^2}{1 - \eta_u + \eta_u^2} =: C_{\mathrm{p}} > 1 \:;
\qquad\quad  C_{\mathrm{p}} \approx  1 + 2 \eta_u^{-1}\,\text{ if }\,
\eta_u \gg 1\:.
\end{equation}
Therefore, in a large curvature phase, the growth function $G(l)$
behaves like
\begin{equation}\label{Glfree}
G(l+1) = G(\li) \prod_{k=\li}^{l} g_k\, >\, G(\li)
\,(C_{\mathrm{p}})^{l+1-\li}\:,
\end{equation}
i.e., we observe exponential growth of $G(l)$ with a given base
$C_{\mathrm{p}}$.

Let us now discuss the growth of $G(l)$ in an era $[\lin,\lout]$. We
define the \textit{era growth factor} $\mathfrak{g}$ as
\begin{subequations}\label{gfera}
\begin{align}
\mathfrak{g} & = \frac{G(\lout+1)}{G(\lin)} %= \prod_{l=\lin}^{\lout} g_l
= \left(\prod_{l=\lin}^{\lout-1} g_l\right) g_{\lout} =
%\frac{1+\uin + (\uin)^2}{1+u_{\lout} + (u_{\lout})^2}
%%% ---->  u_{\lout} instead of \uout
%\frac{1+u_{\lout} + (u_{\lout})^2}{1-u_{\lout} + (u_{\lout})^2}
\frac{1+\uin + (\uin)^2}{\bcancel{1+\uout + (\uout)^2}}
\frac{\bcancel{1+\uout + (\uout)^2}}{1-\uout + (\uout)^2} \:,
\intertext{so that}
\mathfrak{g} & \geq \frac{1+\uin + (\uin)^{2}}{3} \:.
\end{align}
\end{subequations}
%
%where $\uout = u_{\lout}$.
Consider the partition of a sequence into $a=0,1,2,\ldots$ eras. For
the era with number $a$ the era growth factor $\mathfrak{g}_a$ can
be written as
\begin{equation}\label{eragrowthfac}
\mathfrak{g}_a =
%\frac{1+\uinn{a} + (\uinn{a})^2}{1-u_{\lout,a}+(u_{\lout,a})^2}
%%% ---->  u_{\lout} instead of \uout
%= \left(1+\uinn{a}+(\uinn{a})^2\right)
% \frac{(\uinn{a+1})^2}{1+\uinn{a+1}+(\uinn{a+1})^2}
\frac{1+\uinn{a} + (\uinn{a})^2}{1-\uoutt{a}+(\uoutt{a})^2} =
\left(1+\uinn{a}+(\uinn{a})^2\right)
\frac{(\uinn{a+1})^2}{1+\uinn{a+1}+(\uinn{a+1})^2}\:.
%\geq \frac{1+\uinn{a}+(\uinn{a})^2}{3}\:.
\end{equation}

The {\em era growth function\/} $\mathcal{G}$ is defined as the
product of era growth factors. From $\mathcal{G}_{a+1} =
\mathcal{G}_a \mathfrak{g}_a$ %%
%\begin{equation}\label{eragrowrecur}
%\mathcal{G}_{a+1} = \mathcal{G}_a \,\mathfrak{g}_a
%\end{equation}
%
we obtain
\begin{equation*}%\label{eragro}
\mathcal{G}(a) = \mathcal{G}_a =\mathcal{G}_0 \prod_{c=0}^{a-1}
\mathfrak{g}_c = \mathcal{G}_0\left( \prod_{c=0}^{a-1} (\uinn{c})^2
\right) \left[\frac{1+\uinn{0} + (\uinn{0})^2}{(\uinn{0})^2}\:
\frac{(\uinn{a})^2}{1+\uinn{a} + (\uinn{a})^2}\right]\:,
\end{equation*}
where we set $\mathcal{G}_0=1$ by convention. The expression in
brackets is a number in the interval $(1/3,3)$; when $\uinn{0}$ and
$\uinn{a}$ are large the number is close to one.

%Finally, it is of interest to also introduce a continuous growth
%function $G(\tau)$ that describes the situation during
%$l$\raisebox{0.7ex}{\small th} curvature transition:
%%
%\begin{equation}
%G(\tau):=G_{l}g(\tau)\:, \quad {\rm where}\quad
%g(\tau):=\frac{\zeta(\tau)}{\zeta_-}\:
%\end{equation}
%%
%(recall that $\partial_\tau \zeta = 2(1-\Sigma^2) \zeta$).

Let us now compute the expectation values of the growth factors. The
era growth factor $\mathfrak{g}$, cf.~\eqref{gfera}, does not
possess a finite expectation value. This is because the equations
contain $\uin$ ($=\varkappa^{-1}$) and $(\uin)^2$:
\begin{equation*}
\langle \uin \rangle = \int_0^1 w(\varkappa) \varkappa^{-1} d
\varkappa = \infty\,, \quad \langle (\uin)^2 \rangle = \int_0^1
w(\varkappa) \varkappa^{-2} d \varkappa = \infty
\quad\Rightarrow\quad \langle \mathfrak{g}\rangle = \infty\:,
\end{equation*}
where we have used the probability density $w(\varkappa)$ from
Section~\ref{stat}. The same result applies for the growth factor
describing the growth of $G$ in a small curvature phase,
cf.~\eqref{gfcap}. However, the expectation value of the logarithm
of the era growth factor can be estimated straightforwardly:
\begin{equation}\label{expectfrakg}
\langle \log \mathfrak{g} \rangle \geq -\log 3 + \int_0^1
w(\varkappa) \log(1+\varkappa^{-1} + \varkappa^{-2}) d\varkappa
\approx  1.77\:. %1.76522\:.
\end{equation}
By definition, the logarithm of the era growth function
$\mathcal{G}$ is the sum of the logarithms of the era growth
factors,
%$\log \mathcal{G}_a = \sum_{c=1}^a \log \mathfrak{g}_c$,
therefore we expect $\log\mathcal{G}$ to grow according to
\begin{equation*}
\langle \log \mathcal{G}_a \rangle = \left\langle \sum_{c=0}^{a-1}
\log \mathfrak{g}_c \right\rangle\, =\, a\, \langle \log
\mathfrak{g} \rangle \geq 1.77 \,a \:.
\end{equation*}
For further results on growth functions, see
Appendix~\ref{convergenceofsums}.

%%%%%%%%%%%%%%%%%%%%%%%%%%%%%%%%%%%%%%%%%%%%%%%%%%%%%%%%%%%%%%%%%%
\section{Asymptotic shadowing}
\label{asysha}
%%%%%%%%%%%%%%%%%%%%%%%%%%%%%%%%%%%%%%%%%%%%%%%%%%%%%%%%%%%%%%%%%%

%---------------------------------------------------------
\subsection*{Shadowing---asymptotic sequences of orbits}
\label{sha---}
%---------------------------------------------------------

The subsequent considerations constitute the core of our analysis:
we provide evidence that the asymptotic dynamics of solutions of
Einstein's equations which exhibit a generic spacelike singularity
is represented by sequences of Bianchi type I and type II
transitions.

The basic assumption we make is that a solution of Einstein's
equations $\bm{X}(t,x^i)$ that possesses a generic spacelike
singularity is asymptotically silent. We further assume that the
dynamics of $\bm{X}(t,x^i)$ along a timeline associated with a
generic spatial point $x^i$ of such a singularity is asymptotically
local. This entails that the asymptotic behavior of the solution is
described, with an increasing degree of accuracy, by the dynamics on
the SH part of the silent boundary. The SH silent boundary consists
of an infinite number of copies of the finite dimensional SH state
space that is spanned by the state vector $\bm{S} =
(\Sigma_{\alpha\beta},A_\alpha,N_{\alpha\beta})$, where the spatial
coordinates act as the infinite index set. Our assumptions thus lead
to the conjecture that the asymptotic behavior of the solution
$\bm{X}(\tau)$ at a generic spatial point $x^i$ is governed by the
dynamics on the SH state space.

In Section~\ref{prebilliard} we have analyzed one structure on the
SH silent boundary that is expected to be of fundamental importance:
the Kasner circle of equilibrium points $\mathrm{K}^\ocircle$. The
local dynamical systems analysis of $\mathrm{K}^\ocircle$ suggests
that $\E{\alpha}{i}\rightarrow 0$ as $\tau\rightarrow\infty$, which
indicates intrinsic consistency of our considerations. Furthermore,
we are led to the conjecture that there exists a `stable part'
$\bm{S}_{\mathrm{stable}}$ of the state vector $\bm{S}$, whose
variables vanish in the asymptotic limit, i.e.,
$\bm{S}_{\mathrm{stable}}= (N_{12}, N_{23}, N_{31}, A_\alpha)
\rightarrow 0$ as $\tau\rightarrow \infty$. This conjecture is
intimately connected with the assumption that the asymptotic
evolution of a generic solution is largely dominated by Kasner
states, in the sense that the solution spends an increasing amount
of $\tau$-time in a small neighborhood of the (non-flat part of) the
Kasner circle.

Since we expect $\bm{S}_{\mathrm{stable}} \rightarrow 0$, it is the
asymptotic behavior of the remaining `oscillatory variables'
$\bm{S}_{\mathrm{osc}} = (\Sigma_\alpha, R_\alpha, N_1, N_2)$ that
represents the essential asymptotic dynamics of the solution
$\bm{X}(\tau)$. Under the above assumptions the Codazzi
constraints~\eqref{dlcodazzi} imply that
%\begin{equation}
%N_1 R_2 \rightarrow 0 \:, \qquad
%N_1 R_3 \rightarrow 0 \:, \qquad
%N_2 R_1 \rightarrow 0 \:, \qquad
%N_2 R_3 \rightarrow 0\:
%\end{equation}
%
\begin{equation}\label{asyCoda}
N_1 R_2 \rightarrow 0 \:, \qquad N_2 R_2 \rightarrow 0 \:, \qquad
(N_1-N_2) R_3 \rightarrow 0
\end{equation}
as $\tau\rightarrow \infty$, which is to be compared
with~\eqref{codazziap}; in the same limit we obtain that
$\bm{S}_{\mathrm{osc}}$ satisfies the oscillatory dynamical
system~\eqref{unstableeq3}--\eqref{unstableeq10} and the
constraint~\eqref{gausscon}. This suggests that the asymptotic
dynamics of $\bm{S}_{\mathrm{osc}} = (\Sigma_\alpha, R_\alpha, N_1,
N_2)$ is governed by the dynamics on the oscillatory subset
$\mathcal{O}$, which we have shown to be represented by Bianchi type
$\mathrm{I}$, $\mathrm{II}$, $\mathrm{VI}_0$ transitions---frame and
curvature transitions---and Bianchi type $\mathrm{VII}_0$ orbits. In
brief: asymptotically, the state vector $\bm{S}_{\mathrm{osc}}$ of a
solution $\bm{X}$ \textit{shadows} the flow on $\mathcal{O}$.

%The SH Codazzi constraints~\eqref{codazzi}
%imply that
%the oscillatory subset $\mathcal{O}$ decomposes into
%disconnected components of Bianchi type $\mathrm{I}$, $\mathrm{II}$,
%$\mathrm{VI}_0$, and $\mathrm{VII}_0$.

Along the orbit $\bm{S}_{\mathrm{osc}}(\tau)$ the SH Codazzi
constraints~\eqref{codazzi} are satisfied only in the limit
$\tau\rightarrow \infty$; hence the disconnectedness of the Bianchi
components of the oscillatory subset $\mathcal{O}$ is irrelevant for
the solution $\bm{S}_{\mathrm{osc}}(\tau)$. As an immediate
consequence of asymptotic shadowing it thus follows that in the
asymptotic regime $\tau\rightarrow \infty$, the orbit
$\bm{S}_{\mathrm{osc}}(\tau)$ resembles a sequence of Bianchi
transitions and orbits: $\bm{S}_{\mathrm{osc}}(\tau)$ can be
partitioned into a sequence of segments, where each segment is
associated with a heteroclinic orbit on $\mathcal{O}$ (of type I,
II, $\mathrm{VI}_0$, or $\mathrm{VII}_0$). Two subsequent segments
of $\bm{S}_{\mathrm{osc}}(\tau)$ are joined smoothly in a
neighborhood of an equilibrium point of the
Kasner circle.%
\footnote{The partition of $\bm{S}_{\mathrm{osc}}(\tau)$
  into segments involves a certain arbitrariness: which point in the neighborhood
  of $\mathrm{K}^\ocircle$ is chosen to be defined as the end point of one transition and the
  initial point of the subsequent one is a matter of convention, however, asymptotically
  any reasonable convention leads to the natural segmentation into the heteroclinic
  orbits on $\mathcal{O}$.}
As $\tau\rightarrow \infty$ the segments of
$\bm{S}_{\mathrm{osc}}(\tau)$ are approximated to an increasing
degree of accuracy by heteroclinic orbits on $\mathcal{O}$;
simultaneously, the joining of segments occurs increasingly closer
to $\mathrm{K}^\ocircle$; hence $\bm{S}_{\mathrm{osc}}(\tau)$ is
described by a sequence of Bianchi transitions and orbits with an
increasing degree of accuracy. We may call this representation of
the orbit $\bm{S}_{\mathrm{osc}}(\tau)$ in the asymptotic regime an
\textit{asymptotic sequence of $\mathcal{O}$-orbits} $\cA\cS_\cO$
(or, for brevity, simply \textit{asymptotic sequence}). Note that
the concept of asymptotic sequences $\cA\cS_\cO$ gives a precise
meaning to BKL's `piecewise approximations'.

As it will turn out in the following, generic asymptotic sequences
$\cA\cS_\cO$ fall into segments that correspond to Bianchi type I
and II transitions; % and can thus be viewed as concatenations of
%`approximate transitions';
it is suggestive to call asymptotic sequences of this kind
\textit{asymptotic sequences of transitions} $\cA\cS_\cT$. With
slight abuse of notation we write $\cA\cS_\cT =
(\cT_i)_{i\in\mathbb{N}}$, where $\cT_i$ does not denote an exact
transition in present context, but a segment of $\cA\cS_\cO$, which
we can view as an approximate transition.

However, a priori we cannot exclude that Bianchi type
$\mathrm{VI}_0$ transitions (double curvature transitions) and
Bianchi type $\mathrm{VII}_0$ orbits play an important role in the
context of asymptotic sequences $\cA\cS_\cO$. A segment of
$\cA\cS_\cO$ that is approximated by a type
$\mathrm{VI}_0$/$\mathrm{VII}_0$ orbit---an approximate type
$\mathrm{VI}_0$/$\mathrm{VII}_0$ orbit in our nomenclature---is
expected to connect a neighborhood of the point $\mathrm{T}_3$ with
a neighborhood of one of the fixed points on sectors $(312)$ or
$(321)$ on $\mathrm{K}^\ocircle$. Clearly, since the accuracy of the
approximation, i.e., the degree of shadowing, improves as
$\tau\rightarrow \infty$, these neighborhoods are expected to shrink
with increasing $\tau$. However, this does not automatically lead to
an exclusion of type $\mathrm{VI}_0$/$\mathrm{VII}_0$ orbits from
generic asymptotic sequences. Instead the exclusion of type
$\mathrm{VI}_0$/$\mathrm{VII}_0$ orbits follows from a
bootstrap-like argument which we present next.

%We conjecture that the errors appear as random errors in
%this map; since the asymptotic sequences of transitions
%approach exact sequences in the limit $\tau\rightarrow \infty$
%the errors converge to zero in that limit.
%As the ever decreasing errors are essentially random
%the map~\eqref{useqapprox} can be viewed as
%a randomized Kasner map. This randomization justifies the probabilistic
%description of approximate Kasner sequences $(u_l)_{l\in\mathbb{N}}$
%by means of a probability density, see Section~\ref{stat}.

%---------------------------------------------------------
\subsection*{$\bm{R_2\rightarrow 0}$ and
$\bm{N_2\rightarrow 0}$ as $\bm{\tau\rightarrow \infty}$}
\label{R2N2go}
%---------------------------------------------------------

Consider an asymptotic sequence $\cA\cS_\cO$.
Equations~\eqref{unstableeq3}--\eqref{unstableeq10} imply that the
product $(R_1 R_3)$ satisfies
%
%\begin{subequations}\label{RRRB}
\begin{equation*}
\partial_\tau \big(R_1 R_3\big) =
4 (1-\Sigma^2) R_1 R_3 + (\Sigma_1 -\Sigma_3) R_1 R_3 + 2 (R_3^2
-R_1^2) R_2
\end{equation*}
with increasing accuracy as $\tau\rightarrow \infty$. Using the
auxiliary quantity $B$, which is subject to the auxiliary
equation~\eqref{A2eq}, we obtain an identical equation for $(R_2
B)$, i.e.,
\begin{equation*}
\partial_\tau \big(R_2 B\big) =
4 (1-\Sigma^2) R_2 B + (\Sigma_1 -\Sigma_3) R_2 B + 2 (R_3^2 -R_1^2)
R_2\:.
\end{equation*}
%\end{subequations}
%
By choosing appropriate initial data for $B$, it follows that $(R_2
B)$ coincides with $(R_1 R_3)$. Accordingly we obtain
\begin{equation}\label{RRRBA}
R_2 B = R_1 R_3
\quad\text{ or } \quad |R_2| \, \sim \, \big|R_1 R_3\big|\:\,
B^{-1}\:.
\end{equation}

Generically, along an asymptotic sequence of transitions
$\cA\cS_\cT$ the quantity $B$ satisfies $B_i \sim \mathrm{const}\:
G_{l(i)}$ in the asymptotic regime $i\rightarrow \infty$, where
$G_l$ is the growth function of Section~\ref{growth}; this is
because $\cA\cS_\cT$ is approximated by a sequence of transitions
$\cS_\cT$ in the limit $\tau\rightarrow \infty$.
%, i.e., $B$ grows unboundedly as $\tau\rightarrow \infty$.
Along an asymptotic sequence $\cA\cS_\cO$ (which, in contrast to
$\cA\cS_\cT$, might also contain type
$\mathrm{VI}_0$/$\mathrm{VII}_0$ orbits), the behavior of $B$
described by the growth function holds at least piecewise, where the
interruptions are associated with the approximate type
$\mathrm{VI}_0$/$\mathrm{VII}_0$ orbits of $\cA\cS_\cO$; when the
solution shadows a Bianchi type $\mathrm{VI}_0$/$\mathrm{VII}_0$
orbit, a large increase in $B$ ensues, see Section~\ref{growth}.
Since type $\mathrm{VI}_0$/$\mathrm{VII}_0$ orbits thus strengthen
the growth of $B$ in relation to the normal growth described by
$G_{l(i)}$, we conclude that the relation $B_i\sim \mathrm{const}
\,G_{l(i)}$ can be regarded as representing a lower bound for the
growth of $B$.

We can therefore write~\eqref{RRRBA} as $\big|R_2\big|_i \sim \:
\mathrm{const}\:\, \big|R_1 R_3\big|_i\:\, G_{l(i)}^{-1}$ as
$i\rightarrow \infty$, or in simplified notation,
\begin{equation}\label{R2neglect}
|R_2| \sim \: \mathrm{const}\:\, \big|R_1 R_3\big|\:\, G^{-1}\:.
\end{equation}
Since $|R_1 R_3|$ is bounded because of the Gauss
constraint~\eqref{gausscon}, Eq.~\eqref{R2neglect} implies that
$|R_2| \rightarrow 0$ as $\tau\rightarrow \infty$, i.e., generically
$R_2$ is asymptotically suppressed.
%\marginalnote{This is very fast! Because of $|R_1 R_3|$.}
Analogously, $|N_2| \rightarrow 0$ as $\tau\rightarrow \infty$,
which we show in the following.

%To show the asymptotic suppression of $N_2$
%along asymptotic sequences of transitions $\cA\cS_\cT$, i.e.,
%$|N_2| \rightarrow 0$ as $\tau\rightarrow \infty$,
%we first
Consider the product $(N_1 R_3^2)$;
from~\eqref{unstableeq3}--\eqref{unstableeq10} we have
\begin{equation*}%\label{N1R3prov}
\partial_\tau \big(N_1 R_3^2\big) =
2 \left[2 (1-\Sigma^2) -\Sigma^2 - \Sigma_2\right] N_1 R_3^2 - 4 R_1
R_2 R_3 N_1\:.
\end{equation*}
Since, generically, $R_2 \sim \mathrm{const} \, (R_1 R_3)\, G^{-1}$
by~\eqref{R2neglect} we obtain
\begin{equation}\label{N1R32}
\partial_\tau \big(N_1 R_3^2\big) =
2 \left[2 (1-\Sigma^2) -\Sigma^2 - \Sigma_2\right] N_1 R_3^2 - 4
R_1^2 G^{-1} (N_1 R_3^2)\:.
\end{equation}
In the asymptotic regime $\tau\rightarrow \infty$ the last term
in~\eqref{N1R32} vanishes; in fact, by using the results of the
previous sections we show in Appendix~\ref{convergenceofsums} that
the term falls off rapidly enough so that
\begin{equation}\label{N1R3}
\partial_\tau \big(N_1 R_3^2\big) =
2 \left[2 (1-\Sigma^2) -\Sigma^2 - \Sigma_2\right] N_1 R_3^2
\end{equation}
holds asymptotically. Similarly, using the auxiliary quantity $A$
leads to the equation
\begin{equation*}
\partial_\tau \big(N_2 A^2\big) =
2 \left[2 (1-\Sigma^2) -\Sigma^2 - \Sigma_2\right] N_2 A^2
\end{equation*}
for $(N_2 A^2)$. By choosing appropriate initial data for $A$, we
conclude that
\begin{equation*}
|N_2 A^2| \,\sim\, \mathrm{const} \: |N_1 R_3^2|\:;
\end{equation*}
equivalently, by using~\eqref{Ainc}, we can write $|N_2|_i \sim
\mathrm{const}\, |N_1 R_3^2|_i\, G_{l(i)}^{-2}$, or simply,
\begin{equation}\label{N2neglect}
|N_2| \sim\: \mathrm{const} \:\, |N_1 R_3^2|\: G^{-2} \:.
\end{equation}
Since $|R_3^2|$ and $|N_1 -N_2|$ are bounded because of~\eqref{gausscon},
%~\eqref{gausscon}, \eqref{asyCoda},
%\marginalnote{cu: I think this
%is correct but that we have neglected a subtlety here; for type
%VII$_0$ the state space is noncompact, however, that $N_1$ and $N_2$
%can tend to infinity is associated with that they become equal and
%that we obtain the line of fixed points; that they become equal is
%in conflict with the present equation --- we obtain a contradiction,
%and hence we obtain boundedness. I added a reference to the
%asymptotic Codazzi constraint which is of relevance here. Should we
%comment on any of this?}
Eq.~\eqref{N2neglect} implies that,
generically, $|N_2| \rightarrow 0$ as $\tau\rightarrow \infty$.
% note that $N_2$ converges to zero approximately like $R_2^2$.

%---------------------------------------------------------
\subsection*{Exclusion of $\bm{R_2}$ and $\bm{N_2}$}
\label{excluR2}
%---------------------------------------------------------

Here we establish the following result: the statement $(R_2, N_2)
\rightarrow 0$ ($\tau\rightarrow \infty$) does not merely mean that
the amplitudes in $R_2$ and $N_2$ are decreasing, but it also leads
to a complete or stochastic exclusion of the shadowing of orbits
that involve $R_2$ or $N_2$, i.e., $\cT_{R_2}$, $\cT_{N_2}$,
$\cT_{R_2 N_2}$, $\cT_{R_1 R_2 R_3}$, $\cT_{N_1 N_2}$ transitions
and Bianchi type $\mathrm{VII}_0$ orbits.

In Section~\ref{dynosc} and Appendix~\ref{multiple} we have proved
that there are no Bianchi type $\mathrm{VI}_0$ ($\cT_{N_1N_2}$) and
$\mathrm{VII}_0$ orbits with arbitrarily small amplitudes: there
exists $\epsilon > 0$ such that $\max |N_2| \geq \epsilon$ uniformly
for all type $\mathrm{VI}_0$/$\mathrm{VII}_0$ orbits. Since,
generically, $|N_2| \rightarrow 0$ along asymptotic sequences
$\cA\cS_\cO$, this entails that Bianchi type $\mathrm{VI}_0$ and
$\mathrm{VII}_0$ orbits cannot be contained in generic sequences
$\cA\cS_\cO$ in the asymptotic regime, as for sufficiently large
$\tau$, $|N_2|$ satisfies $|N_2| <\epsilon$. Since the shadowing of
Bianchi type $\mathrm{VI}_0$ and $\mathrm{VII}_0$ orbits is thus
completely excluded, asymptotic sequences $\cA\cS_\cO$ are sequences
of transitions $\cA\cS_\cT$, which consist of Bianchi type I/II
transitions only. In brief: a generic $\cA\cS_\cO$ must be an
$\cA\cS_\cT$.
%
%\marginalnote{cu: should we dare insert a comment of the following
%type: (Incidentally, the reasoning here can be formalized and
%adapted to Bianchi types VI$_{-1/9}$ and VIII as pieces in new
%singularity proofs for these models). To me, particularly in the
%compact type VI$_{-1/9}$ case, it seems that we almost get LOCAL
%asymptotic new singularity proofs; assume initial data so that type
%VI, VII transitions cannot take place; the difficulty lies in
%showing that you do not have an infinite repetition of (approximate
%asymptotically more exact) rational $u$-values (now we have the
%`unlikely' comment about this); what do you think?}

Let us now consider the remaining transitions that involve the
excitation of one of the variables $R_2$ or $N_2$, i.e., $\cT_{R_2}$,
$\cT_{N_2}$, $\cT_{R_2 N_2}$, $\cT_{R_1 R_2 R_3}$ transitions.
Consider, for instance, the case of $\cT_{R_2}$ transitions: for a
given $\cT_{R_2}$ transition the quantity $|R_2|$ goes through a
finite maximum $r$ along the transition orbit, which can be computed
via the Gauss constraint~\eqref{gausscon}. However, since $|R_2|
\rightarrow 0$ as $\tau\rightarrow \infty$ by~\eqref{R2neglect},
$|R_2|$ can never reach the value $r$, and hence the given
transition cannot take place when $\tau$ is sufficiently large.

It follows that more and more $\cT_{R_2}$ transitions are excluded
as $\tau\rightarrow \infty$, so that merely transitions in
increasingly smaller neighborhoods of $\mathrm{T}_2$ and
$\mathrm{Q}_2$ are possible; however, it is a priori not clear
whether these transitions `die out' completely after some time, or
if there are infinitely many as $\tau\rightarrow \infty$. Consider,
e.g., a neighborhood of the point $\mathrm{T}_2$: the repeated
occurrence of alternating sequences of $\cT_{N_1}$ and $\cT_{R_2}$
transitions (generating small curvature phases) is quite plausible a
priori, particularly in the light of the statement that small
curvature phases dominate in sequences of transitions $\cS_\cT$ and
thus in the asymptotic dynamics of $\cA\cS_\cT$; see the discussion
in Section~\ref{stat}. However, in the following we establish a
proposition that implies that $\cT_{R_2}$ (and $\cT_{N_2}$)
transitions are excluded \textit{stochastically} for generic
sequences $\cA\cS_\cT$:
\begin{proposition}[Stochastic exclusion of $R_2$ and $N_2$]
\label{subdomprop} A generic asymptotic sequence $\cA\cS_\cT$
contains a finite number of $\cT_{R_2}$, $\cT_{R_1 R_2 R_3}$,
$\cT_{N_2}$, and $\cT_{N_2 R_2}$ transitions only.
\end{proposition}

As for the previous results, the word `generic' is important also in
this context; there might exist sequences $\cA\cS_\cT$ that contain
an infinite number of transitions that involve $R_2$ or $N_2$;
however, these sequences form a set of measure zero in the space of
all sequences. Recall further that asymptotic sequences of
transitions $\cA\cS_\cT$ are the generic case of asymptotic
sequences $\cA\cS_\cO$; hence, in the proposition it is equivalent
to write $\cA\cS_\cO$ instead of $\cA\cS_\cT$.

The validity of the proposition relies on stochastic arguments,
which we present in the remainder of this subsection. For simplicity
we focus on the statement that there is a finite number of
$\cT_{R_2}$ transitions in a generic asymptotic sequence
$\cA\cS_\cT$. The treatment of the other transitions is analogous.

Consider an arbitrary $\cT_{R_2}$ frame transition. Along the
$\cT_{R_2}$ orbit the quantity $|R_2|$ goes through a maximum $r$
that is determined by the Gauss constraint~\eqref{gausscon}; it is
straightforward to express $r$ in terms of the Kasner parameter $u$
that characterizes the $\cT_{R_2}$ transition: in particular, if $u
\gg 1$ (which corresponds to a $\cT_{R_2}$ transition that takes
place in a small neighborhood of $\mathrm{T}_2$), we obtain $r \sim
3 u^{-1}$; if $(u-1) \ll 1$, then the corresponding $\cT_{R_2}$
transition is close to $\mathrm{Q}_2$ and $r \sim (u-1)$;
cf.~Figure~\ref{singletrans}.

By Eq.~\eqref{R2neglect}, along $\cA\cS_\cT$ the quantity $|R_2|$
decreases with time at a rate given by $|R_2|_{l(i)} \propto
G_l^{-1}$; hence, a necessary condition for an (approximate)
$\cT_{R_2}$ transition to take place after the
$l$\raisebox{0.7ex}{\small th} (approximate) curvature transition of
$\cA\cS_\cT$ is that $u$ be sufficiently large (or $(u-1)$ be
sufficiently small):
\begin{equation}\label{occur}
\text{An occurrence of a $\cT_{R_2}$ transition close to } \left\{
\begin{matrix}
\mathrm{T}_2 \\[0.3ex]
\mathrm{Q}_2
\end{matrix}
\right\}\, \text{is only possible if } \left\{
\begin{matrix}
u_l^{-1} < c\, G_l^{-1} \\[0.3ex]
u_l -1 < c\, G_l^{-1}
\end{matrix}
\right\}\:,
\end{equation}
where $c$ is a constant. Based on these considerations we find that
establishing the proposition amounts to proving that for generic
sequences the conditions in~\eqref{occur} %, $u_l^{-1} < c G_l^{-1}$,
are satisfied only for finitely many $l$ as $l\rightarrow \infty$.
For simplicity we focus on the first case of~\eqref{occur}, i.e., we
investigate the condition $u_l^{-1} < c\, G_l^{-1}$; it is
straightforward to adapt the results to the second case.

Since asymptotic sequences $\cA\cS_\cT$ shadow sequences of
transitions $\cS_\cT$, we can adopt the concepts introduced in
Section~\ref{transition}; in particular, to obtain a more
transparent formulation of the problem, we make use of the concept
of eras: we view $\cA\cS_\cT$ as a succession of eras $[(\lin)_a,
(\lout)_a]$ ($a=0,1,2,3,\ldots$), and denote by $\varkappa_a \in
(0,1)$ the value of $u^{-1}$ at the beginning of era $a$, i.e.,
$\varkappa_a = (\uinn{a})^{-1}$. Furthermore, we define $\delta_a$
to be the value of the function on the r.h.s.\ of
condition~\eqref{occur} at the beginning of era $a$, i.e.,
\begin{equation*}
\delta_a = c\, G_{(\lin)_a}^{-1} =c\, G_{(\lout+1)_{a-1}}^{-1} = c\,
\mathcal{G}_{a}^{-1}\:,
\end{equation*}
where $\mathcal{G}_a$ is the era growth function defined in
Section~\ref{growth}. Without loss of generality we assume the
initial value $\delta_0$ of the sequence to be sufficiently small,
i.e., $\delta_0 < \varepsilon$. Based on these concepts, the first
case of~\eqref{occur} translates to the following necessary
condition for the occurrence of $\cT_{R_2}$ transitions:
\begin{equation}\label{eraoccur}
\text{An occurrence of a $\cT_{R_2}$ transition close to
$\mathrm{T}_2$ in era $a$ is only possible if } \varkappa_a <
\delta_a\:.
\end{equation}
To establish the proposition we show that for generic sequences the
condition $\varkappa_a < \delta_a$ is satisfied for a finite number
of eras only.

The arguments we present in the following are stochastic in nature.
Our basic assumption is that we can model the sequence
$(\varkappa_a)_{a\in\mathbb{N}}$ by a random variable $\varkappa$ on
$(0,1)$ with a well-defined probability density. The justification
for this assumption and its origin are discussed in the subsection
`Randomized Kasner sequences' below. For the subsequent computations
we employ the density $w(\varkappa)$ that describes the stochastics
of Kasner sequences; however, the basic qualitative results are
largely independent of the concrete form of the probability density,
see the remarks at the end of this subsection.

Since $\varkappa$ is a random variable, the sequence
$(\varkappa_a)_{a\in\mathbb{N}}$ represents an infinite series of
trials with $\varkappa$. The sequence $(\delta_a)_{a\in\mathbb{N}}$,
on the other hand, represents a sequence $(0,\delta_a)$ of intervals
whose length decreases as $a\rightarrow \infty$. The
condition~\eqref{eraoccur} thus reads: the number of eras that can
contain $\cT_{R_2}$ transitions coincides with the number of `hits'
$\varkappa_a \in (0, \delta_a)$ as $a\rightarrow \infty$.

In Appendix~\ref{hitting} we demonstrate that the behavior of the
sequence $(\delta_a)_{a\in\mathbb{N}}$ is decisive for the `hit
rate': $\delta_a \sim a^{-1}$ leads to a diverging expectation value
for the number of hits as $a\rightarrow \infty$. In our case this
would lead to the conclusion that an infinite number of eras can
contain $\cT_{R_2}$ transitions (although the recurrence intervals
between $\cT_{R_2}$ transitions would become increasingly large). A
sequence $\delta_a \sim a^{-2}$, on the other hand, generates a
finite expectation value; in that case only a finite number of eras
is expected to contain $\cT_{R_2}$ transitions.

The present situation is more complicated since
$(\delta_a)_{a\in\mathbb{N}}$ is not given explicitly, but is
instead determined by a stochastic process. We have the relation
\begin{equation*}
\delta_{a+1} = c\, \mathcal{G}_{a+1}^{-1} = c\, \mathcal{G}_{a}^{-1}
\mathfrak{g}_a^{-1} = \mathfrak{g}_a^{-1}\, \delta_a\:,
\end{equation*}
see Section~\ref{growth}, and by using~\eqref{gfera}
and~\eqref{eragrowthfac},
\begin{equation*}
\delta_{a+1} < \frac{3}{1+\uinn{a}+(\uinn{a})^2} \, \delta_a =
\frac{3 \varkappa_a^2}{1+\varkappa_a + \varkappa_a^2}\,\delta_a\:.
\end{equation*}
Equivalently we can write
\begin{equation*}
\log \delta_{a+1} -\log \delta_a < \log 3 + \log \varkappa_a^2 -
\log (1+\varkappa_a +\varkappa_a^2)\:.
\end{equation*}

Employing the probability density $w(\varkappa)$ for $\varkappa$,
see~\eqref{probdens}, we find an estimate for the expectation value
of $\log \delta_{a+1}/\delta_a$,
\begin{equation*}
\langle \log \delta_{a+1} -\log \delta_a \rangle < \log 3 + \int_0^1
w(\varkappa) \left(\log \varkappa^2 - \log (1+\varkappa
+\varkappa^2)\right) d\varkappa \approx -1.77 < -1\:,
\end{equation*}
see also~\eqref{expectfrakg}. Accordingly, for $a \gg 1$ we expect
\begin{equation*}
\langle \log \delta_a \rangle < -a  + \log \delta_0 < -a +\log
\varepsilon\:.
\end{equation*}
Consequently, the condition~\eqref{eraoccur}, i.e., $\varkappa_a <
\delta_a$ (or, equivalently, $\log\varkappa_a < \log \delta_a$)
translates to the following statement:
\begin{equation*}
\text{An occurrence of a $\cT_{R_2}$ transition close to
$\mathrm{T}_2$ in era $a$ is only possible if } \varkappa_a <
\varepsilon e^{-a}\:.
\end{equation*}

Based on these considerations we can compute the probabilities for
`$N$ hits' (cf. Appendix~\ref{hitting}). For $a\gg 1$ the
probability that $\varkappa\in (0,\varepsilon e^{-a})$ is given by
\begin{equation*}
\mathrm{P}\big(\varkappa \in (0, \varepsilon e^{-a})\big) =
\int_0^{\varepsilon e^{-a}} w(\varkappa) d\varkappa =:
\tilde{\delta}_a \approx \varepsilon w(0) e^{-a}\:.
\end{equation*}
For a series of $A$ trials we find that the probability
$\mathrm{P}^A(0)$ (`no hit', i.e., the probability for the scenario
$\varkappa \not\in [0,\varepsilon e^{-a}]$ $\forall a =1, \ldots,
A$) is given by
\begin{equation*}
\log \mathrm{P}^A(0) = \sum_{a=1}^A \log(1 - \tilde{\delta}_a)
\approx \sum_{a=1}^A \big( - \varepsilon w(0) e^{-a} +
O(\varepsilon^2 e^{-2 a})\big) \approx - \varepsilon w(0) \frac{1-
e^{-A}}{1-e^{-1}}\:.
\end{equation*}
Similarly, the probability for `one hit' is
\begin{equation*}
\mathrm{P}^A(1) = \mathrm{P}^A(0) \sum_{a=1}^A
\frac{\tilde{\delta}_a}{1-\tilde{\delta}_a} \approx  \mathrm{P}^A(0)
\left[\varepsilon w(0)  \sum_{a=1}^A e^{-a} \right] \approx
\mathrm{P}^A(0) \left[ \varepsilon w(0)  \frac{1- e^{-A}}{1-e^{-1}}
\right] \:,
\end{equation*}
and the probability for `two hits' in a series of $A$ trials,
$\mathrm{P}^A(2)$, is given by
\begin{equation*}
\frac{\mathrm{P}^A(2)}{\mathrm{P}^A(0)}= \sum_{a<b}
\frac{\tilde{\delta}_a}{1-\tilde{\delta}_a}\frac{\tilde{\delta}_b}{1-\tilde{\delta}_b}
\approx \frac{1}{2} \left[ \Big(\sum_a
\frac{\tilde{\delta}_a}{1-\tilde{\delta}_a}\Big)^2 - \sum_a
\Big(\frac{\tilde{\delta}_a}{1-\tilde{\delta}_a} \Big)^2 \right] <
\frac{1}{2} \Big[ \varepsilon w(0)  \frac{1- e^{-A}}{1-e^{-1}}
\Big]^2\:.
\end{equation*}
Letting $A\rightarrow \infty$ we obtain
\begin{equation}\label{Pp012}
\mathrm{P}^\infty(0) \approx e^{-p} \approx 1 - p\:,\quad
\mathrm{P}^\infty(1) < p\:,\quad \mathrm{P}^\infty(2) <
\frac{p^2}{2}\:,\quad \ldots\,,
\end{equation}
where $p = \varepsilon w(0) (1-e^{-1})^{-1}$. We conclude that the
expectation value of the number of hits is finite:
\begin{equation*}
\langle \# \text{hits}\rangle = \sum_{i=1}^\infty
i\,\mathrm{P}^\infty(i)  < \sum_{i=1}^\infty  i \frac{p^i}{i!} =
p\,e^p \,<\, \infty
\end{equation*}
Since the `number of hits' corresponds to the number of eras that
contain $\cT_{R_2}$ transitions, we have shown that the number of
eras that contain $\cT_{R_2}$ transitions is expected to be finite
and thus we have established the proposition.

It is important to stress that the arguments of the proof rely on
the assumption that the sequence $\cA\cS_\cT$ is generic. By
stochastic methods it is impossible to exclude that there exist
(non-generic) sequences that contain an infinite number of
$\cT_{R_2}$ (or $\cT_{N_2}$, $\cT_{N_2 R_2}$, $\cT_{R_1 R_2 R_3}$)
transitions.

We conclude this subsection with the observation that the above
results are to a large extent independent of the concrete form of
the probability density $w(\varkappa)$; e.g., it is sufficient to
assume that $w(0)$ is finite to obtain the result~\eqref{Pp012}.
This is essentially due to the rapid decrease of $\delta_a$, which
is an exponential decay that stems from the (era) growth function.

Let us finally summarize the results of this subsection: Bianchi
type $\mathrm{VI}_0$ and $\mathrm{VII}_0$ orbits are excluded from
generic asymptotic sequences $\cA\cS_\cO$, therefore a generic
asymptotic sequence $\cA\cS_\cO$ is an asymptotic sequence of
transitions $\cA\cS_\cT$. Generically sequences $\cA\cS_\cT$ do not
contain transitions that involve an excitation of the variables
$R_2$ and $N_2$, i.e., the transitions $\cT_{N_2 R_2}$, $\cT_{R_2}$,
$\cT_{N_2}$, and $\cT_{R_1 R_2 R_3}$ do not appear. The arguments
leading to the stochastic exclusion of $R_2$ and $N_2$ are based on
the stochastic properties of asymptotic sequences; next we discuss
the stochasticity of asymptotic sequences in more detail.

%---------------------------------------------------------
\subsection*{Randomized Kasner sequences}
\label{randomiK}
%---------------------------------------------------------

In Section~\ref{transition} we have seen that generic sequences of
transitions $\cS_\cT$ are associated with Kasner sequences
$(u_l)_{l\in\mathbb{N}}$, which are determined by the
recursion~\eqref{useq}, where the initial value $u_0 \not\in
\mathbb{Q}$. In Section~\ref{stat} we have presented a stochastic
analysis of Kasner sequences: the collective distribution of all
sequences $(\uinn{j})_{j \in \mathbb{N}}$ (or, equivalently,
$(\varkappa_j)_{j\in\mathbb{N}} = (\uin_j)^{-1}_{j\in\mathbb{N}}$)
possesses a probabilistic description in terms of a probability
density $w$. In the context of asymptotic sequences of transitions
$\cA\cS_\cT$, however, we do not deal with exact Kasner sequences;
in the following we investigate the resulting consequences for the
probabilistic aspects.

An asymptotic sequence of transitions $\cA\cS_\cT$ generates an
approximate Kasner map, i.e.,
\begin{equation}\label{useqapprox}
u_{l+1}  \:\approx \:
\begin{cases}
u_l - 1 & \quad\text{if}\quad u_l >  2 \\
(u_l - 1)^{-1} &  \quad\text{if} \quad 1 < u_l < 2\: .
\end{cases}
\end{equation}
Since asymptotic sequences $\cA\cS_\cT$ shadow sequences $\cS_\cT$
to an increasing degree of accuracy, the errors
in~\eqref{useqapprox} converge to zero as $\tau\rightarrow \infty$.
%
%\marginalnote{jmh: since there are these remarks here, we need less
%at the end of Sec.~\ref{stat}.} Above we have described exact Kasner
%sequences $(u_l)_{l\in\mathbb{N}}$, associated with sequences of
%transitions $\cS_\cT$ on the oscillatory subset $\mathcal{O}$.
%However, exact Kasner sequences are an unattainable idealization. A
%solution that approaches a sequence of transitions generates an
%\textit{approximate Kasner sequence} that only satisfy the Kasner
%map~\eqref{useq} approximately; since the `errors' this introduce
%are, presumably, essentially random in nature, these `physical'
%sequences can be said to satisfy a randomized Kasner map, where the
%errors become smaller with time. Errors are not always a bad thing:
%as noted above, not all exact Kasner sequences can be described by
%means of the probability density $\hat{w}(\varkappa)$, however, due
%to random errors, $\hat{w}(\varkappa)$ applies to approximate Kasner
%sequences.
%
We presume that the nature of these errors is random, which is due
to the fact that the approach $\cA\cS_\cT \rightarrow \cS_\cT$ is
effectively unpredictable. This random element in the approach
$\cA\cS_\cT \rightarrow \cS_\cT$ is the key to a probabilistic
description of an approximate Kasner sequence~\eqref{useqapprox}:
while an exact Kasner sequence might not be described by the
collective probability density $w(\varkappa)$ (or alternatively
$\hat{w}(\varkappa)$), the random errors in~\eqref{useqapprox} smear
out the probability distributions of exact sequences over the entire
collective; we thus expect the sequence
$(\varkappa_j)_{j\in\mathbb{N}}$ associated with an individual
generic randomized asymptotic Kasner sequence to be described by the
probability density $w(\varkappa)$ of Section~\ref{stat}. In brief,
we can regard~\eqref{useqapprox} as a randomized asymptotic Kasner
sequence which admits a probabilistic description in terms of the
density $w(\varkappa)$.

The errors in~\eqref{useqapprox} are not uniform in $u$ (i.e.,
independent of the value of $u$): the Kasner map $u \mapsto u -1$
(or $u\mapsto (u-1)^{-1}$) is relatively robust when $u$ takes a
value that is far from the extremes (which are $u=1$ and
$u=\infty$), i.e., when $u \in (1+\varepsilon, \varepsilon^{-1})$
for some $\varepsilon$; in contrast, the errors in the Kasner map
might be large when $u^{-1}$ or $(u-1)$ is small. To see this we
consider the (approximate) curvature transitions that
generate~\eqref{useqapprox}. When $u \in (1+\varepsilon,
\varepsilon^{-1})$ the associated curvature transition connects two
points on the Kasner circle $\mathrm{K}^\ocircle$ where the
(angular) gradient of $u$ (i.e., the gradient of $u$ along the
Kasner circle) is comparatively small, see Figure~\ref{uplot}. A
perturbation of the curvature transition, i.e., a deviation from the
exact orbit, thus leads to a small error in the Kasner map. In
contrast, when $u > \varepsilon^{-1}$, the associated curvature
transition takes place in a small neighborhood of one of the Taub
points, where the angular gradient of $u$ is large, see
Figure~\ref{uplot}; hence, any deviation from the exact curvature
transition orbit leads to a large error in the $u$-map. This effect
is amplified by the local characteristics in the approach
$\cA\cS_\cT \rightarrow \cS_\cT$: we expect the accuracy of the
shadowing to be smaller in the neighborhood of the Taub points (and
the points $\mathrm{Q}_\alpha$) than elsewhere.
%The neighborhood of the Taub points is dangerous in
%all respects:
The dynamics in the neighborhood of the Taub points is rather
complicated which is due to the fact that these fixed points are not
transversally hyperbolic: the approach to the attractor is expected
to be rather slow, and for fixed points close to the Taub points the
actual dynamics is expected to display relatively large deviations
from the linearized dynamics (more accurately described by `eras of
small oscillations'). It is evident that this leads to an additional
blow-up of errors in the $u$-map for large $u$.

Based on these consideration we expect that the Kasner map is not
well-defined unless $u \in (1+\varepsilon, \varepsilon^{-1})$; only
for this `safe range' of $u$ we can assume that~\eqref{useqapprox}
holds. However, in the asymptotic regime the safe range increases,
i.e., $\varepsilon \rightarrow 0$ as $\tau\rightarrow \infty$, which
is simply because $\cA\cS_\cT \rightarrow \cS_\cT$. Suppose that the
safe range grows rapidly enough. Then we can draw the following
conclusion: the probability that the sequence
$(u_l)_{l\in\mathbb{R}}$ ever leaves the safe range is zero. To see
this, we use the same arguments that led to the stochastic exclusion
of $R_2$ and $N_2$: the `number of hits' $\varkappa_j \in (0,
\epsilon_j)$ (corresponding to the Kasner sequence leaving the safe
range) has a finite expectation value.

It is plausible that the increase of the safe range is in fact an
exponential growth, since it is directly related to the accuracy of
the shadowing. The decay rates of $R_2$ and $N_2$ suggest that the
overall accuracy of the shadowing behaves in the same way, which in
turn leads to the required rapid growth of the safe range. Presuming
the correctness of these considerations we find that generic
asymptotic Kasner sequences are indeed described
by~\eqref{useqapprox}, where the errors converge to zero (uniformly)
as $\tau\rightarrow \infty$.

%%%%%%%%%%%%%%%%%%%%%%%%%%%%%%%%%%%%%%%%%%%%%%%%%%%%%%%%%%%%%%%%%%
\section{The billiard attractor}
%\label{billiardsubset}
\label{billiardattractor}
%%%%%%%%%%%%%%%%%%%%%%%%%%%%%%%%%%%%%%%%%%%%%%%%%%%%%%%%%%%%%%%%%%

In the previous section we have argued that for generic solutions
the variables $R_2$ and $N_2$ are not excited in the asymptotic
regime, i.e., we have established the exclusion of transitions that
involve the variables $R_2$ and $N_2$ from generic asymptotic
sequences of transitions. Accordingly, the essential asymptotic
dynamics of generic solutions $\bm{X}(\tau)$ is described not by the
full oscillatory dynamical
system~\eqref{unstableeq3}--\eqref{unstableeq10}, but instead by a
reduced system that is obtained by setting $R_2 =0$ and $N_2 =0$.

%---------------------------------------------------------
\subsection*{The billiard subset}
\label{billsubse}
%---------------------------------------------------------

We define the \textit{billiard subset} $\mathcal{O}_{\mathcal{B}}$
as the invariant subset $(R_2 = 0) \wedge (N_2 = 0)$ of the
oscillatory subset $\mathcal{O}$, which leaves
$\bm{S}_{\mathrm{billiard}} = (\Sigma_1,\Sigma_2,\Sigma_3, R_1, R_3,
N_1)$ as the state vector.

The system of differential equations on $\mathcal{O}_{\mathcal{B}}$
is given by
\begin{subequations}\label{billiardeqs}
\begin{eqnarray}
\label{billiardeq1}
\partial_{\tau} \Sigma_1 & = &
2(1-\Sigma^2)\Sigma_1 - 2R_3^2 + \textfrac{2}{3}N_1^2 \\
\label{billiardeq2}\partial_{\tau} \Sigma_2 & = &
2(1-\Sigma^2)\Sigma_2 -2(R_1^2 - R_3^2) - \textfrac{1}{3}N_1^2 \\
\label{billiardeq3}\partial_{\tau}\Sigma_3 & = &
2(1-\Sigma^2)\Sigma_3 + 2R_1^2 - \textfrac{1}{3}N_1^2 \\
\label{billiardeq4} \partial_{\tau} R_1 & = & 2(1-\Sigma^2)R_1 +
(\Sigma_2 - \Sigma_3)R_1 \\
\label{billiardeq5}
\partial_{\tau} R_3 & = & 2(1-\Sigma^2)R_3 +
(\Sigma_1 - \Sigma_2)R_3 \\
\label{billiardeq6}
\partial_{\tau} N_1 & = & -2(\Sigma^2 + \Sigma_1)N_1\, ,
\end{eqnarray}
where
\begin{equation*}
\Sigma^2= \textfrac{1}{6}(\Sigma_1^2 + \Sigma_2^2 + \Sigma_3^2 + 2R_1^2
+ 2R_3^2)\, .
\end{equation*}
The Gauss constraint~\eqref{gausscon} reads
\begin{equation}\label{gausscon2}
1-\Sigma^2 - \textfrac{1}{12}N_1^2 = 0
\end{equation}
and the Codazzi constraint~\eqref{codazzi} takes the form
\begin{equation}\label{codazzibilliard}
R_3\,N_1=0 \:.
\end{equation}
\end{subequations}

\begin{figure}[htp]
\psfrag{a}[cc][cc]{$\Sigma_1$} \psfrag{b}[cc][cc]{$\Sigma_3$}
\psfrag{c}[cc][cc]{$\Sigma_2$} \psfrag{d}[cc][cc]{$T_1$}
\psfrag{e}[cc][cc]{$Q_2$} \psfrag{f}[cc][cc]{$T_3$}
\psfrag{g}[cc][cc]{$Q_1$} \psfrag{h}[cc][cc]{$T_2$}
\psfrag{i}[cc][cc]{$Q_3$} \psfrag{j}[cc][cc]{0}
\psfrag{k}[cc][cc]{$\sfrac{\pi}{3}$}\psfrag{l}[cc][cc]{$\sfrac{2\pi}{3}$}
\psfrag{m}[cc][cc]{$\pi$} \psfrag{n}[cc][cc]{$\sfrac{4\pi}{3}$}
\psfrag{o}[cc][cc]{$\sfrac{5\pi}{3}$}\psfrag{p}[cc][cc]{$2\pi$}
\psfrag{q}[cc][cc]{$u$} \psfrag{r}[cc][cc]{$\alpha$}
\psfrag{A}[cc][cc]{$R_3$} \psfrag{B}[cc][cc]{$N_1$}
\psfrag{C}[cc][cc]{$N_1,R_1$} \psfrag{D}[cc][cc]{$R_1$}
\psfrag{E}[cc][cc]{$R_1,R_3$}\psfrag{F}[cc][cc]{$R_3$} \centering
\includegraphics[height=0.45\textwidth]{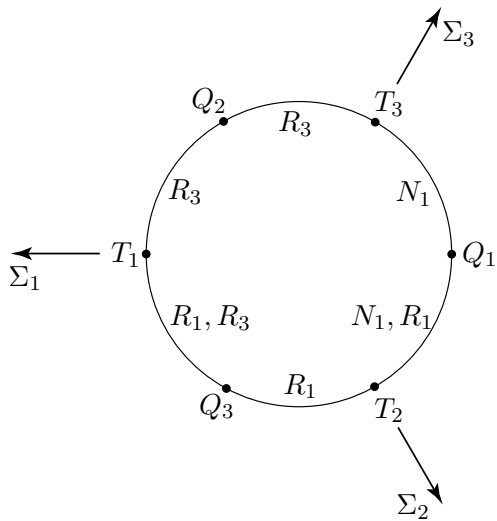}
\caption {Active triggers on $\mathrm{K}^\ocircle$ associated with
the billiard subset; note that $R_3\,N_1=0$.}
\label{billiardtriggers}
\end{figure}

Due to the Codazzi constraint the billiard subset consists of two
invariant components: the component $N_1 = 0$ is the Bianchi type I
subset $\overline{\mathcal{B}}_{R_3 R_1}$, which consists of
$\mathrm{K}^\ocircle \cup \mathcal{B}_{R_1} \cup \mathcal{B}_{R_3}
\cup \mathcal{B}_{R_3 R_1}$; the component $N_1 \neq 0$ is the
Bianchi type II subset, which is given by $\mathcal{B}_{R_1} \cup
\mathcal{B}_{N_1} \cup \mathcal{B}_{R_1 N_1}$. The flow of the
dynamical system on these subsets has been analyzed in
Section~\ref{dynosc}. A depiction of the flow on the boundary
subsets $\mathcal{B}_{R_1}$, $\mathcal{B}_{R_3}$ (single frame
transitions $\cT_{R_1}$, $\cT_{R_3}$) and $\mathcal{B}_{N_1}$
(single curvature transitions $\cT_{N_1}$) is given in
Figure~\ref{singletrans}; for the flow on $\mathcal{B}_{R_3 R_1}$
(double frame transitions $\cT_{R_3 R_1}$) and $\mathcal{B}_{R_1
N_1}$ (mixed curvature transitions $\cT_{R_1 N_1}$) see
Figure~\ref{doubletransition}. The analysis of the stability of the
fixed points on $\mathrm{K}^\ocircle$, as regards
$\mathcal{O}_{\mathcal{B}}$, is summarized in
Figure~\ref{billiardtriggers}.

A sequence of transitions on the billiard subset, which we denote by
$\cB_\cT$, is an infinite concatenation of transition orbits on
$\mathcal{O}_{\mathcal{B}}$. Each transition $\cT_i$ of $\cB_\cT$ is
either a frame transition $\cT_{R_1}$, $\cT_{R_3}$, $\cT_{R_1 R_3}$,
or a curvature transition $\cT_{N_1}$, $\cT_{R_1 N_1}$.
It is understood that the final state of each transition $\cT_i$
(which is a fixed point on $\mathrm{K}^\ocircle$) coincides
with the initial state of the transition $\cT_{i+1}$.
%to get a picture of the sequences of single transitions, see
%Figure~\ref{Kcurvtrans}--\ref{Kframetrans3}
Clearly, billiard sequences $\cB_\cT$ can be viewed as a special
case of the more general sequences of transitions $\cS_\cT$ on
$\mathcal{O}$.

%-------------------------------------------------------
\subsection*{The asymptotic suppression of multiple transitions}
\label{asysuppre}
%-------------------------------------------------------

The considerations of Section~\ref{asysha} can be summarized as
follows: the dynamical evolution of generic solutions $\bm{X}$ is
asymptotically silent and local, and the asymptotic behavior along
generic timelines is governed by the dynamics on the SH silent
boundary. At a spatial point $x^i$, the asymptotic evolution of
$\bm{X}(\tau)$ is thus represented by a solution $\bm{S} =
(\Sigma_{\alpha\beta}, A_\alpha,N_{\alpha\beta})$ of the finite
dimensional SH dynamical system. The stable variables
$\bm{S}_{\mathrm{stable}}= (N_{12}, N_{23}, N_{31}, A_\alpha)$
vanish in the limit $\tau\rightarrow 0$. Among the remaining
oscillatory variables, which are $\bm{S}_{\mathrm{osc}} =
(\Sigma_\alpha, R_\alpha, N_1, N_2)$, two more variables converge to
zero: $R_2$ and $N_2$. This leaves the \textit{billiard variables}
$\bm{S}_{\mathrm{billiard}} = (\Sigma_\alpha, R_1,R_3,N_1)$ as the
variables that capture the essential asymptotic dynamics.

The billiard variables are constrained by the asymptotic condition
\begin{equation*}
N_1 R_3 \rightarrow 0 \quad (\tau\rightarrow \infty)\:,
\end{equation*}
which is the asymptotic Codazzi constraint that ensues
from~\eqref{asyCoda}. In the same limit the variables
$\bm{S}_{\mathrm{billiard}}$ satisfy the billiard
system~\eqref{billiardeq1}--\eqref{billiardeq6} and the Gauss
constraint~\eqref{gausscon2} holds. The asymptotic
evolution of $\bm{S}_{\mathrm{billiard}}$ is thus governed by the
dynamics on the billiard subset $\mathcal{O}_{\mathcal{B}}$; in
other words: the state vectors $\bm{S}_{\mathrm{billiard}}$ of
solutions $\bm{X}(\tau)$ shadow sequences of transitions $\cB_\cT$
on the billiard subset. Accordingly, in the asymptotic regime the orbit
$\bm{S}_{\mathrm{billiard}}(\tau)$ can be partitioned into a
sequence of segments, where each segment is associated with a
transition on $\mathcal{O}_{\mathcal{B}}$. We call this
representation of the orbit $\bm{S}_{\mathrm{billiard}}(\tau)$ an
\textit{asymptotic billiard sequence} $\cA\cB_\cT$.
% (or, for brevity,asymptotic sequence)
Since a billiard sequence $\cB_\cT$ on $\mathcal{O}_{\mathcal{B}}$
is a concatenation of transitions of the types $\cT_{N_1}$,
$\cT_{R_1}$, $\cT_{R_3}$, $\cT_{R_3 R_1}$, $\cT_{N_1R_1}$, an
asymptotic billiard sequence $\cA\cB_\cT$ is a concatenation of
approximate transitions, i.e.,
$\cA\cB_\cT=(\cT_i)_{i\in\mathbb{N}}$, where each $\cT_i$ denotes an
approximate transition in the present context.

In asymptotic billiard sequences $\cA\cB_\cT$, however,
\textit{multiple transitions are suppressed} asymptotically, i.e.,
transitions of the types $\cT_{R_3 R_1}$, $\cT_{N_1R_1}$ do not
occur in the asymptotic regime. In the following we discuss this
statement; however, we focus on the simpler case of $\cT_{R_3 R_1}$
transitions.

From the differential equations~\eqref{billiardeq4}
and~\eqref{billiardeq5} we obtain
\begin{equation*}%\label{RRN}
\partial_\tau \log \frac{R_1}{R_3} = 3 \Sigma_2 \qquad \text{and thus}
\qquad
\log \frac{R_1}{R_3} \propto  3 \int \Sigma_2\, d\tau \:,
\end{equation*}
where we assume for simplicity and without loss of generality that
$R_1, R_3>0$. In an asymptotic billiard sequence $\cA\cB_\cT$ the
variables $R_1$ and $R_3$ (as well as $N_1$) are characterized by
oscillations between zero and finite maxima (which are bounded above
via the Gauss constraint); in particular, $R_1 \not\rightarrow 0$
and $R_3 \not \rightarrow 0$ as $\tau \rightarrow \infty$ (as well
as $N_1 \not \rightarrow 0$). For $\cT_{R_1}$ transitions, while
$R_1$ is of order unity, $R_3$ is small, so that $R_1/R_3$ is large;
conversely, for $\cT_{R_3}$ transitions, $R_1/R_3$ is small.

Consequently, $\log (R_1/R_3)$ and thus $\int \Sigma_2 d\tau$ is a
function that oscillates in the range $(-\infty,\infty)$. The
central property of these oscillations is that the amplitude
increases (rapidly) with increasing $\tau$, which is due to the
increasing accuracy of the shadowing. In the asymptotic regime we
can use a simple model for these oscillations which is based on the
observation that the orbit $\cA\cB_\cT$ spends an increasing amount
of $\tau$-time in a neighborhood of the Kasner fixed points, while
each transition always takes a fixed $\Delta\tau$ (depending on the
transition). We obtain $\int \Sigma_2 d\tau = \sum (\Sigma_2)_i
(\Delta \tau)_i$, where $i \in \mathbb{N}$ consecutively numbers the
Kasner fixed points that the orbit passes, and where, accordingly,
$(\Delta \tau)_i$ is the time the orbit spends in a neighborhood of
the Kasner point $i$. The quantities $(\Sigma_2)_i$ range in
$(-2,2)$; in conformance with our assumptions we expect the
distribution of the $(\Sigma_2)_i$ in $(-2,2)$ to have a
probabilistic description. Therefore, the sum $\sum (\Sigma_2)_i
(\Delta \tau)_i$ can be thought of as arising from a stochastic
process resembling a random walk $\sum s_i$; in the present case,
however, the step size $s_i$ is not fixed, but itself random; most
importantly, the step size increases with the number of steps taken.

The growth of the step size is of crucial importance. Consider an
interval $(-a,a)$ (where $a$ is small compared to the first step
size). If $s_i$ increases slowly with $i$, the probability that
$\sum_{i=1}^n s_i \in (-a,a)$ for some $n$ (recurrence probability)
is one. If the growth is sufficiently fast, however, this
probability is zero; this is case, for instance, if the growth rate
is geometric. The behavior of the quantity $\log (R_1/R_3)$ is
determined accordingly: for every $a$ there exists $\tau_a$ such
that the average step size is large compared to $a$ for all $\tau >
\tau_a$; the probability that $\log (R_1/R_3)$ lies in the interval
$(-a,a)$ for some $\tau > \tau_a$ is expected to vanish; this is
because the step size increases rapidly, which is in turn due to
the expectation that the orbit $\cA\cB_\cT$ shadows the attractor with
a rapidly increasing accuracy (so that $(\Delta \tau)_i$ grows
fast). Consequently, in the asymptotic regime, the probability of
the quantity $\log (R_1/R_3)$ being in any given interval around
zero vanishes. Since, therefore, either $R_1$ is small and $R_3$
large or $R_1$ large or $R_3$ small, our arguments suggest that
multiple transitions are suppressed as $\tau\rightarrow \infty$.

%%
%\marginalnote{How can that be?\\
%Experiment: start with multiple transition.\\
%It might well be the last one! \\
%see also statistics of $\Sigma_\alpha$ in billiard.}
%%

%-------------------------------------------------------
\subsection*{The billiard attractor}
\label{thebillatt}
%-------------------------------------------------------

The asymptotic suppression of multiple transitions in the asymptotic
regime entails that asymptotic billiard sequences $\cA\cB_\cT$ do
not approach general billiard sequences $\cB_\cT$ on the billiard
subset $\mathcal{O}_{\mathcal{B}}$ but sequences on the
\textit{billiard attractor} $\mathcal{O}_{\mathcal{BA}}$.

We define the \textit{billiard attractor} subset
$\mathcal{O}_{\mathcal{BA}}$ as
\begin{equation*}%\label{billiard}
\mathcal{O}_{\mathcal{BA}} =
\mathrm{K}^{\ocircle}\cup\cB_{N_1}\cup\cB_{R_1}\cup\cB_{R_3}\:,
\end{equation*}
i.e., $\mathcal{O}_{\mathcal{BA}}$ is the boundary of the billiard
subset $\mathcal{O}_{\mathcal{B}}$.

An \textit{attractor sequence} of transitions $\cA_\cT$ (or, for
brevity, attractor sequence) is a sequence of transitions on the
billiard attractor $\mathcal{O}_{\mathcal{BA}}$. By definition,
$\cA_\cT$ is an infinite concatenation of single transitions of the
type $\cT_{N_1}$, $\cT_{R_1}$, and $\cT_{R_3}$, i.e., one type of
single curvature transition and two types of single frame
transitions; to get an intuitive picture of attractor sequences we
simply refer to Figure~\ref{Kcurvtrans1}--\ref{Kframetrans3}.

The collection of our results leads to the formulation of the
\textit{dynamical systems billiard conjecture}:
\begin{conjecture}
The asymptotic dynamical evolution of a generic timeline of a
solution of Einstein's vacuum equations (expressed in an Iwasawa
frame) that exhibits a generic spacelike singularity is
characterized as follows:
\begin{itemize}
\item[\rm{(i)}] It is asymptotically silent and local.
\item[\rm{(ii)}] In the asymptotic limit the essential dynamics
is represented by an attractor sequence $\cA_\cT$ on the billiard
attractor $\mathcal{O}_{\mathcal{BA}}$.
\end{itemize}
\end{conjecture}

The dynamical systems billiard conjecture can be viewed as the `dual
formulation' of the cosmological billiard conjecture by Damour,
Hennaux, and Nicolai~\cite{dametal03}.
In Section~\ref{dual} we will establish in
detail the correspondence between the two approaches. In the
remainder of this section we will give a brief summary of our
analysis that has led to the dynamical systems billiard conjecture.

Our previous analysis constitutes a \textit{derivation} of the
dynamical systems billiard conjecture; however, although a
considerable part of our treatment meets the criterion of
mathematical rigor, some of our arguments are heuristic rather than
rigorous (despite their being mathematically convincing); therefore,
the presented derivation of the billiard conjecture does not
represent a rigorous proof but merely a first step toward a rigorous
treatment. Nonetheless, apart from the fact that our analysis
provides strong support for the billiard conjecture, we expect that
many aspects of our considerations are fundamental for a deeper
understanding of the asymptotic dynamics associated with generic
spacelike singularities; in particular we believe that we have
identified several issues that have to be taken into account in the
pursuit of rigor.

Let us therefore recapitulate the main steps in our derivation.
Hereby we are guided by the uncovered hierarchical structure of
(boundary) subsets: the essential asymptotic dynamics can be
restricted successively to subsets of subsets, to boundaries of
boundaries.
\begin{itemize}
\item[$\mathbf{A}$]
Assumption: Asymptotic silence and asymptotically local dynamics.
Asymptotically local dynamics is defined by the requirement that
$E_{\alpha}{}^{i}\rightarrow 0$, $\parb_{\alpha} (\bm{S}, r_\beta,
\Udot_\beta) \rightarrow 0$, $(r_\alpha, \Udot_\alpha) \rightarrow
0$, which leads to a spatial decoupling of the field equations. As a
consequence, a generic solution $\bm{X}(\tau)$ of the Einstein
vacuum equations at a generic spatial point approaches the SH part
of the silent boundary $E_\alpha{}^i = 0$, which is spanned by the
state vector $\bm{S} =
(\Sigma_{\alpha\beta},A_\alpha,N_{\alpha\beta})$.
\item[$\Rightarrow \mathbf{B}$]
Assumption: Asymptotic dominance of the Kasner states in
the asymptotic evolution of solutions.
%the Kasner circle $\mathrm{K}^{\ocircle}$ and the associated stable
%and oscillatory degrees of freedom.
Linear analysis of
the Kasner circle $\mathrm{K}^{\ocircle}$
leads to the conjecture that the variables
$\bm{S}_{\mathrm{stable}}$ vanish in the asymptotic limit $\tau\rightarrow
\infty$, where $\bm{S}_{\mathrm{stable}} = (N_{\alpha\beta},
A_{\alpha})$ ($\alpha \neq \beta$). Hence the essential asymptotic
dynamics is represented by the dynamics on a subset of the SH part
of the silent boundary, the oscillatory subset $\cO$ and its
associated state vector $\bm{S}_{\mathrm{osc}} = (\Sigma_\alpha,
R_\alpha, N_1,N_2)$.
Orbits on $\cO$ are heteroclinic orbits (in general: transitions)
that can be joined to form infinite sequences. These sequences
of orbits/transitions are accessible to a mathematically rigorous
treatment, where stochastic aspects become important.
\item[$\Rightarrow \mathbf{C}$] Assumption: Asymptotic shadowing.
A generic solution $\bm{X}(\tau)$ shadows a sequence of
orbits/transitions on $\mathcal{O}$ with an increasing degree of
accuracy. Our analysis shows that
we have $R_2\rightarrow 0$ and $N_2\rightarrow 0$
along generic orbits $\bm{X}(\tau)$;
this implies that certain types of transitions
that involve the variables $R_2$, $N_2$
are ruled out \textit{a priori} (complete exclusion),
while
it can shown that the probability for the occurrence
of any of the other transitions involving these variables
is zero (stochastic exclusion).
The asymptotic exclusion of $R_2$ and $N_2$ entails
that the essential asymptotic dynamics of $\bm{X}$ is described by
the dynamics on the billiard subset $\cO_\cB$ with its state vector
$\bm{S}_{\mathrm{billiard}} = (\Sigma_1,\Sigma_2,\Sigma_3, R_1, R_3,
N_1)$.
\item[$\Rightarrow \mathbf{D}$] %Assumption:
Asymptotic shadowing of billiard sequences. A generic solution
$\bm{X}(\tau)$ shadows a sequence of transitions on
$\mathcal{O}_{\mathcal{B}}$ (billiard sequence)
with an increasing degree of accuracy.
Our arguments indicate that the variables $(R_1, R_3, N_1)$ of
$\bm{S}_{\mathrm{billiard}}$ cannot be excited simultaneously. This
leads to the conclusion that the generic asymptotic dynamics is
governed by the flow on the boundary of the billiard subset, which
is the billiard attractor $\cO_{\mathcal{BA}}=
\mathrm{K}^{\ocircle}\cup\cB_{N_1}\cup\cB_{R_1}\cup\cB_{R_3}$. The
dynamics on this attractor set is represented by billiard attractor
sequences (consisting of one type of curvature transitions
and two types of frame transitions: $\cT_{N_1}$,
and $\cT_{R_1}$, $\cT_{R_3}$),
which thus form the attractor for generic solutions
$\bm{X}$. In other words: the essential dynamics of a generic
solution $\bm{X}$ is characterized by its approach to an attractor
sequence $\cA_\cT$.
\end{itemize}

In this paper we have given heuristic arguments for
$\mathbf{A}\Rightarrow \mathbf{B}$, but our main focus has been the
other steps. Assuming that generic solutions $\bm{X}$ are asymptotic
sequences of $\mathcal{O}$-orbits (i.e., assuming shadowing of the
flow on $\mathcal{O}$) has inevitably led to the implication
$\mathbf{B}\Rightarrow \mathbf{D}$.
%based on a line of arguments whose foundation --- although
%presumably not a mathematical proof in its formal sense --- is
%mathematically rigorous.
Since stochastic considerations are essential for parts of our
considerations, we obtain only statements for generic solutions. The
possibility exists that there are solutions violating one or the
other properties; however, such solutions are expected to form a set
of measure zero in the space of all solutions.

%%%%%%%%%%%%%%%%%%%%%%%%%%%%%%%%%%%%%%%%%%%%%%%%%%%%%%%%%%%%%%%%%%%%%%%%%%%%%%%%%%%%%%%%%%
\section{`Duality' of Hamiltonian and dynamical systems billiards}
\label{dual}
%%%%%%%%%%%%%%%%%%%%%%%%%%%%%%%%%%%%%%%%%%%%%%%%%%%%%%%%%%%%%%%%%%%%%%%%%%%%%%%%%%%%%%%%%%

In Section~\ref{Hamilton} we have outlined the Hamiltonian approach
to cosmological billiards, where we have followed Damour, Hennaux,
and Nicolai~\cite{dametal03,damnic05}. This approach is based on an
analysis of the Hamiltonian $\mathcal{H}$ in Iwasawa frame
variables, which are: the diagonal degrees of freedom $b^\alpha$,
the off-diagonal variables $\N{\alpha}{i}$ and the conjugate
momenta. An asymptotic Hamiltonian $\mathcal{H}_\infty$ is
constructed from $\mathcal{H}$ by taking the limit $\rho
=\sqrt{-b_\alpha b^\alpha} \rightarrow \infty$ and dropping all
terms except for three terms that are identified as being
`dominant'. The asymptotic Hamiltonian $\mathcal{H}_\infty$ is
assumed to describe the generic asymptotic dynamics of generic
solutions that exhibit a spacelike singularity.

\begin{figure}[ht]
\psfrag{a}[cc][cc]{$\gamma^1$}
\psfrag{b}[cc][cc]{$\gamma^3$}\psfrag{c}[cc][cc]{$\gamma^2$}
\psfrag{j}[cc][cc]{$\Sigma_1$} \psfrag{k}[cc][cc]{$\Sigma_3$}
\psfrag{l}[cc][cc]{$\Sigma_2$} \psfrag{d}[cc][cc]{$T_1$}
\psfrag{e}[cc][cc]{$Q_2$} \psfrag{f}[cc][cc]{$T_3$}
\psfrag{g}[cc][cc]{$Q_1$} \psfrag{h}[cc][cc]{$T_2$}
\psfrag{i}[cc][cc]{$Q_3$}
  \centering
     \subfigure[The cosmological billiard]{
     \label{cosmobilliard}
     \includegraphics[width=0.40\textwidth]{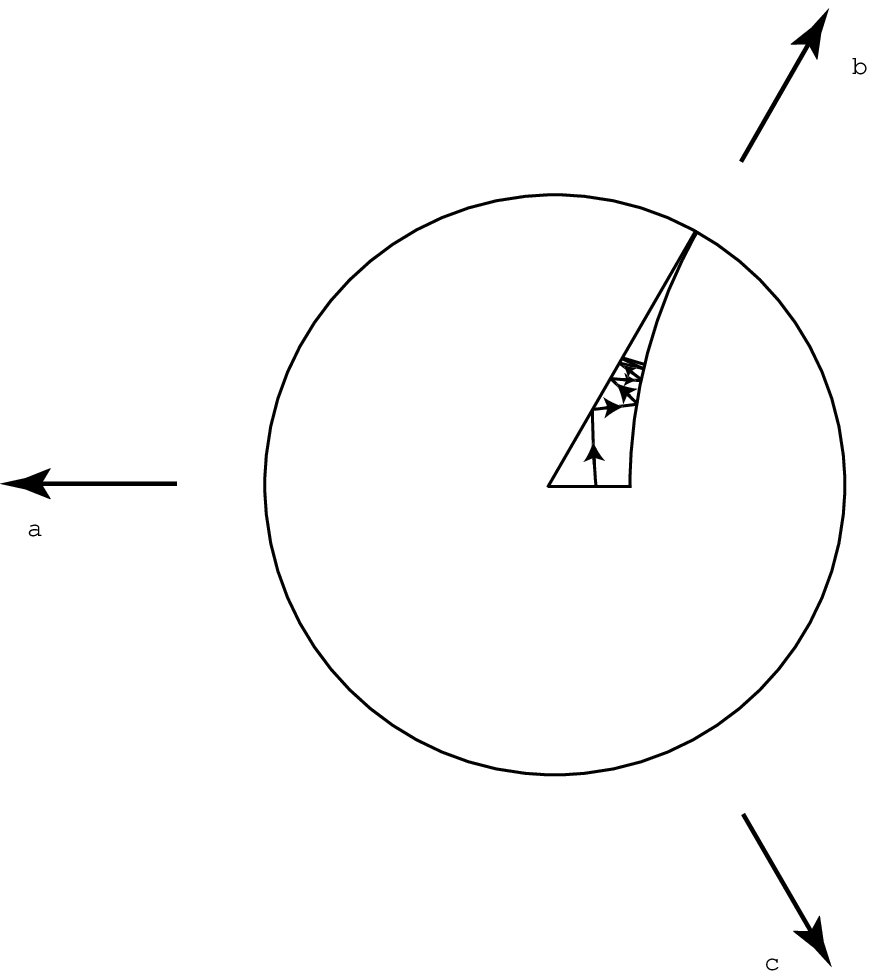}}\qquad
     \subfigure[The billiard attractor]{
     \label{billiardattractorfig}
     \includegraphics[width=0.40\textwidth]{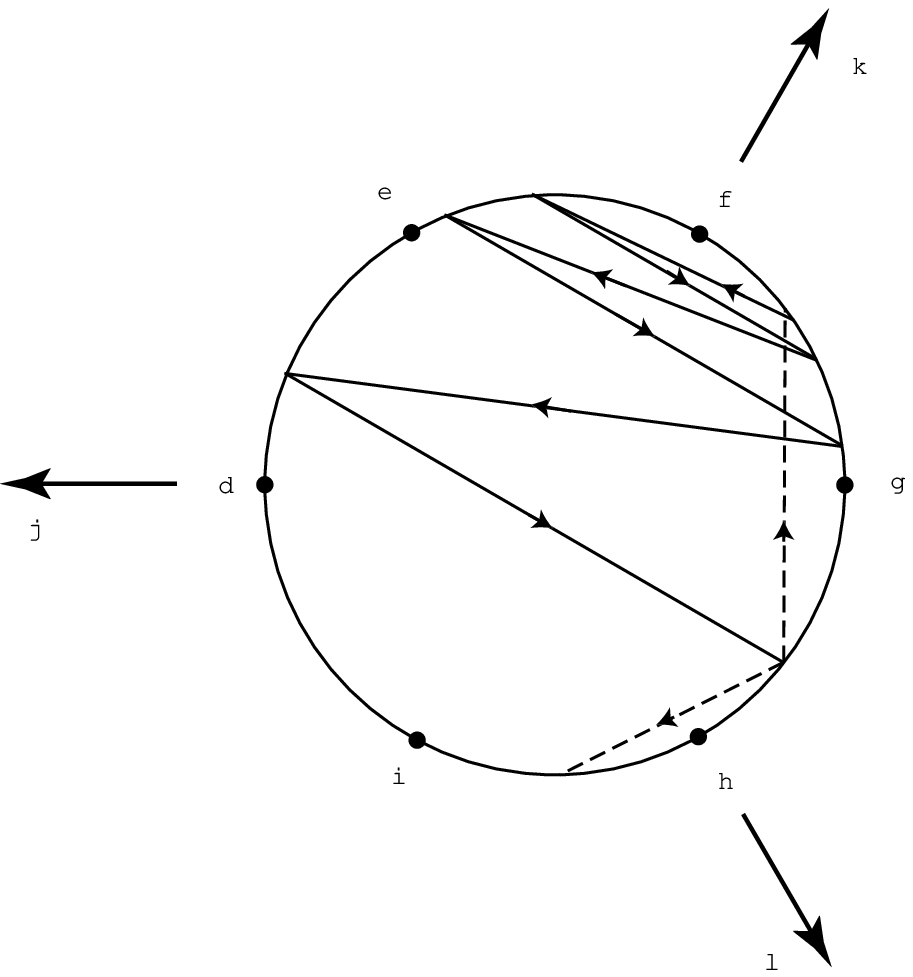}}
  \caption{Figure (a) shows part of an orbit in terms of free Kasner (Fermi frame) motion
           and frame and curvature bounces. This represents a `configuration space' projection
           of the asymptotic dynamics. The disc here represents hyperbolic space. Figure (b)
           shows part of an orbit in terms of single frame and curvature transitions, i.e., it
           shows a part of an attractor sequence. Note that the solution does not quite return
           to any of the Kasner points it has `visited' before. This description
           represents a `momentum space' projection of the asymptotic dynamics. The circle here
           is the Kasner circle $\mathrm{K}^{\ocircle}$. The dashed lines correspond to the two
           possible single transitions that are possible at this stage; which one is realized depends
           on initial data. This corresponds to that free motion in a given direction in Figure (a) may
           either lead to that one hits the wall associated with $R_1$ (the short wall) or $N_1$
       (the curved wall).}
     \label{dualfig}
\end{figure}

A direct consequence from the form of the asymptotic
Hamiltonian~$\mathcal{H}_\infty$ is that the off-diagonal degrees of
freedom are asymptotic constants of the motion, as is a momentum
variable associated with a projection of the diagonal degrees of
freedom---a phenomenon referred to as `asymptotic freezing'; for a
derivation of these results using the present dynamical systems
formalism and the results from the previous sections, see
Appendix~\ref{asympconst}. Therefore, the non-trivial asymptotic
dynamics is encoded in the diagonal degrees of freedom, or, more
precisely, their projections $\gamma^\alpha$: the asymptotic
dynamics at each spatial point is described as a geodesic motion in
a portion of hyperbolic space that is bounded by sharp walls, see
Section~\ref{Hamilton} and~\cite{dametal03,damnic05}. This picture
prompts the terminology `billiard motion' and `cosmological
billiards'; see Figure~\ref{cosmobilliard}.

It is of interest to consider a Hamiltonian that is in a sense a
link between $\mathcal{H}$ and $\mathcal{H}_\infty$: the `dominant'
Hamiltonian $\mathcal{H}_{\mathrm{dom}}$. It is obtained by dropping
the `subdominant' terms of $\mathcal{H}$ before the limit
$\rho\rightarrow \infty$ is taken. Since the asymptotic Hamiltonian
constructed from $\mathcal{H}_{\mathrm{dom}}$ naturally coincides
with ${\cal H}_\infty$, the asymptotic dynamics described by ${\cal
H}$ and ${\cal H}_{\rm dom}$ is the same.

From Section~\ref{Hamilton} and~\cite{dametal03,damnic05} we
obtain the dominant Hamiltonian:
\begin{equation*}
\mathcal{H}_{\rm dom} =  \tilde{N}\:
\left[\textfrac{1}{4} %\sum_{{\alpha},\beta}
\cg^{{\alpha} \beta}
\pi_{\alpha} \pi_{\beta} + \textfrac{1}{2}{\tilde
R}_1^2\,e^{2(b^2-b^3)}+ \textfrac{1}{2}{\tilde R}_3^2\,e^{2(b^1-b^2)} +
\textfrac{1}{8}{\tilde N}_1^2\,e^{-4b^1}\right]\:,
\end{equation*}
where ${\tilde R}_1:=-{\cal P}_3$, ${\tilde R}_3:=-({\cal P}_1 +
n_3\,{\cal P}_2)$, and where ${\tilde N}_1$ is a function of the
variables $n_\alpha$ given by Equation~(\ref{tildeN1}). The
non-trivial dynamics is represented by the diagonal degrees of
freedom $b^\alpha$ (or alternatively by $\rho$ and the projected
variables $\gamma^\alpha$): the dynamics is described as a motion in
a portion of Lorentzian space that is bounded by `exponential
walls'. As the solution approaches the singularity these walls
become increasingly sharp with an infinitely high potential as
limit; we refer to the remarks at the end of this section. The
asymptotic dynamics is thus a free null geodesic motion in
Lorentzian space interrupted by bounces at the walls, which yields
an alternative representation of cosmological billiards.

The asymptotic dynamics as described by Hamiltonian cosmological
billiards can be interpreted in terms of generalized Kasner
solutions: Fermi propagated generalized Kasner solutions appear as
straight lines (geodesics) in flat Lorentzian space in the
$b^\alpha$ description, or, equivalently, as geodesics in hyperbolic
space in the projected $\gamma^\alpha$ description. A free motion
between two bounces at walls thus translates to a (Fermi) `Kasner
epoch', a phase where the solution evolves as a generalized Kasner
solution, see~(\ref{kasnerlineel}). The walls and the associated
bounces, on the other hand, are of two kinds: (i)
frame/centrifugal/symmetry walls and associated bounces; (ii)
curvature walls and associated bounces. Bounces of the former type
merely result in axes permutations of a Kasner solution; bounces of
the latter type are Bianchi type II bounces: the Kasner state
changes in accordance with a change generated by a Bianchi type II
solution.

The Hamiltonian approach emphasizes the dynamics of the
configuration space variables ($b^\alpha$ and the projected
variables $\gamma^\alpha$, respectively); therefore one may say that
the Hamiltonian picture yields a `configuration space'
representation of the asymptotic dynamics. In the following we will
see that the dynamical systems approach can be viewed as a dual
representation, i.e., a `momentum space' representation.

Consider the dominant Hamiltonian from the perspective of the
dynamical systems approach. The terms ${\tilde R}_1 \exp(b^2-b^3)$,
${\tilde R}_3 \exp(b^1-b^2)$, and ${\tilde N}_1 \exp(-2 b^1)$ can be
associated with $R_1$, $R_3$, and $N_1$ by using the
relations~(\ref{R1}) and~(\ref{R3}). Accordingly,
$\mathcal{H}_{\mathrm{dom}}$ contains $\textfrac{1}{2} (R_1^2 +
R_3^2) +\textfrac{1}{8} N_1^2$ modulo a common factor. Recall in
this context that the Hamiltonian is proportional to the Gauss
constraint, cf.~\eqref{gausscon2}. This indicates that there is an
intimate connection between: (i) centrifugal bounces in the
Hamiltonian picture and single frame transitions $\cT_{R_1}$ and
$\cT_{R_3}$ in the dynamical systems picture; (ii) curvature bounces
and single curvature transitions $\cT_{N_1}$.

When we calculate the Hamiltonian equations of ${\cal H}_{\rm
Dom}=0$ associated with the non-trivial degrees of freedom
$b^\alpha$, adopt the appropriate time gauge, and then perform
suitable variable transformations according to the formulas of
Appendix~\ref{relations}, we obtain:
\begin{enumerate}
\item[(i)] The billiard
system, i.e., the differential equations~(\ref{billiardeq1})--\eqref{billiardeq6}
and the Gauss constraint~\eqref{gausscon2}; however, without the
Codazzi constraint $R_3\,N_1=0$.
\item[(ii)] Evolution equations for
$E_\alpha=(E_1{}^1,E_2{}^2,E_3{}^3)$ and the Hubble variable $H$,
with all spatial frame derivatives and all variables $\bm{X}$ (and
$\Udot_\alpha, r_\alpha$) set to zero except $E_\alpha$, $H$, and
$\bm{S}_{\mathrm{billiard}}$.
\end{enumerate}
The unconstrained billiard system (i) forms an independent coupled
system of differential equations. The equations (ii) for $H$ and
$E_\alpha$, on the other hand, decouple from the unconstrained
billiard system. These equations can be regarded as linear equations
with time dependent coefficients, which are provided by the
solutions of the billiard system. (Compare with the treatment of the
auxiliary quantities in Section~\ref{growth}.) The evolution
equations for $E_\alpha$ constitute the lowest order perturbation
into the physical state space of the full equations for $E_\alpha$
in the vicinity of the silent boundary; together with the Hubble
variable $H$ they yield $b^\alpha$ according to
Equation~(\ref{offdiagN}) in Appendix~\ref{relations}. The
off-diagonal degrees of freedom can be treated similarly: the
solutions of the unconstrained billiard system provide the time
dependent coefficients for the lowest order perturbation equations.
In this manner we obtain asymptotic freezing% of the off-diagonal degrees of freedom
, see Appendix~\ref{asympconst}. This result
further strengthens the connection between the Hamiltonian and the
dynamical systems approach.

The asymptotic limit of the dynamics of the unconstrained billiard
system is represented by attractor sequences $\cA_\cT$, see
Figure~\ref{billiardattractorfig}.%
\footnote{A more detailed discussion of this statement will be given elsewhere.}
The duality of the Hamiltonian and the dynamical systems approach
thus becomes apparent: one phase of free motion in the Hamiltonian
billiard picture corresponds to a Kasner fixed point on the Kasner
circle $\mathrm{K}^{\ocircle}$. The underlying reason for this is
that the dynamics has been projected out of the self-similar Fermi
propagated Kasner solutions by means of a conformal normalization
that has yielded scale-invariant variables. The bounces at walls in
the Hamiltonian billiards correspond to motion in the dynamical
systems picture represented by the transition orbits (heteroclinic
orbits) between Kasner points. The curvature transitions $\cT_{N_1}$
are non-scale-invariant solutions and correspond to curvature
bounces in the Hamiltonian billiards; the frame transitions
$\cT_{R_1}$, $\cT_{R_3}$ rotate axes w.r.t.\ a Fermi frame; see
Appendix~\ref{KasnernonFermi}. The transitions appear as straight
lines (in Euclidian space) when projected onto
$\Sigma_\alpha$-space. Note that the transitions and the Hamiltonian
wall bounces yield exactly the same rules for changing Fermi
propagated Kasner states. (However, the Hamiltonian billiard only
gives a consistent picture when we consider generic attractor
sequences, which are characterized by irrational values of the
Kasner parameter $u$. Rational values of the Kasner parameter $u$
leads to motion straight into the sharp corners of the asymptotic
billiard where the dynamics is undefined.)

Changing from the Hamiltonian billiards to dynamical systems
billiards, the notions `bounces at walls' and 'motion along straight
lines' switch places. In the Hamiltonian picture, free motion is
followed by a bounce at a wall, which again gives rise to a phase of
free motion. In the dynamical systems picture, we observe `bounces'
at the fixed points on the Kasner circle $\mathrm{K}^{\ocircle}$; in
fact, since the Kasner circle constitutes the boundary of
$\Sigma_\alpha$-space, it can be viewed as a `wall' in the dynamical
systems picture. Between the bounces we have motion along straight
lines in $\Sigma_\alpha$-space (transitions). In brief: the
Hamiltonian `motion--bounce--motion--bounce' is translated to
`bounce--motion--bounce--motion' in the dynamical systems billiards.

Since the variables $\Sigma_\alpha$ are intimately connected
with the variables $\pi_\alpha$,
it is natural to refer to the projected dynamical
systems picture as a `momentum space' representation of the
asymptotic dynamics, which complements the Hamiltonian configuration
space representation. To compare the two pictures, see
Figure~\ref{dualfig}.

The correspondence between the dynamical systems and Hamiltonian
pictures offers a possibility for mutual support. Let us give a few
examples: (i) As can be inferred from Figures~\ref{smallcurvphase}
and~\ref{dualfig}, the typical `captured' $\cT_{N_1}$-$\cT_{R_3}$
oscillatory behavior (a small curvature phase) corresponds to the
many oscillations in the upper right corner in the Hamiltonian
billiard, see Figure~\ref{cosmobilliard}. (ii)
Figure~\ref{billiardtriggers} indicates that $\cT_{R_1}$ transitions
are less frequent than $\cT_{R_3}$ transitions: $R_1$ and $R_3$ are
unstable in three sectors of $\mathrm{K}^\ocircle$; however, $R_1$
has to compete with $N_1$ in sector $(132)$, which `depletes'
$\cT_{R_1}$ transition at the cost of $\cT_{N_1}$ transitions;
$\cT_{R_3}$ is unaffected by this. The correctness of these
arguments is corroborated by the features of Lobashevski geometry
close to the circular boundary and by the fact that the wall
associated with $\cT_{R_3}$ bounces is longer than the one
associated with $\cT_{R_1}$ bounces. (iii) Consider the Hamiltonian
${\cal H}_{\rm dom}$ and perform a projection of the dynamics onto
spatial constant volume slices in $b^\alpha$-space. The walls are
`exponential walls' whose contours are of a triangular shape where
the corners are smoothed out. Toward the singularity (i.e., with
shrinking volume) the triangle increases in size, where
simultaneously the triangle's sides become increasingly straight and
the corners increasingly sharp (relative to the size of the
triangle), see~\cite{jan01} for similar SH examples. Particularly
the corner shape is sensitive to the additive effects from the
different walls potentials. The increasing size of the potential
contour triangle and the comparatively sharper corners suggests that
it becomes increasingly less likely that the `billiard ball' hits
the corners. This geometrical feature suggests that, statistically,
multiple transitions are asymptotically suppressed, in agreement
with our quantitative analysis.

It is of interest to also compare the dynamical systems picture with
previous (Hamiltonian) treatments of so-called asymptotic velocity
term dominated dynamics; this is done in Appendix~\ref{rel}.

%%%%%%%%%%%%%%%%%%%%%%%%%%%%%%%%%%%%%%%%%%%%%%%%%%%%%%%%%%%%%%%%%%%%%%%%%%%%%%%%%%%%%%%%%%
\section{Models with symmetries in an Iwasawa frame}
\label{sym}
%%%%%%%%%%%%%%%%%%%%%%%%%%%%%%%%%%%%%%%%%%%%%%%%%%%%%%%%%%%%%%%%%%%%%%%%%%%%%%%%%%%%%%%%%%

In this section we give a brief overview on models with symmetries.
Such models are of obvious significance for the study of
cosmological singularities; since the equations simplify we can hope
to obtain more substantial evidence and perhaps even prove theorems
on the asymptotic dynamics of solutions.
This in turn can be expected to shed
new light on the heuristic statements concerning the generic case
without symmetries.

%---------------------------------------------------------
\subsection*{$\bm{G_1}$ models}
\label{G1mo}
%---------------------------------------------------------

By definition, cosmological $G_1$ models possess one spacelike
Killing vector; let us choose this Killing vector to be
$\partial_{x^1}$. The Iwasawa decomposition is compatible with $G_1$
symmetry and we are therefore able to choose an Iwasawa frame that
is symmetry adapted; we set $\parb_\alpha = \E{\alpha}{i}
\partial_{x^i}$, where $\E{\alpha}{i}$ is a lower triangular matrix,
see Section~\ref{confsec}; accordingly,
\begin{align*}
%\begin{equation*}
& \parb_1 = E_1{}^1\partial_{x^1}\,,
& & \parb_2 = E_2{}^1\partial_{x^1}+E_2{}^2\partial_{x^2}\:,
& & \parb_3 = E_3{}^1\partial_{x^1}+E_3{}^2\partial_{x^2} +E_3{}^3\partial_{x^3}\:,
%\end{equation*}
\intertext{where $E_{\alpha}{}^i$ as well as $H$ and ${\cal N}$ are independent of $x^1$.
  Consequently, when $f$ is a function that is independent of $x^1$, it follows
  that}
%
%\begin{equation*}
& \parb_1 f= 0 \:,
& & \parb_2 f = E_2{}^2\partial_{x^2} f\:,
& & \parb_3 f = (E_3{}^2\partial_{x^2} + E_3{}^3\partial_{x^3}) f\:.
%\end{equation*}
\end{align*}
This in turn leads to a decoupling of the $\E{\alpha}{1}$ equations
from the system~\eqref{udotN}-\eqref{theIs}, since $\E{\alpha}{1}
\partial_{x^1}$ disappears from the r.h. sides of the evolution
equations and the constraints. Note also that~\eqref{dl13comts}
and~\eqref{dl13com} do not mix the components $\E{\alpha}{1}$ with
$\E{\alpha}{2}$ and $\E{\alpha}{3}$. In addition, using the
expressions in Appendix~\ref{relations}, the symmetries are seen to
further reduce the state space~(\ref{Xstatespsplit},\ref{Sstatesp});
since
\begin{equation*}
N_{22} = N_{23} = A_1 = \Udot_1 = r_1 = 0\:,
\end{equation*}
the state space associated with the reduced coupled system of equations
is given by
\begin{equation*}
\bm{X}=(E_2{}^2, E_3{}^2,E_3{}^3, \Sigma_\alpha, R_\alpha, N_{1\alpha}, A_2, A_3)\:;
\end{equation*}
in addition $\Udot_2, \Udot_3, r_2, r_3$ are non-zero for most temporal gauge
choices.

Apart from the above generic $G_1$ case there exists special $G_1$
models. The \textit{hypersurface orthogonal} Killing vector case,
also referred to as the polarized case in a $U(1)$ context,
see~\cite{choetal04} and references therein, is given by setting
%the Iwasawa components $\bar{n}_1,\bar{n}_2$ to zero, i.e.,
$E_2{}^1$ and $E_3{}^1$ to zero, so that the metric takes a diagonal
form. Using the expressions in Appendix~\ref{relations}, it follows
that $N_{11}=R_2=R_3=0$. In this case, the sectors $(231)$, $(213)$,
and $(123)$ on the Kasner circle $\mathrm{K}^{\ocircle}$ become
stable, which in turn implies that solutions asymptotically approach
a Kasner state; for proofs, see~\cite{choetal04,choetal06}.

%---------------------------------------------------------
\subsection*{$\bm{G_2}$ models}
\label{G2mo}
%---------------------------------------------------------

Models with two commuting spacelike Killing vectors are known as
$G_2$ models, see, e.g.,~\cite{waiell97}.
%we choose $\partial_{x^1}$ and $\partial_{x^2}$ to be the Killing
%vectors.
An Iwasawa frame that is adapted to the symmetries is obtained by
letting $\parb_1$ and $\parb_2$ be tangent to the group orbits.
Accordingly, $E_\alpha{}^i$ and $H$, ${\cal N}$ are functions that
are independent of $x^1$ and $x^2$ and thus depend on $x^0$ and
$x^3$ only. For any $f = f(x^0,x^3)$ we get
\begin{equation*}
\parb_1 f = 0 \:,\qquad
\parb_2 f = 0 \:,\qquad
\parb_3 f = \E{3}{3} \partial_{x^3} f\:;
\end{equation*}
furthermore, as follows from Appendix~\ref{relations},
\begin{equation*}
N_{22} = N_{13} = N_{23} = N_{33} = A_1=A_2 =\dot{U}_1=\dot{U}_2=r_1=r_2=0\:.
\end{equation*}
In addition one can exploit the remaining freedom in choosing
the frame at a spatial point to obtain $R_2=0$, and then
the Codazzi constraint and the evolution equation for $R_2$
force $R_2$ to be identically zero everywhere~\cite{wai79}.
Using a symmetry adapted Iwasawa frame thus leads to
a decoupling of all $E_\alpha{}^i$ equations except for
the equation for $E_3{}^3$. Accordingly, the reduced state space for
the problem is represented by the state vector
\begin{equation*}
\bm{X} = (E_3{}^3,\Sigma_\alpha, R_1, R_3, N_{11}, N_{12}, A_3)\:;
\end{equation*}
in addition $\Udot_3, r_3$ are non-zero for most temporal gauge choices.
If one in addition chooses the timelike separable area
gauge~\cite{elsetal02}, so that the area density of the $G_2$
symmetry orbits, i.e.,  $\mathcal{A} = (e_1{}^1e_2{}^2)^{-1}$, is a function
of $x^0$ only, then $A_3=r_3$, as follows from~(\ref{dl13com}).

There exists several special classes of $G_2$ models:%
\begin{enumerate}
\item[(i)] {\em Hypersurface-orthogonal\/} models are characterized by the
restriction $N_{11} = R_{3} = 0$.
\item[(ii)] \emph{Orthogonally
transitive} models are obtained by setting $ R_{1} = 0$.
\item[(iii)] The {\em diagonal\/} case is given by $N_{11} = R_{1} = R_{3}= 0$;
this case is also referred to as the polarized case in literature dealing with a
$T^3$ topology~\cite{isemon90,grumon93}.
\item[(iv)] {\em Plane symmetric models\/}
are determined by $N_{11} = N_{12} = R_1 = R_3 = 0$,
$\Sigma_{11}=\Sigma_{22}$.
\end{enumerate}

In~\cite{andetal05} slightly different variables were used to
discuss the general case with $T^3$-topology; these can be obtained
as follows: firstly, note that $\Udot_\alpha = \udot_\alpha/H  -
r_{\alpha}\, , A_{\alpha}= a_{\alpha}/H + r_{\alpha}$, where
$\udot_\alpha$, $a_{\alpha}$ are the acceleration and spatial
connection coefficients associated with ${\bf g}$, respectively.
Secondly, to identify the present frame with the one used
in~\cite{andetal05} we permute both frame and coordinate indices
from the present ones to the ones used in~\cite{andetal05} according
to $1\rightarrow 3$, $2 \rightarrow 1$, $3 \rightarrow 2$. Then, in
addition, we make the few scalings and notational changes that are
needed in order to obtain the variables
$(E_1{}^1,r,\Udot,\Sigma_+,\Sigma_-,\Sigma_\times,\Sigma_2,N_-,N_\times)$
used in~\cite{andetal05} in terms of the present ones; $E_3{}^3$ is
replaced with $E_1{}^1$, which was used in~\cite{andetal05}, while
the other variables are related as follows:
\begin{subequations}\label{2plus1vartot}
\begin{align}
\label{2plus1var}
r & = r_3=A_3\:, & \Udot & =  r_3 + \Udot_3\:, &    &  \\
\Sigma_{+} & = \textfrac{1}{2}(\Sigma_{11} + \Sigma_{22})\:, &
\Sigma_{-} & =  \textfrac{1}{2\sqrt{3}}(\Sigma_{11}-\Sigma_{22})\:, &
\Sigma_{\times} & = -\textfrac{1}{\sqrt{3}}R_3\:,\\
\label{2plus1varf}
\Sigma_2 & = -\textfrac{1}{\sqrt{3}}R_1\:, & N_{-} & =
\textfrac{1}{2\sqrt{3}}N_{11}\:, &  N_{\times} & =
\textfrac{1}{\sqrt{3}}N_{12}\:.
\end{align}
\end{subequations}

For our present purposes it is of interest to note that numerical
experiments performed in the general $G_2$ case in~\cite{andetal05}
gave evidence for the claim that multiple transitions become
increasingly rare and that the billiard attractor gives a correct
generic asymptotic description. However, the results in the present
paper suggest that it would be useful to numerically check our more
detailed statements about asymptotic suppression and decay rates (as
well as freezing, as discussed in Appendix~\ref{asympconst}); the
same also holds for the next class of models.

%---------------------------------------------------------
\subsection*{Generic Bianchi type VI$_{-1/9}$ models}
\label{genexpB}
%---------------------------------------------------------

The evolution and constraint equations for SH cosmologies are given
by setting all terms involving spatial derivatives to zero, which
includes setting $\dot{U}_{\alpha}=0=r_{\alpha}$, see also
Sections~\ref{confsec} and~\ref{asympsil}. This implies that the
conformal Hubble normalization becomes equivalent with the usual
Hubble normalization used in, e.g.,~\cite{waiell97}. While the Iwasawa
gauge is compatible with the symmetry adapted frames of Bianchi
types I--VII, since they admit a subgroup of two spacelike commuting
Killing vectors, it is incompatible with the symmetry adapted frames
of Bianchi types VIII and IX. The most general model with a symmetry
adapted
Iwasawa frame is the general Bianchi type $\text{VI}_{-1/9}$ model,
also sometimes referred to as the exceptional type $\text{VI}_{h}$
model.

The asymptotic dynamics of generic Bianchi cosmologies of type
$\text{VI}_{-1/9}$ has been studied in \cite{hewetal03}. The
compatibility of these models with the Iwasawa gauge is responsible
for their particular significance in the context of generic
inhomogeneous cosmologies: the general type $\text{VI}_{-1/9}$ case
is the only SH vacuum case where models possess the same attractor toward
the singularity as generic inhomogeneous cosmologies in an Iwasawa
frame; in particular they exhibit oscillatory asymptotic behavior. The
type $\text{VI}_{-1/9}$ models admit a $G_2$ subgroup; adapting the
spatial Iwasawa frame to the three-dimensional symmetry surfaces so
that the first two vectors are adapted to the $G_2$ subgroup implies
that the same quantities as in the $G_2$ case become zero, however,
the extra symmetry forces some additional quantities to be zero as
well. The Codazzi constraint with $\alpha=2$ yields
$R_1(N_{12}-3A_3)=0$, but, in contrast to other type VI$_h$ models,
$h=-1/9$ implies, with the present conventions, that $N_{12}=3A_3$;
hence $R_1$ is not forced to vanish since the Codazzi constraint is
automatically satisfied, see, e.g.,~\cite{waiell97}; the general
type $\text{VI}_{-1/9}$ case is thus characterized by nonzero
off-diagonal shear components $R_1$ and $R_3$. It follows that we
can choose
\begin{equation*}
\bm{S} = (\Sigma_{11},\Sigma_{22},R_1,R_3,N_{11},A_3)\:,
\end{equation*}
as the reduced state space variables.

Using the variables~\eqref{2plus1vartot} with the opposite signs%
\footnote{This is because the opposite sign was used
in~\cite{waiell97,hewetal03} for the definition of $R_\alpha$.}
for $\Sigma_2$ and $\Sigma_{\times}$,
and $r_\alpha=\Udot_\alpha =0$, yields the
system given in~\cite{hewetal03}, where a numerical investigation
indicated that multiple transitions become increasingly rare and
that the billiard attractor %(global, because of the SH assumption)
gives a correct generic asymptotic description, in agreement with
the present general analysis.

%%%%%%%%%%%%%%%%%%%%%%%%%%%%%%%%%%%%%%%%%%%%%%%%%%%%%%%%%%%%%%%%%%%%%%%%%%%%%%%%%%%%%%%%%%
\section{Gauge considerations}
\label{gauge}
%%%%%%%%%%%%%%%%%%%%%%%%%%%%%%%%%%%%%%%%%%%%%%%%%%%%%%%%%%%%%%%%%%%%%%%%%%%%%%%%%%%%%%%%%%

It is a matter of course that there exists many useful spatial frame
choices; while our present considerations are based on the choice of
an Iwasawa frame, other prominent gauges are: the frame used in
UEWE~\cite{uggetal03}; the $SO(3)$ choice used by Benini and Montani
as the starting point for a Misner/Chitr\'e billiard
analysis~\cite{benmon04}; the Fermi choice ($R_\alpha=0$). Each of
these gauges exhibit some computational advantages over the others
in special contexts, and each has its physical merits: e.g., Fermi
observers associated with Fermi frames do not get dizzy. But it is
only in special situations, when one considers
an important class of spacetimes that
share a special, often global, feature, that one can argue that a
particular frame choice is preferred. It is probably safe to
conclude that no single frame choice will be optimal as regards all
features one may be interested in; e.g., Iwasawa frames are
incompatible with the symmetry adapted SH frames of Bianchi type
VIII and IX, which are historically the prime examples of
`Mixmaster' dynamics.

The asymptotic causal structure associated with asymptotic
silence indicates that local and quasi-local aspects, and not global
ones, are most important when trying to understand the physics of
generic spacelike singularities in GR. Since this suggests that there
is no preferred choice of frame, it is natural that results
should be translated to a frame independent description in order to
separate gauge features from physics. In particular, the attractor
should be characterized within the framework of a spatial frame
invariant description of the union of the silent Kasner and Bianchi
type II subsets; %., which holds for any computationally useful spatial
%frame choice;
it follows from this that we are interested in
quantities that do not involve $R_\alpha$.

Let us therefore characterize this subset in a frame invariant way.
It is described by the equations
\begin{equation*}
G^{ij} = \Udot^{2} = r^{2} = A^{2} =
N_{\alpha\beta}\Sigma_{\gamma\sigma}\Sigma^{\beta\sigma}(N^{\alpha\gamma} -
\Sigma^{\alpha\gamma}) = D_{N} = \Delta_{N}=0\:,
\end{equation*}
where we have introduced the following quantities:
\begin{equation*}
G^{ij}:= \delta^{\alpha\beta}\,
E_{\alpha}{}^{i}\,E_{\beta}{}^{j}\, ,\qquad  D_{N}:=
\det(N_{\alpha\beta})\, ,\qquad
\Delta_{N}:=\textfrac{1}{2}\,[N_{\alpha\beta}N^{\alpha\beta} -
(N_{\alpha}{}^{\alpha})^{2}]\:.
\end{equation*}
The quantity $G^{ij}$ is the conformal Hubble-normalized contravariant
three-metric%
\footnote{Incidentally, this
  suggests that it is this object that is natural to use if one wants
  to pursue a metric approach to the Einstein field equations and generic spacelike
  singularities.}%
, see \cite{rohugg05};
if $N^{\alpha\beta}$ is of rank two, then $D_{N} =
0$; if $N^{\alpha\beta}$ is of rank one (as it is for Bianchi type
II), then $D_{N} = \Delta_{N} = 0$. The quantity
$N_{\alpha\beta}\Sigma_{\gamma\sigma}\Sigma^{\beta\sigma}(N^{\alpha\gamma} -
\Sigma^{\alpha\gamma})$ is the square of the expression
$\epsilon_{\alpha}{}^{\beta\gamma}\,N_{\beta\delta}\,
\Sigma_{\gamma}{}^{\delta}$, which is associated with the asymptotic
Codazzi constraints.

To explore the frame invariant description of the attractor it is of
interest to consider the following Weyl spatial and spacetime scalar
invariants
\begin{align*}
%\label{2plus1varA}
\mathcal{E}^2 &
=\textfrac{1}{6}\,\mathcal{E}_{\alpha\beta}\mathcal{E}^{\alpha\beta}\:,
& \mathcal{H}^2 &
=\textfrac{1}{6}\,\mathcal{H}_{\alpha\beta}\mathcal{H}^{\alpha\beta}\:,\\
%\label{weyl}
\mathcal{W}_1  & =  \frac{C_{abcd}C^{abcd}}{48H^4} =
\mathcal{E}^2 - \mathcal{H}^2\:, & \mathcal{W}_2 & =
\frac{C_{abcd}{}^{*}C^{abcd}}{48H^4} =
\textfrac{1}{3}\,\mathcal{E}_{\alpha\beta}\mathcal{H}^{\alpha\beta}\:,
\end{align*}
where $\{\mathcal{E}_{\alpha\beta},\mathcal{H}_{\alpha\beta}\}=
\{E_{\alpha\beta},H_{\alpha\beta}\}/H^2$ are the electric and the
magnetic part of the Weyl curvature tensor. (Here, $E_{\alpha\beta},
H_{\alpha\beta}$ refer to the orthonormal frame components of the
Weyl tensor of ${\bf g}$ while
$\mathcal{E}_{\alpha\beta},\mathcal{H}_{\alpha\beta}$ are the
orthonormal frame components of the Weyl tensor of ${\bf G}$, or,
equivalently, the conformal orthonormal frame components of the Weyl
tensor of ${\bf g}$).

The attractor consists of a sequence of Kasner states connected by
Bianchi type II solutions (i.e., attractor sequences).
To formulate this in terms of curvature frame invariants we proceed
as follows: we first give the electric and magnetic parts of the
Weyl tensor for the $\mathcal{B}_{N_1}$ subset:
\begin{align*}% \label{weylcomp}
\mathcal{E}_{11} & =  2 + (1-\Sigma_1)\Sigma_1 -
\textfrac{10}{9}(1-\Sigma^2)\:, & \mathcal{H}_{11} & =
-\textfrac{3}{2}N_1\Sigma_1\:,  \\
\mathcal{E}_{22} & =  2 + (1-\Sigma_2)\Sigma_2 -
\textfrac{22}{9}(1-\Sigma^2)\:, & \mathcal{H}_{22} & =
\textfrac{1}{2}N_1\,(\Sigma_2 + 2\Sigma_1)\:,\\
\mathcal{E}_{33} & =  2 + (1-\Sigma_3)\Sigma_3 -
\textfrac{22}{9}(1-\Sigma^2)\:, & \mathcal{H}_{33} & =
\textfrac{1}{2}N_1\,(\Sigma_3 + 2\Sigma_1)\:.
\end{align*}
The equations~(\ref{Zeqs}) then allow us to express the above
equations in terms of the frame invariant $\zeta$, and subsequently
we can compute the curvature scalars in terms of $\zeta$, i.e., we
obtain a frame invariant description of the scalars. However, the
resulting expressions are quite complicated and we will refrain from
giving them explicitly. The Kasner states yield $\mathcal{W}_1 =
27u^2(1+u)^2/(1 + u +u^2)^3$ and $\mathcal{W}_2=0$; $\mathcal{W}_1$
is monotonically decreasing in $u$ with the range $(0,4)$ with
$\mathcal{W}_1=0$ when $u=\infty$, i.e., for the Taub state, while
$\mathcal{W}_1$ reaches its maximum value $\mathcal{W}_1=4$ for the
non-flat LRS Kasner state $u=1$.

The scalars can be expressed implicitly in $\tau$-time for a Bianchi
type II transition since it is possible to integrate~(\ref{Zeqs});
we obtain
\begin{equation*}%\label{tauzeta}
\tau(\zeta) =
\ln\left[(\zeta-\zeta_{-})^{\frac{1}{4u\zeta_{-}}}
(\zeta_{+}-\zeta)^{\frac{-1}{4u\zeta_{+}}}\zeta^{-1/6}\right]\,.
\end{equation*}

\begin{figure}[hp]
\psfrag{a}[cc][cc]{$\mathcal{H}^2$}
\psfrag{b}[cc][cc]{$\mathcal{E}^2$}
\psfrag{c}[cc][cc]{$\mathcal{W}_1$}
\psfrag{d}[cc][cc]{$\mathcal{W}_2$} \psfrag{A}[cc][cc]{$\Sigma_1$}
\psfrag{B}[cc][cc]{$\Sigma_3$} \psfrag{C}[cc][cc]{$\Sigma_2$}
\psfrag{D}[cc][cc]{$T_1$} \psfrag{e}[cc][cc]{$Q_2$}
\psfrag{f}[cc][cc]{$T_3$} \psfrag{g}[cc][cc]{$Q_1$}
\psfrag{h}[cc][cc]{$T_2$} \psfrag{i}[cc][cc]{$Q_3$} \centering
\subfigure[A typical attractor sequence.]{
 \label{weylorbit}
  \includegraphics[height=0.45\textwidth]{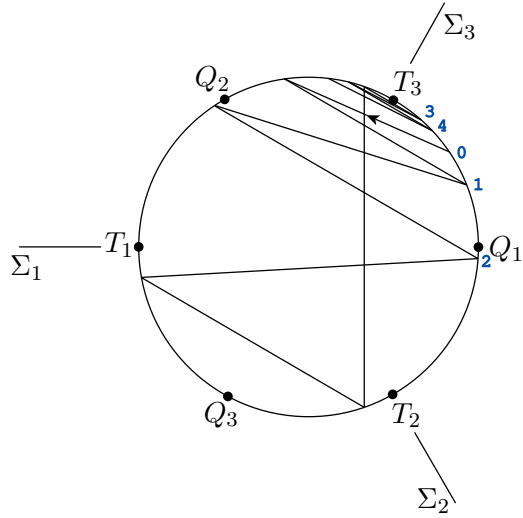}}\\
  \subfigure[$\mathcal{E}^2 - \mathcal{H}^2$]{
  \label{E2H2weyl}
  \includegraphics[height=0.35\textwidth]{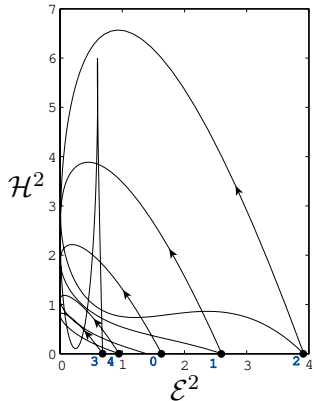}}\qquad
  \subfigure[$\mathcal{W}_1 - \mathcal{W}_2$]{
  \label{W1W2weyl}
  \includegraphics[height=0.35\textwidth]{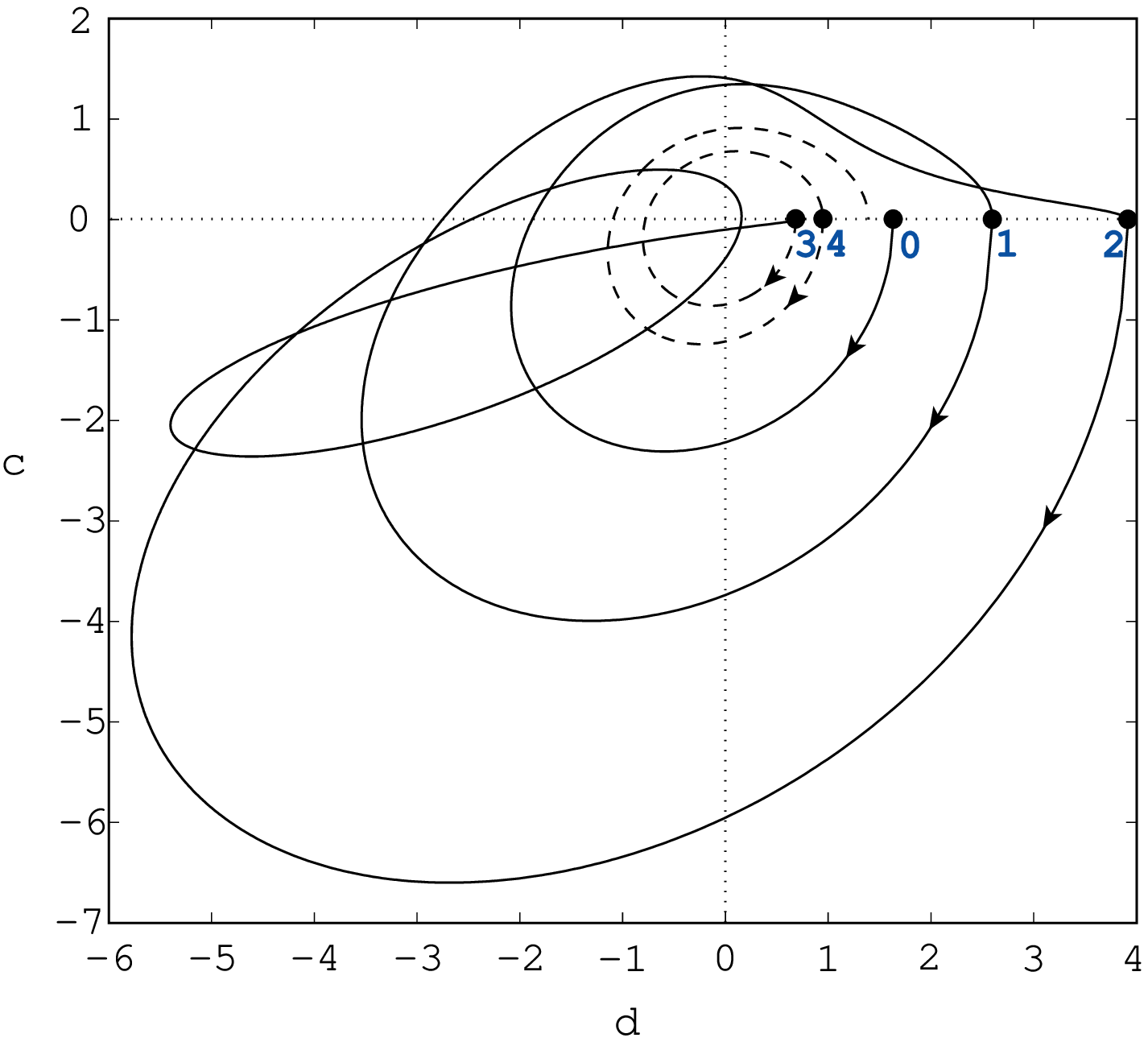}}\qquad
\caption{In Figure~\ref{weylorbit} a typical attractor sequence
is shown, which is associated a sequence of Kasner epoch
characterized by the values of $u$ given by
$\{u_0,u_1,u_2,u_3,u_4\} = \{3.18,2.18,1.18,(1.18-1)^{-1}=5.56,4.56\}$.
Figures~\ref{E2H2weyl} and~\ref{W1W2weyl} display the corresponding
Weyl scalar properties (note that the scalars are constant during
frame transitions); the bullets pinpoint the Kasner epochs. The
dashed line in~\ref{W1W2weyl} indicates a new era and the arrows
show the direction of increasing time toward the singularity.}
\label{weylfigs}
\end{figure}

In Figure~\ref{weylorbit} we plot a typical attractor sequence; note
that a change of era occurs. We then plot the $\mathcal{E}^2$,
$\mathcal{H}^2$ and $\mathcal{W}_1$, $\mathcal{W}_2$ scalars against
each other in Figs.~\ref{E2H2weyl} and~\ref{W1W2weyl}, and obtain a
frame independent description of the same orbit in terms of
curvature properties.

Apart from spatial gauge considerations, it is of interest to ask
oneself: to what degree do asymptotic silence and local dynamics
for a generic spacelike singularity depend on the temporal gauge one
is using? Note that asymptotic silence has been defined
gauge invariantly, and hence the issue is whether a generic spacelike
singularity is asymptotically silent (which our results indicate)
and to what extent asymptotic local dynamics is a temporally gauge
robust feature for a generic spacelike singularity. The asymptotic
causal structure associated with asymptotic silence suggests
considerable temporal gauge robustness, and we expect a large class
of time choices to be compatible with asymptotic local dynamics.

We first note that if one considers a foliation associated with a
time choice, and if it is possible to choose a parametrization of
this foliation so that ${\cal N}$ is bounded and greater (or
smaller) than zero asymptotically toward the singularity, then the
billiard attractor is a local attractor in the full state space; the
difference between using such a time variable and the local
reparametrization that uses $\tau$ only amounts to an unimportant
conformal factor on the right hand sides of the billiard equations
for each fixed $x^i$. We then note that the billiard attractor is a
local attractor if one uses {\em any\/} temporal choice compatible
with the asymptotic local dynamics condition, which, in particular,
implies that the timelike congruence is asymptotically conformally
geodesic since $\Udot_\alpha\rightarrow 0$ (which includes the
synchronous choice used by BKL). However, presumably it is possible
to choose a time gauge such that e.g., $\Udot_\alpha \not\rightarrow
0$, which is hence non-compatible with asymptotic local dynamics,
even though the spacetime may exhibit a singularity for which there
exists a large class of time choices for which the dynamics is
asymptotically local.

%clear to what extent recurring spike formation, which we will discuss next, is temporally
%gauge robust or not.

%%%%%%%%%%%%%%%%%%%%%%%%%%%%%%%%%%%%%%%%%%%%%%%%%%%%%%%%%%%%%%%%%%%%%%%%%%%%%%%%%%%%%%%%%%
\section{Concluding remarks}
\label{concl}
%%%%%%%%%%%%%%%%%%%%%%%%%%%%%%%%%%%%%%%%%%%%%%%%%%%%%%%%%%%%%%%%%%%%%%%%%%%%%%%%%%%%%%%%%%

In this paper we have considered the dynamics of solutions of the
Einstein equations in the asymptotic limit toward a generic
spacelike singularity: the asymptotic behavior of solutions is
represented by the billiard attractor and the associated attractor
sequences. We have derived the billiard attractor based on the
dynamical systems formulation of the Einstein equations in the
conformal Hubble-normalized Iwasawa frame approach; in
addition we have established the `duality' of the presented
framework with the Hamiltonian approach to cosmological billiards of
Damour, Hennaux, and Nicolai~\cite{dametal03}.

The cornerstones of our derivation of the billiard attractor have
been (i) the identification of a hierarchy of state spaces and
invariant subsets; in the asymptotic limit, dynamics is restricted
to subsets of subsets, to boundaries of boundaries, descending from the
full state space via the SH silent boundary and the oscillatory
subset down to the billiard attractor; (ii) a thorough analysis of
sequences of transitions; in particular, we have introduced the
concepts of small and large curvature phases that appeared
prominently in our considerations; (iii) stochastic analysis; we
have investigated the probabilistic nature of Kasner sequences and
we have discussed the effects leading to randomization; randomized
Kasner sequences and randomized asymptotic attractor sequences are
fundamental concepts in our analysis; (iv) the computation of decay
rates; in connection with probabilistic aspects we have given the
decay rates of several quantities; this decay has turned out to be
the underlying reason for the restriction of the dynamics to subsets
and boundaries of subsets. The presented derivation of the billiard
conjecture is not mathematically rigorous, but depends on arguments
that are merely heuristic---despite their being convincing. However,
we are positive that in any endeavor aimed at obtaining proofs,
several concepts and methods introduced in this paper will play a
prominent role. Let us elaborate on this.

Generalizing the concept of an era has played an important role in
our analysis. The partitioning of sequences into small and large
curvature phases has shown that small curvature phases dominate over
large curvature phases from a stochastic point of view; hence, in
the asymptotic evolution of solutions the phases when the solution
is close to the Taub solution are crucial---a fact that might not
have been appreciated enough previously. Note that the division into
small and large curvature phases has featured prominently in the
more technical computations, see Appendix~\ref{convergenceofsums}
for details.

Asymptotic (attractor) sequences give a precise meaning to the
notion of `piece-wise approximations' by BKL. The dynamics of a
generic timeline of a solution with a generic spacelike singularity
is expected to asymptotically shadow the attractor, and hence the
associated orbit can be partitioned into segments where each segment
can be approximated with an increasing degree of accuracy by a
heteroclinic orbit on the attractor. However, due to `errors'
associated with the approach toward the attractor, the solution does
not exactly follow the heteroclinic orbit structure on the
attractor; instead these increasingly small errors lead to an
approximation that might be subsumed under the notion of a
randomized sequence of heteroclinic orbits.

Randomization constitutes an important ingredient in several
contexts. The underlying structures that allowed one to obtain
mathematical proofs about the attractor of the diagonal Bianchi type
IX models in a Fermi propagated frame~\cite{rin00,rin01} are
specific for these models and are not available in other cases;
since it is such cases that are relevant for the present general
scenario, the diagonal Bianchi type IX models are misleading. Our
present analysis suggests that one has to know the (asymptotic)
history of a solution to unravel its asymptotic features; this
causes a dilemma since this requires that one finds the solution,
which seems unlikely. However, randomization makes it possible to
stochastically examine the cumulative effects of small and large
curvature phases and this allows one to estimate decay rates and
give a description of what is going to happen generically. In our
opinion statistical analysis will be an essential new ingredient in
future proofs.

In this paper we have obtained decay rates associated with
asymptotic suppression and freezing (see Appendix~\ref{asympconst})
which will hold in the neighborhood of the billiard attractor as
well. Since decay rates are likely to be ingredients in any proof,
it is important that numerics not only check that the dynamics of a
generic spatial point approaches the attractor, but that the
attractor is approached in the way described by the decay rates;
therefore, these results offer new input for numerical
investigations.

As we have seen, the asymptotic dynamics is inevitably restricted to
subsets of subsets, to boundaries of boundaries, which underlines
the importance of the hierarchical structure of invariant sets
associated with asymptotically local SH dynamics. However, there
also exists a hierarchy of invariant subsets that is based on the
number of Killing vectors the spacetime admits. This symmetry-based
hierarchy is of interest since it is associated with different
levels of technical difficulty; taking highly symmetric, and thus
technically simple, cases as a starting point, one can analytically
and numerically explore, step-by-step, the cases of decreasing
symmetry by making use of the gained experience; since several
structures that are relevant to the non-symmetric case already
appear in cases of special symmetries, this is likely to lead the
way to the generic case. The present work shows how the
Hubble-normalized Iwasawa frame approach naturally provides a
framework for dealing with this type of hierarchy; the formulation
comprises different symmetry levels; note, however, that the Iwasawa
decomposition is natural only when one has commuting spacelike
Killing vectors, and hence it is not appropriate for Bianchi types
VIII or IX. The simplest model with an oscillatory singularity as
described by the present formalism is the general Bianchi type
$\mathrm{VI}_{-1/9}$ model. The next level of difficulty as regards
oscillatory singularities is the general case with two commuting
spacelike Killing vectors; then the general case with one spacelike
Killing vector; and finally the general case with no Killing
vectors. Incidentally, by imposing geometric restrictions on the
Killing vectors such as hypersurface orthogonality, see,
e.g.,~\cite{waiell97,lim04}, one can obtain relatively simple
classes of models with non-oscillatory singularities. Exploiting
this outlined hierarchy of symmetries might prove to be the key to a
complete understanding of the nature of generic spacelike
singularities.

A dynamical system that is of particular interest in our analysis is
the system~\eqref{billiardeqs}; when we do not impose the Codazzi
constraint~(\ref{codazzibilliard}) we refer to this system as the
unconstrained billiard system, see also the discussion in
Section~\ref{dual} in connection with the `dominant' Hamiltonian. We
expect the unconstrained billiard system to possess the same generic
asymptotic behavior as the full system $\bm{X}$ in the neighborhood
of an asymptotically silent singularity; hence, in our approach, it
is this system that is of central importance for the asymptotic
dynamics, rather than, e.g., the Bianchi type IX system. We will
provide a thorough analysis of this system and its properties
elsewhere.

Another interesting issue
we have put little emphasis on in
%we have only addressed implicitly in
the present paper is the question of consistency. It is suggestive
to take the billiard attractor and the associated billiard
conjecture as a starting point for further developments by inserting
attractor sequences $\cA_\cT$ into the relevant equations; using
similar reasoning as done here for the suppression of variables, we
will be able to derive more substantial justification for the claim
that $E_\alpha{}^i \rightarrow 0$ (a major ingredient for asymptotic
silence) and $\bm{S}_{\mathrm{stable}}\rightarrow 0$ when
$\tau\rightarrow\infty$ (cf.\ steps $\mathbf{A}$ and $\mathbf{B}$ in
Section~\ref{billiardattractor}). In addition we will obtain further
information on the approach of solutions toward the asymptotic
limit. We will pursue issues of this kind in future work.

In this paper we have been concerned with the vacuum GR case,
however, the approach and methods we have developed will be
applicable to the case when matter is included and also to other
theories (see, e.g.,~\cite{dametal03,damnic05} and references
therein). As regards possible sources this naturally suggests a
study of the influence of matter on singularities. As a simple
example we refer to~\cite{heiugg06}, which gives an indication that,
e.g., Vlasov matter behaves differently in some respects than
perfect fluids. Of particular interest is the question of structural
stability of generic spacelike singularities, especially since
generic singularities
seem to uncover the essential properties of matter. There are
indications that a classification of the influence of matter on
generic singularity structure naturally rests on (i) energy
conditions and (ii) whether the effective propagation speed is less
than the speed of light or not. There are also indications that
suggest a subclassification based on the behavior of matter in the
case of speed of light propagation: e.g., massless scalar fields and
electromagnetic fields influence the generic spacelike singularity
in different ways; does this motivate a subclassification based on
spin? In the sublight case it seems to be natural to base a
subclassification on the question of what features are affected by
matter and what features are not, e.g., the Hubble-normalized
energy-momentum tensor may go to zero, which leads to an asymptotic
description of the geometry by vacuum solutions, but this does not
necessarily mean that, e.g., the Hubble-normalized rotation of a
perfect fluid tends to zero. Another issue is how, and if, matter
influences the connection between generic spacelike singularities
and weak null singularities in asymptotically flat spacetimes,
see~\cite{limetal06}. Questions and issues like these lead to a
variety of possible research projects that could make our
understanding more substantial.

In this paper we have assumed asymptotic silence and asymptotic
local dynamics, and our work has provided results and evidence that
support the consistency of this scenario for generic timelines.
However, this does not mean that there could not exist interesting
phenomena associated with special timelines, indeed, we believe that
a set of measure zero of timelines exhibit spike formation and
recurring `spike transitions'~\cite{andetal05} that are associated
with non-local dynamics. Even so, it still seems that asymptotic
silence prevails and that asymptotic local dynamics play an
important role---remarkably, spike transitions seem to be governed
by variations of the Kasner map, which is associated with
asymptotically local dynamics, hinting at further more deeply hidden
structures. Asymptotic spike formation is associated with the
unstable variables $N_1$, $R_1$, $R_3$ going through a zero at a
spatial point (where a zero in $N_1$ yields a `true spike', seen in
the curvature, while zeroes in $R_1$, $R_3$ yield false spikes,
i.e., gauge effects not seen in the curvature; presumably, zeroes in
$N_2$ and $R_2$ do not play a similar role since these variables are
generically suppressed in the asymptotic limit). There are hints that
spikes may interfere---destructively and constructively. Although
uncertain, constructive spike interference seems to dominate, which,
if correct, leads to asymptotic `spike cascading'~\cite{lim06}.
There are many other unresolved spike issues as well: are there
spikes that undergo infinitely many recurring spike transitions?
Where do spikes form and in what way---how does spike interference
work? Do spikes asymptotically form a dense set? Can spikes leave
observational imprints in, e.g., the cosmic microwave background?
These question are clearly interesting in themselves, moreover, a
clarification of some of these issues is of considerable interest in
the context of eventual generic singularity proofs.

Ultimately one might ask oneself the question why generic
singularities should be studied at all in a classical GR context?
Firstly, there exists a regime between the Planck era and the GUT
era where GR is expected to hold and where the approach toward the
singularity is presumably described by the dynamics toward a generic
singularity (recall that one of the points of inflation is to `erase
the effects of initial data' and that before this erasure a
singularity is presumably generic according to this line of
reasoning). Secondly, black hole formation is associated with
initial data reflecting the complexities of the real universe; one
would hence also in this case expect generic spacelike singularities
to play a role before one enters the Planck regime. Thirdly, the
formation of generic singularities is associated with considerable
structure, even in the case of spike formation: can this structure
be used to asymptotically quantize gravity where it needs to be
quantized, namely in the ultra-strong gravitational field in the
neighborhood of a generic spacelike singularity?\footnote{For the
exploitation of some of these structures in the context of
quantization of special models, see,
e.g.,~\cite{ashetal93,ashetal93b}.}

%%%%%%%%%%%%%%%%%%%%%%%%%%%%%%%%%%%%%%%%%%%%%%%%%%%%%%%%%%%%%%%%%%%%%%%%
\subsection*{Acknowledgments}
It is a pleasure to thank Lars Andersson, Henk van Elst, Woei Chet
Lim, and John Wainwright for many helpful and stimulating
discussions. CU is supported by the Swedish Research Council.
%%%%%%%%%%%%%%%%%%%%%%%%%%%%%%%%%%%%%%%%%%%%%%%%%%%%%%%%%%%%%%%%%%%%%%%

%%%%%%%%%%%%%%%%%%%%%%%%%%%%%%%%%%%%%%%%%%%%%%%%%%%%%%%%%%%%%%%%%%%%%%%%%%%%%%%%%%%%%%%%%%
%%%%%%%%%%%%%%%%%%%%%%%%%%%%%%%%%%%%%%%%%%%%%%%%%%%%%%%%%%%%%%%%%%%%%%%%%%%%%%%%%%%%%%%%%%
%%%%%%%%%%%%%%%%%%%%%%%%%%%%%%%%%%%%%%%%%%%%%%%%%%%%%%%%%%%%%%%%%%%%%%%%%%%%%%%%%%%%%%%%%%

%%%%%%%%%%%%%%%%%%%%%%%%%%%%%%%%%%%%%%%%%%%%%%%%%%%%%%%%%%%%%%%%%%%%%%%%%%%%%%%%%%%%%%%%%%
%%%%%%%%%%%%%%%%%%%%%%%%%%%%%%%%%%%%%%%%%%%%%%%%%%%%%%%%%%%%%%%%%%%%%%%%%%%%%%%%%%%%%%%%%%
%%%%%%%%%%%%%%%%%%%%%%%%%%%%%%%%%%%%%%%%%%%%%%%%%%%%%%%%%%%%%%%%%%%%%%%%%%%%%%%%%%%%%%%%%%

%%%%%%%%%%%%%%%%%%%%%%%%%%%%%%%%%%%%%%%%%%%%%%%%%%%%%%%%%%%%%%%%%%%%%%%%%%%%%%%%%%%%%%%%%%
%%%%%%%%%%%%%%%%%%%%%%%%%%%%%%%%%%%%%%%%%%%%%%%%%%%%%%%%%%%%%%%%%%%%%%%%%%%%%%%%%%%%%%%%%%
%%%%%%%%%%%%%%%%%%%%%%%%%%%%%%%%%%%%%%%%%%%%%%%%%%%%%%%%%%%%%%%%%%%%%%%%%%%%%%%%%%%%%%%%%%

%%%%%%%%%%%%%%%%%%%%%%%%%%%%%%%%%%%%%%%%%%%%%%%%%%%%%%%%%%%%%%%%%%%%%%%%%%%%%%%%%%%%%%%%%%
%%%%%%%%%%%%%%%%%%%%%%%%%%%%%%%%%%%%%%%%%%%%%%%%%%%%%%%%%%%%%%%%%%%%%%%%%%%%%%%%%%%%%%%%%%
%%%%%%%%%%%%%%%%%%%%%%%%%%%%%%%%%%%%%%%%%%%%%%%%%%%%%%%%%%%%%%%%%%%%%%%%%%%%%%%%%%%%%%%%%%

%%%%%%%%%%%%%%%%%%%%%%%%%%%%%%%%%%%%%%%%%%%%%%%%%%%%%%%%%%%%%%%%%%%%%%%%%%%%%%%%%%%%%%%%%%
%%%%%%%%%%%%%%%%%%%%%%%%%%%%%%%%%%%%%%%%%%%%%%%%%%%%%%%%%%%%%%%%%%%%%%%%%%%%%%%%%%%%%%%%%%
%%%%%%%%%%%%%%%%%%%%%%%%%%%%%%%%%%%%%%%%%%%%%%%%%%%%%%%%%%%%%%%%%%%%%%%%%%%%%%%%%%%%%%%%%%

\begin{appendix}
\label{bappe}

%%%%%%%%%%%%%%%%%%%%%%%%%%%%%%%%%%%%%%%%%%%%%%%%%%%%%%%%%%%%%%%%%%%%%%%%%%%%%%%%%%%%%%%%%%
\section{Iwasawa variables and useful equations} \label{relations}
%%%%%%%%%%%%%%%%%%%%%%%%%%%%%%%%%%%%%%%%%%%%%%%%%%%%%%%%%%%%%%%%%%%%%%%%%%%%%%%%%%%%%%%%%%

In this section we derive in some detail the connection between the
Hamiltonian Iwasawa variables of Section~\ref{Hamilton} and the
conformal Hubble-normalized variables of Section~\ref{confsec}. Let
us begin by recalling that for an Iwasawa frame
\begin{subequations}\label{coframeEE}
\begin{align}
\sEE{\alpha}{i} & = \exp(-b^\alpha) \N{\alpha}{i}\:,
&
\sE{\alpha}{i} & = \exp(b^\alpha) \bN{i}{\alpha}
& \qquad (\text{no summation over $\alpha$})\:, \\
\intertext{where $\N{\alpha}{i}$ and $\bN{i}{\alpha}$ are the upper
triangular matrices given by~\eqref{IwasawaNdef}
and~\eqref{inverseoffdiag}, respectively; the off-diagonal
components of $\N{\alpha}{i}$ ($\bN{i}{\alpha}$) are denoted by
$n_\alpha$ ($\bar{n}_{\alpha}$). Therefore, due to~\eqref{framedef},
we obtain for the conformal frame components:} \label{coframerel}
\EE{\alpha}{i} & = H \exp(-b^\alpha) \N{\alpha}{i}\:, &
\E{\alpha}{i} & = H^{-1} \exp(b^\alpha) \bN{i}{\alpha} & \qquad
(\text{no summation})\:.
\end{align}
\end{subequations}
In the following we use the notation $E_{\alpha} =
\E{\alpha}{\alpha}$ (no sum over $\alpha$) and analogously
$e_{\alpha} = \sE{\alpha}{\alpha}$. Note that the frame variables
$e_\alpha$ have been chosen to be positive, from which it follows
that also $E_\alpha>0$, since we take the cosmological model to be
expanding toward the future and hence $H>0$.
Equation~\eqref{coframeEE} can be inverted easily:
\begin{equation}\label{offdiagN}
b^{\alpha} = \log (e_\alpha) = \log \left(H\,E_{\alpha}\right)\:,
\qquad\quad
\bar{n}_1 = \frac{\E{2}{1}}{E_{2}}\:, \quad\,
\bar{n}_2 = \frac{\E{3}{1}}{E_{3}}\:, \quad\,
\bar{n}_3 = \frac{\E{3}{2}}{E_{3}}\:.
\tag{\ref{coframeEE}${}^\prime$}
\end{equation}

In order to make contact between the Hamiltonian variables and the
conformal Hubble-normalized variables we first consider the Hubble
scalar $H$. Since $g_{i j} = \sEE{\alpha}{i} \sEE{\beta}{j}
\delta_{\alpha\beta}$, we have
\begin{equation*}
\sqrt{g} = \exp(-b^1 -b^2 -b^3) = \exp\left(-\sum\nolimits_\alpha b^\alpha\right)\:;
\end{equation*}
accordingly, we obtain for the derivative (w.r.t.\ coordinate time $x^0$):
\begin{equation}\label{sqrtgder}
\frac{\partial}{\partial x^0} \,\sqrt{g} = - \sqrt{g}
\left( \frac{\partial}{\partial x^0} \sum\nolimits_\alpha b^\alpha \right).
\end{equation}
Since the derivative of $\sqrt{g}$ is proportional to the expansion $\theta$
according to
\begin{equation*}
\frac{1}{\sqrt{g}} \: \frac{\partial}{\partial x^0}\sqrt{g} =
N \theta = 3 N H\:,
\end{equation*}
this implies that
\begin{equation}\label{Hinb}
H = -\frac{1}{3} \frac{1}{N} \frac{\partial}{\partial x^0} \sum\nolimits_\alpha b^\alpha \:.
\end{equation}
Using the relation~\eqref{def} between $\dot{b}^\alpha$ and $\pi_{\alpha}$,
where we recall that $N = \tilde{N} \sqrt{g}$,
we obtain a representation of $H$ in terms of the momenta, i.e.,
\begin{equation}\label{Hinmomenta}
H = \frac{1}{12} \frac{1}{\sqrt{g}} \sum\nolimits_\alpha \pi_\alpha = \frac{1}{12}
\left(\exp\sum\nolimits_\alpha b^\alpha\right)   \sum\nolimits_\beta \pi_\beta \:.
\end{equation}

Equation~\eqref{Hinmomenta} suggests that we define a quantity
$\Lambda$,
\begin{equation}\label{LambdainHam}
\Lambda:= 2 H \sqrt{g} = \frac{1}{6} \sum\nolimits_\alpha
\pi_\alpha\:,
\end{equation}
which we will use frequently in the following. In terms of the frame
components, $\Lambda$ can be expressed as
\begin{equation}\label{Lambdainconf}
\Lambda = 2 H \sqrt{g} = 2\,(H^2\,E_1\,E_2\,E_3)^{-1}\:,
\end{equation}
since $\sqrt{g} = (e_1 e_2 e_3)^{-1} = H^{-3} (E_1 E_2 E_3)^{-1}$.

Inverting the commutator equations~\eqref{commuta} yields the
conformal Hubble-normalized variables in terms of $\cn$ and
$\E{\alpha}{i}$ and their derivatives:
\begin{subequations}
\begin{align}
\label{comm}
\Udot_{\alpha}  & =  \parb_{\alpha}\log\cn\:,
& q & = \textfrac{1}{3}\,E^{\alpha}{}_{i}\,\parb_{0}E_{\alpha}{}^{i}\:,\\
\label{comm1}
R^{\alpha} & =  \textfrac{1}{2}\,\epsilon^{\alpha}{}_{\beta}{}^{\gamma}\,
E^{\beta}{}_{i}\,\parb_{0}E_{\gamma}{}^{i}\:, & \Sigma_{\alpha\beta} &
=  - E^{\gamma}{}_{i}\,
\parb_{0}E_{\langle\alpha}{}^{i}\,\delta_{\beta\rangle \gamma}\:,\\
\label{comm2} A_{\alpha} & =
\textfrac{1}{2}\,E^{\beta}{}_{i}\,\parb_{\alpha}E_{\beta}{}^{i} -
\textfrac{1}{2}\,\EE{\beta}{i}\, \parb_{\beta} E_{\alpha}{}^{i}\:, & N^{\alpha\beta} & =
E^{(\alpha}{}_{i}\,\epsilon^{\beta)\gamma\delta}\,
\parb_{\gamma}E_{\delta}{}^{i}\:.
\end{align}
\end{subequations}

Inserting~\eqref{coframerel} into the equation~\eqref{comm} for $q$ yields
\begin{equation}\label{qinthings}
q = -\frac{\parb_0 H}{H} + \frac{1}{3} \parb_0 \sum\nolimits_\alpha b^\alpha\:.
\end{equation}
Since $\partial_{x^0} = H N \parb_0$, Equation~\eqref{Hinb} takes the simple
form $\parb_0 \sum_\alpha b^\alpha = -3$, which in turn implies
$\parb_0 \sqrt{g} = 3 \sqrt{g}$ from~\eqref{sqrtgder}.
Using the expression~\eqref{Hinmomenta} to compute $\parb_0 H$
we eventually obtain
\begin{equation}\label{qinpi}
q = 2 - \frac{1}{6} \Lambda^{-1} \parb_0 \sum\nolimits_\alpha \pi_\alpha\:.
\end{equation}
Note that Equation~\eqref{qinthings} reproduces
the dimensional equation $\parb_0 H = -(1+q) H$ for $H$.

Inserting~\eqref{coframerel} into~\eqref{comm1} yields
\begin{subequations}\label{RbisSigma}
\begin{align}
R_1 & = -\textfrac{1}{2} e^{b^3-b^2} \parb_0 n_3 \:, &
R_2 & = \textfrac{1}{2} e^{b^3-b^1}
\left( -n_3 \parb_0 n_1 + \parb_0 n_2 \right) \:, &
R_3  & = -\textfrac{1}{2} e^{b^2-b^1} \parb_0 n_1\:, \\
\Sigma_{23}  & = + \textfrac{1}{2} e^{b^3-b^2} \parb_0 n_3  \:, &
\Sigma_{13} & = \textfrac{1}{2} e^{b^3-b^1}
\left( -n_3 \parb_0 n_1 + \parb_0 n_2 \right) \:, &
\Sigma_{12}  & = + \textfrac{1}{2} e^{b^2-b^1} \parb_0 n_1
\end{align}
for $R_\alpha$ and $\Sigma_{\alpha\beta}$ ($\alpha\neq\beta$), and
\begin{equation}
\Sigma_{11} = \frac{1}{3} \parb_0 \sum_\alpha b^\alpha - \parb_0 b^1
\:,\qquad \Sigma_{22} = \frac{1}{3} \parb_0 \sum_\alpha b^\alpha -
\parb_0 b^2 \:,\qquad \Sigma_{33} = \frac{1}{3}
\parb_0 \sum_\alpha b^\alpha - \parb_0 b^3
\end{equation}
\end{subequations}
for the diagonal elements of $\Sigma_{\alpha\beta}$. The main
observation is that the off-diagonal components of
$\Sigma_{\alpha\beta}$ are given by the Fermi rotation parameters
$R_\alpha$ according to
\begin{equation}\label{Iwasawagauge}
( \Sigma_{23},  \Sigma_{31}, \Sigma_{12}) = (-R_1,R_2,-R_3)\:,
\end{equation}
which is a fundamental property of the Iwasawa gauge.

In order to express to r.h.\ sides of~\eqref{RbisSigma} in terms
of the momenta $\pi_\alpha$ and $\mathcal{P}_\alpha$
we write~\eqref{momentarelinv} in the form
\begin{subequations}\label{momentarelinvII}
\begin{align} \label{defII}
\parb_0 b^{\alpha} & = \frac{1}{\Lambda} \sum_{\beta} \cg^{{\alpha}
\beta}\pi_{\beta}\,, &
\parb_0 n_2 & = \frac{2}{\Lambda} \, e^{2(b^1-b^2)}\,\left(n_3\,{\cal
P}_1 + (e^{2(b^2-b^3)} + n_3^2)\,{\cal P}_2\right) ,\\
\label{def2II}
\parb_0 n_1 & = \frac{2}{\Lambda} \,e^{2(b^1-b^2)}\,
\left({\cal P}_1 + n_3\,{\cal P}_2\right) ,&
\parb_0 n_3 &
= \frac{2}{\Lambda} \, e^{2(b^2-b^3)} \, {\cal P}_3 \,,
\end{align}
\end{subequations}
where we use $\tilde{N}^{-1} \partial_{x^0} = (\Lambda/2) \parb_0$.
A straightforward computation then results in
\begin{subequations}\label{confinHami}
\begin{align}
\label{Sigmaalphapi}
\Sigma_\alpha &= \Sigma_{\alpha\alpha} = 2 - \Lambda^{-1}\,\pi_{\alpha}\:, &
& (\text{no sum}) \\
\label{R1}
R_1 &= -\Lambda^{-1}\,\exp(b^2-b^3)\,{\cal P}_3\:, &
& R_2 = \Lambda^{-1}\exp(b^1-b^3)\,{\cal P}_2\:, & \\
\label{R3} R_3 &=  -\Lambda^{-1}\,\exp(b^1 - b^2)(n_3\,{\cal P}_2 + {\cal P}_1)\:,
\end{align}
\end{subequations}
where $\Lambda$ is given in terms of the Hamiltonian variables by~\eqref{LambdainHam}.

Inversion of~\eqref{confinHami} yields
\begin{subequations}
\begin{align}
\label{pirel}
\pi_{\alpha} &= \Lambda\,[ 2 - \Sigma_{\alpha}]\:, & &
\tag{\ref{Sigmaalphapi}${}^\prime$} \\
{\cal P}_1 & = -\Lambda\,R_3\, E_2/E_1 +\bar{n}_3{\cal P}_2\:,
& {\cal P}_2 & = \Lambda\,R_2\, E_{3}/E_{1} \:, & &\qquad
\tag{\ref{R1}${}^\prime$}\\
{\cal P}_3 & = -\Lambda\,R_1 \, E_{3}/E_{2}\:,
\tag{\ref{R3}${}^\prime$}
\end{align}
\end{subequations}
where $\Lambda$ is given by~\eqref{Lambdainconf}.

To obtain the conformal Hubble-normalized variables $A_\alpha$ and
$N_{\alpha\beta}$ in terms of the Hamiltonian variables we proceed
analogously. Based on~\eqref{comm2} we find
%
%\begin{subequations}
\begin{align*}
N_\alpha &= \Lambda^{-1}\,\exp{(-2b^\alpha)}\,{\tilde N}_\alpha  &
& (\text{no summation over $\alpha$})\:, \\
N_{\alpha\beta} &= \Lambda^{-1}\,\exp{(-(b^\alpha +
b^\beta))}\,f_{\gamma -}  &
& (\alpha \neq \beta \neq \gamma)\:, \\
A_\alpha - r_\alpha &= \Lambda^{-1}\,\exp{(-(b^\alpha +
b^\beta))}\,f_{\gamma +}  &
& (\alpha \neq \beta \neq \gamma)\:,
\end{align*}
%\end{subequations}
%
where
\begin{subequations}
\begin{align}
\label{tildeN1}
{\tilde N}_1 &= 2(
\bptl_2\bar{n}_3-\bptl_3\bar{n}_1+
(\bar{n}_1+1)(\bar{n}_1\bptl_1\bar{n}_3-\bptl_2\bar{n}_3))\:, &   & \\
{\tilde N}_2 &= -2\bptl_1\bar{n}_3\:, &   & \\
{\tilde N}_3 &= 0\:, &   & \\
f_{1\pm} & =   \bptl_1(b^3 \pm b^2)\:, &
\bptl_1 &= \partial_1\:, \\
f_{2\pm} & =  \bptl_2(b^3 \pm b^1) \mp \bptl_1\bar{n}_1\:, &
\bptl_2 &= \bar{n}_1\partial_1 + \partial_2\:, \\
f_{3\pm} & =  \bptl_3(b^1 \pm b^2) \mp \bptl_2\bar{n}_3 +
\bar{n}_1\bptl_1\bar{n}_3 - \bptl_1\bar{n}_2\:, & \bptl_3 &=
\bar{n}_2\partial_1 + \bar{n}_3\partial_2 + \partial_3\:;
\end{align}
\end{subequations}
as usual we denote the diagonal elements of $N_{\alpha\beta}$ by $N_\alpha$.
In particular we see that
\begin{equation*}
N_3 = 0
\end{equation*}
in the Iwasawa gauge.

Finally we note that the momentum conjugate to $\lambda = \log \rho$
reads
\begin{equation*}
\pi_{\lambda} =
\Lambda\,\log[(e_1)^{(2-\Sigma_1)}(e_2)^{(2-\Sigma_2)}(e_3)^{(2-\Sigma_3)}]\:.
\end{equation*}
where $e_\alpha=H E_\alpha=\exp (b^\alpha)$.

We conclude this section by investigating the equations for the
conformal spatial frame components $\E{\alpha}{i}$ and the
consequences thereof in some detail; these equations will be useful
in Appendix~\ref{asympconst}. From~\eqref{ccomts0}
and~\eqref{dl13comts} we see
\begin{equation*}
\delzero \E{\alpha}{i} = F_\alpha^{\weg\beta} \E{\beta}{i}\:, \quad\text{ where }\quad
F_\alpha^{\weg\beta} = q \delta_{\alpha}^{\weg\beta} -
\Sigma_\alpha^{\weg\beta} -\epsilon_{\alpha\weg\gamma}^{\wweg\beta} R^\gamma\:.
\end{equation*}
In Iwasawa gauge~\eqref{Iwasawagauge} we obtain
\begin{subequations}
\begin{align}\label{delzeroEetc}
& \delzero E_1\; = (q - \Sigma_{1}) E_{1} & & \delzero E_2\; = (q -
\Sigma_{2}) E_{2} & & \delzero E_3\; = (q - \Sigma_{3}) E_{3}\:,
\end{align}
where $E_1$, $E_2$, $E_3$, and $\Sigma_1$, $\Sigma_2$, $\Sigma_3$,
are again the diagonal components of $\E{\alpha}{i}$ and
$\Sigma_{\alpha\beta}$, and
\begin{align}
& \delzero \E{2}{1} = 2 R_3 E_{1} + (q - \Sigma_{2}) \E{2}{1} &
& \delzero \E{3}{2} = 2 R_1 E_{2} + (q - \Sigma_{3}) \E{3}{1} \\
&\delzero \E{3}{1} = -2 R_2 E_{1} + 2 R_1 \E{2}{1} + (q - \Sigma_{3}) \E{3}{1}\:,
\end{align}
\end{subequations}
for the non-zero off-diagonal components of $E_\alpha{}^i$.
%Recall from Section~\ref{confsec} that $q = 2 \Sigma^2
%+( I_q)$, where $\Sigma^2 = \frac{1}{6} \Sigma_{\alpha\beta}
%\Sigma^{\alpha\beta}$.
Using that
\begin{equation*}
\delzero H = -(1+q) H
\end{equation*}
we find for the derivatives of $e_\alpha = H E_\alpha$:
\begin{equation}\label{edelzero}
\delzero e_1 = -(1+\Sigma_1) e_1
\qquad
\delzero e_2 = -(1+\Sigma_2) e_2
\qquad
\delzero e_3 = -(1+\Sigma_3) e_3\:.
\end{equation}
Furthermore, from
\begin{equation*}
\sqrt{g} = H^{-3} \left(E_{1} E_{2} E_{3} \right)^{-1}\:,
\end{equation*}
we obtain
\begin{equation*}
\delzero \sqrt{g} = 3 \sqrt{g}
\end{equation*}
from a direct computation based on~\eqref{delzeroEetc};
therefore,
\begin{equation}\label{Lambdaderiv}
\delzero \Lambda = \delzero (2 H \sqrt{g})  = (2 -q ) (2 H \sqrt{g}) = (2 - q) \Lambda  \:.
\end{equation}
Evidently, this equation is consistent with~\eqref{LambdainHam} and~\eqref{qinpi}.

%%%%%%%%%%%%%%%%%%%%%%%%%%%%%%%%%%%%%%%%%%%%%%%%%%%%%%%%%%%%%%%%%%%%%%%%%%5
\section{Deriving the generalized Kasner line element of BKL}
\label{Kasner}
%%%%%%%%%%%%%%%%%%%%%%%%%%%%%%%%%%%%%%%%%%%%%%%%%%%%%%%%%%%%%%%%%%%%%%%%%%5

The Kasner subset on the silent boundary $\E{\alpha}{i} =0$ is characterized
by $\Udot_\alpha = r_\alpha=0$ (SH condition) and
$0=1-\Sigma^2 = A_{\alpha} = N_{\alpha\beta}$.
We specialize to the Kasner circle
$\mathrm{K}^{\ocircle}$ by setting $R_\alpha=0$ which corresponds to
setting $\Sigma_{\alpha\beta} = 0$ ($\alpha \neq \beta$).
Therefore, for every point on $\mathrm{K}^{\ocircle}$,
$\Sigma_{\alpha\beta}= \mathrm{diag}[\hat{\Sigma}_{1},\hat{\Sigma}_{2},\hat{\Sigma}_{3}]$,
where the shear variables
$\hat{\Sigma}_{\alpha}$ ($\alpha =1 \ldots 3$) are temporally constant
spatially dependent functions;
we set $\mathrm{diag}[{\hat \Sigma}_{11},{\hat
\Sigma}_{22},{\hat \Sigma}_{33}]= \mathrm{diag}[3p_1-1,3p_2-1,3p_3-1]$.

In order to derive the leading asymptotic expressions for $H$
and $E_\alpha{}^i$ we first note that $q =2$ which is due to~\eqref{defofmuchsig2q}.
Equations~\eqref{Hubblenorm} and~\eqref{ccomts0},~\eqref{dl13comts} yield
\begin{equation*}
H = {\hat H}\,e^{3\tau}\, ,\quad E_ \alpha{}^i = {\hat
E}_\alpha{}^i\,e^{-3(1-p_\alpha)\tau}\, \quad (\rm{no\,\, sum\,\,
over}\,\, \alpha)\, ,
\end{equation*}
which leads to
\begin{equation*} e^\alpha{}_i = {\hat e}^\alpha{}_i\,e^{-3p_\alpha\,\tau}\, \quad
(\rm{no\,\, sum\,\, over}\,\, \alpha)\, ;
\end{equation*}
recall that hatted objects denote temporally constant spatially dependent functions.

In the synchronous gauge of BKL, i.e., $N=1$, $\cn=H$,
equation~(\ref{ttaurel}) leads to
\begin{equation*}
t = \hat{t} + \textfrac{1}{3}\,{\hat H}^{-1} e^{-3\tau}\:.
\end{equation*}
In addition, BKL also choose the synchronous time coordinate to be a
simultaneous bang time function. In the present derivation this
amounts to setting ${\hat t}=0$ as an initial value condition, which
can be accomplished by making a coordinate transformation.
Consequently we obtain $t \propto e^{-3\tau}$. With suitable
redefinitions, i.e., $\hat{e}^{1}{}_{i} \propto l_i$,
$\hat{e}^{2}{}_{i} \propto m_i$, $\hat{e}^{3}{}_{i} \propto n_i$,
and $a=t^{p_1}$, $b=t^{p_2}$, $c=t^{p_3}$, the `generalized Kasner
line element' of BKL ensues:
\begin{equation}\label{kasnerlineel}
ds^2 = -dt^2 + [a^2\,l_i\,l_j + b^2\,m_i\,m_j +
c^2\,n_i\,n_j]\,dx^idx^j\:.
\end{equation}
Thus, instead of ad hoc assuming the above line element, and
inserting it into the field equations in order to analyze its
consistency, we have now derived it as the lowest order perturbation
of $\mathrm{K}^{\ocircle}$ on the silent boundary.

Finally, note that an equivalent derivation can be performed in a
Fermi frame instead of an Iwasawa frame: we could start with
$R_\alpha=0$, which would lead to a 4-dimensional ellipsoid of
Kasner equilibrium points characterized by temporally constant
values ${\hat \Sigma}_{\alpha\beta}$, and then make a temporally
constant rotation and diagonalize this matrix.

%%%%%%%%%%%%%%%%%%%%%%%%%%%%%%%%%%%%%%%%%%%%%%%%%%%%%%%%%%%%%%%%%%%%%%%%%%5
\section{Kasner solutions in a rotating frame}
\label{KasnernonFermi}
%%%%%%%%%%%%%%%%%%%%%%%%%%%%%%%%%%%%%%%%%%%%%%%%%%%%%%%%%%%%%%%%%%%%%%%%%%5

It is customary to represent a generalized Kasner solution in a
Fermi frame with constant diagonal shear variables:
$\mathrm{diag}[\hat{\Sigma}_{1},\hat{\Sigma}_{2},\hat{\Sigma}_{3}] =
\mathrm{const}$. However, when one considers a representation in a
frame that rotates w.r.t.\ the Fermi frame, one obtains
time-dependent non-diagonal shear variables
$\Sigma_{\alpha\beta}(\tau)$ that are related to
$\hat{\Sigma}_{\alpha}$ through time-dependent rotations
$O^\alpha_\beta(\tau)$, i.e.,
\begin{equation*}
\Big(O^\alpha_{\alpha^\prime} \, \Sigma_{\alpha\beta}
\,O^\beta_{\beta^\prime}  \Big)_{\alpha^\prime, \beta^\prime} =
\mathrm{diag}\big(\hat{\Sigma}_1,\hat{\Sigma}_2,\hat{\Sigma}_3\big)
= \mathrm{diag}\big( 3 p_1 -1, 3 p_2 -1 ,3 p_3 -1 \big) =
\mathrm{const}\:,
\end{equation*}
%
%cf.~\eqref{Kasnercircledef2};
where the quantities
$(\hat{\Sigma}_1,\hat{\Sigma}_2,\hat{\Sigma}_3)$ can be regarded as
the time-independent eigenvalues of $\Sigma_{\alpha\beta}(\tau)$.
Accordingly, $\Sigma_{\alpha\beta}(\tau)$ generates time-independent
frame invariants: the linear invariant reduces to the trace-free
shear condition, i.e., $\mathrm{tr}
\:\big(\Sigma_{\alpha\beta}(\tau)\big) = \Sigma_1(\tau) +
\Sigma_2(\tau) + \Sigma_3(\tau) = \hat{\Sigma}_1 + \hat{\Sigma}_2 +
\hat{\Sigma}_3 = 0$; the quadratic invariant yields the Gauss
constraint, i.e.,
\begin{equation*}
\Sigma^{\alpha\beta}\Sigma_{\alpha\beta} =
\Sigma_1^2 + \Sigma_2^2 + \Sigma_3^2 + 2R_1^2 + 2R_2^2 + 2R_3^2 =
{\hat \Sigma}_1^2 + {\hat \Sigma}_2^2 + {\hat \Sigma}_3^2 = 6\: .
\end{equation*}
Finally, the conserved cubic invariant is given by
\begin{equation*}
\det \Sigma_{\alpha\beta}(\tau) = \hat{\Sigma}_1 \hat{\Sigma}_2 \hat{\Sigma}_3 = \mathrm{const}\:.
\end{equation*}
This invariant can also be expressed in terms of a Weyl scalar. The
magnetic Weyl tensor is identically zero for Kasner, but the
electric part is non-trivial and thus one obtains one non-zero
quadratic Weyl scalar, related to $\det(\Sigma_{\alpha\beta})$
according to
\begin{equation*}
\mathcal{W}_1 = \frac{C_{abcd}C^{abcd}}{48H^4} =
2 - \det(\Sigma_{\alpha\beta}) = \mathrm{const}\:.
\end{equation*}
Note that $C_{abcd}C^{abcd}$ is the Kretschmann scalar associated
with the metric ${\bf g}$ while the quantity
$C_{abcd}C^{abcd}/H^4$ is the
Kretschmann scalar connected with ${\bf G}$.

Since $\det\Sigma_{\alpha\beta} = \hat{\Sigma}_1 \hat{\Sigma}_2
\hat{\Sigma}_3$ and $(\hat{\Sigma}_1, \hat{\Sigma}_2,
\hat{\Sigma}_3) \in \mathrm{K}^{\ocircle}$, $\det\Sigma_{\alpha
\beta}$ is a constant in the range $[-2,2]$. Conversely, each value
of $\det\Sigma_{\alpha \beta}$ from the interval $(-2,2)$ generates
a unique ordered triple $\hat{\Sigma}_\alpha < \hat{\Sigma}_\beta <
\hat{\Sigma}_\gamma$ ($\alpha\neq\beta\neq\gamma\neq\alpha$), while
$\det\Sigma_{\alpha \beta} = \pm 2$ leads to the LRS points
$\Sigma_\alpha = \pm 2$, $\Sigma_\beta =  \Sigma_\gamma = \mp 1$
($\alpha\neq\beta\neq\gamma\neq\alpha$).

The value of $\det\Sigma_{\alpha\beta}$ thus characterizes a Kasner
state frame invariantly and uniquely; however, it is usually more
convenient to use the \textit{Kasner parameter} $u$ instead, which is
defined implicitly through
\begin{equation}\label{ufidef}
\det(\Sigma_{\alpha\beta})= 2 - \frac{27u^2(1+u)^2}{(1+u+u^2)^3}=
\mathrm{const}\, ,\qquad u\in [1,\infty]\:,
\end{equation}
where $\det\Sigma_{\alpha\beta}$ is a monotonically increasing
function of $u$. The Kasner parameter $u \in [1,\infty]$
parametrizes the one-parameter set of Kasner states $(p_1,p_2,p_3)$
according to
\begin{equation}\label{ueqA}
p_\alpha=\frac{-u}{1+u+u^2}\, ,\qquad p_\beta=\frac{1+u}{1+u+u^2}\,
,\qquad p_\gamma = \frac{u(1+u)}{1+u+u^2}\:,
\end{equation}
for sector $(\alpha,\beta,\gamma)$ of the Kasner circle, i.e.,
$p_\alpha < p_\beta < p_\gamma$ ($\hat{\Sigma}_\alpha <
\hat{\Sigma}_\beta < \hat{\Sigma}_\gamma$); here $u \in (1,\infty)$.
It is easy to check that $p_\alpha + p_\beta + p_\gamma = 1$ and
$p_\alpha^2 + p_\beta^2 + p_\gamma^2 = 1$. The value $u=1$
represents the points $\mathrm{Q}_1, \mathrm{Q}_2, \mathrm{Q}_3$,
i.e., the three equivalent representations of the non-flat LRS
Kasner solution; $u=\infty$ defines the Taub points $\mathrm{T}_1,
\mathrm{T}_2, \mathrm{T}_3$
and thus the Taub solution.%
\footnote{BKL
  define the Kasner parameter $u$ according to
  $p_1 = -u/(1+u+u^2)$, $p_2 = (1+u)/(1+u+u^2)$, $p_3 = u(1+u)/(1+u+u^2)$,
  cf.~\eqref{ueqA}; here,
  the order of the Kasner exponents is fixed as
  $p_1\leq p_2 \leq p_3$ by definition. We find it more natural to
  use the frame independent definition~\eqref{ufidef}, and permute the ordering
  of $p_\alpha$ according to the sector one considers when dealing
  with frame dependent matters.}

The considerations of this section simplify the analysis of the
silent Kasner subset $\mathcal{K}$, see the subsection `The silent
Kasner subset' in Section~\ref{dynosc}. Since a solution of
$\mathcal{K}$ is a representation of a Kasner solution in an Iwasawa
frame, properties such as $u = \mathrm{const}$ easily follow.

%%%%%%%%%%%%%%%%%%%%%%%%%%%%%%%%%%%%%%%%%%%%%%%%%%%%%%%%%%%%%%%%%%%%%%%%%%5
\section{Multiple transitions}
\label{multiple}
%%%%%%%%%%%%%%%%%%%%%%%%%%%%%%%%%%%%%%%%%%%%%%%%%%%%%%%%%%%%%%%%%%%%%%%%%%5

Here we give the proofs of some statements made in
Section~\ref{dynosc} concerning multiple transitions; as an
alternative to using frame invariants we will use elementary methods
from the theory of dynamical systems.

\textit{Multiple frame transitions} are solutions of~\eqref{unstableeqs}
on the silent Kasner subset $\mathcal{K}$,
%characterized by $(N_1, N_2) \equiv (0,0)$,
for which at least two of the variables $(R_1,R_2,R_3)$
do not vanish identically.
The equations are
\begin{subequations}\label{frametranssys}
\begin{align}
\partial_{\tau} \Sigma_1 & = - 2 R^2_2 - 2 R^2_3 &
\partial_{\tau} \Sigma_2 & = - 2 R^2_1 + 2 R^2_3 &
\partial_{\tau} \Sigma_3 & = 2 R^2_1 + 2 R^2_2  \\
\partial_{\tau} R_1 & = (\Sigma_2 - \Sigma_3) R_1 + 2 R_2 R_3 &
\partial_{\tau} R_2 & = (\Sigma_1 - \Sigma_3) R_2 &
\partial_{\tau} R_3 & = (\Sigma_1 - \Sigma_2) R_3 - 2 R_1 R_2
\end{align}
\end{subequations}
together with $\Sigma^2 = 1$. The set $R_2 = 0$ is an invariant
subset of~\eqref{frametranssys} (while $R_1 = 0$ and $R_3 = 0$ are
not invariant), hence possible classes of multiple frame transitions
are $\cT_{R_3 R_1}$ (which satisfy $R_1 \not\equiv 0$, $R_2 \equiv
0$, $R_3 \not\equiv 0$) and $\cT_{R_1 R_2 R_3}$ (where all three
functions are different from zero).

First, let $R_2 = 0$, i.e., we consider $\cT_{R_3 R_1}$ transitions.
In this case, $R_1 =0$ and $R_3 =0$ are invariant subspaces; we
consider, without loss of generality, the state space
defined by $R_1 >0$ and $R_3 >0$; since $\Sigma^2 =1$, the
closure of this state space is compact.
The functions $\Sigma_1$ and $\Sigma_3$ are monotone on the state space,
and hence the monotonicity principle%
    \footnote{The monotonicity principle essentially states that
      if there exists
      a function that is monotone along the orbits of a state space
      with compact closure, then the $\alpha$/$\omega$-limit set
      of every orbit is contained on the boundary of the state space;
      see~\cite{waiell97,fjaetal06} and references therein.%
    }
implies that the $\alpha$-limit and the $\omega$-limit of each orbit
must lie on the boundaries $R_1 =0$ (i.e., on
$\overline{\mathcal{B}}_{R_3}$) or $R_3 = 0$ (i.e., on
$\overline{\mathcal{B}}_{R_1}$) of the state space. Using the known
structure of the flow on the boundaries (which contain the single
transition orbits) we conclude that the $\alpha$-/$\omega$-limit
must be a source/sink on $\mathrm{K}^{\ocircle}$. The fixed points
in four of the sectors of $\mathrm{K}^{\ocircle}$ are saddles; the
fixed points in sector $(321)$ are sources, since both $R_1$ and
$R_3$ belong to the unstable subspaces, see Figure~\ref{Ktrigg}; the
points in sector $(123)$ are sinks. Thus, $\cT_{R_3R_1}$ orbits
originate in sector $(321)$ and end in sector $(123)$, see
Figure~\ref{TR1R3f}. (Recall that, in this paper, `time' is directed
toward the past singularity; the nomenclature `$\alpha$-limit set'
and `$\omega$-limit set' is used in accordance with the chosen
time-direction.)

Second, let $R_2 \neq 0$, i.e., we consider $\cT_{R_1 R_2 R_3}$
transitions; without loss of generality we assume $R_2 > 0$. Since
$\Sigma_1$ and $\Sigma_3$ are monotone functions, the monotonicity
principle yields that the $\alpha$-/$\omega$-limit of every orbit
must lie on the boundary $R_2 = 0$. Applying the previous result on
the limit sets of $\cT_{R_3R_1}$ transitions we see that the
$\alpha$-limit must be an equilibrium point in sector $(321)$ and
the $\omega$-limit a point in sector $(123)$.

For \textit{mixed frame/curvature transitions}
$\cT_{N_1 R_1}$ the system~\eqref{unstableeqs} reduces to
\begin{subequations}\label{mixedcurvsys}
\begin{align}
\label{mixedcurv1}
\partial_{\tau} \Sigma_1  & = 2(1-\Sigma^2)\Sigma_1 + \textfrac{2}{3} N_1^2 &
\partial_{\tau} N_1 & = -2 (\Sigma^2 + \Sigma_1) N_1 \\
\partial_{\tau} R_1 & =  2(1-\Sigma^2) R_1 + (\Sigma_2 - \Sigma_3) R_1  & & \\
\partial_{\tau} \Sigma_2  & = 2(1-\Sigma^2)\Sigma_2 - 2 R^2_1 - \textfrac{1}{3} N_1^2 &
\partial_{\tau} \Sigma_3  & = 2(1-\Sigma^2)\Sigma_3 + 2 R^2_1  + \textfrac{1}{3} N_1^2
\end{align}
\end{subequations}
together with $(1-\Sigma^2) = (1/12) N_1^2$; the remaining variables
vanish identically. Without loss of generality we consider the case
$N_1 >0$ and $R_1 >0$. By~\eqref{mixedcurvsys} the function
$(\Sigma_1 + 4)$ is monotone on the (relatively compact) state
space, and hence the monotonicity principle guarantees that the
$\alpha$- and the $\omega$-limit of every orbit resides on the
boundaries $N_1 =0$ or $R_1 = 0$. Using the structure of the flow on
the boundaries (as given by the single curvature transitions of
Figure~\ref{singletrans}), we find that the $\alpha$-limit of each
orbit is a fixed point on sector $(132)$, and that the
$\omega$-limit is a point on sector $(213)$ or $(231)$ of
$\mathrm{K}^{\ocircle}$, see Figure~\ref{TN1R1f}.

From~\eqref{mixedcurvsys} we can derive the validity of the Kasner
map~\eqref{BKLMap} explicitly. The equations~\eqref{mixedcurv1} form
a two-dimensional decoupled system for $\Sigma_1$ and $N_1$ by means
of the Gauss constraint; the underlying reason for this is that
$\Sigma_1$ and $N_1$ are invariant under the frame rotations in the
$\langle e_2, e_3 \rangle$-plane that are induced by $R_1$. Since
the subsystem for $\Sigma_1$ and $N_1$ is identical to the one for
the single curvature transition case $\cT_{N_1}$, the respective
solutions are identical; in particular, as in~\eqref{Zeq},
\begin{equation*}
\Sigma_1 = 4 + (1 +u^2) \zeta \:,
\end{equation*}
where $\zeta$ is defined as in~\eqref{Zeq2} and $u = u_-$. This
implies that $(\Sigma_1)_+$ is identical to its counterpart in the
$\cT_{N_1}$ case (provided that the initial values $(\Sigma_1)_-$
are the same). Note, however, that $\Sigma_2$ and $\Sigma_3$ differ
from the functions given in~\eqref{Zeq}, since $R_1 \neq 0$ changes
the evolution of these quantities. In particular we find that
$(\Sigma_2)_+^{\cT_{N_1R_1}} = (\Sigma_3)_+^{\cT_{N_1}}$ and
$(\Sigma_3)_+^{\cT_{N_1R_1}} = (\Sigma_2)_+^{\cT_{N_1}}$, which
reflects a relative rotation of the axes. However, in terms of the
Kasner parameter $u$ the two final states are indistinguishable,
which is a consequence of the frame invariance of $u$; hence
$u_+$ is given in terms of $u_-$ by the Kasner map~\eqref{BKLMap}.

The equations on the silent Bianchi type $\mathrm{VI}_0$/$\mathrm{VII}_0$
subsets are given by
\begin{equation*}
\partial_{\tau} \Sigma_\alpha = 2 (1-\Sigma^2) \Sigma_\alpha + {}^3\mathcal{S}_{\alpha\alpha}
\qquad
\partial_{\tau} N_{a} = -2(\Sigma^2 + \Sigma_{a})N_{a}\:,
\end{equation*}
$\alpha = 1\ldots 3$, $a=1,2$\,;
here, the quantities ${}^3\mathcal{S}_{\alpha\alpha}$ are polynomials in $N_1$ and $N_2$
given by~\eqref{someabbrevs}.
The Gauss constraint~\eqref{gausscon} reads
\begin{equation}\label{gaussconVII}
\Sigma_1^2 + \Sigma_2^2 + \Sigma_3^2 + \frac{1}{2} (N_1 - N_2)^2 = 6 \:.
\end{equation}
Since ${}^3\mathcal{S}_{33} = -(1/3) (N_1 - N_2)^2$ we obtain
\begin{equation}\label{2S3dec}
\partial_\tau (2 -\Sigma_3) = \frac{1}{6} (N_1 - N_2)^2 (2-\Sigma_3) \:,
\end{equation}
i.e., $(2 - \Sigma_3)$ is a strictly monotonically increasing function
unless $N_1 = N_2$.

The Bianchi type $\mathrm{VI}_0$ subset is given by $(N_1 > 0)
\wedge (N_2 < 0)$ (or, equivalently, by the reversed inequalities).
Hence, the Gauss constraint~\eqref{gaussconVII} leads to a compact
state space given by $\Sigma_1^2+\Sigma_2^2 +\Sigma_3^2 + (1/2) N^2
< 6$, where $N_1 = N$ and $N_2 = N - \sqrt{12(1 - \Sigma^2)}$. The
function $(2-\Sigma_3)$ is strictly monotone, since $(N_1 - N_2) >
0$ everywhere. It thus follows from the monotonicity principle that
the $\alpha$-/$\omega$-limits of orbits must reside on the
boundaries of the state space, i.e., on $\mathrm{K}^\ocircle \cup
\mathcal{B}_{N_1} \cup  \mathcal{B}_{N_2}$. Using the known
structure of the flow on the Bianchi type II subsets, see
Figure~\ref{singletrans}, and the resulting local properties of the
fixed points on $\mathrm{K}^\ocircle$, we find that the
$\alpha$-limit of each orbit is the fixed point $\mathrm{T}_3$ on
$\mathrm{K}^\ocircle$ and that the $\omega$-limit is a point on
sector $(312)$ or $(321)$.

The Bianchi type $\mathrm{VII}_0$ subset is given by $(N_1 > 0)
\wedge (N_2 > 0)$ (or, equivalently, by the reversed inequalities).
In this case, the Gauss constraint does not enforce a compact state
space. The derivative of the function $(2-\Sigma_3)$ is not positive
when $N_1 = N_2$, see~\eqref{2S3dec}; however, $3 \partial_\tau^3
(2-\Sigma_3) |_{N_1 = N_2} = 4 (\Sigma_1 -\Sigma_2)^2 N_1^2
(2-\Sigma_3)$, and hence $(2 - \Sigma_3)$ is a strictly
monotonically increasing function except on the LRS subset $N_1 =
N_2$ and $\Sigma_1 = \Sigma_2$, which yields that
$(\Sigma_1,\Sigma_2,\Sigma_3) = (-1,-1,2)$ or
$(\Sigma_1,\Sigma_2,\Sigma_3) = (1,1,-2)$. The latter is an
invariant subset with $\partial_\tau N_a = -4 N_a$, which represents
the non-flat LRS Kasner solution in a Bianchi type $\mathrm{VII}_0$
symmetry foliation. The set $(\Sigma_1,\Sigma_2,\Sigma_3) =
(-1,-1,2)$, on the other hand, is a one-dimensional set of fixed
points, parametrized by $N_1 = N_2 = \mathrm{const} > 0$, which we
denote by $\mathcal{L}_3^+$; each fixed point represents the
Minkowski spacetime in a Bianchi type $\mathrm{VII}_0$ symmetry
foliation, see~\cite[pp.\ 130, 133]{waiell97}. It can be proved,
see~\cite[Theorem 5]{rin03}, that the $\alpha$-limit of every orbit
in the Bianchi type $\mathrm{VII}_0$ state space is one of the fixed
points on $\mathcal{L}_3^\pm$. The $\omega$-limits are points on
sectors $(312)$ or $(321)$ of $\mathrm{K}^\ocircle$.

For each individual Bianchi type $\mathrm{VI}_0$ or $\mathrm{VII}_0$
orbit, the quantity $|N_1|$ (and, equivalently, $|N_2|$) goes
through a maximum value. It is of interest to note that there exists
a uniform bound, i.e., there exists $\epsilon > 0$ such that
$\max_\tau |N_1(\tau)| \geq \epsilon$ (and $\max_\tau |N_2(\tau)|
\geq \epsilon$) uniformly for all Bianchi type
$\mathrm{VI}_0$/$\mathrm{VII}_0$ orbits. In order to see this,
assume that the assertion is false; then there exists for all
$n\in\mathbb{N}$ a Bianchi type $\mathrm{VI}_0$ (Bianchi type
$\mathrm{VII}_0$) orbit $\mathfrak{T}_n$ such that $\max_\tau
|N_1(\tau)| < 1/n$, i.e., $|N_1(\tau)| <1/n$ $\forall \tau$. In
particular, $|N_1(\tau_n)| < 1/n$, where $\tau_n$ is such that
$\Sigma_3(\tau_n) = -1$. Since the state space is compact (or, in
Bianchi type $\mathrm{VII}_0$, the intersection of the state space
with the set $|N_1| \leq 1$), without loss of generality the limit
$\lim_{n\rightarrow \infty} \Sigma_\alpha(\tau_n)$
exists---otherwise we go over to a subsequence. If the limit is
neither of the points $\mathrm{Q}_1$, $\mathrm{Q}_2$, then
$(\Sigma_{\alpha}, N_1,N_2)(\tau_n)$ converges to a point on a
$\cT_{N_2}$ single curvature transition orbit. Hence, for
sufficiently large $n$, the corresponding orbit $\mathfrak{T}_n$
shadows this $\cT_{N_2}$ transition. However, the flow on the
boundary of $\mathcal{B}_{N_1 N_2-}$ ($\mathcal{B}_{N_1N_2+}$)
forces $\mathfrak{T}_n$ to also shadow the $\cT_{N_1}$ transition
preceding/succeeding it. Along these $\cT_{N_1}$ transitions we have
$|N_1| \not< (1/n)$ for large $n$, which is a contradiction. The
argument is analogous when $\lim_{n\rightarrow \infty}
\Sigma_\alpha(\tau_n)$ is one of the points $\mathrm{Q}_1$,
$\mathrm{Q}_2$; in this case we employ the saddle structure of these
fixed points to show that for sufficiently large $n$, the orbit
$\mathfrak{T}_n$ shadows a $\cT_{N_1}$ transition along which $|N_1|
\not< (1/n)$, which is again the desired contradiction.

%%%%%%%%%%%%%%%%%%%%%%%%%%%%%%%%%%%%%%%%%%%%%%%%%%%%%%%%%%%%%%%%%%%%%%%%%%
\section{Behavior of an auxiliary quantity}
\label{Bbehave}
%%%%%%%%%%%%%%%%%%%%%%%%%%%%%%%%%%%%%%%%%%%%%%%%%%%%%%%%%%%%%%%%%%%%%%%%%%

In this appendix we consider Eq.~\eqref{A2eq}, i.e.,
\begin{equation}\label{A3eq}
\frac{\partial}{\partial \tau} B = 2 (1-\Sigma^2) B + 2 R_3^2 - 2
R_1^2\:;
\end{equation}
we show that, generically, $B_i \sim \,\mathrm{const}\:G_{l(i)}$
as $i\rightarrow \infty$, cf.~\eqref{Bincgen}, along a sequence
of transitions $\cS_\cT$. (In Section~\ref{growth} this statement
was shown to be true for sequences $\cS_\cT$ that do not contain
$\cT_{N_2}$ and $\cT_{N_2 R_2}$ transitions, which is equivalent to
assuming $N_2 \equiv 0$.)

For the following it is convenient to use a continuous time variable
along $\cS_\cT$; we introduce `sequence time' $\lambda$ by
$\lambda = (1/2) + (\arctan \tau)/\pi$, i.e., $\tau = \tan
\left[(2\lambda -1) \pi/2\right]$; accordingly, the time interval
$\lambda \in (i,i+1)$ (for $i\in\mathbb{N}$) corresponds to $\tau$
ranging in $(-\infty,\infty)$ during the $i$\raisebox{0.7ex}{\small
th} transition. By using sequence time we define the growth function
$G$ by
\begin{equation*}%\label{gengrofun}
G(\lambda): \qquad \partial_\lambda G = 2 (1 -\Sigma^2)
\frac{d\tau}{d\lambda}\, G \:, \qquad G(0) = 1\:.
\end{equation*}
By construction, $G(\lambda)$ coincides with $G_{l(i)}$ as defined
in Section~\ref{growth} for $\lambda =i$; this is because $G_l =
\prod_{k=0}^{l-1} g_k$, where the growth factors $g_k$ are defined
as the ratio $\zeta_+/\zeta_-$ (where $\zeta_\pm =
\lim_{\tau\rightarrow \pm\infty} \zeta$) that arises from the
equation $\partial_\tau \zeta = 2(1-\Sigma^2) \zeta$ for the
$k$\raisebox{0.7ex}{\small th} curvature transition, see~(\ref{geq})
and Section~\ref{growth}.

The derivative of $B$ is closely related to derivatives of
$\Sigma_2$; in sequence time $\lambda$ we find
\begin{equation}\label{A3Sigma2}
\frac{\partial}{\partial \lambda} B =
\begin{cases}
\frac{\partial}{\partial \lambda} (\Sigma_2 -2) + 2
(1-\Sigma^2)\frac{d\tau}{d\lambda}
\big[B - (\Sigma_2 -2)\big] & \text{when } N_2 =0\\[0.5ex]
\frac{\partial}{\partial \lambda} (\Sigma_2 +4) + 2
(1-\Sigma^2)\frac{d\tau}{d\lambda} \big[B - (\Sigma_2 +4)\big]&
\text{when } N_1 =0 \:.
\end{cases}
\end{equation}
Integrating~\eqref{A3eq} thus amounts to integrating the two
equations in~\eqref{A3Sigma2} alternately: as long as $N_2 =0$ we
use the first equation; whenever there is a $\cT_{N_2}$ or a
$\cT_{N_2 R_2}$ transition the second equation is employed. For
$k\in\mathbb{N}$, let $\lambda_k \in \mathbb{N}$ be a sequence of
times that represents the `switches' between the two equations
in~\eqref{A3Sigma2}, i.e., $N_2 \equiv 0$ in
$(\lambda_k,\lambda_{k+1})$ for even $k$ and $N_1 \equiv 0$ in
$(\lambda_k,\lambda_{k+1})$ for odd $k$. Integration
of~\eqref{A3Sigma2} and a somewhat cumbersome iteration yields
\begin{equation*}
B(\lambda) = (\Sigma_2 -2)(\lambda) + 6 G(\lambda) \left[
\sum_{j=1}^k (-1)^j \frac{1}{G(\lambda_j)}\right] + G(\lambda) [B(0)
- (\Sigma_2 -2)(0)]
\end{equation*}
for $\lambda \in (\lambda_k,\lambda_{k+1})$ with $k$ even (and a
similar result for odd $k$). By the Leibniz criterion the sum
$\sum_{j=1}^k (-1)^j G(\lambda_j)^{-1}$ converges as $k\rightarrow
\infty$. Moreover, $|\sum_{j>k}  (-1)^j G(\lambda_j)^{-1}| <
G(\lambda_{j+1})^{-1}$, hence $G(\lambda) \sum_{j>k}  (-1)^j
G(\lambda_j)^{-1}$ is bounded. Since also $(\Sigma_2 -2)(\lambda)$
is bounded, we can replace the sum with an infinite series and collect
the remaining terms in a term labeled $b(\lambda)$,
\begin{equation*}
B(\lambda) = G(\lambda) \left[B(0) +6  \sum_{j=1}^\infty (-1)^j
\frac{1}{G(\lambda_j)} -(\Sigma_2 -2)(0) \right] + b(\lambda) \:;
\end{equation*}
here $b(\lambda)$ is bounded (and oscillatory). Consequently,
generically, $B(\lambda) \sim \,\mathrm{const}\: G(\lambda)$ holds
asymptotically, or, in different notation, $B_i \sim
\,\mathrm{const}\: G_{l(i)}$.

%%%%%%%%%%%%%%%%%%%%%%%%%%%%%%%%%%%%%%%%%%%%%%%%%%%%%%%%%%%%%%%%%%%%%%%%%%
\section{Convergence of sums}
\label{convergenceofsums}
%%%%%%%%%%%%%%%%%%%%%%%%%%%%%%%%%%%%%%%%%%%%%%%%%%%%%%%%%%%%%%%%%%%%%%%%%%

Consider a sequence of transitions $\cS_\cT$ on the oscillatory
subset $\mathcal{O}$; in our nomenclature, $\cS_\cT$ denotes
a generic sequence, i.e., a sequence that does not
contain any double curvature transitions, see Section~\ref{transition};
therefore, the sequence $\cS_\cT$ is associated with a Kasner
sequence $(u_l)_{l\in\mathbb{N}}$, where
$l\in\mathbb{N}$ is the index that consecutively numbers the
curvature transitions of $\cS_\cT$.

In this appendix we first analyze the series
\begin{equation}\label{convseries}
\sum_l G_l^{-1} |(\Delta\Sigma_2)|_l \:,
\end{equation}
where $G_l$ is the growth function and $|(\Delta\Sigma_2)|_l$
denotes the sum of the absolute changes in $\Sigma_2$ occurring for
the frame transitions between the $(l-1)$\raisebox{0.7ex}{\small th}
and the $l$\raisebox{0.7ex}{\small th} curvature transition. Series
of the type~\eqref{convseries} arise in connection with
the asymptotic suppression
of the variable $N_2$, see Section~\ref{asysha} and the discussion
below, and in the context of the
asymptotic freezing of the Hamiltonian variables
$n_i$ and $\mathcal{P}_i$, see Section~\ref{asympconst}. In the
following we prove that the series~\eqref{convseries} converges. To
establish this result we analyze contributions of the sum in large
and small curvature phases separately.

First, we consider~\eqref{convseries} for a large curvature phase
(which we assume to be
associated with a sufficiently large value of $\eta_u$):
$l\in [l_{\mathrm{i}}, l_{\mathrm{f}}]$. Obviously,
$|(\Delta\Sigma_2)|_l$ is uniformly bounded (by $2 \sqrt{3}$), so
that $G_l^{-1} |(\Delta\Sigma_2)|_l  \leq \mathrm{const}\:
G_l^{-1}$. Furthermore, in a large curvature phase, $g_l >
C_{\mathrm{p}}
>1$, cf.~\eqref{glfree}, so that $G_{l+1}^{-1} = g_l^{-1} G_l^{-1} <
G_l^{-1} C_{\mathrm{p}}^{-1}$ for all $l\in[l_{\mathrm{i}},
l_{\mathrm{f}}]$. Hence, for a large curvature phase, each term in
the sum can be bounded by a term of a geometric series.
%, $G_l^{-1} |(\Delta\Sigma_2)|_l \leq \mathrm{const}\:
%G_{l_{\mathrm{i}}}^{-1} C_{\mathrm{p}}^{-(l-l_{\mathrm{i}})}$.

Second, we consider~\eqref{convseries} for a small curvature phase;
since small curvature phases dominate over large curvature phases in
the probabilistic description of sequences, see Section~\ref{stat},
the subsequent considerations are of central signifiance. Recall
that the prototype of a small curvature phase is the alternating
sequence of $\cT_{N_1}$ and $\cT_{R_3}$ transitions in a
neighborhood of the point $\mathrm{T}_3$, which is characteristic
for billiard sequences $\cB_\cT$. The following considerations are
adapted to small curvature phases of this kind; however, the results
hold for general small curvature phases with obvious minor
modifications. Let $l\in [l_{\mathrm{in}}, l_{\mathrm{out}}]$ be a
small curvature phase (where we assume $\uin > \eta_u+1$ to obtain a
phase consisting of at least two curvature transitions):
\begin{equation}\label{convcaptur}
\begin{split}
\sum_{l = l_{\mathrm{in}}+1}^{l_{\mathrm{out}}+1} G_l^{-1}
|(\Delta\Sigma_2)|_l & = \sum_{l = \lin+1}^{\lout+1}
G_{l_{\mathrm{in}}}^{-1} \Big[\prod_{j=\lin}^{l-1}
g_j\Big]^{-1}|(\Delta\Sigma_2)|_{l}  \\
& = G_{\lin}^{-1} \sum_{l = \lin+1}^{\lout+1}
\left(\frac{1+ u_l +u_l^2}{1+\uin+(\uin)^2}\right) \left(\frac{3 ( 1
+ 2 u_l)}{1 + u_l + u_l^2}\right) \\
& = G_{l_{\mathrm{in}}}^{-1}
\sum_{k=1}^{L} \frac{3 (1 + 2 (\uin - k))}{1 + \uin + (\uin)^2} =
G_{l_{\mathrm{in}}}^{-1} \frac{3 L (2 \uin
-L)}{1 + \uin + (\uin)^2} \\
& \approx G_{l_{\mathrm{in}}}^{-1} \frac{3 ((\uin)^2 - \eta_u^2)}{1
+ \uin + (\uin)^2} \,<\, 3 \:G_{l_{\mathrm{in}}}^{-1} \, < \,
\mathrm{const}\:\,G_{l_{\mathrm{in}}}^{-1} \:,
\end{split}
\end{equation}
% % number on last line:
%\begin{align}\label{convcaptur}
%\nonumber \sum_{l = l_{\mathrm{in}}+1}^{l_{\mathrm{out}}+1} G_l^{-1}
%|(\Delta\Sigma_2)|_l & = \sum_{l = \lin+1}^{\lout+1}
%G_{l_{\mathrm{in}}}^{-1} \Big[\prod_{j=\lin}^{l-1}
%g_j\Big]^{-1}|(\Delta\Sigma_2)|_{l}  \\
%\nonumber
%& = G_{\lin}^{-1} \sum_{l = \lin+1}^{\lout+1}
%\left(\frac{1+ u_l +u_l^2}{1+\uin+(\uin)^2}\right) \left(\frac{3 ( 1
%+ 2 u_l)}{1 + u_l + u_l^2}\right) \\
%\nonumber
%& = G_{l_{\mathrm{in}}}^{-1}
%\sum_{k=1}^{L} \frac{3 (1 + 2 (\uin - k))}{1 + \uin + (\uin)^2} =
%G_{l_{\mathrm{in}}}^{-1} \frac{3 L (2 \uin
%-L)}{1 + \uin + (\uin)^2} \\
%& \approx
%G_{l_{\mathrm{in}}}^{-1} \frac{3 (u_{l_{\mathrm{in}}}^2 -
%\eta_u^2)}{1 + \uin + (\uin)^2} \,<\, 3 \:G_{l_{\mathrm{in}}}^{-1}
%\, < \, \mathrm{const}\:\,G_{l_{\mathrm{in}}}^{-1} \:.
%\end{align}
%
%%\begin{align}\label{convcaptur}
%%\nonumber
%%\sum_{k= i_0}^{ i_0 + k_{\mathrm{exit}}} G_k^{-1}
%%|(\Delta\Sigma_2)|_k & = G_{i_0}^{-1} \sum_{k = i_0}^{ i_0 +
%%k_{\mathrm{exit}}} \left(\frac{1+ u_k
%%+u_k^2}{1+u_{i_0}+u_{i_0}^2}\right) \left(\frac{3 ( 1 + 2 u_k)}{1 +
%%u_k + u_k^2}\right) = G_{i_0}^{-1} \sum_{k=0}^{k_{\mathrm{exit}}}
%%\frac{3 (1 + 2 (u_{i_0} - k))}{1 + u_{i_0} + u_{i_0}^2}
%%\:= \\
%%& = G_{i_0}^{-1} \frac{3 (u_{i_0}^2 - \eta_u^2)}{1 + u_{i_0} +
%%u_{i_0}^2} < 3 G_{i_0}^{-1} \:.
%%\end{align}
%
where we have used~\eqref{grofacscp} for the growth factors $g_j$
and~\eqref{DeltaSigma2s} for $|(\Delta\Sigma_2)|_l$.

From~\eqref{Glfree} we see that $G_{l_{\mathrm{in}}}^{-1} <
G_{l_{\mathrm{in}}-1}^{-1} C_{\mathrm{p}}^{-1}$; furthermore,
$G_{l_{\mathrm{out}}+2}^{-1} < G_{l_{\mathrm{in}}}^{-1} (1-\eta_u
+\eta_u^2)/(1 + \uin + (\uin)^2)$, cf.~\eqref{gfcap}, so that
$G_{l_{\mathrm{out}}+2}^{-1} < G_{l_\mathrm{in}}^{-1}
C_{\mathrm{p}}^{-1}$. This leads to the important conclusion that
the sum associated with a small curvature phase can be treated on an
equal footing with a single term in a large curvature phase, i.e.,
\begin{align*}
\sum_l  G_l^{-1} |(\Delta\Sigma_2)|_l & = \:\,  \ldots \,+
(\Delta\Sigma_2)|_{l_{\mathrm{in}}-1} G_{l_{\mathrm{in}}-1}^{-1} +
(\Delta\Sigma_2)|_{l_{\mathrm{in}}} G_{l_{\mathrm{in}}}^{-1}\: + \\
& \qquad\quad +\:
\left[\sum_{l = \lin+1}^{\lout+1} G_l^{-1} |(\Delta\Sigma_2)|_l \right]+
(\Delta\Sigma_2)|_{l_{\mathrm{out}}+2} G_{l_{\mathrm{out}}+2}^{-1} +\,\ldots  \\[1ex]
& < \mathrm{const} \: \left( \ldots + G_{l_{\mathrm{in}}-1}^{-1} +
G_{l_{\mathrm{in}}}^{-1}+ G_{l_{\mathrm{in}}}^{-1}
+G_{l_{\mathrm{out}}+2}^{-1} +\ldots   \right) \\[0.5ex]
& < \mathrm{const} \: \left( \ldots + G_{l_{\mathrm{in}}-1}^{-1} +
G_{l_{\mathrm{in}}-1}^{-1} C_{\mathrm{p}}^{-1} +
G_{l_{\mathrm{in}}-1}^{-1} C_{\mathrm{p}}^{-1} +
G_{l_{\mathrm{in}}-1}^{-1} C_{\mathrm{p}}^{-2} + \ldots \right)\:.
\end{align*}
Therefore,
\begin{equation*}
\sum_l G_{l}^{-1} |(\Delta\Sigma_2)|_l \leq \mathrm{const}\:
\sum_{k} C_{\mathrm{p}}^{-k} < \infty\:,
\end{equation*}
which proves the asserted convergence of the series.

%Note that the sum~\eqref{convseries}
%is only piecewise defined
%for a sequence of transitions that contains
%double curvature transitions, since the growth factor
%is infinite for $\cT_{N_1 N_2}$.

For asymptotic sequences of transitions $\cA\cS_\cT$ the above
considerations apply in the asymptotic regime when $\cA\cS_\cT$
shadows sequences on $\mathcal{O}$ and thus consists of approximate
transitions $(\cT_i)_{i\in\mathbb{N}}$. The convergence result also
holds analogously for asymptotic sequences $\cA\cS_\cO$: since the
possible (approximate) Bianchi type $\mathrm{VI}_0$ and
$\mathrm{VII}_0$ orbits are associated with a very large increase in
$G$, convergence in the sum is strengthened.

The established convergence of~\eqref{convseries} has direct
applications in our discussion of asymptotic freezing in
Appendix~\ref{asympconst}. The convergence result is also the basis
for the proof of the assertion made in Section~\ref{asysha} that
Eq.~\eqref{N1R32} simplifies to~\eqref{N1R3} along an asymptotic
sequence of transitions $\cA\cS_\cT$; this will been shown in the
following:

Integration of~\eqref{N1R32}, i.e.,
\begin{equation}\label{N1R32again}
\partial_\tau \big(N_1 R_3^2\big) =
2 \left[2 (1-\Sigma^2) -\Sigma^2 - \Sigma_2\right] N_1 R_3^2 - 4
R_1^2 G^{-1} (N_1 R_3^2)\:,
\end{equation}
yields
\begin{equation*}
\log \big(N_1 R_3^2\big) = 2 \int \left( 2 (1-\Sigma^2) -\Sigma^2 -
\Sigma_2\right)d\tau - 4 \int R_1^2 G^{-1} d\tau + \mathrm{const}\:.
\end{equation*}
In the asymptotic regime, the integrals can be written as sums over
the (approximate) transitions of the asymptotic sequence; the second
integral contributes only for transitions involving $R_1$. We can
make the estimate
\begin{equation}\label{R12G-1}
\int R_1^2 G^{-1} d\tau \leq \sum_l G_{l+1}^{-1} \int_{\tau_l}^{\tau_{l+1}} R_1^2
d\tau\:,
\end{equation}
where the limits of the integral on the r.h.s.\ denote the
$\tau$-time that elapses during the frame transitions between
curvature transition number $l$ and $l+1$. In the asymptotic regime,
the integrals of this type can be approximated by using exact
transitions; in particular, there exists a uniform upper bound,
which suffices to prove convergence in large curvature phases in
complete analogy to the considerations above.

The variable $R_1$ is non-zero along $\cT_{R_1}$ transitions; along
other single transitions it vanishes. Note that multiple transitions
involving $R_1$ can be ignored in the present context, since we
focus on small curvature phases, where multiple transitions do not
appear. For a $\cT_{R_1}$ transitions we have $-2 R_1^2 =
\partial_\tau \Sigma_2$, hence
$-2 \int_{\tau_l}^{\tau_{l+1}} R_1^2 d\tau =
(\Sigma_2)_+ -(\Sigma_2)_-$. Therefore, since
$|2 \int_{\tau_l}^{\tau_{l+1}} R_1^2| \leq |\Delta\Sigma|_l$,
the sum in~\eqref{R12G-1} can be estimated by a sum
of the type~\eqref{convseries}; we thus obtain convergence and
\begin{equation*}
\log \big(N_1 R_3^2\big) = 2 \int \left( 2 (1-\Sigma^2) -\Sigma^2 -
\Sigma_2\right)d\tau + \mathrm{const}^\prime\:
\end{equation*}
as $\tau\rightarrow \infty$. Consequently, the term $-4 R_1^2 G^{-1}
(N_1 R_3^2)$ in~\eqref{N1R32again} can be neglected as
$\tau\rightarrow \infty$; this leads to the simplified
equation~\eqref{N1R3} and thus establishes the claim.

To prove asymptotic freezing of $\pi_\lambda$ in
Section~\ref{asympconst}, we need to consider series of a slightly
more general type than~\eqref{convseries}. Consider a sequence of
transitions $\cS_\cT$ and the associated series
\begin{equation}\label{convseries2}
\sum_l G_l^{-a} u_l^{-c}\:,
\end{equation}
where $a>0$, $c>0$ (and for simplicity $2 a - c > -1$); note in
this context that the series
\begin{equation*}
\sum\limits_{l} G_l^{-1} \log G_l\: \frac{1}{u_l}\:,
\qquad
\sum\limits_{l} G_l^{-1} \frac{\log u_l}{u_l}\:\:,
\end{equation*}
see~\eqref{someserlog}, can be estimated by a series
of the type~\eqref{convseries2} (with $a = 1 -\epsilon$, $c = 1-\epsilon$).
In the following we prove convergence of the series~\eqref{convseries2}.

In analogy to~\eqref{convcaptur}, we obtain during a small curvature
phase
\begin{equation*}
\sum_{l = l_{\mathrm{in}}+1}^{l_{\mathrm{out}}+1} G_l^{-a}
(u_l)^{-c} =
G_{l_{\mathrm{in}}}^{-a} \sum_{l =\lin+1}^{\lout+1}
\Big(\frac{1+ u_l +u_l^2}{1+\uin+(\uin)^2}\Big)^a
(u_l)^{-c} < (1+\epsilon)
G_{l_{\mathrm{in}}}^{-a} \sum_{l = l_{\mathrm{in}}+1}^{l_{\mathrm{out}}+1}
\frac{(u_l)^{2 a -c}}{(\uin)^{2 a}} \:,
\end{equation*}
where we have used an estimate of the type $(1+ u + u^2)<(1+\epsilon)
u^2$ for some $\epsilon > 0$ (which is small when $\eta_u$ is
large). Further, by estimating the sum through
the associated integral, i.e.,
\begin{equation*}
\sum_{l = l_{\mathrm{in}}+1}^{l_{\mathrm{out}}+1} (u_l)^{2 a -c}
=
\sum_{k = 1}^{L}
(\uin - k)^{2 a -c}
\leq \mathrm{const}\:
\int_{\eta_u}^{\uin} u^{2 a -c} d u\:,
\end{equation*}
we arrive at
\begin{equation}
\label{capturedab}
\sum_{l = l_{\mathrm{in}}+1}^{l_{\mathrm{out}}+1} G_l^{-a}
(u_l)^{-c}
< \mathrm{const}\: G_{l_{\mathrm{in}}}^{-a}
\frac{1}{2 a -c+1} (\uin)^{1-c} = \mathrm{const}
\:G_{l_{\mathrm{in}}}^{-a} (\uin)^{1-c}\:.
\end{equation}
Modulo the large curvature phase terms whose sum converges
straightforwardly, and which hence can be suppressed, the
series~\eqref{convseries2} can be written as
\begin{equation*}
\sum_l G_l^{-a} (u_l)^{-c} = \sum_j \sum_{l =
l_{\mathrm{in},j}+1}^{l_{\mathrm{out},j}+1} G_l^{-a}
(u_l)^{-c}\:,
\end{equation*}
where $[(\lin)_j, (\lout)_j]$ denotes the $j$\raisebox{0.7ex}{\small
th} small curvature phase. Based on the estimate~\eqref{capturedab}
we therefore obtain
\begin{equation}\label{sumofcaptured}
\sum_l G_{l}^{-a} (u_l)^{-c} < \mathrm{const}\: \sum_j
\left(G_{(\lin)_j}\right)^{-a} \left(\uinn{j}\right)^{1-c}\:,
\end{equation}
where $\uinn{j}$ is the initial value of $u$ for the small curvature
phase number $j$, which begins with transition $l_{\mathrm{in},j}$.
%For brevity, we denote $u_{l_{\mathrm{in},j}}$ by $u^{\mathrm{in}}_j$.

Regarded as a random variable, $\uin = \varkappa^{-1}$ is associated
with the probability density $\hat{w}(\varkappa)$, see~\eqref{hatw}.
Therefore, $\uin =\varkappa^{-1}$ does not possess a finite
expectation value, cf.~Section~\ref{stat}; however,
$(\uin)^{1-c}=\varkappa^{c-1}$ does, since
\begin{equation}\label{uinpower}
\langle(\uin)^{1-c}\rangle \, =\, \int_0^{\eta_u^{-1}}
\hat{w}(\varkappa)\varkappa^{c-1} d\varkappa \sim \hat{w}(0)
\frac{1}{c} (\eta_u^{-1})^c \sim \,\frac{1}{c}\: \eta_u^{1-c} \:.
\end{equation}
It follows that the expectation value of the sum
in~\eqref{sumofcaptured} exists, i.e.,
\begin{equation*}
\left\langle \sum_{j=1}^{n} \left(G_{(\lin)_j}\right)^{-a}
\left(u^{\mathrm{in}}_j\right)^{1-c}\right \rangle = \sum_{j=1}^{n}
\left(G_{(\lin)_j}\right)^{-a}\, \frac{\eta_u^{1-c}}{c}\:.
\end{equation*}
Hence, since $G_{(\lin)_j}$ increases geometrically, i.e.,
$G_{(\lin)_{j+1}} > C_{\mathrm{p}} \,G_{(\lin)_j}$, the limit
$n\rightarrow \infty$ exists. We conclude that the
series~\eqref{sumofcaptured} and thus~\eqref{convseries2} converges.

%%%%%%%%%%%%%%%%%%%%%%%%%%%%%%%%%%%%%%%%%%%%%%%%%%%%%%%%%%%%%%%%%%%%%%%%%%
\section{Hitting intervals stochastically}
\label{hitting}
%%%%%%%%%%%%%%%%%%%%%%%%%%%%%%%%%%%%%%%%%%%%%%%%%%%%%%%%%%%%%%%%%%%%%%%%%%

Let us introduce a continuous random variable $\varkappa \in [0,1]$
with probability density
\begin{equation*}
w: [0,1]\ni \varkappa \,\mapsto\, w(\varkappa) \in \mathbb{R}\:,
\qquad w(0) > 0\:.
\end{equation*}
A simple example is a uniformly distributed random variable, i.e.,
$w(\varkappa) \equiv 1$.

Now consider a sequence $(\delta_n)_{n\in\mathbb{N}}$ where
$\delta_n \in (0,1)$ $\forall n$ and where $\delta_n \rightarrow 0$
($n\rightarrow \infty$). The sequence $(\delta_n)_{n\in\mathbb{N}}$
generates a sequence of intervals $[0,\delta_n]$ whose length
decreases as $n\rightarrow \infty$.

The probability that $\varkappa\in [0,\delta_n]$ is given by
\begin{equation}\label{hitprobability}
\mathrm{P}\big(\varkappa \in [0,\delta_n]\big) = \int_0^{\delta_n}
w(\varkappa) d\varkappa =: \tilde{\delta}_n \approx w(0) \delta_n\:,
\end{equation}
where the approximation holds for $\delta_n \ll 1$. (Evidently, the
result is exact for a uniformly distributed random variable.) We
denote the event $\varkappa\in [0,\delta_n]$ as a `hit'; the
associated probability $\mathrm{P}\big(\varkappa \in
[0,\delta_n]\big) =\tilde{\delta}_n$ is the `hit probability'.

Consider a series of $n=1,\ldots, N$ trials. The probability that no
hits occur during $N$ trials, i.e., $\varkappa \not\in [0,\delta_n]$
$\forall n =1, \ldots, N$ (`no hits'), is given by
\begin{equation*}
\mathrm{P}^N(0) = \prod_{n=1}^N (1- \tilde{\delta}_n)\:;
\end{equation*}
analogously, the probability for `one hit' is
\begin{equation*}
\mathrm{P}^N(1) = \sum_{n=1}^N \tilde{\delta}_n \prod_{j=1,j\neq
n}^N (1-\tilde{\delta}_j) = \Big(\prod_{n=1}^N (1-\tilde{\delta}_n)\Big)
\sum_{i=1}^N \frac{\tilde{\delta}_i}{1-\tilde{\delta}_i} =
\mathrm{P}^N(0) \sum_{i=1}^N
\frac{\tilde{\delta}_i}{1-\tilde{\delta}_i} \:.
\end{equation*}
Finally, the probability for `$N$ hits' is the product $\prod_{n=1}^N
\tilde{\delta}_n$.

As a simple example let $\delta_n = \epsilon n^{-1}$ with $\epsilon
\ll 1$. We obtain
\begin{equation*}
\log \mathrm{P}^N(0) = \sum_{n=1}^N \log(1 - \tilde{\delta}_n)
\approx \sum_{n=1}^N \big( - w(0) \delta_n + O(\delta_n^2)\big) =
-\epsilon w(0)\sum_{n=1}^N \frac{1}{n} + O(\epsilon^2)\:;
\end{equation*}
therefore $\log \mathrm{P}^N(0) \rightarrow -\infty$ as
$N\rightarrow \infty$, and $\mathrm{P}^N(0) \rightarrow 0$ as
$N\rightarrow \infty$ (independently of $\epsilon$). Similarly,
\begin{equation*}
\mathrm{P}^N(1) = \mathrm{P}^N(0) \sum_{i=1}^N
\frac{\tilde{\delta}_i}{1-\tilde{\delta}_i} \approx \exp\left(-
\epsilon w(0)\sum_{n=1}^N \frac{1}{n}\right) \: \left[\epsilon w(0)
\sum_{n=1}^N \frac{1}{n} \right]\:;
\end{equation*}
therefore $\mathrm{P}^N(1) \rightarrow 0$ as $N\rightarrow \infty$
(independently of $\epsilon$). The analogous result holds for the
general case: one can show that $\mathrm{P}^N(i) \rightarrow 0$ as
$N\rightarrow \infty$ for all $i\in\mathbb{N}$. We conclude that for
an infinite series of trials the probability of getting a finite
number $i$ of hits is zero, $\mathrm{P}^\infty(i) = 0$. We expect an
infinite number of hits as $N\rightarrow \infty$ (although the
intervals between two hits are expected to become increasingly
large).

As a second simple example let $\delta_n = \epsilon n^{-2}$ with
$\epsilon \ll 1$. We obtain
\begin{equation*}
\log \mathrm{P}^N(0) = \sum_{n=1}^N \log(1 - \tilde{\delta}_n)
\approx \sum_{n=1}^N \big( - w(0) \delta_n + O(\delta_n^2)\big) =
-\epsilon w(0)\sum_{n=1}^N \frac{1}{n^2} + O(\epsilon^2)\:;
\end{equation*}
in this case, $\log \mathrm{P}^N(0) \rightarrow -c$ as $N\rightarrow
\infty$ (with $c>0$), hence $\mathrm{P}^N(0) \rightarrow e^{-c} > 0$
as $N\rightarrow \infty$. Analogously, one can show that
$\mathrm{P}^N(i) \rightarrow \mathrm{const} > 0$ as $N\rightarrow
\infty$ for all $i \in \mathbb{N}$; in addition, $\mathrm{P}^N(i)
\leq \mathrm{P}^N(0) (c^i/i!)$. Accordingly, for an infinite series
of trials, the probability of getting a finite number $i$ of hits is
finite, i.e., $\mathrm{P}^\infty(i) > 0$, where
$\mathrm{P}^\infty(i) \rightarrow 0$ as $i\rightarrow \infty$. An
important conclusion is that the expectation value of the number of
hits is finite,
\begin{equation*}
\langle \# \text{hits}\rangle = \sum_{i=1}^\infty
i\,\mathrm{P}^\infty(i)  \,<\, \infty\:.
\end{equation*}
Note that the expectation value exists because
$\mathrm{P}^\infty(i)$ falls off sufficiently rapidly as
$i\rightarrow \infty$.

\begin{remark}
For the above computations we have made use of the approximation
$\tilde{\delta}_n \approx w(0) \delta_n$,
cf.~\eqref{hitprobability}. Alternatively, we could have used an
estimate of the type $\tilde{\delta}_n \leq \|w\|_{\infty}
\delta_n$.
\end{remark}

%%%%%%%%%%%%%%%%%%%%%%%%%%%%%%%%%%%%%%%%%%%%%%%%%%%%%%%%%%%%%%%%%%%%%%%%%%%%%%%%%%%%%%%%%%
\section{Asymptotic constants of the motion}
\label{asympconst}
%%%%%%%%%%%%%%%%%%%%%%%%%%%%%%%%%%%%%%%%%%%%%%%%%%%%%%%%%%%%%%%%%%%%%%%%%%%%%%%%%%%%%%%%%%

The analysis of Damour et al.~\cite{dametal03} indicates that the
off-diagonal Iwasawa frame variables $n_i$, their conjugate momenta
$\mathcal{P}_i$, and the momentum $\pi_\lambda$ conjugate to
$\lambda = \log \rho$, see Section~\ref{Hamilton}, are asymptotic
constants of motion; i.e., these quantities converge to functions
that only depend on the spatial variables (so-called asymptotic
freezing). In this appendix we will derive these results from a
dynamical systems perspective.

%-----------------------------------------------------------------------
\subsection*{Asymptotic freezing of $\bm{n_i}$ and $\pmb{\mathcal{P}}_{\bm{i}}$}
%\subsection*{Asymptotic freezing of $\bm{n}_{\bm{i}}$ and $\bm{\mathcal{P}}_{\bm{i}}$}
%%$\bm{n}_{\bm{i}}$ and $\bm{\mathcal{P}}_{\bm{i}}$}
%-----------------------------------------------------------------------

We begin by establishing asymptotic freezing of the quantities
$\barn_i$ and $\mathcal{P}_i$. In the conformal Hubble normalized
variables these quantities are given by
\begin{subequations}\label{momentas}
\begin{align}
\bar{n}_1 & = \frac{\E{2}{1}}{E_{2}}\:, \qquad\quad \bar{n}_2 =
\frac{\E{3}{1}}{E_{3}}\:, \qquad\quad \bar{n}_3 =
\frac{\E{3}{2}}{E_{3}}\\
\mathcal{P}_1 & = -2H{\sqrt g}\, R_3 \,\frac{E_2}{E_1} +
{\bar n}_3\, \mathcal{P}_2\, ,\qquad
\mathcal{P}_2 = 2 H {\sqrt g}\, R_2 \,\frac{E_{3}}{E_{1}}\, ,\qquad
\mathcal{P}_3 = - 2H{\sqrt g}\, R_1 \, \frac{E_{3}}{E_{2}}\:,
\end{align}
\end{subequations}
where we have used the notation
\begin{equation*}
E_1 := E_1{}^1\, ,\quad E_2 := E_2{}^2\, ,\quad E_3 := E_3{}^3\, ;
\end{equation*}
see Appendix~\ref{relations} for details.
Since $n_1 = -\bar{n}_1$, $n_2 =-\bar{n}_2 + \bar{n}_1 \bar{n}_3$, $n_3 = -\bar{n}_3$,
cf.~\eqref{inverseoffdiag}, asymptotic freezing of $\bar{n}_i$ implies that
also $n_i$ converge to constants.

Consider the quantities $\barn_i$. From the formulas for
$\parb_0 \E{\alpha}{i}$, see Appendix~\ref{relations}, it is straightforward
to compute
\begin{equation*}
\parb_0 \barn_1 = 2 R_3 \frac{E_1}{E_2} \:,\qquad
\parb_0 \barn_2 = -2 R_2 \frac{E_1}{E_3} + 2 R_1 \frac{\E{2}{1}}{E_3} \:,\qquad
\parb_0 \barn_3 = 2 R_1 \frac{E_2}{E_3} \:.
\end{equation*}
Let us now consider a sequence of transitions (or a billiard sequence)
that is approximated by an attractor sequence $\cA_\cT$
in the asymptotic regime $\tau\rightarrow \infty$.
Recall that $\cA_\cT$
is an infinite concatenation of $\cT_{N_1}$, $\cT_{R_1}$, and $\cT_{R_3}$
transitions on the billiard attractor $\mathcal{O}_{\mathcal{BA}}$, see
Section~\ref{billiardattractor}.
As usual, for notational simplicity we drop the distinction between
exact transitions and approximate transitions.
To lowest order, the evolution equations for $\barn_i$ along $\cA_\cT$
are given by
\begin{equation}\label{delzerons}
\partial_\tau \bar{n}_1= -2 R_3 \frac{E_{1}}{E_{2}} \:,\qquad \partial_\tau
\bar{n}_2 = -2\,\bar{n}_1 \, R_1 \frac{E_{2}}{E_{3}}\:,\qquad \partial_\tau
\bar{n}_3 = -2 R_1 \frac{E_{2}}{E_{3}}\:.
\end{equation}
Let us first focus on the evolution of $\barn_1$.
We rewrite the evolution equation as
\begin{equation}\label{n1bar2}
\partial_\tau \bar{n}_1= -2 R_3^2 \left(\frac{E_{1}}{E_{2} R_3}\right)\:,
\end{equation}
and investigate the behavior of the quantity $E_1/(E_2 R_3)$. Along
$\cA_\cT$ we obtain
\begin{equation}\label{E1oE2R3}
\frac{\partial}{\partial \tau} \Big( \frac{E_1}{E_2 R_3} \Big) =
-2(1 -\Sigma^2)\,\Big( \frac{E_1}{E_2 R_3} \Big)\:;
\end{equation}
therefore, $E_1/(E_2 R_3)$ behaves like the inverse of the auxiliary
quantity $A$, see~(\ref{A1eq}). Accordingly,
$E_1/(E_2 R_3)$ decreases like the inverse of the growth function, i.e.,
\begin{equation}\label{E1E2R3}
\Big( \frac{E_1}{E_2 R_3}\Big)_i = G_{l(i)}^{-1}\,\Big(
\frac{E_1}{E_2 R_3}\Big)_0 \:,
\end{equation}
see~\eqref{Ainc}. Inserting this result into~\eqref{n1bar2} we are
able to explicitly compute the change in $\barn_1$ for each
transition: $\Delta \barn_1$ is zero (in the lowest order
approximation) for each $\cT_{R_1}$ and each $\cT_{N_1}$ transition
and
\begin{equation}\label{barn1Delta}
(\Delta \barn_1)_i = - \Big( \frac{E_1}{E_2 R_3} \Big)_i \:|\Delta
\Sigma_2|_i = - \mathrm{const}\: G_{l(i)}^{-1} \:|\Delta \Sigma_2|_i
\:,
\end{equation}
when $\cT_i$ is a $\cT_{R_3}$ frame transition; $(\Delta\Sigma_2)_i$
denotes the change in $\Sigma_2$ during this transition;
here we have used that $E_1/(E_2 R_3)$ is constant during $\cT_i$
and that $\partial_\tau \Sigma_2 = 2 R_3^2$.

To establish asymptotic freezing of $\barn_1$ we must show that the
summation of all $(\Delta\barn_1)_i$ is finite (where the sum is
over all $\cT_{R_3}$ frame transitions of $\cA_\cT$). Let
$\mathcal{I}_{R_3} = \{i\,|\, \cT_i \text{ is } \cT_{R_3}\:\}$; the
series
\begin{equation}\label{barn1sum}
\sum\limits_{i\in \mathcal{I}_{R_3}} (\Delta\barn_1)_i
= \mathrm{const}\: \sum\limits_{i\in \mathcal{I}_{R_3}} G_{l(i)}^{-1} |\Delta\Sigma_2|_i
%\leq \mathrm{const} \: \sum_l G_l^{-1} |\Delta\Sigma_2|_l
\end{equation}
can be rewritten as
\begin{equation}\label{barn1sumagain}
\sum\limits_{l} G_l^{-1} |\Delta\Sigma_2|_l < \infty \:,
\end{equation}
where $l$ is the running index that consecutively numbers the
curvature transitions in $\cA_\cT$;  $|(\Delta\Sigma_2)|_l$ then
denotes the sum of the absolute changes in $\Sigma_2$ occurring for
the frame transitions between the $(l-1)$\raisebox{0.7ex}{\small th}
and the $l$\raisebox{0.7ex}{\small th} curvature transition.
The series~\eqref{barn1sumagain} is proved to converge
in Appendix~\ref{convergenceofsums}; hence~\eqref{barn1sum} converges.
Finiteness of the sum~\eqref{barn1sum} entails that $\barn_1$
converges to a constant value (which depends on the spatial variables),
i.e., $\barn_1$ exhibits asymptotic freezing.

The arguments used to establish asymptotic freezing of
$\barn_2,\barn_3$ are analogous. Instead of $E_1/(E_2 R_3)$ one can
use the quantity $E_2/(E_3 R_1)$, which again exhibits a behavior of
the type~\eqref{E1E2R3}. In addition, the already established
asymptotic constancy of $\barn_1$ can be employed in the equation
for $\barn_2$.

\begin{remark}
Asymptotic freezing does not depend on the assumption that the
solution converges to a attractor sequence (although this leads to
simplifications in the computations). Analogously, one can show
asymptotic freezing in the case that the asymptotic sequence
converges to an arbitrary sequence of transitions
on the billiard subset $\mathcal{B}$ (including double transitions).
\end{remark}

To establish asymptotic freezing of $\mathcal{P}_i$ we employ the
equation~\eqref{E1oE2R3} and the analogous equations for the quantities
$E_1/(E_3 R_2)$ and $E_2/(E_3 R_1)$: obviously,
for the inverse quantities $E_2 R_3/E_1$, $E_3
R_2/E_1$, $E_3 R_1/E_2$, the evolution equations are of the form
\begin{equation*}
 \partial_\tau A = 2(1-\Sigma^2)A\, .
\end{equation*}
Furthermore we note that the evolution equation for the quantity
$\Lambda = 2 H \sqrt{g}$ is
\begin{equation}\label{2Hgder}
\partial_{\tau} \Lambda = -2(1-\Sigma^2)\Lambda\:,
\end{equation}
see Appendix~\ref{relations}. From these equations it is immediate
that
\begin{equation*}
\partial_\tau \mathcal{P}_i =0\:,
\end{equation*}
where we have used that $\barn_3$ converges to a constant
in the equation for $\mathcal{P}_1$.
It follows that $\mathcal{P}_i$ are constant in the asymptotic regime
so that asymptotic freezing holds trivially.

%\begin{remark}
%Strictly speaking we have proved that $\mathcal{P}_i =
%\mathrm{const}$ (and $\barn_i \rightarrow \mathrm{const}$) for
%solutions $\cA\cS_\cT$ of the unconstrained billiard system.
%Asymptotic freezing follows when we invoke the conjecture that we
%expect the asymptotic dynamics of $\mathbf{X}(x^0,x^i)$ to coincide
%with the asymptotic dynamics of the unconstrained system, where the
%error this entails is expected to essentially exponentially
%decrease.
%\end{remark}

\begin{remark}
Based on the results of Appendix~\ref{relations}
we are able to compute the following ratios:
\begin{equation*}
\frac{R_1R_3\Lambda}{R_2}=\frac{{\cal P}_3(n_3{\cal P}_2 + {\cal
P}_1)}{{\cal P}_2}\:, \qquad
\frac{N_1R_3^2}{N_2\Lambda^2}=\frac{{\tilde N}_1(n_3{\cal P}_2 +
{\cal P}_1)^2}{{\tilde N}_2}\:, \qquad
\frac{E_1\Lambda}{E_2R_3}=-\,\frac{1}{n_3{\cal P}_2 + {\cal P}_1}\:,
\end{equation*}
where r.h.\ sides, and hence also the l.h.\ sides, are asymptotic
constants of the motion. Since the quantity $\Lambda$
satisfies~\eqref{Lambdaderiv}, it behaves like the auxiliary
quantity $A$ (or like $B$ in the generic case); we thus
find---\textit{a posteriori}---the asymptotic behavior of $R_2$ and
$N_2$, in consistency with the results of Section~\ref{asysha}.
\end{remark}

%-----------------------------------------------------------------------
\subsection*{Asymptotic freezing of $\bm{\pi_\lambda}$}
%\label{freeze2}
%-----------------------------------------------------------------------

The momentum $\pi_\lambda$ conjugate to the variable $\lambda$
is given by
\begin{equation*}
\pi_{\lambda} = \Lambda\,
\log\left[(e_1)^{(2-\Sigma_1)}(e_2)^{(2-\Sigma_2)}(e_3)^{(2-\Sigma_3)}\right]\:,
\end{equation*}
see Appendix~\ref{relations}; recall that $e_\alpha$ is defined by
$e_\alpha = H E_\alpha$ and that $\Lambda = 2 H \sqrt{g}$. The
derivative of the variables $e_\alpha$ reads
\begin{equation}\label{parte}
\partial_\tau e_\alpha = (1+\Sigma_\alpha) e_\alpha\:,
\end{equation}
cf.~\eqref{edelzero} in Appendix~\ref{relations}. Using these
formulas and Eqs.~\eqref{billiardeq1}--\eqref{billiardeq3},
and~\eqref{gausscon2} for $\partial_\tau \Sigma_\alpha$,
and~\eqref{2Hgder} for $\partial_\tau \Lambda$, we find that the
derivative of $\pi_\lambda$ is given by
\begin{subequations}\label{pilamder}
\begin{equation}
\Lambda^{-1}\partial_{\tau}\pi_{\lambda} =
\log[(e_1)^{X_1}(e_2)^{X_2}(e_3)^{X_3}] + 6(1-\Sigma^2) +
2R_1^2+2R_3^2 \:,
\end{equation}
where
\begin{equation} X_1= -12(1-\Sigma^2) + 2R_3^2\, ,\qquad X_2 = 2(R_1^2-R_3^2)\,
,\qquad X_3 = -2R_1^2\:.
\end{equation}
\end{subequations}

In order to establish asymptotic freezing of $\pi_\lambda$
we consider a sequence of transitions
that is approximated by an attractor sequence $\cA_\cT$
in the asymptotic regime $\tau\rightarrow \infty$
and investigate~\eqref{pilamder} along this sequence of
$\cT_{N_1}$, $\cT_{R_1}$, and $\cT_{R_3}$ transitions.
We begin
by analyzing the behavior of certain functions in terms of the
growth function $G$; we first note that
\begin{equation*}
\Lambda = 2 H \sqrt{g} = \Lambda_0\, G^{-1}\:, \quad \text{ so that }\quad
\Lambda_i = \Lambda_0 \,G_{l(i)}^{-1}\:.
\end{equation*}
From~\eqref{billiardeqs} and~\eqref{parte} we obtain the following equations:
\begin{subequations}
\begin{align}
&\partial_\tau \left[ (1-\Sigma^2) e_1^4 \right] =
4 (1-\Sigma^2)\, \left[ (1-\Sigma^2) e_1^4 \right]\:, \qquad \\[1ex]
& \partial_\tau \left[ R_1 \frac{e_3}{e_2}\right] =
2 (1-\Sigma^2)\, \left[  R_1 \frac{e_3}{e_2}\right] \:,\qquad
\partial_\tau \left[ R_3 \frac{e_2}{e_1}\right] =
2 (1-\Sigma^2)\, \left[  R_3 \frac{e_2}{e_1}\right] \:;
\end{align}
\end{subequations}
hence the relations
\begin{equation*}
\left[ (1-\Sigma^2) e_1^4 \right] \propto G^2 \:,\qquad \left[ R_1
\frac{e_3}{e_2}\right] \propto G \:,\qquad \left[ R_3
\frac{e_2}{e_1}\right] \propto G
\end{equation*}
describe the behavior of these functions along $\cA_\cT$ in terms of
growth function $G$.
On the basis of these formulas Eq.~\eqref{pilamder}
simplifies: For a $\cT_{R_3}$ frame transition, \eqref{pilamder}
reduces to
\begin{equation}\label{partialpilamR3}
\partial_{\tau}\pi_{\lambda} =
\Lambda \left[ 2 R_3^2 \log \frac{e_1}{e_2} +2 R_3^2 \right]
=
2 \Lambda_0 \, G^{-1} \, R_3^2\, \Big[ - \log G + \log R_3 +\mathrm{const} \Big]\:.
\end{equation}
The fact that growth function $G$ is constant along the $\cT_{R_3}$
transition makes it possible to integrate~\eqref{partialpilamR3}.
Since $\partial_\tau \Sigma_2 = 2 R_3^2$ we obtain
that
\begin{equation}\label{someR3ints}
2 \int R_3^2 \,d\tau = \Delta \Sigma_2 \:,
\quad \text{and}\quad
2 \int R_3^2 \log R_3\, d\tau = \Delta \Sigma_2 \,(\log \Delta\Sigma_2 -1)\:,
\end{equation}
where we have used that $R_3^2 = [\Sigma_2 - (\Sigma_2)_-] [(\Sigma_2)_+-\Sigma_2]$
for the second integral; as usual, $\Delta\Sigma_2 = (\Sigma_2)_+ -(\Sigma_2)_-$,
where $(\Sigma_2)_\pm$ denotes the values of $\Sigma_2$ at the endpoints of
the transition.
Using~\eqref{someR3ints}
we find that
\begin{equation*}
\Delta \pi_\lambda = \Lambda_0 \, G^{-1}\, \Delta\Sigma_2\,
\Big[-\log G + \log \Delta\Sigma_2 + \mathrm{const}\Big]\:,
\end{equation*}
or, in index notation, where we assume that $\cT_i$ is the $\cT_{R_3}$ transition
under consideration:
\begin{equation}\label{delpilam1}
(\Delta \pi_\lambda)_i = \Lambda_0 \, G_{l(i)}^{-1}\, (\Delta\Sigma_2)_i\,
\Big[-\log G_{l(i)} + \log (\Delta\Sigma_2)_i + \mathrm{const}\Big]\:;
\end{equation}
here, the constant is independent of $i$.
%Note that if $\cT_{R_3}$ is a transition in a small curvature phase we have
%$\Delta\Sigma_2 = 3(1+ u_l +u_l^2)^{-1} (1+ 2 u_l) = 6 u_l^{-1} + O(u_l^{-2})$.
If $\cT_i$ is a $\cT_{R_1}$ transition instead of a $\cT_{R_3}$, the result is analogous.

In the case of a $\cT_{N_1}$ curvature transition, \eqref{pilamder} reduces to
\begin{equation}\label{pilamcurv}
\partial_{\tau}\pi_{\lambda} = 6 \Lambda (1-\Sigma^2)
[-2\log e_1+1] =
6 \Lambda_0 \,G^{-1} (1-\Sigma^2) \Big[ -\log G +
\textfrac{1}{2} \log (1-\Sigma^2) +\mathrm{const} \Big]\:.
\end{equation}
The integral of $2(1-\Sigma^2)$ yields the logarithm of the growth factor $g = \zeta_+/\zeta_-$
associated with the transition $\cT_{N_1}$, i.e.,
\begin{equation*}
2 \int (1-\Sigma^2) d\tau = \int \frac{d\zeta}{\zeta} = \log \frac{\zeta_+}{\zeta_-} = \log g\:.
\end{equation*}
The integral of $2(1-\Sigma^2)\log (1-\Sigma^2)$ is more involved; using~\eqref{Zeq2} we find
\begin{equation}\label{zetalogzetadzeta}
2 \int (1-\Sigma^2) \log (1-\Sigma^2) d\tau =
\left(\log \frac{3}{\zeta_-\zeta_+}\right) \log g +
\int_{\zeta_-}^{\zeta_+} \frac{1}{\zeta} \log [(\zeta-\zeta_-)(\zeta_+-\zeta)] d\zeta \:.
\end{equation}
If the Kasner parameter $u$ ($ = u_-$) of the $\cT_{N_1}$ transition is large
(i.e., if $\cT_{N_1}$ belongs to a small curvature phase),
then~\eqref{zetalogzetadzeta} simplifies by using
the approximations
\begin{equation*}
g \approx 1 + \frac{2}{u}\:,\qquad
\zeta_+ -\zeta_- \approx \frac{6}{u^3}\:;
\end{equation*}
in particular
\begin{equation*}
\int_{\zeta_-}^{\zeta_+} \frac{1}{\zeta} \log [(\zeta-\zeta_-)(\zeta_+-\zeta)] d\zeta
\,\approx\, -\frac{12}{u} \left[ \log u - \frac{\log 6 -1}{3}\right]\:,
\end{equation*}
so that
\begin{equation*}
2 \int (1-\Sigma^2) \log (1-\Sigma^2) d\tau \,\approx\,
\frac{2}{u} \left[ -2 \log u +\log 12 -2\right]
\end{equation*}
for $u\gg 1$. Since $G$ is not constant along the $\cT_{N_1}$ orbit,
Eq.~\eqref{pilamcurv} cannot be integrated explicitly. However, by
making the estimate $G \geq G_-$, where $G_-$ is the initial value
of the growth function, we obtain
\begin{equation*}
|\Delta\pi_\lambda| \leq 3 \Lambda_0\, G_-^{-1} \left[ \log G_-\,
\log g + \mathrm{const}\: \log g  + \Big|\int (1-\Sigma^2) \log
(1-\Sigma^2) d\tau\Big|  \right]\:,
\end{equation*}
which reduces to
\begin{equation*}
|\Delta\pi_\lambda| \leq 6 \Lambda_0\, G_-^{-1} \frac{1}{u}
\Big[ \log G_- + \log u + \mathrm{const} \Big]
\end{equation*}
in the case $u\gg 1$. In index notation, when we assume that the
$\cT_{N_1}$ transition under consideration is the
$i$\raisebox{0.7ex}{\small th} transition of the sequence, i.e.,
$\cT_{i} = \cT_{N_1}$, we have
\begin{equation}\label{delpilam2}
|\Delta\pi_\lambda|_i \leq 6 \Lambda_0 \, G_{l(i)}^{-1}\, \frac{1}{u_{l(i)}}
\Big[ \log G_{l(i)} + \log u_{l(i)} + \mathrm{const} \Big] \:,
\end{equation}
where the estimate holds in a small curvature phase (with sufficiently large $u_l$).

To establish asymptotic freezing of $\pi_\lambda$ we must show that
the summation of all $(\Delta\pi_\lambda)_i$ is finite (where the
sum is over all transitions of $\cA_\cT$), i.e.,
\begin{equation}\label{pilamsumfin}
\sum\limits_{i} |\Delta\pi_\lambda|_i < \infty \:.
\end{equation}
Finiteness of this sum entails that $\pi_\lambda$
converges to a constant value (depending on the spatial variables),
i.e., $\pi_\lambda$ exhibits asymptotic freezing.

We merely state the main arguments of the proof and omit the
computational details:
in analogy to the considerations of Appendix~\ref{convergenceofsums},
we may focus on the behavior of~\eqref{pilamsumfin} in small
curvature phases (i.e., the alternating phases
of $\cT_{R_3}$ and $\cT_{N_1}$ transitions in the neighborhood of
$\mathrm{T}_3$), since in large curvature phases, the treatment
of the sum is simple.
Since $|\Delta \Sigma_2| \sim u^{-1}$ when $u \gg 1$,
see~\eqref{DeltaSigma2s},
the equations~\eqref{delpilam1} and~\eqref{delpilam2}
are of the same type and lead to sums of the type
\begin{equation}\label{someserlog}
\sum\limits_{l} G_l^{-1} \log G_l\: \frac{1}{u_l}  \qquad \text{and}
\qquad
\sum\limits_{l} G_l^{-1} \frac{\log u_l}{u_l}\:,
\end{equation}
where $l$ is the running index that consecutively numbers
the curvature transitions in $\cA_\cT$.
Finiteness of~\eqref{someserlog} is proved in
Appendix~\ref{convergenceofsums}, hence
finiteness of~\eqref{pilamsumfin} ensues and
we obtain asymptotic freezing of $\pi_\lambda$.

%%%%%%%%%%%%%%%%%%%%%%%%%%%%%%%%%%%%%%%%%%%%%%%%%%%%%%%%%%%%%%%%%%%%%%%%%%5
\section{AVTD singularities and the dynamical systems approach}
\label{rel}
%%%%%%%%%%%%%%%%%%%%%%%%%%%%%%%%%%%%%%%%%%%%%%%%%%%%%%%%%%%%%%%%%%%%%%%%%%5

In this appendix we discuss the connection between
work on AVTD singularities and the dynamical systems approach.
There exist no rigorous results for inhomogeneous models with
oscillatory singularities, but there are some for inhomogeneous
models with so-called AVTD singularities. In this appendix we
address how work in this latter field relates to the present
approach.

The term velocity dominated singularities was first introduced by
Eardley et al.~\cite{earetal72}. Isenberg, Moncrief and co-workers
later used the approach of Eardley et al. to obtain rigorous
results, see~\cite{isemon90} and references in~\cite{ber02}. Let us
here follow Isenberg and Moncrief~\cite{isemon90}. They use the
standard initial value problem as their starting point, the spatial
metric and the extrinsic curvature being the dependent variables.
They then consider the system of differential equations one obtains
if one drop all spatial derivatives, except in the Codazzi
constraints, and sets the three-curvature to zero. By doing so they
obtain a velocity term dominated (VTD) ODE system of evolutionary
equations constrained algebraically by the Gauss constraint; the
solution of this system can subsequently be inserted into the
Codazzi constraints yielding a VTD-solution. A spacetime is said to
be asymptotically velocity term dominated (AVTD) if it is a solution
to Einstein's equations such that it, w.r.t. some appropriate norm,
asymptotically approach a solution to the VTD equations, see p87, 88
in~\cite{isemon90} for details. If furthermore the asymptotic limit
yields a singularity this singularity is said to be AVTD.

If one uses a conformal Hubble-normalized orthonormal frame, then
the equations for the VTD ODE system are the Bianchi type I
evolution equations and the Gauss constraint on the silent
boundary---and the equations for $E_\alpha{}^i$ and the Hubble
variable $H$, with all spatial frame derivatives dropped and all
variables set to zero except for $\Sigma_{\alpha\beta}$ (and
possible matter variables), i.e., the lowest order perturbations of
the evolution equations for $E_\alpha{}^i$ and $H$ w.r.t. the silent
Bianchi type I subset; this therefore yields the generalized Bianchi
type I solutions, which in the vacuum case are the generalized
Kasner solutions. The VTD treatment of the Codazzi constraints
corresponds to inserting these solutions into the lowest non-zero
perturbation of the constraints w.r.t. the silent Bianchi type I
subset (i.e., one sets all variables to zero except for
$E_\alpha{}^i,\Sigma_{\alpha\beta}$, and possible matter variables,
in the constraints).

In the vacuum case AVTD singularities occur for special spacetimes,
notably ones with symmetries, as discussed in Section~\ref{sym}. In
cases with one or two commuting spacelike Killing vectors, this
naturally leads to an Iwasawa frame representation, and hence to a
fairly simple correspondence between the present Iwasawa based
conformally Hubble-normalized variables and variables that has been
used in previous work in this area (thus the AVTD starting point is
obtained by plugging in the Kasner circle $\mathrm{K}^{\ocircle}$
into the lowest order approximation for $E_\alpha{}^i$ and $H$, as
done in Appendix~\ref{Kasner}).

Arguably the most impressive result about AVTD singularities was
obtained by Andersson and Rendall~\cite{andetal05} for massless
scalar fields and stiff perfect fluids. In this case the Bianchi
type I equations on the silent boundary yields the generalized
Jacobs solutions, and it turns out that an open subset of these
solutions are stable and hence attract an open set of solutions. It
should, however, be pointed out that the approach of Andersson and
Rendall is naturally related to a Fermi propagated frame rather than
an Iwasawa frame. Stiff perfect fluid spacetimes has also been
investigated in terms of the UEWE dynamical systems approach by
Coley and Lim~\cite{collim05}.

We note that within the present framework AVTD singularities are
more appropriately geometrically described as singularities that are
associated with (Hubble) {\em asymptotic\/} (Hubble) {\em conformal
spatial flatness\/}.

To study when singularities are not AVTD Moncrief and coworkers have
introduced the ``Method of Consistent Potentials"
(MCP)~\cite{grumon93,beretal98}, and references in~\cite{ber02}. In
this case one inserts the VTD solutions (the generalized Kasner
solutions in the vacuum case) into the full equations (or rather,
into the Hamiltonian for the full equations); if all terms are
decaying the spacetime is AVTD, if not it is conjectured to have
local Mixmaster dynamics. We note that this corresponds to the
linear perturbation of the Kasner circle $\mathrm{K}^{\ocircle}$ in
the present context, as can be seen by inserting the generalized
Kasner solutions into the relations given in
Appendix~\ref{relations} and comparing with the linearized
$\mathrm{K}^{\ocircle}$ result. Here, however, this result is seen
in the context of the full conformally normalized state space
picture.

We note that the Hamiltonian approach using the `dominant'
Hamiltonian ${\cal H}$ suggests a direct generalization of the
VTD/AVTD approach, or equivalently, the billiard subset and its
lowest order perturbation into the physical state space (with the
ODE solution, yielding a lowest order perturbation, inserted into
the constraints), see Section~\ref{dual}. An advantage of the
present approach is that it naturally splits the problem into two
parts and hence offers a possibility of achieving more modest goals
than cracking the whole problem: it may be possible to prove some
statements that are connected with the silent boundary only---it may
not be necessary to also prove things about perturbations thereof.

%Moreover, in this paper we have
%derived, albeit using some heuristic arguments, the structures that
%are believed to be of relevance asymptotically for generic behavior
%of generic singularities.

\end{appendix}

\label{bibbegin}

\end{document}